\documentclass[apj,iop,revtex4]{emulateapj}
\usepackage[utf8]{inputenc}
\usepackage[varg]{txfonts}
\usepackage{url}
\usepackage{colordvi}
\usepackage{xcolor}
\usepackage[utf8]{inputenc}
\usepackage{natbib}
\usepackage{blindtext}
\usepackage{pdflscape}
\usepackage{pdflscape,array,booktabs}
\usepackage{lscape}
\usepackage{afterpage}
\usepackage{longtable}
\usepackage{rotating}
\usepackage{float}
\usepackage{threeparttable}  
\usepackage{multirow}
\usepackage{graphics}
\usepackage{epstopdf}
\usepackage[title]{appendix}
\usepackage{enumitem}
\bibpunct{(}{)}{;}{a}{}{,}
\RequirePackage{color}

\begin{document}
\nocite{*} \title{Optically-faint massive Balmer Break Galaxies at
  z$>$3 in the CANDELS/GOODS fields}

\author{Bel\'en Alcalde Pampliega\altaffilmark{1,2}, 
 Pablo G. P\'erez-Gonz\'alez\altaffilmark{3,1}, 
 Guillermo Barro\altaffilmark{4},
 Helena Dom\'inguez S\'anchez\altaffilmark{5},  
 M. Carmen Eliche-Moral\altaffilmark{6},  
 Nicol\'as Cardiel\altaffilmark{1},
 Antonio Hern\'an-Caballero\altaffilmark{1}
 Luc\'ia Rodriguez-Muñoz\altaffilmark{7},
 Patricia S\'anchez Bl\'azquez\altaffilmark{8},
 Pilar Esquej\altaffilmark{9}}

\altaffiltext{1}{Departamento de F\'isica de la Tierra y Astrof\'isica, Faultad de CC F\'isicas, Universidad Complutense de Madrid E-2840 Madrid, Spain.}
\altaffiltext{2}{Isaac Newton Group of Telescopes (ING), Apto. 321, E-38700 Santa Cruz de la Palma, Canary Islands, Spain.}
\altaffiltext{3}{Centro de Astrobiolog\'{\i}a (CAB, INTA-CSIC), Carretera de Ajalvir km 4, E-28850 Torrej\'on de Ardoz, Madrid, Spain}
\altaffiltext{4}{Astronomy Department,Department of Physics, University of the Pacific, Stockton, CA 95211, USA.}
\altaffiltext{5}{Department of Physics and Astronomy, University of Pennsylvania, Philadelphia, PA 19104, USA.}
\altaffiltext{6}{Instituto de Astrof\'isica de Canarias, Calle V\'ia L\'actea, s/n E-38205, La Laguna, Tenerife, Spain.}
\altaffiltext{7}{Dipartimento di Fisica e Astronomia, Universit\`{a} di Padova, vicolo dell'Osservatorio 2, I-35122 Padova, Italy.}
\altaffiltext{8}{Departamento de F\'isica Te\'orica, Universidad Aut\'onoma de Madrid, Cantoblanco, 28049, Spain.}
\altaffiltext{9}{Herschel Science Centre, ESA, Villafranca del Castillo, Apartado 78, E-28691 Villanueva de la Ca\~nada, Spain.}

%19 Instituto de Astrof\'ısica, Pontifica Universidad de Chile, Av. Vicuna Mackenna 4860, 782-0436 Macul, Santiago, Chile

\shorttitle{Optically faint massive BBGs at z$>$3}
\shortauthors{B. Alcalde Pampliega et al.}	
%\slugcomment{B. Alcalde, \email{belenalc@ucm.es}}

\date{Accepted for publication in The Astrophysical Journal 26/03/2019}

\begin{abstract}

  We present a sample of 33 Balmer Break Galaxies (BBGs) selected as
  HST/F160W dropouts in the deepest CANDELS/GOODS fields
  ($H\gtrsim27.3$~mag) but relatively bright in {\it Spitzer}/IRAC
  ($[3.6],[4.5]<24.5$~mag), implying red colors (median and quartiles:
  $\langle H-[3.6]\rangle=3.1^{3.4}_{2.8}$\,mag). Half of these BBGs
  are newly identified sources. Our BBGs are massive ($\langle
  \log(\rm{M}/\rm{M}_\sun)\rangle=10.8^{11.0}_{10.4}$) high redshift
  ($\langle z\rangle=4.8^{5.1}_{4.4}$) dusty ($\langle
  \rm{A(V)}\rangle=2.0^{2.0}_{1.5}$~mag) galaxies.  The SEDs of half
  of our sample indicate that they are star-forming galaxies with
  typical specific SFRs 0.5-1.0~Gyr$^{-1}$, qualifying them as main
  sequence (MS) galaxies at $3<z<6$. One third of those SEDs indicates
  the presence of prominent emission lines (H$\beta$+$[OIII]$,
  H$\alpha$$+$[NII]) boosting the IRAC fluxes and red colors.
  Approximately 20\% of the BBGs are very dusty
  ($\rm{A(V)}\sim2.5$~mag) starbursts with strong mid-to-far infrared
  detections and extreme SFRs ($\rm{SFR}>10^{3}\,\rm{M}_\sun/yr$) that
  place them above the MS. The rest, 30\%, are post-starbursts or
  quiescent galaxies located $>2\sigma$ below the MS with
  mass-weighted ages older than 700~Myr. Only 2 of the 33 galaxies are
  X-ray detected AGN with optical/near-infrared SEDs dominated by
  stellar emission, but the presence of obscured AGN in the rest of
  sources cannot be discarded.  Our sample accounts for 8\% of the
  total number density of $\log(\rm{M}/\rm{M}_\sun)>10$ galaxies at
  $z>3$, but it is a significant contributor (30\%) to the general
  population of red $\log(\rm{M}/\rm{M}_\sun)>11$ galaxies at $4<z<6$.
  Finally, our results point out that 1 of every 30 massive
  $\log(\rm{M}/\rm{M}_\sun)>11$ galaxies in the local Universe was
  assembled in the first 1.5~Gyr after the Big Bang, a fraction that
  is not reproduced by state-of-the-art galaxy formation simulations.
\end{abstract}
%Massive red galaxies:BBG


%\maketitle

\section{Introduction}
\label{sec:Intro}

Understanding how galaxies form and evolve is one of the central
challenges of modern astronomy. In the current $\Lambda$CDM paradigm,
dark matter halos are the primary structures which provide seeds for
gas collapse and allow the baryonic growth of galaxies. Dark matter
halos assemble primarily in a hierarchical manner, with low-mass halos
forming early and merging to produce more massive halos as they move
down to lower redshifts \citep{Kauffmann1993, Reed2003}. Consequently,
the most massive galaxies are expected to appear in massive halos at
lower redshifts following a similar hierarchical assembly. However,
observational studies suggest that massive galaxies
($\log(\rm{M}/\rm{M}_\sun)>11$) in the local universe form rapidly in
strong bursts of star formation at early times
\citep[e.g.,][]{perez2008stellar,
  mancini2009searching,2010ApJ...709..644I,Caputi2011,2011ApJ...739...24B,
  Ilbert2013,Muzzin2013,Tomczak2014,Grazian2015}. Similarly, many
surveys have identified a substantial population of massive galaxies
at redshifts up to $z\sim4$, when the universe was only 1.5 Gyr old
\citep{Mobasher2005,Wiklind2008,caputi2012nature}.Although
  some of those studies present evidence for evolved stellar
  populations or suppressed star-formation \citep{fontana2009fraction,
    2014ApJ...783L..14S, 2014ApJ...787L..36S, 2014ApJ...794...68N},
  the existence of fully quiescent galaxies at very high redshifts is
  still controversial \citep{2016ApJ...830...51S,
    Hill2017,Glazebrook2017,Marsan2017,Simpson2017,Schreiber2018}.
Determining when the first massive galaxies emerged and characterizing
the evolution of their number density is particularly important to
improve our picture of galaxy evolution and to constrain galaxy
formation models. A major complication to address these questions is
gathering a complete, robust and un-biased census of massive galaxies
up to the highest redshifts possible.

The advent of deep surveys with the WFC3 camera on the Hubble Space
Telescope (HST) has significantly expanded our census of distant
galaxies up to $z\sim10$ \citep[e.g.,][]{2015ApJ...803...34B,
  Oesch2018}.  However, the use of near infrared (NIR) observations
implies that the sample selection at $z\gtrsim3$ is based on the
rest-frame UV emission of the galaxies, which is particularly
sensitive to the effects of dust attenuation and biased towards the
detection of blue systems. Consequently, while UV-based
selection techniques, such as the Lyman break dropout
\citep[][LBG]{madau1996high} or the search for Lyman $\alpha$ emitters
(LAEs), are very effective at identifying blue, (typically) low-mass
star-forming galaxies \citep{steidel2003lyman,giavalisco2004rest},
these methods are strongly biased against red, dusty, or evolved
galaxies which typically make most of the massive galaxy population at
mid-to-high redshifts
\citep[e.g.,][]{2011ApJ...739...24B,2014ApJ...787L..36S, Caputi2015}.
Thus, rest-frame UV-selected samples at high-redshift are likely
incomplete, missing massive red galaxies which could potentially be
identified with observations at longer wavelengths.

At $z>3$, the strongest spectral features in the rest-frame optical
continuum of galaxies, the 4000~$\AA$ and Balmer breaks, are shifted
redward of $\sim$1.5~$\mu$m, thus making mid-to-far IR (or even radio)
observations essential to identify the presence of massive galaxies.
A number of different selection techniques based on ``extremely'' red
mid-IR colors
\citep[e.g.,][]{fontana2006galaxy,rodighiero2007unveiling,
  1998ARA&A..36..189K,huang2011four,caputi2012nature, Nayyeri2014,
  2016ApJ...816...84W, Schreiber2016} and/or bright far-IR or
sub-millimeter emission \citep{casey2012,Riechers2013,Vieira2013} have
successfully identified a {\it hidden} population of massive galaxies
at redshifts $z\gtrsim3$ which are missing from even the deepest HST
surveys. The extremely red colors of these galaxies can indicate
either a heavily obscured burst of star formation (usually accompanied
by a strong far-IR emission from the heated dust), or a
  quiescent, passively evolving galaxy
  \citep[e.g.,][]{barros2014lbg}. Overall, the intrinsically faint
optical-to-NIR fluxes of these galaxies, typically coupled with high
dust obscurations, make the modeling of their spectral energy
distributions (SEDs) very challenging \citep[see,
e.g.,][]{Schaerer2013,deBarros2014}.  Consequently, the inferred
redshifts, stellar population properties and star-formation rates
(SFRs) are quite uncertain
\citep[e.g.,][]{Michalowski2010,daCunha2015}.

A way forward to overcome the mentioned limitations when studying
massive galaxies at high redshift is to gather large unbiased samples
from the deepest cosmological surveys carried out in the mid-IR. These
surveys must be very deep and cover relatively wide areas, since $z>3$
massive galaxies are relatively scarce systems \citep[the typical
number density is $\sim0.1\,\rm{arcmin}^{-2}$;][]{caputi2012nature}.
This methodology would provide more robust, statistically significant,
constraints on the overall properties of the oldest massive galaxies,
counting with the contribution of red mid-IR detected sources to the
massive end of the $z>3$ galaxy population. Simultaneously, 
  by using mid-IR deep surveys we can also characterize the
incompleteness of our current mass-limited samples, that are typically
based on NIR selections with HST. In addition, follow up observations
of {\it bona fide} candidates to massive high-z galaxies in new
spectral ranges (e.g., with ALMA or the upcoming JWST) can alleviate
the SED fitting limitations and provide more precise values of their
redshifts and stellar population properties, which are essential to
have a complete view of the very early phases of massive galaxy
formation.

In this context, in the present paper we aim at obtaning a
(more) complete sample of massive galaxies at $z\gtrsim3$. In order
to achieve this goal, we focus our analysis on the search and
characterization of mid-IR bright, near-IR faint galaxies that might
have been missed in the HST-based, near-IR selected catalogs
presented by the CANDELS \citep{2013ApJS..207...24G} and 3D-HST
\citep{2014ApJS..214...24S} surveys. In particular, we present the
results of an IRAC 3.6+4.5$\,\mu$m selection and multi-wavelength
analysis of a sample of red massive galaxies at $z>3$, i.e., probing
the massive galaxy population formed roughly between the 1st and 2nd
Gyr in the lifetime of the Universe.  Hereafter, we will call these
objects Balmer Break Galaxies (BBGs).  Our sample of BBGs has been
built by searching for extremely red objects in the Spitzer 3.6 and
4.5$\,\mu$m IRAC images that are not detected in the F160W CANDELS
deep observations carried out over the 330 arcmin$^{2}$ in
the GOODS-N and GOODS-S fields.  At $3<z<7$, the wavelengths in the
3-5~$\mu$m range (and, therefore, the 3.6 and 4.5$\,\mu$m IRAC bands)
are a robust proxy for a deep and roughly constant stellar mass cut
\citep{fontana2006galaxy, 2010ApJ...709..644I,2009ApJ...707.1387C},
allowing us to built a mass-complete sample of BBGs. After
  presenting our method to identify these IRAC-selected, NIR-faint
  massive galaxies at $z>3$, we compare their colors, redshifts and
  other stellar properties to those of NIR faint, $H-[3.6]$ color
  selected and mass-selected galaxies in the CANDELS catalog to
  compare and characterize the regions of the redshift - stellar mass
  parameter space that are being populated by our newly identified
  BBGs.  Lastly, we study the SFRs and stellar populations properties
  of all the BBGs (which are hard to constrain due to their
  intrinsically faint nature in all but the mid-IR bands), and we
  analyze the role of these galaxies in the context of galaxy
  evolution, especially at the high-mass end.

This paper is organized as follows: in \S\ref{sec:data}, we present
the data set available in the GOODS fields. The procedure followed to
select our sample of IRAC-bright BBGs and the methods applied for
searching for counterparts in other bands are presented in
\S\ref{sec:sampleselection} and \S\ref{sec:SEDsfit}, respectively. In
\S\ref{sec:SEDsfit} we describe our estimations of the photometric
redshifts and stellar properties derivation together with the SED
fitting procedure.  In \S\ref{sec:Pysicalprop} we present our results
and discuss on the properties of different subsamples of BBGs,
including a comparison with the literature. Finally, in
\S\ref{sec:results}, we summarize our findings and present the
conclusions. Appendix~\ref{A:PhotUncertainty} contains a
  detailed description of the photometric measurements in optical and
  NIR bands. Appendix~\ref{A:CANDELSample} describes the analysis of
  the comparison samples used throughout the paper. And
  Appendix~\ref{A:SEDs} shows the spectral energy distributions and
  postage stamps of all the BBGs presented in this work.

We adopt a cosmology with H$_{0}$=\mbox{70 km\,s$^{-1}$\,Mpc$^{-1}$},
$\Omega_{\rm M}$=0.3, and $\Omega_{\Lambda}=0.7$. All the magnitudes
refer to the AB system \citep{Oke&Gunn1983}. The IMF is assumed to be
that presented in Chabrier (2003).

\section{Multi-wavelength dataset}
\label{sec:data}

This work presents the search and analysis of a sample of BBG
candidates in two of the deepest cosmological fields, namely GOODS-N
(RA$=12^{h}36^{m}55^{s}$ DEC$=+62^o14'15"$) and GOODS-S
(RA$=3^{h}32^{m}31^{s}$ DEC$=-27^o48'54"$)
\citep{2004ApJ...600L..93G}. In the following we describe the
multi-band datasets available in these fields and that we have used in
our analysis.  We limit our search for BBGs to the area surveyed by
CANDELS (see \S2.2), which is also covered by other surveys probing
wavelengths from the UV to the far-IR and mm. In total, we work with a
sky region of $330\,\rm{arcmin}^2$ ($160\,\rm{arcmin}^2$ in GOODS-N
and $170\,\rm{arcmin}^2$ in GOODS-S).

\subsection{Spitzer/IRAC data}
\label{ssec:data_irac}
Our BBG candidate search is primarily based on deep mid-IR images
taken in the GOODS fields by the \textit{Spitzer} Infrared Array
Camera (IRAC) from 3.6 to 8.0~$\mu$m. Here we make use of the deepest
multi-epoch mosaics in these regions which are based on observations
from the GOODS/IRAC survey \citep{dickinson2003great} and the
\textit{Spitzer} Extended Deep Survey (SEDs, \citealp{ashby2013seds}).
The four IRAC bands centered at 3.6~$\mu$m, 4.5~$\mu$m, 5.6~$\mu$m,
and 8.0$\,\mu$m have a 5$\sigma$ limiting magnitudes of 26.1, 25.5,
23.5, and 23.4~mag, respectively.

\subsection{CANDELS HST WFC3 near-IR data}
\label{ssec:data_nIR}
Deep NIR imaging of the GOODS fields was obtained with the HST/WFC3
camera as part of the the Cosmic Assembly Near-Infrared Deep
Extragalactic Legacy Survey (CANDELS; \citealt{grogin2011candels},
\citealt{koekemoer2011candels}).  Here we make use of the publicly
available F105W, F125W ($J$), F140W (only in GOODS-S) and F160W ($H$)
mosaics as well as the $H$-band selected galaxy catalogs in both
fields presented in \citet{2013ApJS..207...24G} for GOODS-S and Barro
et al. (2019, in prep.) for GOODS-N. These catalogs include UV-to-NIR
multi-band photometry as well as photometric redshifts and stellar
population properties based on the fitting of the SEDs. See also
\citet{2013ApJS..206...10G} for more details on the data reduction and
the creation of the catalogs. Note that the GOODS fields are the only
2 out of the 5 CANDELS fields that have the deepest layer of NIR
observations, reaching a 5$\sigma$ sensitivity limit of 27.6~mag
\citep{grogin2011candels}.

We also use $K$-band imaging taken with the VLT/HAWK-I
instrument in GOODS-S field (HUGS survey; \citealt{Fontana2014HUGS}).
Similarly, in GOODS-N we use $K$-band imaging from CFHT/WIRCam
\citep{Kajisaw2011}. The 5$\sigma$ depths of these datasets
are 26.3 and 24.7~mag, respectively.

\subsection{HST/ACS and SHARDS optical data}
\label{ssec:data_ACS_SH}
In addition to the near and mid-IR imaging described above, we make
use of the deep optical mosaics taken with the HST/Advanced Camera for
Surveys (ACS) in the GOODS fields as part of the GOODS and CANDELS
surveys. There are publicly available mosaics in 5 bands: F435W ($b$),
F606W($v$), F775W ($i$), F850W ($z$), and F814W. They reach 5$\sigma$
point-source sensitivity limits of 28.5, 28.8, 28.1, 27.6 and 28.4~mag
\citep{2004ApJ...600L..93G}.

Furthermore, we also use imaging data from the GTC Survey for High-z
Absorption Red and Death Sources (SHARDS; \citealt{perez2013shards})
which consists of observations in 25 contiguous medium-band
(R$\sim$50) filters covering the spectral range 500-950~nm, reaching
an AB magnitude of 27.0 at least at the 3$\sigma$ level.

Apart from the publicly available single-band mosaics, we have also
created two stacked images by combining either all the HST bands
(optical and NIR) or all the SHARDS medium-bands. The goal of building
these stacks is to increase the limiting depth of our search for
optical counterparts to our mid-IR bright, near-IR faint (or even
undetected) BBGs.

\subsection{Far-IR an sub-mm data}
\label{ssec:data_fIR}
The GOODS fields have also been observed in the far IR by {\it
  Spitzer} and Herschel as part of the GOODS
\citep{dickinson2003great}, the GOODS-Herschel
\citep{2011A&A...533A.119E} and the PACS Evolutionary Probe (PEP;
\citealp{2011A&A...532A..49B}; \citealp{2011A&A...532A..90L}) surveys.
Here we make use of the Spitzer/MIPS- 24 and 70 $\mu$m mosaics
presented in \cite{perez2008stellar}, and the Herschel PACS- 100 and
160 $\mu$m, and SPIRE- 250, 350 and 500 $\mu$m catalogs described in
\citet{2011A&A...533A.119E} and \citet{2013A&A...553A.132M}. The
5$\sigma$ limits of these surveys are 30(30)~$\mu$Jy,
1.2(1.2)~$\mu$Jy, 1.7(1.5)~mJy, 3.6 (3.2)~mJy, 9 (8)~mJy, 12(11)~mJy,
and 13(11)~mJy for 24, 70, 100, 160,250, 350, and 500~$\mu$m in
GOODS-N (GOODS-S).

We also search for additional far-IR data in the following surveys:
SCUBA \citep{2005ApJ...635..853B,Pope2005} in GOODS-N, and LABOCA
\citep{Weis2009}, LESS and ALMA follow up ALESS
\citep{Hodge2013,2013MNRAS.432....2K} in GOODS-S. At millimeter
wavelengths, AzTEC 1.1mm
\citep{2008MNRAS.391.1227P,2011MNRAS.410.2749P}, MAMBO 1.2mm
\citep{2005ApJ...635..853B}, and GISMO 2mm \citep{2014ApJ...790...77S}
surveys are available in GOODS-N. These surveys reach sensitivities of
2-5\,mJy corresponding to L(IR) of $\gtrsim10^{12}$ for $z\sim4$.

\subsection{X-Ray}
\label{ssec:data_xray}
We have used X-ray data from the Chandra 2 Ms source catalog by
\citet{2003AJ....126..539A}, covering the entire surveyed region of
the F160W mosaic in GOODS-N, and 4 MS catalog from \cite{Xue2011} in
GOODS-S. The on-axis sensitivity limits in soft$/$hard bands are of
$2.5\times10^{-17}/1.4\times10^{-16}$\,erg cm$^{-2}$ s$^{-1}$ and
$9.1\times10^{-18}/5.5\times10^{-17}$\,erg cm$^{-2}$ s$^{-1}$ in 2 Ms
and 4 Ms respectively. These fluxes correspond to X-ray luminosities
$\rm{L}_X\rm{(2-10 keV)}>10^{43}$~erg/s ($\rm{L}_X\rm{(2-10
  keV)}>10^{44}$~erg/s) for $z>3$ ($z>4$), according to
\cite{Ueda2014}; see also \cite{Padovani2017}.

\section{Selection of BBGs at $z>3$}
\label{sec:sampleselection}

\begin{figure}[t]
  \begin{center} 
     \includegraphics[width=1.\linewidth]{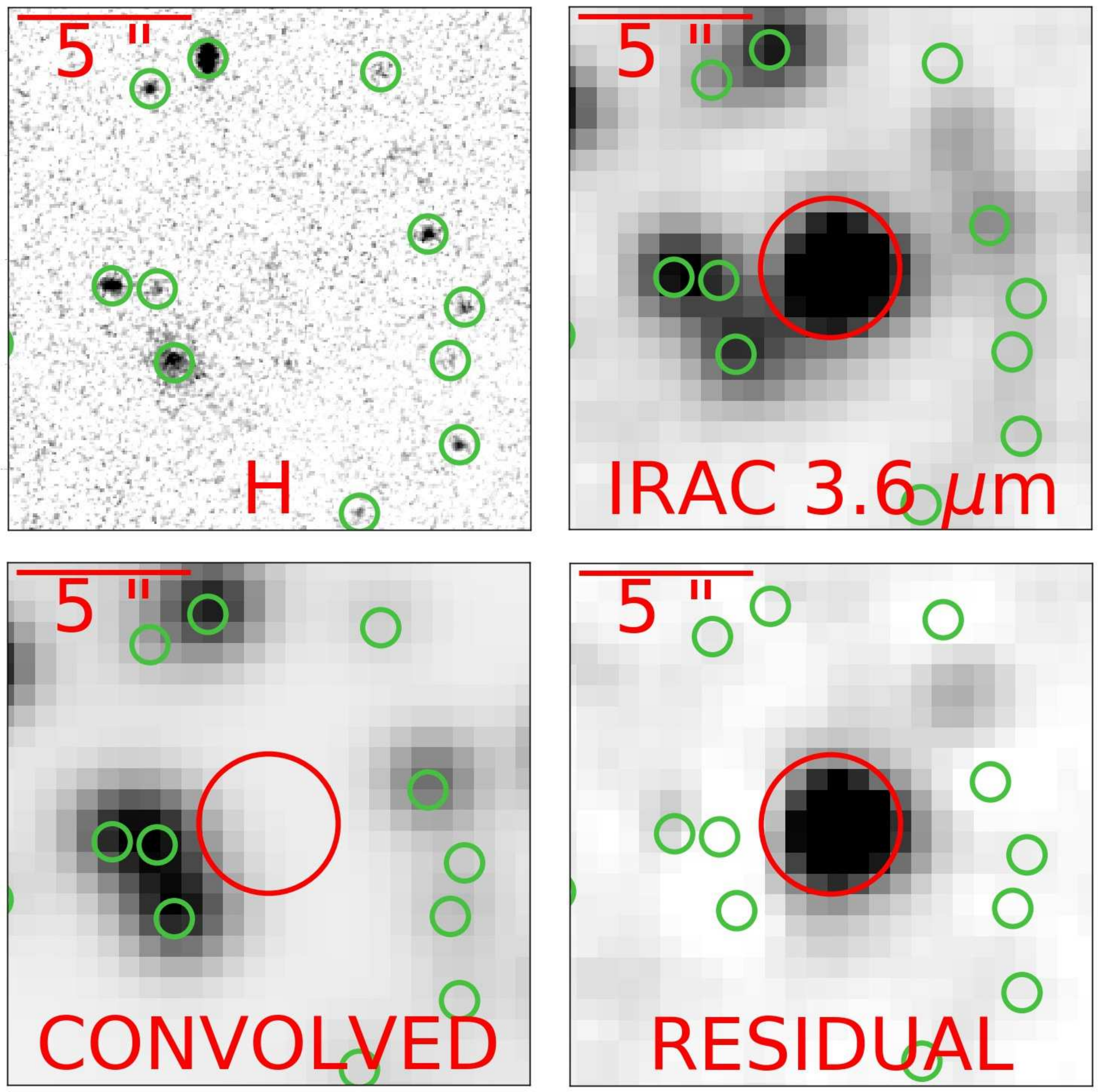} 
  \end{center}
  \caption{ Example of the deconvolution process followed in the IRAC
    3.6 and 4.5~$\mu$m images (see Barro et al. 2019, in prep.).
    \textit{Top left:} Cutout of the $H-$band image centered in the
    position of our source GDN\_BBG02 (see Table~\ref{tab:photprop}).
    Its position is also marked in all panels with a $2\arcsec$ radius
    red circle. $H$-band detected sources from the CANDELS public
    catalog (Barro et al.  2019, in prep.; see also the 3D-HST catalog
    in \citealt{2014ApJS..214...24S}) are encircled in green.
    \textit{Top right}: IRAC 3.6~$\mu$m image showing all the $H$-band
    sources detected in the region. \textit{Bottom left:} TFIT
    ``template'' image built by convolving the HST image to the
    3.6~$\mu$m resolution, and then scaling all $H$-band detected
    sources to reproduce the flux measured in the original IRAC image.
    \textit{Bottom right:} residual image obtained by subtracting the
    scaled ``template'' from the IRAC science frame. Note how the
    $H$-band dropout appears clearly defined in this residual image
    (jointly with another fainter dropout to the NW beyond our IRAC
    magnitude cut).}
  \label{fig:convolution}
\end{figure}

In this Section, we describe the selection technique used to identify
candidates to massive high redshift galaxies. Briefly, candidates are
identified by searching for relatively bright IRAC sources which have
no optical/NIR counterparts in very deep HST imaging (i.e., dropouts).
This technique is an extension of the classical ERO (Extremely Red
Objects; see e.g., \citealt{mccarthy2004eros}) or DRG \citep[Distant
Red Galaxy;][]{franx2003significant,brammer2007density} methods used
to select red galaxies at $z\sim2$, and has been used in different
papers to identify massive galaxies at higher redshifts
(\citealt{huang2011four}, \citealt{caputi2012nature};
\citealt{stefanon2015stellar}; \citealt{2016ApJ...816...84W}).
All these methods use a single-color selection threshold to
identify extremely red galaxies at $z\gtrsim3$ with strong breaks
around the Balmer/4000~$\AA$ rest-frame region. At these redshifts,
the Balmer/4000~\AA\, break lies in between the HST $F160W$ and the
Spitzer IRAC 3.6~$\mu$m. Therefore, we use the $H-[3.6]$ color to
search for BBGs. Using a single color to identify a spectral feature
often exhibits degeneracies. In our case, a strong Balmer or
4000~\AA\, break can be explained by either evolved stellar
populations or younger stellar populations with significant dust
obscuration (see, e.g., \citealt{barros2014lbg}). Throughout the
paper we will use the name Balmer Break Galaxies to refer to all the
candidates to massive high redshift galaxies identified by our color
selection, regardless or their intrinsic stellar ages. In
Section~\ref{sec:Pysicalprop}, we infer redshifts and stellar
population properties from the fitting of their SEDs, and discuss
their typical ages and dust attenuations.  We will show that the
average redshift distribution of the BBGs peaks at $z\gtrsim4$, in
agreement with the prediction from the color selection. Overall, our
sample of sources identified with a red $H-[3.6]$ color exhibits
intermediate ages of $\sim$1~Gyr (with strong Balmer breaks) and
relatively high dust attenuations (A(V)$\gtrsim$1.5~mag).  Only a
handful of BBGs have ages consistent, within the large
uncertainties, with having more evolved stellar populations. We
note, however, that at $z\gtrsim4$, where most of our BBGs seem to
lie, there is little room for stellar populations older than 1~Gyr
(the age of the Universe $z=4$ is around 1.5~Gyr), so the observed
colors are due, most probably, to Balmer jumps rather than
4000~\AA\, breaks (justifying our nomenclature).  See
\citet{Dunlop2013} and references therein for a discussion about
Balmer Break Galaxies at high redshift.

\subsection{The F160W dropout search}
\label{ssec:data_search}
Our selection technique is based on two conditions. BBG candidates are
required to be bright in the first two channels of IRAC, $[3.6]$ and
$[4.5]\leq24.5$~mag, and they must be undetected (dropouts) in the
HST/F160W ($H\gtrsim27$~mag) mosaics (according to publicly available
catalogs). The search for dropouts in F160W relies on the multi-band
catalogs published by \citet{2013ApJS..207...24G} and Barro et al.
(2019 in prep.) for the CANDELS GOODS-S and GOODS-N regions,
respectively (and we also checked the catalogs published by the 3D-HST
team, \citealt{2014ApJS..214...24S}). The IRAC photometry presented in
these works for $H$-band detected sources is based on a PSF-matching
technique, TFIT \citep{laidler2007tfit}, which is also used and
described extensively in \cite{2013ApJS..206...10G}.  Briefly, TFIT is
used to generate a model of the IRAC image by convolving the high
spatial resolution HST/F160W mosaic with the appropriate PSF
transformation kernel. Then, the fluxes on the resulting ``template''
image are scaled to those of the galaxies in the IRAC frame on a
galaxy-by-galaxy basis, taking into account the contamination by
neighboring sources. The individual scaling factors provide
PSF-matched H-[3.6] colors for all $H$-band detected galaxies.
Lastly, TFIT subtracts the scaled ``template'' image from the original
IRAC mosaic creating a residual frame which is used to verify the
quality of the source extraction and flux measurements. In our case,
we use these residual images to search for potential $H$-band dropouts
with bright IRAC magnitudes.  Figure~\ref{fig:convolution} illustrates
this procedure highlighting the detection of a BBG candidate.

\subsection{Masking and BBG candidate extraction}

Before searching for BBG candidates in the residual IRAC image, we
performed a series of iterative masking procedures to smooth the
image. This cleaning procedure is necessary because the IRAC residual
image often contains spurious flux coming from saturation artifacts as
well as from the wings and cores of bright sources which are not
properly subtracted. This problem is typically caused by slight
changes (at the 5\% flux level) in the IRAC's point-spread function
(PSF) along the mosaic.

We applied three different cleaning masks to the residual IRAC images.
First, to avoid detecting $H$-band bright sources, we created a mask
including pixels above a threshold flux (6$\times10^{-3}$ and
3$\times10^{-2}$ Jy per pixel in GOODS-N and GOODS-S respectively) in
the convolved images, which is equivalent to masking $H$-band bright
pixels.  Secondly, we masked the regions contaminated by the brightest
([3.6]$<$20) stars in the field using circular masks with
magnitude-dependent radii:

\begin{equation}
r = -21.8\times H + 380.8, \rm{ for } 14<H<16,
\end{equation}
\begin{equation}
r= -4.6\times H + 122.4, \rm{ for } 16<H<20,
\end{equation}
\noindent with $r$ expressed in arcsec.

Lastly, we masked artifacts (negative fluxes in the convolved images)
that appeared as a result of the TFIT convolution process.  These 3
masks were applied to the IRAC residual image, replacing the affected
pixels by the median background calculated in a $1\arcmin$ region
around each source.

\begin{figure}[t]
  \begin{center}
      \includegraphics[width=1.\linewidth]{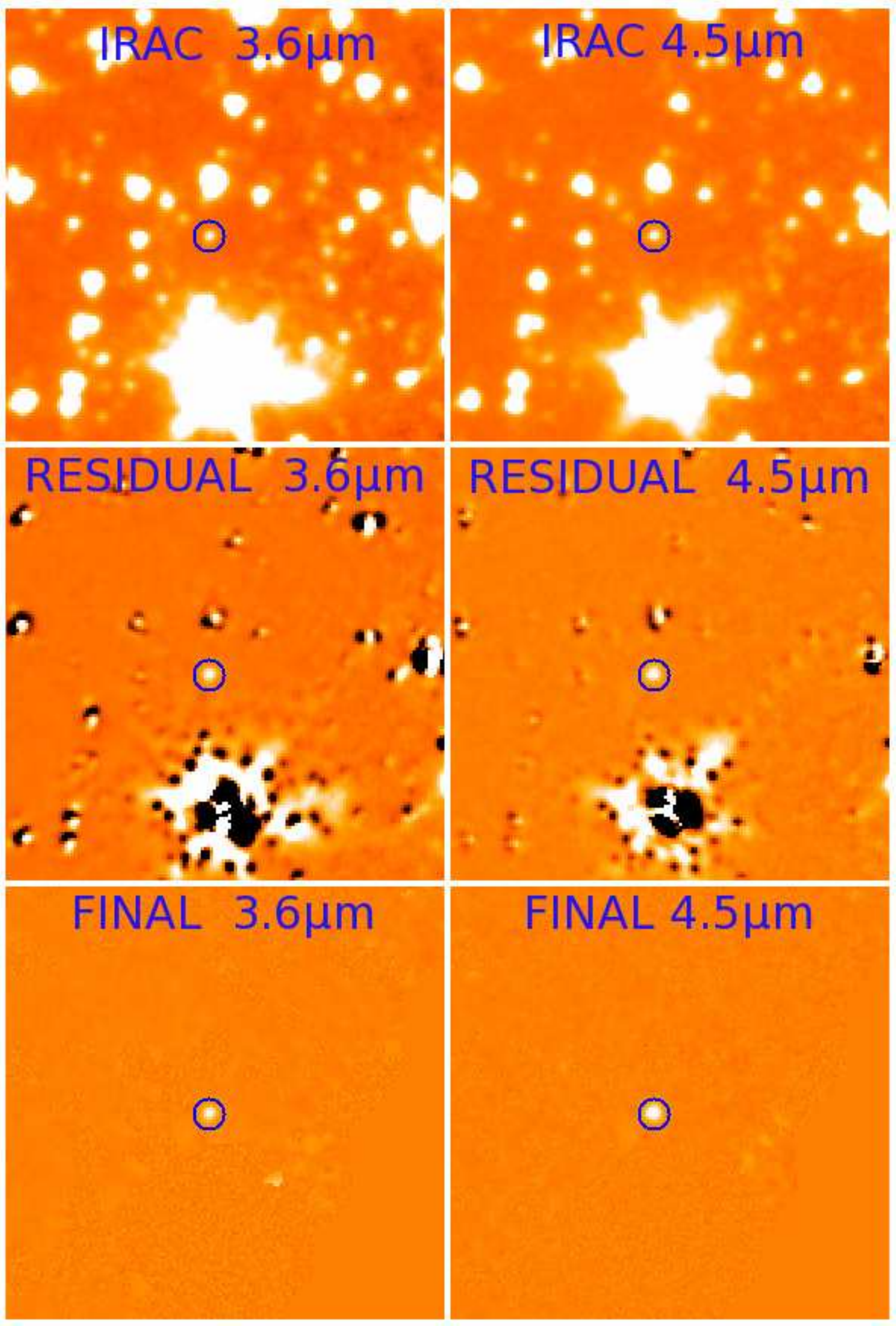} 
\end{center}
\caption{Example of the detection of one of the BBGs as an $H$-band
  dropout using our technique, which combines convolution of $H$-band
  images to the PSF of IRAC images using TFIT, followed by a cleaning
  algorithm and a scaling on a galaxy-by-galaxy basis. The left and
  right columns show this procedure on the IRAC 3.6~$\mu$m and
  4.5~$\mu$m residual images, respectively. The
  $40\arcsec\times40\arcsec$ cutouts are centered in the position of
  GDN\_BBG03 (blue circle).  In each column, the \textit{top panels}
  show the original IRAC image.  The \textit{second row} presents the
  residual image obtained by subtracting the ``template'' built by
  TFIT from the IRAC science frame. The \textit{third row} shows the
  final residual images after masking artifacts.}
 \label{fig:detection}
\end{figure}

We also applied a mathematical morphology (MM) method to the regions
around $H$-band bright sources to avoid extra flux arising from their
wings.  We iteratively generated one-pixel-width outlines applying
dilation
\citep{lea1989algorithm,maccarone1996fuzzy,lybanon1994segmentation},
subtracted the median flux and added the median background value to
each contour.  Figure~\ref{fig:detection} exemplifies our cleaning
procedure showing the environment of one of the $H$-band dropouts in
the raw and residual IRAC images, as well as in the final cleaned
image.

\subsection{Dual detection in IRAC 3.6 and 4.5~$\mu$m}
\label{ssec:IRACdetection}

After applying the masks to the IRAC residual images we used
SExtractor \citep{bertin1996sextractor} to search for BBG candidates.
We required the BBG candidate to be detected both in IRAC 3.6~$\mu$m
and 4.5~$\mu$m. Thus, we run SExtractor separately on the two cleaned
residual images and then we cross-matched the resulting catalogs
within a $1\arcsec$ radius, keeping only the sources in common. The
dual 3.6+4.5~$\mu$m detection provides a more robust selection and
further reduces the impact of spurious detections around artifacts.

In addition, we required all BBG candidates to be IRAC-bright ($[3.6]$
and $[4.5]\leq24.5$~mag) and not included in the $H$-band selected
CANDELS catalogs. To account for the latter, we removed all the
sources in the dual 3.6+4.5~$\mu$m catalog with a F160W counterpart in
any of the GOODS-N or GOODS-S CANDELS catalogs (and also 3D-HST
catalogs presented in \citealt{2014ApJS..214...24S}) identified within
a search radius of $0.5\arcsec$. Therefore, all our final selected
sources are, in principle, dropouts in the $H$-band.

\afterpage{
%\begin{landscape}
\begin{turnpage}
\begin{table*}
\centering
\begin{threeparttable}  
\small
\vspace*{0.cm}
  \caption{Observed properties of our sample of BBGs at $z>$3.}
    \hspace*{-500pt}\begin{tabular}{>{\raggedright}p{0.6cm}cccccccccccc}
    \toprule
 & ID & RA        & DEC        & F160W            & IRAC CH1          & IRAC CH2    & IRAC CH3          & IRAC CH4       & MIPS 24     & Comments\tnote{$^a$}  \\  
  & &	J2000      &   J2000     &   (AB mag)        &   (AB mag)         &   (AB mag)    &   (AB mag)         &   (AB mag)    & ($\mu$Jy)       \\ \hline \\
 1 & GDN$\_$BBG01 & 189.23577500 &  62.20206944 & $>$27.73              &  24.51  $\pm$  0.14 & 23.76 $\pm$  0.07  &  23.57  $\pm$  0.12 &   23.09  $\pm$  0.09    &  69.9 $\pm$ 6.0  & FIR             \\  
 2 & GDN$\_$BBG02 & 189.30782083 &  62.30743889 & 26.51 $\pm$  0.16   &  21.77  $\pm$  0.04 & 22.00 $\pm$  0.04  &  21.39  $\pm$  0.04 &   20.91  $\pm$  0.04    &  69.1 $\pm$ 6.2  & FIR, SMG,       W16                 \\  
 3 & GDN$\_$BBG03 & 189.18331250 &  62.32746389 & 27.25 $\pm$  0.25   &  22.86  $\pm$  0.05 & 22.32 $\pm$  0.04  &  21.82  $\pm$  0.04 &   21.41  $\pm$  0.04    &  29.5 $\pm$ 5.8  &                  W16                 \\  
 4 & GDN$\_$BBG04 & 189.43552083 &  62.29016111 & $>$27.20             &  24.03  $\pm$  0.09 & 23.42 $\pm$  0.06  &  22.92  $\pm$  0.07 &   22.61  $\pm$  0.07    &        \nodata       &                          \\  
 5 & GDN$\_$BBG05 & 189.14454583  & 62.10413611 & $>$26.47   &  24.06  $\pm$  0.10 & 23.55 $\pm$  0.07  &  23.11  $\pm$  0.08 &   22.22  $\pm$  0.06    &  36.1 $\pm$ 7.3  &                                    \\  
 6 & GDN$\_$BBG06 & 189.08689167 &  62.29081389 & 27.59 $\pm$  0.18   &  24.98  $\pm$  0.21 & 24.39 $\pm$  0.13  &  24.08  $\pm$  0.18 &   23.55  $\pm$  0.12    &        \nodata       &                                    \\  
 7 & GDN$\_$BBG07 & 189.10292917 &  62.31471944 & 26.11 $\pm$ 0.16   &  24.65  $\pm$  0.16 & 24.34 $\pm$  0.13  &  \nodata                &   \nodata                   &        \nodata       &                    ME                \\  
 8 & GDN$\_$BBG08 & 189.39490000 &  62.31689167 &  27.13 $\pm$ 0.16  &  23.95  $\pm$  0.09 & 23.75 $\pm$  0.07  &  23.33  $\pm$  0.10 &   23.03  $\pm$  0.09    &        \nodata       &                  W16                 \\  
 9 & GDN$\_$BBG09 & 189.40792083 &  62.21698889 & $>$27.49              &  24.43  $\pm$  0.13 & 24.19 $\pm$  0.11  &  \nodata                &   \nodata                   &        \nodata       &                                    \\  
 10& GDN$\_$BBG10 & 189.02397083 &  62.22303333 &  $>$27.31   &  24.01  $\pm$  0.09 & 23.68 $\pm$  0.07  &  23.39  $\pm$  0.10 &   23.02  $\pm$  0.09    &        \nodata       &                  W16                 \\  
 11& GDN$\_$BBG11 & 189.10607083 &  62.15669722 & $>$27.08              &  24.10  $\pm$  0.10 & 23.91 $\pm$  0.09  &  23.0   $\pm$  0.08 &   22.44  $\pm$  0.06    &        \nodata       & X-Ray,                      \\  
 12& GDN$\_$BBG12 & 188.96070833 &  62.18147500 & $>$27.22             &  24.46  $\pm$  0.14 & 24.21 $\pm$  0.12  &  \nodata                &    \nodata                  &        \nodata       &                            \\  
 13& GDN$\_$BBG13 & 189.25702917 &  62.25032778 &  $>$26.45          &  23.79  $\pm$  0.08 & 23.82 $\pm$  0.08  &  23.96  $\pm$  0.17 &   23.14  $\pm$  0.09    &  21.3 $\pm$ 5.2  &                  W16                 \\  
 14& GDN$\_$BBG14 & 189.14461667 &  62.11762500 & 26.43 $\pm$  0.20   &  24.37  $\pm$  0.12 & 24.39 $\pm$  0.13  &  24.22  $\pm$  0.21 &   23.51  $\pm$  0.11    &        \nodata       &                           ME \\  
 15& GDN$\_$BBG15 & 189.16360000 &  62.12178333 & 26.34 $\pm$  0.10   &  24.25  $\pm$  0.11 & 24.18 $\pm$  0.11  &  \nodata                &   \nodata                   &        \nodata       &                    B15,    ME \\  
 16& GDN$\_$BBG16 & 189.42832083 &  62.26589167 & $>$27.00              &  23.52  $\pm$  0.07 & 23.61 $\pm$  0.07  &  23.27  $\pm$  0.09 &   22.31  $\pm$  0.06    & 301.0 $\pm$ 8.4  &  FIR,             W16                 \\  
 17& GDN$\_$BBG17 & 189.38280000 &  62.33746667 & 26.49 $\pm$  0.11   &  24.78  $\pm$  0.18 & 23.97 $\pm$  0.09  &  \nodata                &   \nodata                   &       \nodata        &                    B15,     ME \\ \hline  \\
 18& GDS$\_$BBG01 &  53.13474583 & -27.90747222 & 26.40 $\pm$  0.11   &  23.36  $\pm$  0.08 & 23.69 $\pm$  0.10  &  22.58  $\pm$  0.07 &   22.45  $\pm$  0.08    &        \nodata       &                  W16                 \\  
 19& GDS$\_$BBG02 &  53.19989167 & -27.90467500 & 26.78 $\pm$  0.17   &  22.40  $\pm$  0.05 & 22.23 $\pm$  0.05  &  21.58  $\pm$  0.04 &   21.13  $\pm$  0.04    &  52.2 $\pm$ 4.9  &  FIR, SMG, X-Ray, B15,  W16              \\  
 20& GDS$\_$BBG03 &  53.04758333 & -27.86863611 & 26.15 $\pm$  0.11   &  23.39  $\pm$  0.08 & 23.30 $\pm$  0.07  &  22.48  $\pm$  0.07 &   22.36  $\pm$  0.07    &        \nodata       &                  W16                 \\  
 21& GDS$\_$BBG04 &  53.21421250 & -27.85935833 & $>$27.18              &  24.42  $\pm$  0.17 & 24.57 $\pm$  0.19  &  24.68  $\pm$  0.39 &   \nodata                   &        \nodata       &                                    \\  
 22& GDS$\_$BBG05 &  53.04210833 & -27.84253333 & $>$27.38              &  23.99  $\pm$  0.13 & 24.12 $\pm$  0.14  &  23.01  $\pm$  0.10 &   22.65  $\pm$  0.09    &        \nodata       &                                    \\  
 23& GDS$\_$BBG06 &  53.03234167 & -27.83515833 & 26.73 $\pm$  0.16   &  24.51  $\pm$  0.17 & 24.40 $\pm$  0.17  &  \nodata                &    \nodata                  &        \nodata       &                                    \\  
 24& GDS$\_$BBG07 &  53.11909167 & -27.81397778 & 26.71 $\pm$  0.10   &  23.65  $\pm$  0.10 & 23.48 $\pm$  0.08  &  22.81  $\pm$  0.09 &   22.61  $\pm$  0.09    &        \nodata       &                  W16,        ME \\  
 25& GDS$\_$BBG08 &  53.16725833 & -27.71545278 & $>$27.68              &  24.23  $\pm$  0.16 & 23.95 $\pm$  0.12  &  22.99  $\pm$  0.10 &   23.00  $\pm$  0.11    &        \nodata       &                   H11             \\  
 26& GDS$\_$BBG09 &  53.06088750 & -27.71842778 & 27.10 $\pm$  0.12   &  23.83  $\pm$  0.11 & 23.66 $\pm$  0.09  &  22.93  $\pm$  0.09 &   22.33  $\pm$  0.07    &  24.4 $\pm$ 6.0  &  FIR,             H11 ,W16             \\  
 27& GDS$\_$BBG10 &  53.13275833 & -27.72019444 & $>$27.02              &  23.79  $\pm$  0.11 & 23.73 $\pm$  0.10  &  22.65  $\pm$  0.08 &   22.23  $\pm$  0.07    &        \nodata       &                 H11 ,W16             \\  
 28& GDS$\_$BBG11 &  53.08477083 & -27.70801944 & $>$ 27.51    &  23.67  $\pm$  0.10 & 23.21 $\pm$  0.07  &  22.38  $\pm$  0.06 &   22.01  $\pm$  0.06    &        \nodata       &                  H11 ,W16             \\  
 29& GDS$\_$BBG12 &  53.19106667 & -27.69395833 & 26.02 $\pm$  0.09   &  24.01  $\pm$  0.13 & 23.83 $\pm$  0.11  &  23.08  $\pm$  0.11 &   22.68  $\pm$  0.09    &        \nodata       &                                    \\  
 30& GDS$\_$BBG13 &  53.02082083 & -27.69909722 & $>$27.09              &  24.02  $\pm$  0.13 & 23.79 $\pm$  0.10  &  22.92  $\pm$  0.09 &   22.49  $\pm$  0.08    &        \nodata       &                  W16                 \\  
 31& GDS$\_$BBG14 &  53.12760833 & -27.70668611 & 25.72 $\pm$  0.07   &  22.79  $\pm$  0.06 & 22.68 $\pm$  0.05  &  22.28  $\pm$  0.06 &   21.86  $\pm$  0.06    &        \nodata       &                  W16  ,ME               \\  
 32& GDS$\_$BBG15 &  53.04935000 & -27.75788611 & 26.61 $\pm$  0.17   &  24.11  $\pm$  0.14 & 23.73 $\pm$  0.10  &  22.85  $\pm$  0.09 &   22.69  $\pm$  0.09    &        \nodata       &                                  \\  
 33& GDS$\_$BBG16 &  53.19653333 & -27.75699444 & $>$27.24          &  23.86  $\pm$  0.12 & 23.81 $\pm$  0.11  &  22.78  $\pm$  0.08 &   22.71  $\pm$  0.09    &        \nodata       &                  B15,W16     \\ \hline                                                                                  
    \bottomrule
    \end{tabular}\hspace*{-5pt}%
     \begin{tablenotes}
          \item[a] H11: Reported in \cite{huang2011four}, B15: Reported in \cite{2015ApJ...803...34B}, W16: Reported in \cite{2016ApJ...816...84W} \\
          SMG: Source detected at wavelengths longer than 850$\mu$m. \\
          FIR: Source detected by PACS and$/$or  SPIRE. \\
          X-Ray: Source detected in X-rays. \\
          ME: Multiple counterparts or extended morphology in the HST individual bands or stacks.
      \end{tablenotes}
  \label{tab:photprop}
\end{threeparttable} 
\end{table*}%
\end{turnpage}
%\end{landscape}
}

Finally, we visually inspected all the remaining BBG candidates in the
IRAC residual images to remove any possible remaining artifacts due to
contamination from bright or/and nearby objects. We also discarded 3
sources that qualified as dropouts but were found to lie at $z<2$
while in this paper we are interested in $z>3$ galaxies. The final
catalog contains 33 \textit{bona fide} BBG candidates (17 in GOODS-N
and 16 in GOODS-S). By sample construction, our sources are relatively
bright in the first 2 IRAC channels and extremely faint (dropouts)
bluewards of 2~$\mu$m. The coordinates and magnitudes (see next
Section) of these sources are given in Table~\ref{tab:photprop}.

\section{SEDs, photometric redshifts and stellar population properties of the IRAC BBG candidates}
\label{sec:SEDsfit}

In this section we describe the multi-wavelength characterization of
the BBG candidates presented in the previous section. This
characterization consists on the measurement of the SED of each source
using the data described in \S\ref{sec:data}. We also discuss the
estimate of their photometric redshifts and stellar population
properties based on the fitting of those multi-band SEDs to stellar
population synthesis models.

\subsection{Multi-band photometry from near-to-far IR data and deep
  optical stacks}
\label{ssec:multiphot}

We measured multi-wavelength photometry for the 33 IRAC-selected BBGs
following the methods described in depth in \cite{perez2008stellar}
and \citet{2011ApJS..193...13A}. All 33 BBGs are in fact detected in
the IRAC catalogs of \cite{perez2008stellar} for the GOODS regions.
We chose not to adopt the SEDs of \cite{perez2008stellar} for the
BBGs. Instead, we repeated the same photometric procedure to take
advantage of the new and deeper mosaics in the region (see \S2, for
more details).

First, we measured the photometry in the four IRAC bands using the
residual IRAC images provided by TFIT and described in the previous
section. The use of these ``cleaned'' images reduces the flux
contamination due to bright neighbors, stars or image artifacts. For
each object, we considered several aperture radii, ranging from
$0.75\arcsec$ to $2\arcsec$, in order to to maximize the SNR of the
measurement and reduce the contamination from nearby sources. We
applied the appropriate aperture correction to the measurement for
each radius.  The typical scatter between measurements using different
apertures are 0.13, 0.16, 0.16 and 0.15\,mag in IRAC 3.6, 4.5, 5.8 and
8.0~$\mu$m.  The new photometry is fully consistent with the values in
\cite{perez2008stellar}. By definition of the sample, our BBGs are
detected in the first two IRAC bands, and $\sim80\%$ of them are
detected in the four IRAC channels.

Redwards of the IRAC channels ($\lambda>8\,\mu$m), we
measured the photometry in the {\it Spitzer}/MIPS and Herschel/PACS
and SPIRE far-IR bands. As discussed in \S~\ref{ssec:SFR}, these
fluxes can be used to characterize the dust emission properties of the
BBGs.  A total of 8 sources ($\sim$25\% of the sample) were detected
in MIPS24, and 5 of those MIPS sources were detected by PACS, and 4 of
them by SPIRE. We also searched for counterparts in the submillimeter
catalogs available in the GOODS fields (see \S~\ref{sec:data}). Among
the 8 MIPS emitters, 3 BBGs were detected at 850~$\mu$m and one of
them is also detected at 1200~$\mu$m.  Additionally, 2 BBGs were
detected in X-rays.

Last, we focus on the more difficult task of trying to characterize
the SEDs bluewards of $\lambda\lesssim3.6\mu$m and measuring the
photometry in the optical, \mbox{near-}, \mbox{mid-} and far-IR bands.
We find that $\sim40$\% of the BBGs are detected in the $K$-band
images, presenting magnitudes in the range K$\sim$24.6-26.7~mag
($\rm{SNR}>5$). All 33 BBGs were required to be undetected in the
CANDELS (and 3D-HST) F160W catalogs. However, in principle there could
still be a weak flux in the image that was missed by the CANDELS
source extraction procedure. In order to constrain that potential
signal, we first searched for weak detections of the BBG candidates in
the deep (9-band) HST and (25-band) SHARDS stacks described in
\S~\ref{ssec:data_ACS_SH}, using a $1\arcsec$ search radius around the
IRAC positions. Interestingly, we found that 11 out of 16 ($\sim$70\%)
and 9 out of the 17 ($\sim$50\%) BBG candidates are detected in the
GOODS-S or GOODS-N HST stacks, respectively. In addition, 6 of the
latter ($\sim$35$\%$) are also detected in the SHARDS stack (only
available in GOODS-N). Out of our 33 sources, 7 are exclusively
detected in the IRAC data. These IRAC-only 7 sources present
magnitudes around 24.0-24.5 mag, $\sim$0.3~mag fainter than the rest.

For all the BBG candidates we measured accurate positions using the
following method. First, we tweaked the World Coordinate System (WCS)
solution for the IRAC images locally taking as a reference the F160W
CANDELS mosaic. For this purpose, we aligned the IRAC and HST images
using detected galaxies in a $30\arcsec$ circle around each BBG
candidate. Typically, this implies corrections in the WCS of the IRAC
images by offsets smaller than $0.2\arcsec$. The rms of the comparison
between source centroids calculated in HST and IRAC images is 0.3".
For the BBGs with counterparts in the HST stack, we adopted the more
robust and reliable astrometric positions based on the HST imaging.
Using these positions, we measured the fluxes in the F160W
mosaic. For the galaxies with extended or multiple-knots morphology
(as inferred from the stacks), we used elliptical apertures with
their semimajor axis raging between 0.6 and 0.9”. For the rest of
the sample, we used apertures of $0.4\arcsec$ radius which maximized
the SNR in the HST images for point-like ultra faint sources. We
also considered larger apertures and checked the consistency of our
results regarding the photometric procedure. Please refer
to appendix \ref{A:PhotUncertainty} for a detailed description of
the method used in this work to derive consistent and reliable
photometry in the optical and near-IR HST bands.

We obtained weak but reliable fluxes ($H=25.7-27.6$;
$\rm{SNR}>5$) for 17 out of the 33 galaxies.  This clearly indicates
that many BBG candidates are (weakly) detected in the F160W mosaics,
but they were missed in the CANDELS (and 3D-HST) catalogs due to the
lower completeness level of the selection procedure at faint
magnitudes. Finally, in all the cases where we could not recover a
positive flux in our measurements, we measured upper limits.  The
upper flux limit was computed as 5$\sigma$ of the sky noise measured
in an empty region around the source with the same aperture size as
that used for the photometric measurement, and taking into account
pixel-to-pixel noise correlations \citep{perez2008stellar}.

In summary, most BBGs have fluxes in 5-6 optical-to-mid-IR bands. The
complete SEDs were then used to estimate photometric redshifts and
characterize the stellar emission (see next subsections).  Around
$20\%$ of the sample also has far-IR detections that were used to
analyze the dust emission and (obscured) SFR properties (see
\S\ref{ssec:SFR}).

\subsection{Photometric redshifts}
\label{ssec:SEDphotoz}

\begin{figure}[t]
  \begin{center}
    \includegraphics[width=1.\linewidth]{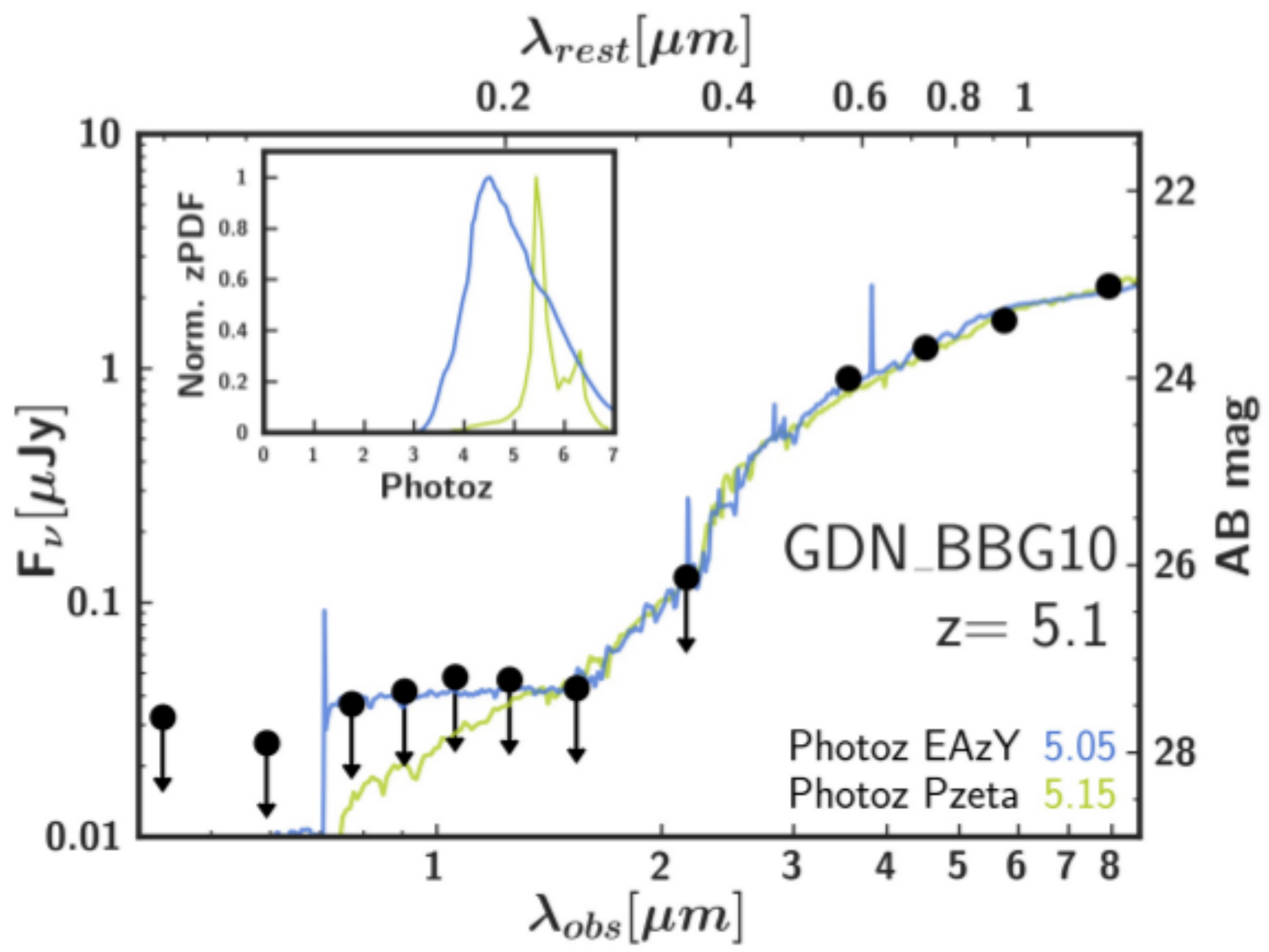}
  \end{center}	
  \caption{SED of one of our sources in GOODS-N (black circles,
    including measured fluxes and 5$\sigma$ upper limits), jointly
    with the best-fitting models used by our two photo-z codes (EAzY
    in blue and pzeta in green). The inset shows the probability
    distribution functions (zPDF) and the legend gives the two
    redshift estimates obtained with each code, as well as the final
    assumed value.}
  \label{fig:sedfitting_photoz}
\end{figure}

We estimated photometric redshifts for the 33 BBGs using two different
codes, namely pzeta (\citealt{perez2008stellar}) and EAzY
\citep{brammer2008eazy}.  Both codes estimate the redshift by fitting
the SEDs in the spectral range where the emission is most probably
dominated by stellar emission, i.e., at wavelengths bluer than
rest-frame $\sim3\,\mu$m, using a limited set of galaxy and AGN
templates.  Figure~\ref{fig:sedfitting_photoz} illustrates the fitting
procedure with the two codes and shows the resulting photometric
redshift probability distribution functions (zPDFs).

Although the majority of the sources are extremely faint at short
wavelengths ($\lambda<3.6\,\mu$m), we used the low SNR data points
(both from individual images and stacks) to impose upper limits and
better constrain the photo-z. Moreover, we visually inspected the
results for each galaxy to verify that the fit relies primarily on the
high SNR data and that the best-fit template was realistic. When
necessary, we re-run the code excluding the lowest SNR data
(SNR$\lesssim$5) or any template which was not consistent with
being undetected at longer wavelengths or with all the upper
limits.  In general, the best-fit photo-z from both codes and the
zPDFs are consistent. Thus, we adopted the result with the most
conservative uncertainty (i.e., the broader zPDF). In the cases where
the results did not agree or none of the solutions properly fitted the
SED, we analyzed the zPDF and adjusted the permitted redshift range
interval to obtain the most reliable value.

Obtaining reliable photometric redshifts obviously benefits
from having high quality photometry in many bandpasses. Given the
difficulties in deriving reliable fluxes for BBGs at short
wavelengths, we repeated the whole photo-z estimation process for
each galaxy with the 3 different sets of SEDs explained in appendix
\ref{A:PhotUncertainty}.  Hence, we investigated the error in
redshift estimate including the contribution of the photometric
uncertainties. Relatively small (considering the type of very faint
galaxies that we are dealing with) photometric redshift differences
were found when using the different photometric approaches,
typically below $\delta z=0.2$. In any case, we checked the
consistency of the results presented in the following sections using
the zPDFs for the 3 SED types presented in Appendix~A. These
different zPDFs were also used to estimate uncertainties in those
results, for example, in the photometric redshift distribution of
our sample of BBGs.

\subsection{Stellar population properties}  
\label{sec:Stellarpop}

Based on the best-fit photometric redshift computed in the previous
section, we fitted again the SEDs of the BBGs using stellar population
synthesis models in order to characterize their stellar masses, dust
attenuations and mass-weighted ages (t$_{m}$).  We performed this fit
using two different codes, namely \texttt{synthesizer}
\citep{perez2008stellar} and \texttt{FAST}
\citep{2009ApJ...700..221K}. We used the same set of stellar
population and dust modeling assumptions for both codes. We considered
the stellar population models from \citet{bruzual2003stellar},
assuming a delayed exponential star formation history (SFR(t)
  $\propto t e^{-t/\tau}$), a \citet{2003PASP..115..763C} IMF, and we
adopted the \citet{2000ApJ...533..682C} dust attenuation law. Each
stellar population model is characterized by four parameters:
timescale $\tau$, age $t$, metallicity $Z$, and dust attenuation
$A(V)$. We assumed solar metallicity and we allowed the other three
free parameters to vary within the ranges presented in
Table~\ref{table:sps}. Figure~\ref{fig:sedfitting} shows an example of
the fitting procedure and the resulting stellar population properties.

Overall, the results for stellar masses, dust attenuations and
mass-weighted ages using the two codes are roughly consistent within
the typical uncertainties (e.g., $\sim0.3$~dex for the stellar mass).
Nonetheless, we further verified the robustness of the estimated
parameters by analyzing possible degeneracies (clusters of likely
solutions) in the full parameter space using a Monte Carlo algorithm
included in \texttt{synthesizer} (see e.g.,
\citealt{2016MNRAS.457.3743D} for more details). Briefly, we generated
1000 random variations of the SED of each galaxy assuming Gaussian
photometric errors, and we explored the resulting set of best-fit
parameters in the age vs. $\tau$ space. The 1000 Monte Carlo (MC)
particles typically form 1-3 clusters of solutions. In most cases
there is always one cluster with a much higher likelihood (defined
from the number of MC particles belonging to that cluster).  However,
a small fraction of the galaxies ($\sim$15$\%$) exhibit two or more
solutions with similar likelihood. In those cases we compared each set
of results with those from FAST and we visually inspected the best-fit
result for each cluster to identify the most reliable solution.  We
also used the MC simulations to assign uncertainties to the best-fit
parameters based on the 68$\%$ probability contours around the median
result for each cluster. Given the faintness of BBGs, the
uncertainties in their derived physical parameters are relatively
large.  The statistical effect of the uncertainties in the
photometric redshifts on the stellar masses was also considered.
Stellar mass probability distribution functions (smPDF) for each
galaxy were constructed from the zPDFs described in
Section~\ref{ssec:SEDphotoz}.  These smPDFs have been used to
estimate uncertainties of the results presented in the following
sections.

\begin{center}
\normalsize	
  \begin{threeparttable}
  \caption{\label{table:sps} Parameter space (star formation
    timescale, age, dust attenuation, and metallicity) allowed in the
    stellar population synthesis fitting procedure.}
\begin{tabular}{lccc}
%  \begin{table} 	
  \hline
  \raisebox{-1ex}{Parameter} &{range} & {units} & {step} \\[-1ex]
  \\\hline
  \hline\\[0ex]
  & $8.0-10.0$ & log(yr) & 0.1 dex\\[-2ex]
  \raisebox{2ex}{Timescale $(\tau)$}
  & $6.0-9.5$ & log(yr) & 0.1 dex\\[-2ex]
  \raisebox{2ex}{Age $(t)$}
  & $0.0-4.0$ & mag & 0.1 mag \\[-2ex]
  \raisebox{2ex}{Dust attenuation A(V)\tnote{$^a$}}
  & $1.0$ & $Z_{\odot}$ & fixed \\[-2ex]
  \raisebox{2ex}{Metallicity $(Z)$ }\\
  \hline
\end{tabular} 
     \begin{tablenotes}
          \item[a] For non-IR emitters we only allowed dust attenuation to vary between 0 and 2 magnitudes.          
     \end{tablenotes}
\end{threeparttable}   
\end{center}
%\end{table}

\subsection{Star formation rates}  
\label{ssec:SFR}

We computed SFRs for the BBGs using the most reliable SFR tracer
available for each galaxy. This approach is similar to the SFR
``ladder'' method described in \citet{2011ApJ...738..106W}. In brief,
we rely in the IR-based SFR estimates for galaxies detected at
mid-to-far IR wavelengths, and we used the best fitting SPS model 
to estimate the SFR for the rest. As shown in \citet{2011ApJ...738..106W} 
the agreement between these estimates for galaxies with a moderate 
attenuation (faint IR fluxes) ensures a continuity between SFR indicators.

\begin{figure}[t]
\begin{center}
   \includegraphics[width=1.\linewidth]{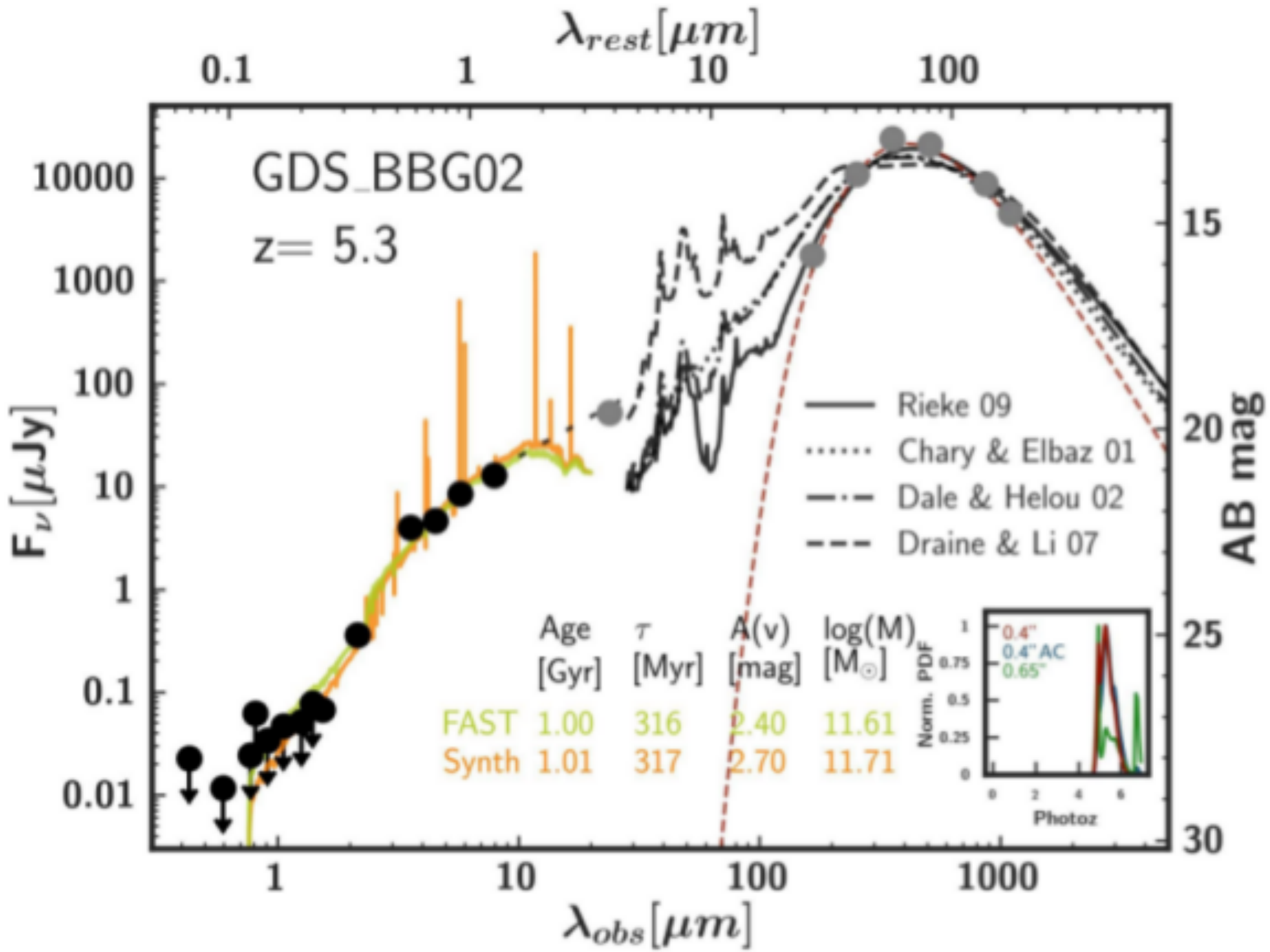}
\end{center}
\caption{SED fitting example for one of our BBG candidates. Bands
  (most probably) dominated by stellar emission are plotted with black
  dots, and bands probing the dust emission are plotted with grey
  dots. The inset shows the photo-z zPDFs from \texttt{EAzY}
 for the 3 different sets of SEDs obtained with distinct
 photometric approaches (red, blue and green). The best fit and
  the inferred parameters from \texttt{synthesizer} and \texttt{FAST}
  are shown in orange and green respectively.  The best-fitting
  solutions for different dust emission libraries
  \citep{Rieke2009,Chary2001,dale2002infrared,draine2007infrared} used
  to characterize MIR/FIR and sub-mm fluxes are shown with different
  black lines. The best dust emission model is the one shown with a
  solid black line.  The red dashed line corresponds to the modified
  black-body model fitting the data above 20~$\mu$m.}
\label{fig:sedfitting}
\end{figure}

For IR-detected galaxies ($\sim$25$\%$ of the sample) the total SFRs,
SFR$_{IR+UV}$, are computed from a combination of IR and rest-frame UV
luminosities (uncorrected for attenuation) applying the
\cite{1998ARA&A..36..189K} equation normalized to a
\cite{2003PASP..115..763C} IMF:
\begin{equation}
 \rm{SFR}_{UV+IR}=1.1 \times 10^{-10} (\rm{L}_{IR} +3.3\times \rm{L}_{UV}) \,\, [\rm{M}_\odot~\rm{yr}^{-1}]
\end{equation}
The UV luminosity traces the (typically small) fraction of ionizing
photons that are not absorbed by the dust. The IR luminosity is
determined from the fitting of the available MIR/FIR and sub-mm data
to 4 dust emission models
\citep{Rieke2009,Chary2001,dale2002infrared,draine2007infrared},
following the methods described in
\cite{perez2008stellar,perez2010improving} and
\cite{2011ApJS..193...13A}. For each IR-detected BBG we used the
L$_{IR}$ of the model that more accurately fitted the data for the
estimate of SFR$_{UV+IR}$. In all cases, the typical scatter of
L$_{IR}$ estimations based on different template libraries (including
the fit to a modified black body) is below 0.1~dex. See
Figure~\ref{fig:sedfitting} and Appendix~\ref{A:SEDs} for some examples
of the IR SED fitting.

For IR undetected galaxies ($\sim$75$\%$ of the sample), we used the
SFR values inferred from the SED fit. This is also the case for MIPS
emitters at $z\gtrsim$5, for which MIPS 24~$\mu$m shifts out of the
rest-frame mid-IR region ($\lambda=$4-20~$\mu$m) where dust emission
models are not defined and would produce highly uncertain L$_{IR}$
values due to the large extrapolation involved. For these two types of
sources (28 BBGs in total) we used the SFR averaged over the
last 100~Myr of the Star Formation History (SFH). We adopted this
measurement over an L$_{UV}$-based value corrected for extinction due
to the faintness of the BBGs. Indeed, the rest-frame UV is redshifted
into the ACS and WFC3 bands, where our galaxies are extremely faint or
even undetected.  Consequently, UV slope measurements cannot be
performed properly or would be highly uncertain. While some BBGs
exhibit marginal optical detections, the SED-based SFRs can be
computed uniformly for all galaxies and they include the F160W fluxes
when available.

In summary, we estimate SFRs using a variety of methods for
all BBGs.  The reliability of these SFR estimations is, in general,
not high, mostly because of the faintness of our galaxies, which is
especially extreme in the rest-frame UV (given that BBGs were
selected as red objects missed by the deepest optical and
near-infrared surveys). Nevertheless, the SFRs for the several FIR
emitters in our sample (5 out of 33) are more robust, given that the
dust emission is very well constrained. For the sources with only a
MIPS mid-IR detection or undetected, the SFRs, based on SED fitting,
should be considered as a lower limit, since typically they are
significantly smaller than the SFRs calculated from dust emission
probed by the FIR data. We also remark here that the SFRs for the
galaxies in the comparison samples have been estimated in a similar
way, so our results about the BBGs relative to the known
(catalogued) populations of massive galaxies at high-z are robust.
Further observations capable of measuring emission lines or fainter
MIR/FIR fluxes are needed for better accuracy in the SFR analysis.

\section{Physical properties of BBGs}
\label{sec:Pysicalprop}

\begin{figure*}[t]
  \begin{center}
    \includegraphics[width=1.\textwidth]{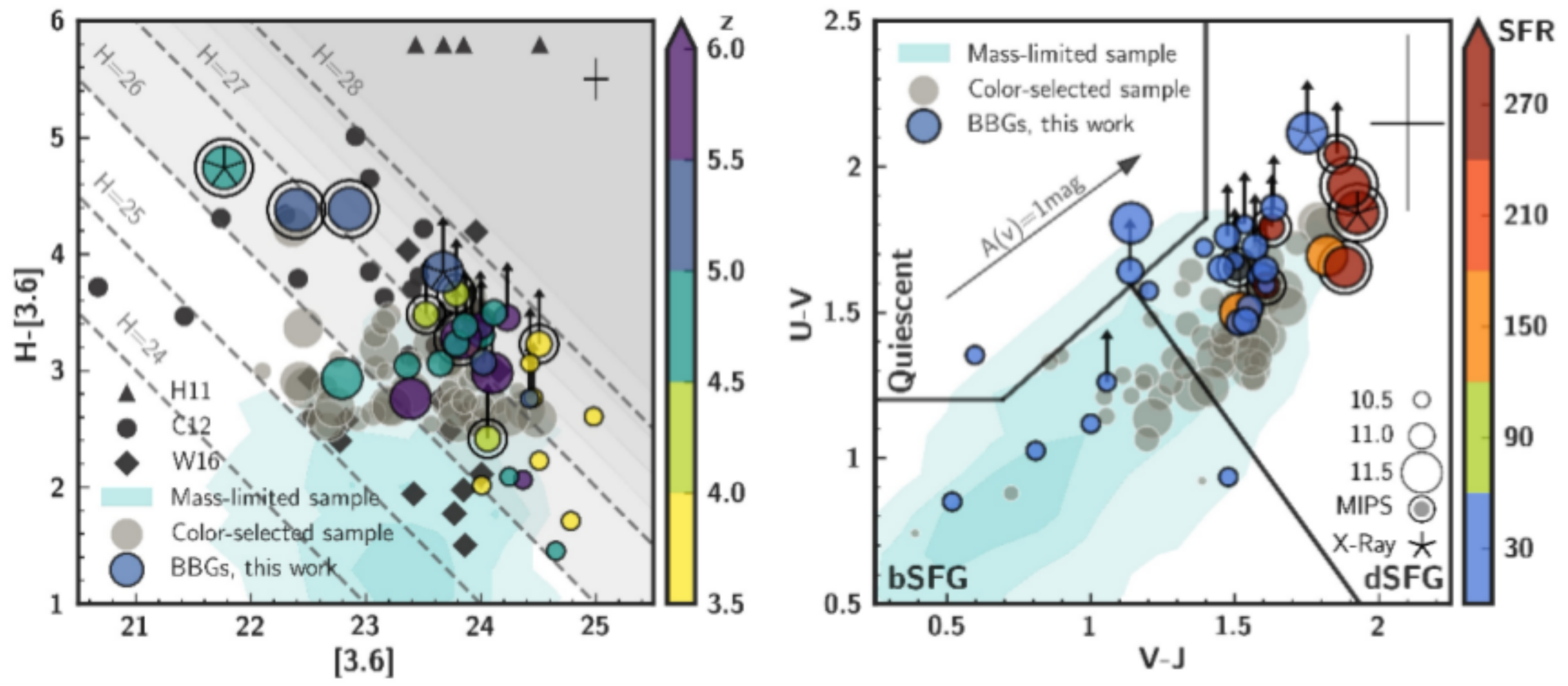}    
  \end{center}
  \caption{\textit{Left panel:} Observed-frame
    $H-[3.6]$ color plotted versus the observed [3.6] magnitude for
    our sample of BBGs, color-coded by their photometric redshift and
    scaled in size as a function of their stellar mass (legend shown
    in right panel). The CANDELS color-selected sample is plotted in
    grey with the size also scaled according to their masses. The 0.3,
    1 and 2~$\sigma$ distribution of the mass-limited sample
    are shown by the light blue density regions. The dark filled
    triangles represent the four massive galaxies at $z>4.5$ found by
    \cite{huang2011four}, dark filled circles show the extremely red
    sources from \cite{caputi2012nature} and dark diamonds correspond
    to the $H$-band dropouts reported by \citet{2016ApJ...816...84W}.
    Lines (dashed) of constant H values are also shown. The darker
    background indicates the regions of lower completeness for
    $H$-band \citep{2013ApJS..207...24G}.  Individual error bars are
    not plotted for clarity but the average values for our sample of
    BBGs are shown in the top-right corner of each panel. $H$-band
    upper limits ($H-[3.6]$ lower limits) are also shown.
    \textit{Right panel:} Rest-frame $U-V$ vs. $V-J$ color-color plot,
    where BBGs are color-coded by SFR and scaled by stellar mass.  SFR
    lower limits are shown in dark grey.  Sources from the CANDELS
    color-selected sample scaled by mass are also shown. The 0.3, 1
    and 2~$\sigma$ distribution of the mass-limited sample
    are shown by the light blue areas. MIPS detected galaxies are
    surrounded with another circle while X-ray detected galaxies are
    highlighted with an star inside the symbol.  The
    \cite{Whitaker2011} upper boundary (black wedge) separates
    quiescent galaxies (top left) from star-forming galaxies (SFGs,
    bottom). The black diagonal line denotes an additional criterion
    proposed in our work (perpendicular to the attenuation vector) to
    separate blue (bSFG) from dusty (dSFG) star-forming galaxies. The
    plot also includes a 1~mag attenuation vector (which assumes a
    \cite{2000ApJ...533..682C} law). $H$-band upper limits ($U-V$
    lower limits) are also shown.}
		\label{fig:col_mag_ios}
\end{figure*}

Here we analyze the distribution of observed colors, photometric
redshifts, and stellar population properties of the 33 BBGs using the
results from the UV-to-FIR SED fitting techniques described in
\S~\ref{sec:SEDsfit}. We will also compare BBGs with two samples
constructed with the CANDELS GOODS-S and GOODS-N $H$-band selected
catalogs presented in \citet{2013ApJS..207...24G} and Barro et
  al. (2019): a mass-limited and a color-selected sample. The
mass-limited sample is composed by massive
($\rm{M}>10^{10}\,\rm{M}_{\odot}$) galaxies at $z>3$. We
remark that this is a sample constructed with a simple cut in
redshift and mass rather than a mass complete sample. The
color-selected sample, aimed to reproduce our BBG selection, is
composed by red ($H-[3.6]>2.5$~mag) faint ($H>25$~mag) galaxies.
 These 2 samples are characterized in Appendix~B. In
addition, we compare our sample of BBGs with the samples of red
galaxies of similar nature presented in \citet[][H11]{huang2011four},
\citet[][C12]{caputi2012nature}, and
\citet[][W16]{2016ApJ...816...84W}.

\subsection{Observed IR colors and photometric redshifts}
\label{ssec:SEDphotozprop}

\begin{table}
\setlength\tabcolsep{4pt}
 \begin{center} 
   \caption{Statistical properties of the different samples. Median
     values,1$^\mathrm{st}$ and 3$^\mathrm{rd}$ quartiles of their
     redshift, magnitudes, colors and masses are shown.}
 \begin{tabular}{lcccccc}
  \normalsize
 \\\hline \\
 &{z} &{H}& {[3.6]}& {[4.5]} &{H$-$[3.6]}&{M}\\[-2ex]
 \raisebox{3ex}{Sample} 
 &{ } &{mag} & {mag}& {mag} & {mag}&{M$_{\odot}$}\\[0ex]
 \\\hline
 \hline\\
 &$4.8^{5.1}_{4.4}$& $27.1^{27.3}_{26.5}$&$24.0^{24.4}_{23.7}$&$23.8^{24.1}_{23.6}$&$3.1_{2.8}^{3.4}$& $10.8_{10.4}^{11.1}$\\[-2ex]
 \raisebox{3ex}{BBGs}
  &$3.8^{4.7}_{3.3}$ & $24.9^{25.7}_{24.2}$&$23.3^{23.8}_{22.8}$&$23.2^{23.8}_{22.6}$&$1.6^{2.1}_{1.1}$& $10.4_{10.1}^{10.6}$\\[-2ex]
 \raisebox{3ex}{Mass-lim.$^a$}\tnote{a}
 &$4.7^{5.3}_{4.1}$&$26.5^{26.8}_{25.9}$&$23.6^{23.9}_{23.1}$&$23.4^{23.9}_{22.7}$&$2.8^{3.1}_{2.6}$& $10.8_{10.4}^{11.1}$\\[-2ex]
 \raisebox{3ex}{Color-sel.$^b$} \tnote{$^b$}
 &$5.7$ &$29.8^{30.2}_{29.5}$&$23.8^{24.0}_{23.6}$ &$23.5^{23.7}_{23.4}$&$ 6.1^{6.6}_{5.5}$&\nodata\\[-2ex]
 \raisebox{3ex}{H11$^c$} \tnote{$^c$}
 &$4.1_{3.4}^{4.9}$&$26.9_{26.5}^{27.5}$&$23.2_{22.4}^{23.5}$&$22.5_{22.1}^{23.13}$&$ 3.7_{3.5}^{4.2}$& \nodata \\[-2ex]
 \raisebox{3ex}{C12$^d$} \tnote{$^d$}
 &$4.7^{5.7}_{4.2}$&$26.0_{25.4}^{26.7}$&$23.7_{23.2}^{23.9}$&$23.4_{22.9}^{23.6}$&$ 2.6_{2.1}^{3.0}$& \nodata\\[-2ex]
 \raisebox{3ex}{W16$^e$} \tnote{$^e$}\\
 \hline
 \label{table:medianprop}
 \end{tabular} 
	 \begin{tablenotes}
	 	  \item [] 	The statistics have been calculated by assuming 5$\sigma$ values for the $H$-band undetected galaxies.
		   \item[a] $^a$ CANDELS mass-limited sample.
	       \item[b] $^b$ CANDELS color-selected sample.
	       \item[c] $^c$ H11: \cite{huang2011four} .
	       \item[d] $^d$ C12: \cite{caputi2012nature}.
	       \item[e] $^e$ W16: \cite{2016ApJ...816...84W}. \\
	 \end{tablenotes}
 \end{center} 
\end{table}

The left panel of Figure\,\ref{fig:col_mag_ios} shows the $H-[3.6]$
vs.  $[3.6]$ color-magnitude diagram for our 33 BBGs color-coded by
redshift. For this one and the rest of figures in the
  following sections, we use our fiducial photometry. For reference,
  we describe how the different photometric methods described in
  Appendix~\ref{A:PhotUncertainty} affect the results.

As demonstrated in Appendix~\ref{A:CANDELSample}, a red
$H-[3.6]\gtrsim2$~mag color is a good proxy to identify
massive red galaxies (dusty or evolved) at $z>3$ (BBGs).
These galaxies present strong Balmer (or D4000) breaks, which produce
the red colors (sometimes in combination with dust attenuation).
Thus, $H$-band dropouts in the CANDELS catalogs
($H(5\sigma)\sim27$~mag) with bright [3.6] magnitudes are excellent
high redshift BBG candidates. However, as discussed in
\S~\ref{sec:SEDsfit}, some of our BBGs have weak $H$-band detections.
They are undetected in the CANDELS catalogs mostly due to the
increasing catalog incompleteness at $H>25$~mag (cf.  Figure 4 in
\citealt{2013ApJS..207...24G}). As a result, a small fraction
($\sim20\%$) of our BBGs have faint $H$-band magnitudes but colors
slightly bluer than $H-[3.6]\sim2.5$~mag (see the bottom-right corner
of the color-magnitude diagram in Figure\,\ref{fig:col_mag_ios}). Note
that the bluest BBGs have faint IRAC magnitudes, $[3.6]=24-25$.
Likewise, one can also identify the opposite situation, i.e., galaxies
with relatively bright H and IRAC magnitudes but red
H-[3.6]$\gtrsim2.5$ colors, that are also good BBG candidates.  This
is the case for the grey dots in Figure\,\ref{fig:col_mag_ios}, which
depict the color-selected sample (see Appendix~\ref{A:CANDELSample}
for more details). The distribution of the color-selected sample in
the diagram shows that our BBG dropout criterion is essentially a
faint-end extension of a color and $H$-band limited sample.
Table~\ref{table:medianprop} summarizes the average properties of our
sample of BBGs and the galaxies selected by color from the CANDELS
catalog. Our BBGs are typically galaxies with a very faint or
  inexistent detection in the $H$-band, $\langle H
  \rangle\sim26.5$~mag (for the 60\% of sources with detections in the
  CANDELS data). These galaxies are remarkably bright in IRAC,
  $\langle [3.6] \rangle\sim24$~mag, which converts them in very red
  sources, $\langle H-[3.6] \rangle\sim3.1$~mag. For reference, these
  statistical properties change by 0.3-0.5~mag when considering the
  other 2 photometric methods described in
  Appendix~\ref{A:PhotUncertainty}; for example, for the photometric
  apertures of size $r=0.65\arcsec$, our sample of BBGs presents
  $\langle H \rangle\sim26.7$~mag and $\langle H-[3.6] \rangle
  \sim2.8$~mag. Comparing our BBGs with other samples, we find that the
CANDELS color-selected sample has $\sim0.4$~mag brighter $[3.6]$ and
$H$-band magnitudes and a slightly bluer $H-[3.6]$ color. We note that
both samples have very similar photometric redshift distributions
peaking at $z\sim5$ (see discussion below).

Figure~\ref{fig:col_mag_ios} also compares the 33 BBGs to other
samples of red, massive high-z candidates (black symbols) from the
works of H11, C12, and W16. In particular, H11 and W16 identified 4
and 16 BBG candidates, respectively, in the GOODS fields. Our
selection method recovers all these galaxies (as indicated in Table\,
\ref{table:medianprop}). Our photometry is consistent with these works
in the IRAC bands ($\langle\Delta[3.6]\rangle\sim0.06$~mag) and
slightly brighter in $H$-band, particularly relative to H11 ($\langle
\Delta~H_{\rm H11} \rangle\sim2.67$~mag). This difference suggests
that our forced photometric measurement is very effective at
recovering flux for faint $H$-band sources. This difference translates
to H11 reporting much redder colors for the galaxies in common with
our sample. We also find significant differences in some cases
  ($\langle \Delta~H_{\rm W16} \rangle\sim0.69$~mag) between our
  $H$-band magnitudes and those reported by W16. Overall, the mean
color of our BBGs (H-[3.6]$\sim$3~mag) is similar to the values reported
in W16 and C12.  However, the C12 sample is slightly redder
($\sim$0.4~mag), but we must also take into consideration that their
sample was built in a field with shallower HST and IRAC data (the UDS,
see \citealt{2013ApJS..206...10G}). This translates to the C12 sample
being around 1~mag brighter at 3.6 and 4.5~$\mu$m than our BBGs.

\begin{figure}[t]
\begin{center}
    \includegraphics[width=1.\linewidth]{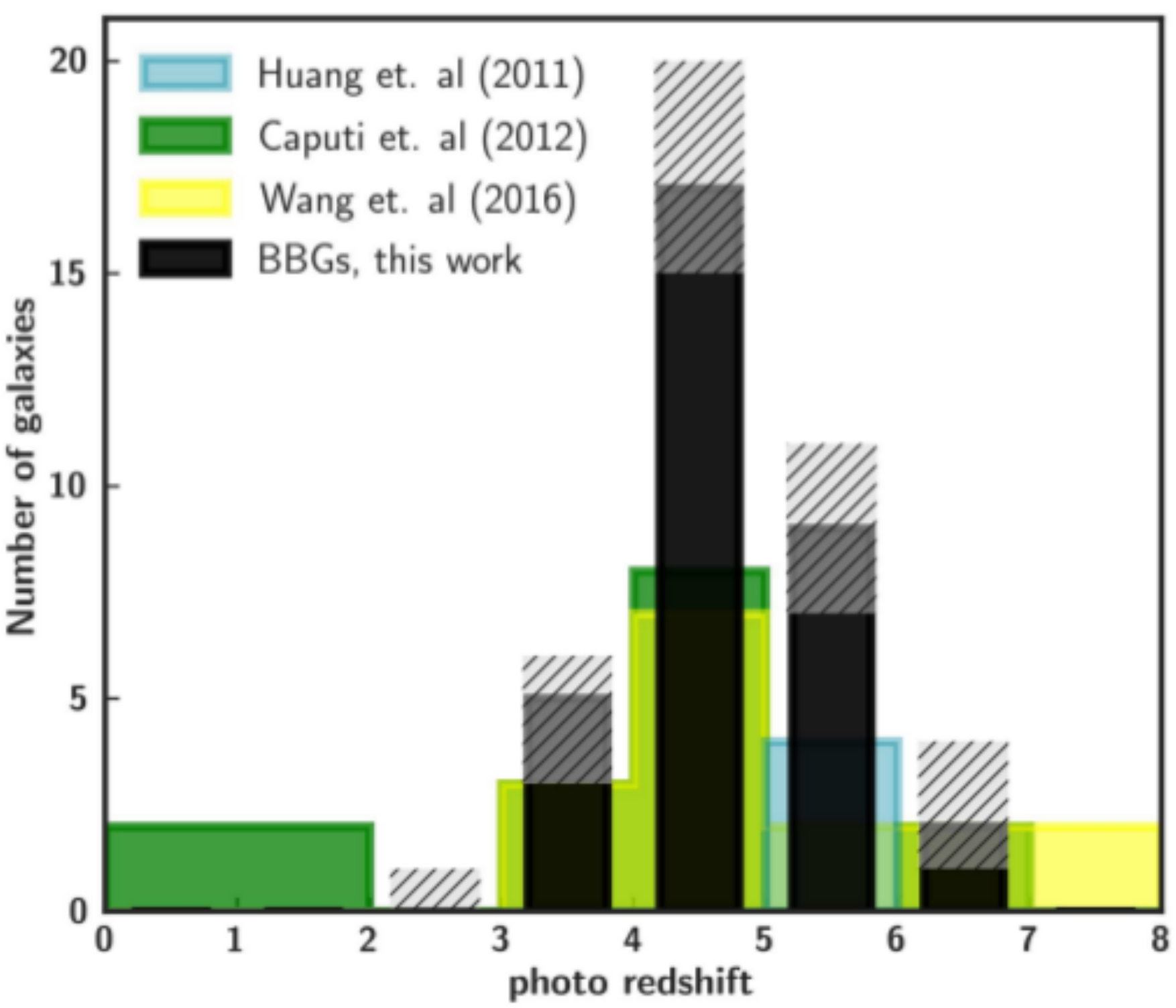}
\end{center}
\caption{Photometric redshift distribution of our sample of BBGs,
  together with previous samples of similar galaxies reported in the
  literature and explained in the text. BBGs are shown with
    black bars and the uncertainty (see text for details) is
    highlighted with hatching black lines. The photo-z distribution
  of galaxies reported in \cite{caputi2012nature} and
  \cite{2016ApJ...816...84W} are shown by green and yellow histograms
  respectively.  The 4 sources presented in \cite{huang2011four}, all
  assumed to be at $z\sim5.7$, are shown in light blue. Photo-z
  statistics for each sample are presented in Table 3.}
\label{fig:zhist_lit}
\end{figure}

Table\,\ref{table:medianprop} also indicates a remarkably good
agreement in the average photometric redshift values for all the
color-selected BBG samples. Typically, these mid-IR bright red
galaxies lie at $\langle z\rangle\sim4-6$.  The redshift distributions
of our galaxies and those in H11, C12, and W16 are shown in
Figure\,\ref{fig:zhist_lit}. The uncertainties in the photo-z
  histogram for our BBGs was derived as the $\pm1$ sigma deviation in
  each bin arising from building 1000 photo-z histograms based on our
  BBGs zPDF for the 3 different sets of SEDs (refer to appendix~\ref
  {A:PhotUncertainty} for more details). While a few BBGs are found
at $z<4$, the majority of the sample is skewed to higher redshifts
$z\gtrsim4$ ($80\pm1$\% of the sample lies at $4<z<6$). We
  remark that, as we anticipated at the beginning of this Section, the
  red colors of our BBGs strongly point to a $z>3$ redshift. As shown
  in this histogram, and considering the uncertainties in the
  optical/NIR photometry, a small fraction ($<5$\%) of the total
  sample would present photometric redshifts below $z=3$.

\subsection{Rest-frame UVJ colors, stellar masses and dust attenuations}

The right panel of Figure\,\ref{fig:col_mag_ios} shows a $UVJ$ diagram
for our sample of BBGs, color-coded by SFR and sized by mass. We
compare our BBGs with the CANDELS color-selected and
mass-limited samples. The black lines indicate the quiescent
region (upper-left), and the low and high extinction star-forming
regions, bSFG (lower-left) for blue star-forming galaxies, and dSFG
for dusty systems (upper-right) (see e.g., \cite{2011ApJ...739...24B}
or \cite{2012ApJ...745..179W} for a detailed discussion of the age and
extinction patterns in the $UVJ$ diagram).

Figure\,\ref{fig:col_mag_ios} shows that the BBGs overlap with the
most extinguished, dustier galaxies in the CANDELS color-selected
sample. Note also that the color-selected sample identifies redder,
more massive dusty or evolved galaxies than a pure
mass-limited sample with no color constraints (located in the
shaded region).  This is highlighted by the blue regions which show
the 0.3, 1 and 2$\sigma$ color distributions of massive
($M>10^{10}\rm{M}_{\odot}$) galaxies at $z>3$ drawn from the CANDELS
catalogs (see Appendix~\ref{A:CANDELSample} for more details). The
predominantly dust-obscured nature of the BBGs is further confirmed by
the strong mid-to-far IR detections of the galaxies in the upper right
corner of the UVJ diagram ($\sim$30\% of the BBGs in the dSFG region),
which imply also large SFRs (see next section).

Nonetheless, about $\sim20\%$ of the BBG sample exhibits blue $UVJ$
colors at the opposite extreme of the dSFGs. These galaxies are also
among the bluest ($H-[3.6]\lesssim2.5$~mag) and brightest
($H\sim26$~mag) of the sample, located at the bottom-right corner of
the color-magnitude diagram on the left panel of Figure\,\ref{fig:col_mag_ios} 
(see discussion in Section \ref{ssec:Stacks}). This suggests that they are included in
the otherwise red sample of $H$-band dropouts due to incompleteness in
the CANDELS catalog.

\begin{figure*}[t]
  \begin{center}
      \includegraphics[width=1.\linewidth]{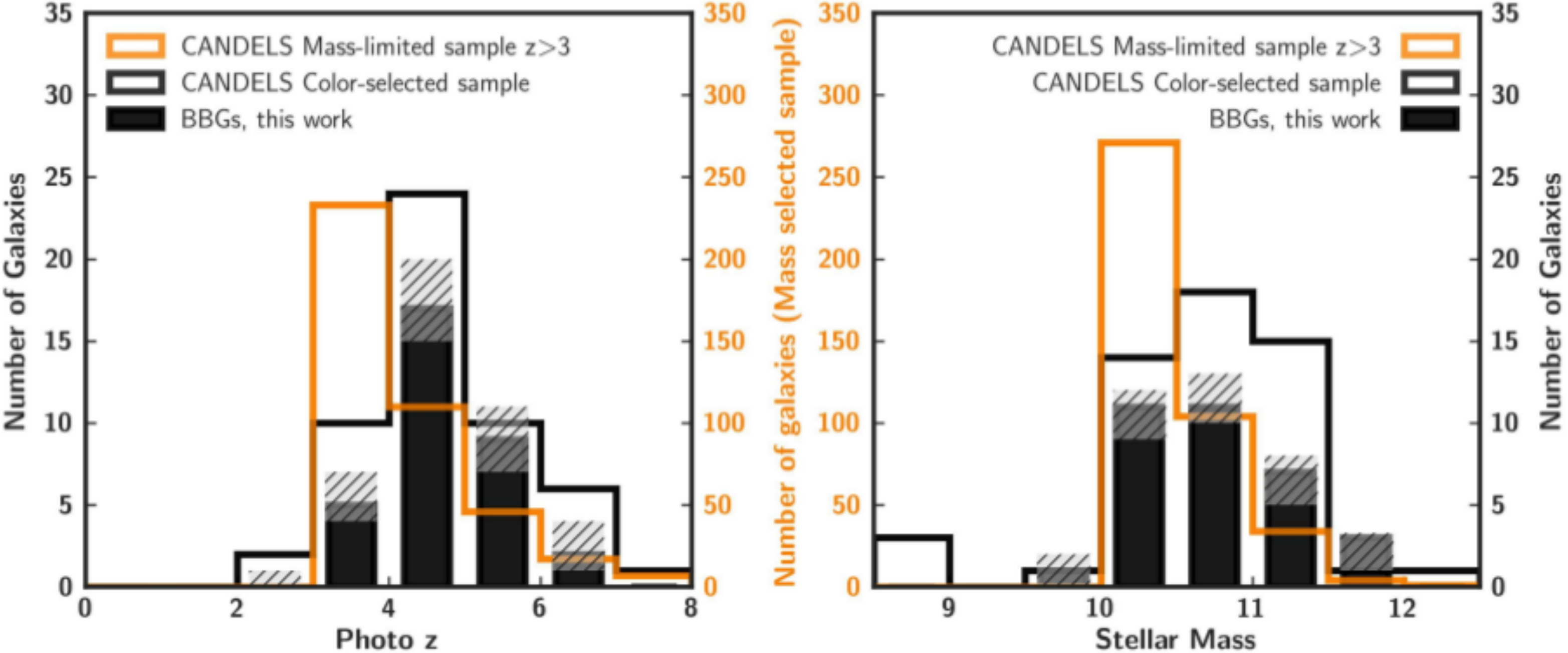}  
  \end{center}
  \caption{ Photometric redshift (\textit{left panel}) and stellar
    mass (\textit{right panel}) distributions of our sample of BBGs
    compared to those of the color-selected and mass-limited
    CANDELS samples explained in the text (see also
    Appendix~\ref{A:CANDELSample}). The photo-z distribution of the
    mass-limited and color-selected samples are shown by
    black and orange line histograms respectively. BBGs are
    shown with black bars and their uncertainties are
      highlighted with hatching black lines. Given the substantially
    higher number of galaxies in the mass-limited sample, a
    different axis has been used for it, as indicated by the orange
    labels.}
  \label{fig:z_mass_hist_sample}
\end{figure*}

Interestingly, Figure\,\ref{fig:col_mag_ios} also indicates
  that there are few quiescent galaxies both in the mass-limited and
  BBGs samples and none in the color-selected and BBG sample. We
remark, however, that several galaxies count with lower limits in
$U-V$, so their colors are consistent with the quiescence locus. The
lack of quiescent galaxies at $z>3$ (and we remind the reader that we
are more efficient in detecting galaxies at $4<z<6$) is consistent
with recent works that have identified several massive quiescent
galaxies at $z=3-4$ (e.g.,
\citealt{2014ApJ...783L..14S,2015ApJ...808L..29S}).  These quiescent
galaxies have lower redshifts and therefore brighter $H$-band
magnitudes and bluer $H-[3.6]$ colors than our BBGs. For example, the
6 quiescent galaxies identified by \cite{2014ApJ...783L..14S} in
GOODS-S have a median color of $H-[3.6]=2.2$~mag, and a redshift of
$z=3.7$, whereas 80\% our BBGs are at $z>4$. Altogether, the
  color distribution of the BBGs suggests that it is increasingly more
  difficult to find bona fide fully quenched galaxies beyond
  $z\sim4$. Possibly this is because, at such high redshifts, even
the most evolved galaxies did not have time to reach a mass-weighted
age older than $\rm{t}\sim1$~Gyr, which is the approximate threshold
for a single stellar population to make it into the quiescent region
of the $UVJ$ diagram (see Figure~\ref{fig:UVJ_CANDELS} in the
Appendix).

\begin{figure*}[t]
  \begin{center}
     \includegraphics[width=1.\linewidth]{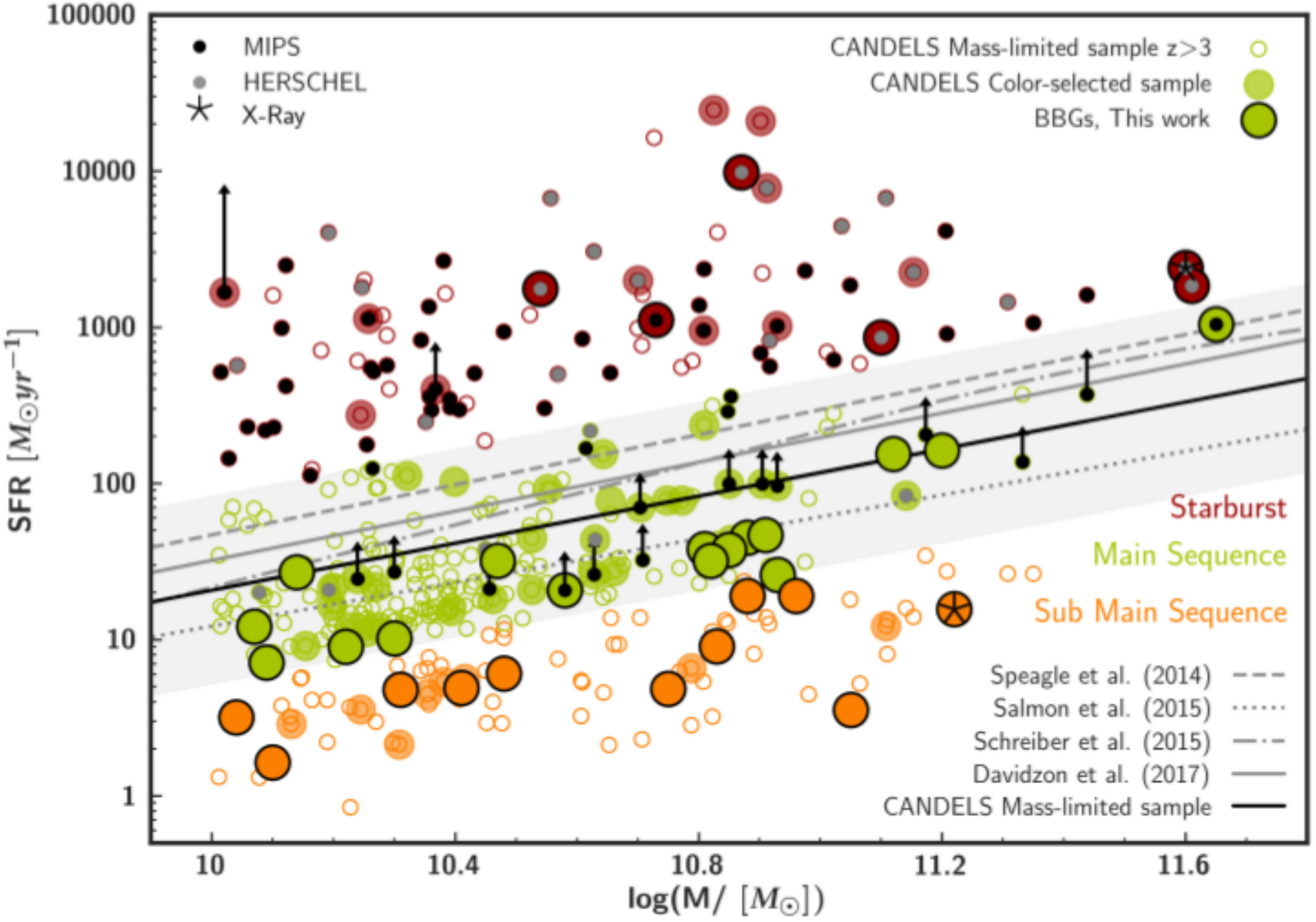}   
  \end{center}
  \caption{SFR vs stellar mass plane for the CANDELS comparison
    samples (color-selected ---filled symbols--- and
    mass-limited ---open symbols---) and the BBGs (filled
    symbols enclosed by a black circle) reported in this work, color
    coded according to their position with respect to the main
    sequence: starburst galaxies are shown in deep-red, MS galaxies in
    green, and sub-MS in orange.  The MS at $z=4$ according to
    \cite{2014ApJS..214...15S}, \cite{salmon2015relation},
    \cite{2015A&A...575A..74S} and \cite{Davidzon2017} are shown with
    different grey lines (see the legend in the panel).  The MS
    inferred for the CANDELS mass-limited sample is shown
    with a black solid line. The grey-shaded region delimits the
    2$\sigma$ area around the MS.  MIR/FIR emitters are marked with an
    enclosed black/grey circle.  The galaxies with MIPS detections but
    no IR-derived SFRs (because they lie at $z>5$, see text for
    details) are shown as lower limits.  X-rays emitters are also
    highlighted with a star symbol.}
    \label{fig:SFR_M}
\end{figure*}

Figure\,\ref{fig:z_mass_hist_sample} shows the photometric redshift
and stellar mass distributions of the BBGs compared to the CANDELS
color- and mass-limited samples. As shown in the previous
section, an $H$-band dropout selection in the GOODS/CANDELS field
(implying an ``extremely'' red $H-[3.6]$ color) identifies galaxies at
$z\gtrsim4$ (Figure\,\ref{fig:zhist_lit}), whereas the CANDELS color
and, especially, mass-limited samples have a more pronounced
tail at $z\sim3$. The stellar mass histograms confirm the intuition we
had when examining the $UVJ$ diagram in Figure\,\ref{fig:col_mag_ios}:
both the color-selected and BBG samples are biased towards the
identification of more massive galaxies than those from the bluer,
mass-limited sample.  This correlation between color, mass
and dust attenuation is fully consistent with previous results at
lower redshifts (e.g., \citealt{2011ApJ...739...24B},
\citealt{2016ApJ...830...51S}, \citealt{2017A&A...601A..63W},
\citealt{2017arXiv171005489F}). Typically, our BBGs have stellar
masses around or above $\log(\rm{M}/\rm{M}_\sun)=11$.  In absolute
terms, the BBG sample recovers a number of massive obscured galaxies
similar to that of the CANDELS color-selected sample and one tenth of
the fraction of mass-limited galaxies (but the latter is more
biased towards lower redshifts). This implies that the completeness
level of the deepest $H$-band selected catalogs (e.g., CANDELS,
3D-HST) is not very high and they are missing a significant fraction
(around 40\%) of massive red galaxies at $z>4$ (see discussion in
\S\ref{quant}).

\subsection{BBGs and the star formation main sequence}
\label{ssec:SFH}

Figure~\ref{fig:SFR_M} shows the SFR vs. stellar mass diagram for the
BBGs and the CANDELS color-selected and mass-limited samples
(main statistical properties given in Table~\ref{table:sfr}). The grey
lines shows the star-forming ``main sequence'' (MS; see, e.g.,
\citealt{noeske2007star}, \citealt{2011A&A...533A.119E}) at $z=4$
(from \citealt{2014ApJS..214...15S}, \citealt{salmon2015relation},
\citealt{2015A&A...575A..74S}, and \citealt{Davidzon2017}). The MS
illustrates the known correlation between SFR and mass for typical
star-forming galaxies that has been shown to exist even up to $z=7-8$
\citep{stark2009evolutionary, 2013ApJ...763..129S,salmon2015relation}.
The MS inferred for the CANDELS mass-limited sample ($\langle
\rm{M}\rangle\sim10^{10.3}\,\rm{M}_{\odot}$ and $\langle
\rm{z}\rangle\sim3.8$; black line) is consistent with the MSs from the
literature, in spite of the typically high uncertainties in the SFR
and stellar mass estimations for these redshifts. The grey region
shows 2$\sigma$ around the MS ($\sigma\sim0.3$~dex). The color code
indicates galaxies in the MS (green) and those above and below the MS
2$\sigma$ band, namely, starburst (deep-red) and sub-MS (orange)
galaxies.
 
Despite the overall high scatter around the MS at
$\rm{M}\lesssim10^{11.2}\,\rm{M}_{\odot}$, the few galaxies at the
high mass end, $\rm{M}\gtrsim10^{11.4}\,\rm{M}_{\odot}$, are located
above the MS.  At lower masses
$\rm{M}\lesssim10^{10.5}\,\rm{M}_{\odot}$, however, all the BBGs are
either in the MS or below. This result is fully consistent with the
inferences from the $UVJ$ diagram in Figs.\,\ref{fig:col_mag_ios} and
\ref{fig:UVJ_SFRsamples}, i.e., their red intrinsic colors are mainly
due to the presence of strong bursts of obscured star formation, with
a few galaxies being consistent with harbouring more evolved stellar
populations.

\begin{figure*}[t]
  \begin{center}
         \includegraphics[width=1.\linewidth]{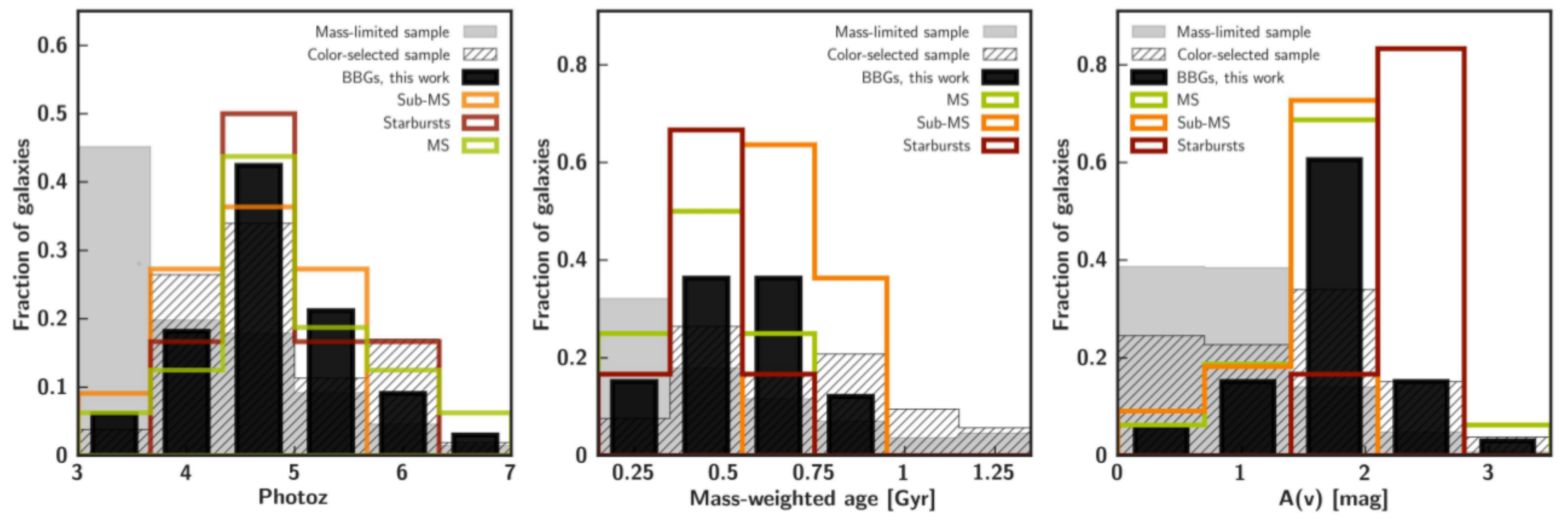}   
  \end{center}
  \caption{Distributions of the redshift (\textit{left panel}),
    mass-weighted age (\textit{middle panel}), and attenuation
    (\textit{right panel}) for the different star formation
    subsamples. The black filled histograms represent the fraction of
    BBGs while the distribution of the mass-limited and
    color-selected samples are drawn with a filled and hatched grey
    histogram respectively.  Starbursts, MS and sub-MS BBGs are shown
    in deep-red, green and orange respectively. Each histogram is
    normalized to its sum.}
  \label{fig:hito_MSamples}
\end{figure*}

This result is further confirmed by Figure~\ref{fig:hito_MSamples},
which shows the mass-weighted age and attenuation histograms for the
three samples. The distributions for the BBGs are clearly skewed
towards higher attenuations, up to A(V)~$\sim3$~mag for some
starbursts, in contrast with the mass-limited sample,
typically characterized by attenuations around A$_{V}\sim1$~mag or
below. BBGs show older mass-weighted ages (typically older than
300~Myr) than the bluer, mass-limited sample (which peaks
below 300~Myr). In line with the general location of BBGs in these
histograms, the color-selected sample is built up by marginally older
and less extincted galaxies.  The average mass-weighted age of our
BBGs is 0.5~Gyr, slightly older than the typical age for the
mass-limited sample (0.3~Gyr) and similar or younger to the
average for the color-selected sample (0.7~Gyr). These young
ages point out that our color and magnitude selection was
preferentially identifying the Balmer and not the 4000~\AA\, break
(typical of more evolved populations, $>1$~Gyr). This is the
justification for the choice of name (BBGs) for the new population
of high-z massive galaxies presented in this paper. The oldest
ages can be found among the sub-MS BBGs, which can be as old as
0.9~Gyr.  The redshift distribution is roughly homogeneous for
the three subsamples of BBGs divided by star formation activity,
peaking at $z\sim4.5$, although starbursts dominate at higher
redshifts and subMS galaxies at lower redshits with a flatter
distribution between $z\sim3.5$ and $z\sim4.5$.  This could be
identified as the assembly of the first quiescent massive galaxies,
which would be actively forming stars in the MS, or even very
actively forming stars in the starburst region at $z\gtrsim4.5$, and
may start to evolve more passively or even to quench by $z\sim4$
(after more than $\sim1$~Gyr).

Concerning attenuations, typical attenuations are around
2~mag, and as large as 3.4~mag for starbursting BBGs. Both the
mass-limited and color-selected galaxies present
significantly smaller attenuations, around 1~mag.

The most relevant conclusion out of this comparison is that the
magnitude-limited F160W samples based on CANDELS data miss $\sim15\%$
of the massive ($M>10^{10.5}\,\rm{M}_{\odot}$) galaxies at
$z\gtrsim4$. The fraction increases as we move to higher masses (see
\S\ref{quant}). In relative terms, the impact of adding missed
BBGs to a mass-limited sample is larger on the starburst region. We
also remark that $\sim20\%$ of the total number of BBGs, $\sim30\%$ if
we only consider $\rm{M}\geq10^{10.5}\,\rm{M}_{\odot}$, are
starbursts. If we combine the mass-limited sample and our BBGs, these
fractions are 25\% and 35\%, respectively. These figures are much
larger than those found at $z\sim2-3$, $2-4\%$ for
$\rm{M}\gtrsim10^{10}\,\rm{M}_{\odot}$
\citep{2011ApJ...739L..40R,2015A&A...575A..74S}. Our fraction is
slightly larger than the $15\%$ reported by \cite{2017ApJ...849...45C}
for a sample of galaxies at $z\sim4-5$ and $\langle \rm{M}\rangle
\sim10^{9.5}\,\rm{M}_{\odot}$. 

Interestingly, very few of our BBGs lie far below the MS to be
considered as completely quiescent. If we consider the average
redshift of our sample, $\langle z\rangle=4.8$, corresponding to a Universe age of
1.2~Gyr, there is not much time for a massive galaxy to completely
quench unless its star formation is very short and starts very early.
This result is consistent with what we found with the $UVJ$ diagram.
For convenience, we show another version of it in
Figure\,\ref{fig:UVJ_SFRsamples}, but this time the BBGs and
color-selected galaxies are color-coded by star formation subsamples. 
We remind the reader again that the $U-V$ color for our sample has 
a relatively high uncertainty (0.2~mag), given that our sources are
extremely faint in the bands probing the $U$-band.  Therefore, some of
our BBGs are consistent with being quiescent (see next subsection), or
at least have moved to a post-starburst phase. 

In any case, 4 sub-MS BBGs (GDN\_BBG08, GDN\_BBG09, GDN\_BBG12 and
GDS\_BBG14), present mass-weighted ages around $\sim0.9$~Gyr (with
times from the start of their SFH around $\rm{t}\sim1.5$~Gyr; see
Table~\ref{tab:stellarprop}), which translates to $z\gtrsim18$ for the
onset of their star formation.

\begin{figure}[h]
  \begin{center}
     \includegraphics[width=1.\linewidth]{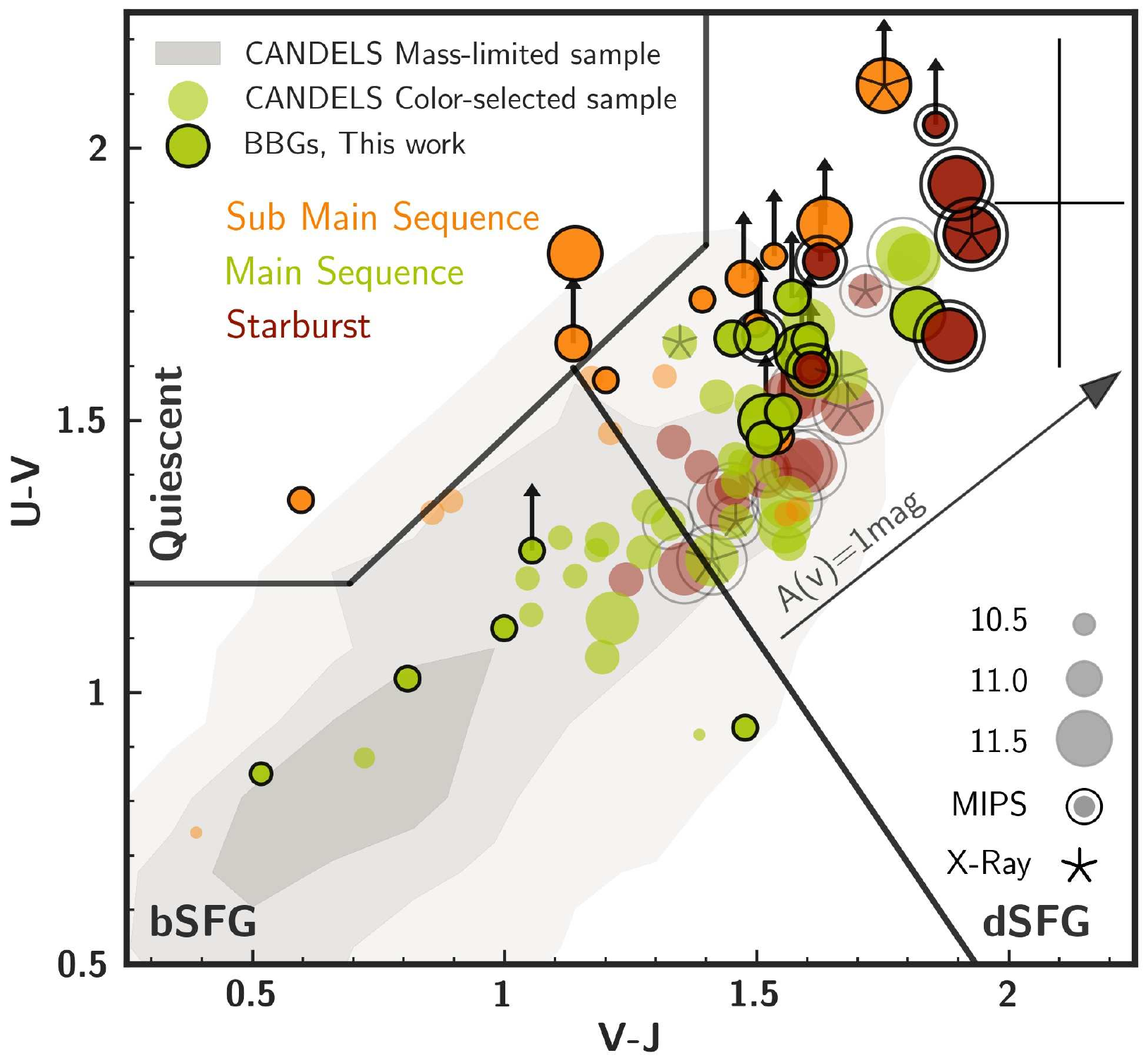}   
     \caption{Rest-frame $U-V$ vs. $V-J$ colors for BBGs and the
       CANDELS color-selected comparison sample color-coded by star
       formation subsamples (with respect to the MS, see the color
       code in Figure\,\ref{fig:SFR_M}) and sized by stellar mass.
       $H$-band non detections are shown as $U-V$ lower limits. MIPS
       detections are marked with an open circle and X-ray detected
       galaxies are highlighted with a star inside the symbol. The
       0.3, 1 and 2~$\sigma$ distributions of the mass-limited sample
       are shown by the light grey areas.  The \citet{Whitaker2011}
       upper boundary (black wedge) separates quiescent galaxies (top
       left) from SFGs (bottom). The black diagonal line denotes an
       additional criterion to separate blue (bSFG) from dusty (dSFG)
       star-forming galaxies.  The 1~mag attenuation vector computed
       assuming a \cite{2000ApJ...533..682C} reddening law is also
       shown. Error bars are not plotted for clarity but the average
       values are shown in the top-right corner.}
       \label{fig:UVJ_SFRsamples}
 \end{center}   
\end{figure}

\subsection{Stacked SEDs of the BBGs}
\label{ssec:Stacks}
In this section we further compare the colors of BBGs and galaxies in
the color-selected and mass-limited samples grouped by their
location relative to the MS. Given the intrinsically faint fluxes of
BBGs at nearly all wavelengths, here we analyze averaged rest-frame
SEDs of multiple BBGs to obtain a better and finer sampling of their
average SEDs, and we compare them to the other 2 samples of (brighter)
galaxies. The SEDs are normalized at 0.7~$\mu$m rest-frame, which
roughly corresponds to the IRAC~3.6 and~4.5~$\mu$m bands, where BBGs
are brightest. We only include photometric points with reliable
detections (SNR~$>3$) at wavelengths shorter than IRAC. We also
removed from the stacks a small number of galaxies with unconstrained
SFRs or redshifts $z\gtrsim6$, which cannot be properly normalized at
0.7~$\mu$m rest-frame if they are not detected beyond 4.5~$\mu$m.
Sources from the mass and color-selected samples with uncertain IRAC
photometry ($<5\sigma$ in [3.6], [4.5]) were also removed.
Furthermore, we visually inspected their HST images to remove
potentially blended objects or sources dominated by a central
point-like emission (AGN candidates).

\begin{figure*}[t]
  \begin{center}
        \includegraphics[width=1.\textwidth]{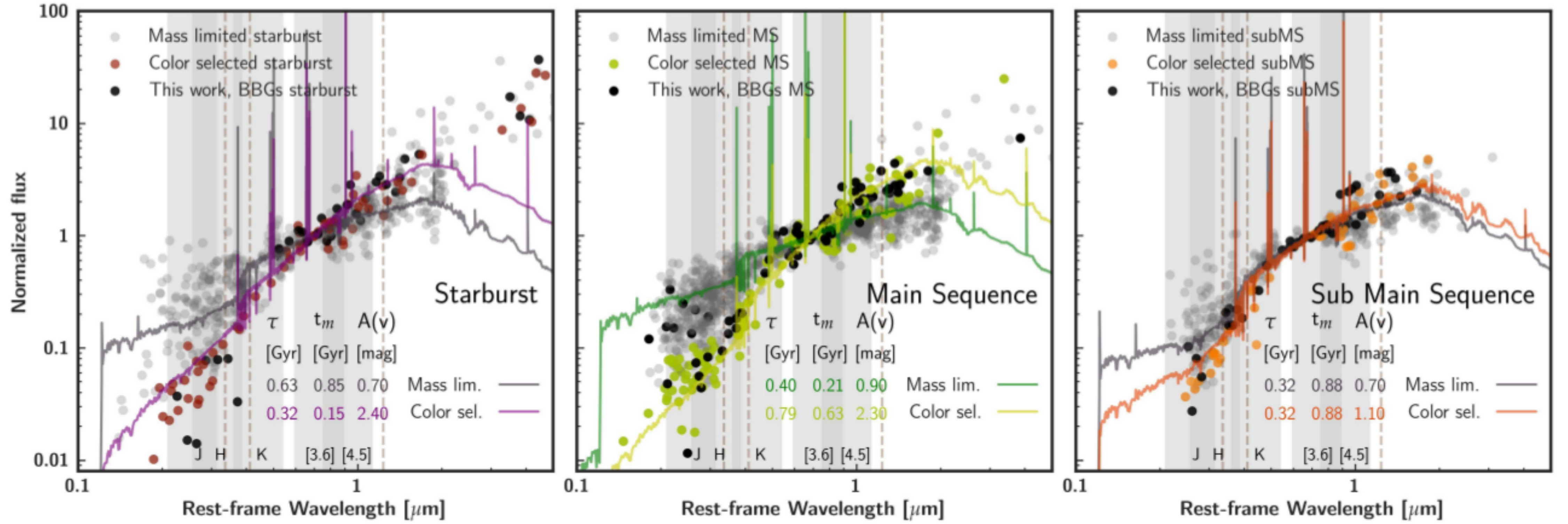}
  \end{center}
  \caption{Rest-frame average SEDs for the CANDELS comparison samples
    and our BBG sources grouped by the different galaxy types
    according to their position with respect to the MS in the SFR vs M
    plane (see Figure~\ref{fig:SFR_M}). SEDs are normalized at 0.7
    $\mu$m rest-frame (observed photometry between the IRAC $[3.6]$
    and $[4.5]$ selection bands). Sources with uncertain ($SNR<5$)
    photometry were excluded from the stacks. \textit{Left panel:}
    rest-frame SEDs of starburst galaxies from the mass-limited
    sample (grey), the color-selected sample (color symbols), and the
    BBGs reported in this work (black). We also show two best-fit 
    stellar population models fitted to the median
    photometry for the mass-limited and color-selected 
    subsamples. The mass-weighted age and the attenuation of these
    fitted models are given in the panel. Grey-shaded regions
    delimit the rest-frame wavelength ranges probed by the observed
    $J$, $H$, $K$, 3.6~$\,\mu$m and 4.5~$\mu$m bands for $z=4$. Brown
    dashed lines mark the location of rest-frame $UVJ$ bands.
    \textit{Middle panel:} the same as the left panel but for the Main
    Sequence galaxies. \textit{Right panel:} the same as in the other
    two panels, but for sub-MS galaxies.}
  \label{fig:MultiSED}
\end{figure*}

\begin{table}
\setlength{\tabcolsep}{0.3em}
\begin{threeparttable} 
	\caption{\label{table:sfr} Statistical properties of starburst, MS and sub-MS BBG, color-selected and mass-limited samples. 
	Median values, 1$^{st}$ and  3$^{rd}$ quartiles of their redshift, masss-weighted age, extinction, SFR, sSFR and Mass are 
	shown. The percentage of sources belonging to each star formation mode are also shown. Sources with SFR lower limits have not 
	been taken into account.}
	\begin{center} {\scriptsize
	\begin{tabular}{llccccccc}
	\hline\\

	\multicolumn{2}{l}{Sample}& {}&z& {t$_m$} & {A(V)} & {SFR} & {sSFR} & {Mass} \\[0.5ex]\cmidrule{4-9}
	\multicolumn{2}{l}{}& {}& &{[Gyr]} & {[mag]} & {[M$_{\odot}$yr$^{-1}$]} & {[Gyr$^{-1}$]} & {[M$_{\odot}$]} \\[-2.ex]
	\\\hline
	\hline\\
	{Mass-limited}&  All&{}&  $3.8_{3.3}^{4.8}$  &$ 0.3_{0.2}^{0.6}$ & $0.7_{0.4}^{1.3}$ & $26_{13}^{124}$   & $1.2_{0.6}^{6}$ & $10.4_{10.2}^{10.6}$ \\ [1.ex]
	& Starburst    & 25\%  &  $3.8_{3.2}^{5.2}$  & $0.3_{0.1}^{0.5}$ & $1.0_{0.2}^{1.6}$ &$ 737_{356}^{1622}$  &$ 27_{12}^{63}$ &$ 10.5_{10.3}^{11.0}$  \\ [1.ex] 
	& MS           & 56\%  & $ 4.1_{3.3}^{4.8}$  & $0.2_{0.2}^{0.5}$ &$ 0.7_{0.4}^{1.1}$ &$ 23_{15}^{41}$   &$ 1.0_{0.7}^{1.7}$ &   $10.3_{10.1}^{10.5}$\\ [1.ex] 
	& SubMS        & 19\%  &  $3.5_{3.3}^{4.0}$  & $0.7_{0.3}^{1.0}$ & $0.5_{0.2}^{1.0}$ &$ 5_{4}^{11}$    &$ 0.2_{0.1}^{0.2}$ & $ 10.3_{10.1}^{10.5}$ \\ [1.ex]
	\hline\\
	{Color-selected}& All&{}& $ 4.7_{4.1}^{5.3}$   & $0.7_{0.4}^{0.8}$& $1.5_{0.7}^{2.0}$  & $ 67_{12}^{236}$    &  $1.3_{0.6}^{10}$   &  $10.8_{10.5}^{11.1} $ \\ [1.ex]
	& Starburst     & 22\% &  $ 5.0_{4.6}^{5.6}$   & $0.7_{0.5}^{0.8}$& $1.7_{1.6}^{1.9}$  & $ 1077_{478}^{2185}$ &  $46_{16}^{143} $   & $11.0_{10.8}^{11.1} $   \\ [1.ex] 
	& MS            & 58\% &  $ 4.8_{4.4}^{5.3}$   &$ 0.5_{0.3}^{0.8}$& $1.3_{0.7}^{2.0} $ &  $44_{19}^{90} $   &  $1.2_{0.6}^{1.4} $  & $ 10.8_{10.5}^{11.1} $  \\ [1.ex] 
	& SubMS         & 20\% &   $3.9_{3.7}^{4.0}$   &$ 0.8_{0.4}^{1.0}$& $0.8_{0.3}^{2.0} $ &  $5_{3}^{6} $   &  $0.2_{0.1}^{0.2} $  & $ 10.3_{10.2}^{10.5} $  \\ [1.ex]
	\hline\\
	{BBGs}& All & {}     & $ 4.8_{4.3}^{5.1}$  & $0.5_{0.5}^{0.6}$ & $2.0_{1.5}^{2.0}$ & $26_{9}^{154}$    & $0.5_{0.2}^{1.1}$ & $10.8_{10.4}^{11.0}$ \\ [1.ex]
	& Starburst   & 19\% & $ 4.7_{4.4}^{5.2}$  & $0.5_{0.5}^{0.5}$ & $2.5_{2.4}^{2.6}$ & $1803_{1275}^{2258}$  & $14_{6}^{43}$  & $11.0_{10.9}^{11.6}$  \\ [1.ex] 
	& MS          & 48\% & $ 4.8_{4.5}^{5.3}$  & $0.5_{0.4}^{0.6}$ & $1.9_{1.6}^{2.0}$ &  $32_{19}^{45}$   & $0.6_{0.5}^{1.0}$ &  $10.8_{10.3}^{11.0}$  \\ [1.ex] 
	& SubMS       & 33\% & $ 4.7_{4.0}^{5.0}$  &$ 0.7_{0.7}^{0.8}$ & $1.7_{1.4}^{2.0}$ &  $5_{4}^{12}$    & $0.2_{0.1}^{0.2}$ & $10.8_{10.4}^{11.0}$  \\ [1.ex] 	
\hline\\
	\end{tabular}
  }
  \end{center}
\end{threeparttable}   
\end{table}

Figure~\ref{fig:MultiSED} shows the individual and averaged SEDs
(constructed by fitting stellar population models to the stacked
photometry) in the sub-MS, MS and starburst regions. Quite
reassuringly, we find that BBGs exhibit very similar average SEDs to
the color-selected sample in all 3 regions, confirming again that the
former are the faint-end extension of the latter. The comparison in
the starburst region is particularly revealing, as both samples show
clear detections at wavelengths longer than
$\lambda_{\rm{r-f}}>2$~$\mu$m. This is what would be expected for dust
emission in heavily enshrouded galaxies.  Indeed, dust emission is
starting to dominate the integrated SED at
$\lambda_{\rm{r-f}}\sim2$~$\mu$m and by $\lambda_{\rm{r-f}}>3$~$\mu$m
dust emits more than 50\% of the integrated light for $\sim$35\% of
sources.  In addition, the best-fit stellar population model to the
stack for these starbursts indicates a large attenuation,
A(V)$\sim2$\,mag. The comparison in the MS region shows that some BBGs
have slightly bluer colors than the color-selected sample at short
wavelengths $\lambda_{\rm{r-f}}<400$~nm. These galaxies are the
smaller mass, bluer ($H-[3.6]\lesssim2$) BBGs discussed in the
previous sections, which are indeed more similar to the overall bluer
stack of the mass-limited sample. This is confirmed by the
$UVJ$ diagram (Figure\,\ref{fig:UVJ_SFRsamples}) where the few green
circles characterized by bluer $UVJ$ colors lie in the region with the
highest density of galaxies from the mass-limited sample.
Most of the BBG sample in this MS region, however, is characterized by
a red SED reproduced by a stellar population with very similar A(V)
compared to the starburst sample, but with considerably older age. We
note that these two quantities are highly degenerated in the models
fitting the stacked photometry, but should the attenuation be smaller,
the age should be older, so there must be a real difference between
the starburst and MS sub-samples. This is also confirmed by the lower
fraction of MIPS emitters in the MS (100\% of the starburst BBG are
detected in the mid-/far-IR and 9\% for the MS BBGs).  Finally, there
are very few BBGs or color-selected galaxies in the sub-MS region to
infer any statistically significant result.  However, as expected,
both populations appear to have redder SEDs than those in the
mass-limited sample, which favors older (or more extincted)
galaxies.

\begin{figure}[t]
  \begin{center}
     \includegraphics[width=1.\linewidth]{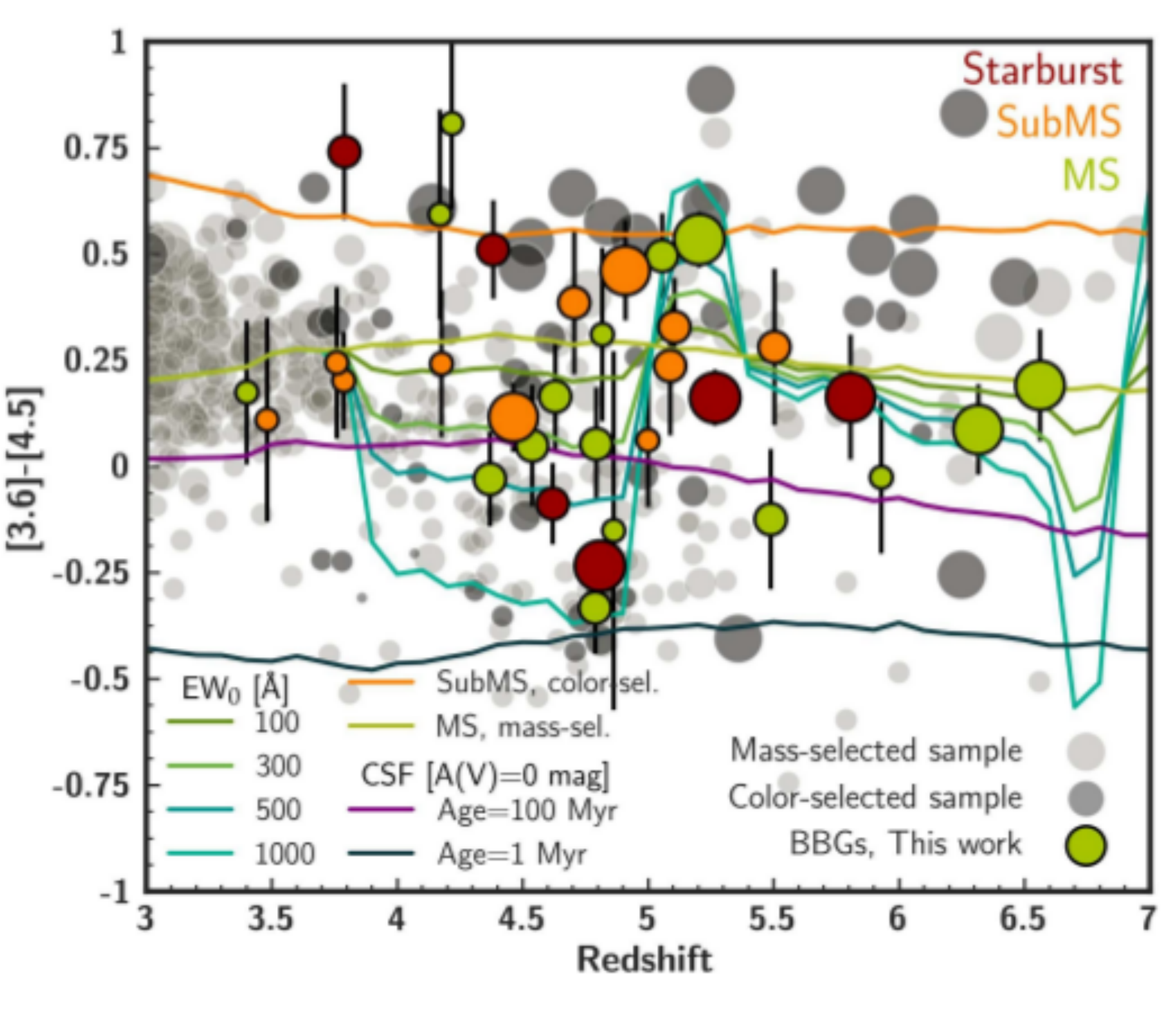}   
  \end{center}
  \caption{IRAC $[3.6]-[4.5]$ colors vs. redshift for our BBGs
    compared to the mass-limited and color-selected samples, shown by
    light and dark grey circles, respectively. BBGs are color-coded by
    star formation subsamples and all the galaxies are scaled by
    stellar mass. The green and orange lines show the colors of the MS
    and subMS best-fitting templates (see Figure~\ref{fig:MultiSED}
    and main text for details). The effect of including emission lines
    (H$\beta$+$[OIII]$, H$\alpha$$+$[NII]) of different EWs in the MS
    best-fitting template is shown by different blue and green lines
    (see legend). For example, H$\alpha$$+$[NII] at $z\sim4.5$
    produces bluer $[3.6]-[4.5]$ colors (when the lines are in the
    3.6~$\mu$m IRAC channel), and redder for $z\sim5.2$ (when the same
    lines enter the 4.5~$\mu$m band).  Similar effects could be
    expected for other templates. The colors for no extinction,
    constant SFH starburst of 1 and 100~Myr are shown in dark blue and
    purple, respectively.  }
    \label{fig:Elines}
\end{figure}

In addition, the bluest BBGs could have been selected due to the
presence of a strong (i.e., high EW) emission line (such as H$\alpha$
at $z\sim4.5$) in the $[3.6]$ band (see e.g.,
\citealt{2012ApJ...761...85K}, \citealt{2016MNRAS.460.3587M} and
\citealt{2014ApJ...784...58S}). Flux contamination by a strong
emission line in the IRAC bands would lead to a brighter $[3.6]$
magnitude and a redder $H-[3.6]$ color for a galaxy with relatively
low stellar mass, which would push it into the color selection region
for BBGs. We show this effect in Figure~\ref{fig:Elines},
where we depict IRAC colors $[3.6]-[4.5]$ vs. redshift for our BBGs
compared to the mass-limited and color-selected sample. Typically,
the BBGs present colors around $[3.6]-[4.5]=0.2$~mag (which are
consistent with the average stellar population models shown in
Figure~\ref{fig:MultiSED} for main sequence systems). Remarkably,
half a dozen galaxies are characterized by colors as blue as
$[3.6]-[4.5]=-0.4$~mag, and a similar number presents very red colors,
$[3.6]-[4.5]=+0.5$~mag. The blue colors can be attributed to the
presence of prominent emission lines, more specifically,
$H\alpha+[NII]$ entering the 3.6~$\mu$m filter at $z=4-5$ and
presenting a equivalent width $EW_0\sim300-1000$~\AA\, (see, e.g.,
\citealp{2014ApJ...784...58S,smit2015}).  The blue IRAC colors could
also be reproduced with very young stellar populations. As an
example, in the plot we show the colors expected for constant SFH
starburst with ages 1 and 100~Myr and no extinction. We note,
however, that those models would not be compatible with the very red
colors observed for $H-[3.6]$. Concerning the galaxies with very red
IRAC colors, some of them could be explained with $H\alpha+[NII]$
entering the 4.5~$\mu$m filter at $z\sim5.2$, but also with a very
red model dominated by stellar continuum (such as the one shown in
Figure~\ref{fig:MultiSED} corresponding to a sub-MS galaxy). Some
examples of this type of galaxy with strong emission lines are
GDN\_BBG07, and also GDN\_BBG15 and GDN\_BBG17, which were cataloged
as LBGs by \citet{2015ApJ...803...34B}. The average stellar mass
($\log\rm{M}/\rm{M}_\sun$) and their quartiles for bSFG (6/33) are
$10.2^{10.1}_{10.1}$, less massive compared to dSFG (25/33),
$10.8^{11.0}_{10.6}$. Remarkably, these galaxies present similar
[3.6] and [4.5] magnitudes compared to dSFGs, but they are not
detected at wavelengths longer than 5~$\mu$m. This would be
consistent with their SED being flatter, corresponding to blue
sources with emission lines.

In Figure \,\ref{fig:UVJ_SFRsamples} we showed that the dSFGs
region (containing 73\% of the BBGs) is populated by mainly
starburst and MS galaxies (which are also located within the bSFG
region). Only three subMS galaxes lie in the quiescent region and
the rest are located in the dSFG region while very close to the
boundary.  The few galaxies consistent with being evolved or in a
post-starburst state at $z\gtrsim4$ are mainly located in the sub-MS
or MS regions.  As discussed in the previous sections, it is
difficult to identify these candidates reliably because most BBGs
and color-selected galaxies are located near the quiescence
boundary, and uncertainties in the observed photometry and the
photometric redshifts can easily scatter them in or out of the dead
galaxy region. In Figure~\ref{fig:MultiSED} we marked the pivot
points of the $UVJ$ colors and the rest-frame wavelength ranges probed
by the observed $JHK$, 3.6~$\,\mu$m, and 4.5~$\mu$m bands to show the
main challenges in characterizing the rest-frame colors at
$z\gtrsim3$ with our current datasets, especially for red
sources. Indeed, the rest-frame $U$-band is probed by the $H-$band
where our BBGs are, by definition, undetected or very faint.
Consequently, those galaxies undetected in $H-$band were considered to
have $U-V$ lower limits. In addition, the rest-frame $J$-band lies in
the IRAC 5.8$\,\mu$m band, which is significantly shallower
($\sim2$~mag) than 3.6$\,\mu$m or 4.5$\,\mu$m.  

Summarizing our results in this section, the vast majority of
BBGs correspond to dSFGs, a smaller fraction to bSFGs, and only 3
BBGs are identified as quiescent galaxies. A few other BBGs are
still marginally consistent with being quiescent based on the $UVJ$
diagram due to the relatively large photometric uncertainties.
Particularly, other quiescent galaxies might be those having older
mass-weighted ages (Figure~\ref{fig:hito_MSamples}) and no far-IR
detections.

%
%\begin{table}
%  \caption{\label{table:sfr}Median sSFR for starbursts, MS and sub-MS  BBG, color-selected and mass-limited samples. 
%    The percentage of sources belonging to each star formation mode are also shown. Sources with sSFR lower limits have not
%    been taken into account.} 
%\begin{center} {\small
%\begin{tabular}{lccccc}
%\hline\\
%&{} &{sSFR [Gyr$^{-1}$]} & {} \\[1ex] \cmidrule{2-4}
%&{Starburst} &{MS} & {sub-MS} \\[-2.5ex]
%\raisebox{1.5ex}{Sample} 
%\\\hline
%\hline\\
%& 46 (25\%) & 1 (56\%)&0.2 (19\%)\\[-2ex]
%\raisebox{3ex}{mass-limited}
%& 27 (22\%) & 1 (58\%)&0.2 (20\%)\\[-2ex]
%\raisebox{3ex}{Color-selected}
%& 14 (19\%) &0.6 (48\%) &0.2 (33\%)\\[-2ex]
%\raisebox{3ex}{BBGs, this work}
%\\\hline
%\end{tabular} }
%\end{center}
%\end{table}

\subsection{AGN in the BBG sample}
\label{ssec:AGNs}
\begin{figure}[t]
  \begin{center}
     \includegraphics[width=1.\linewidth]{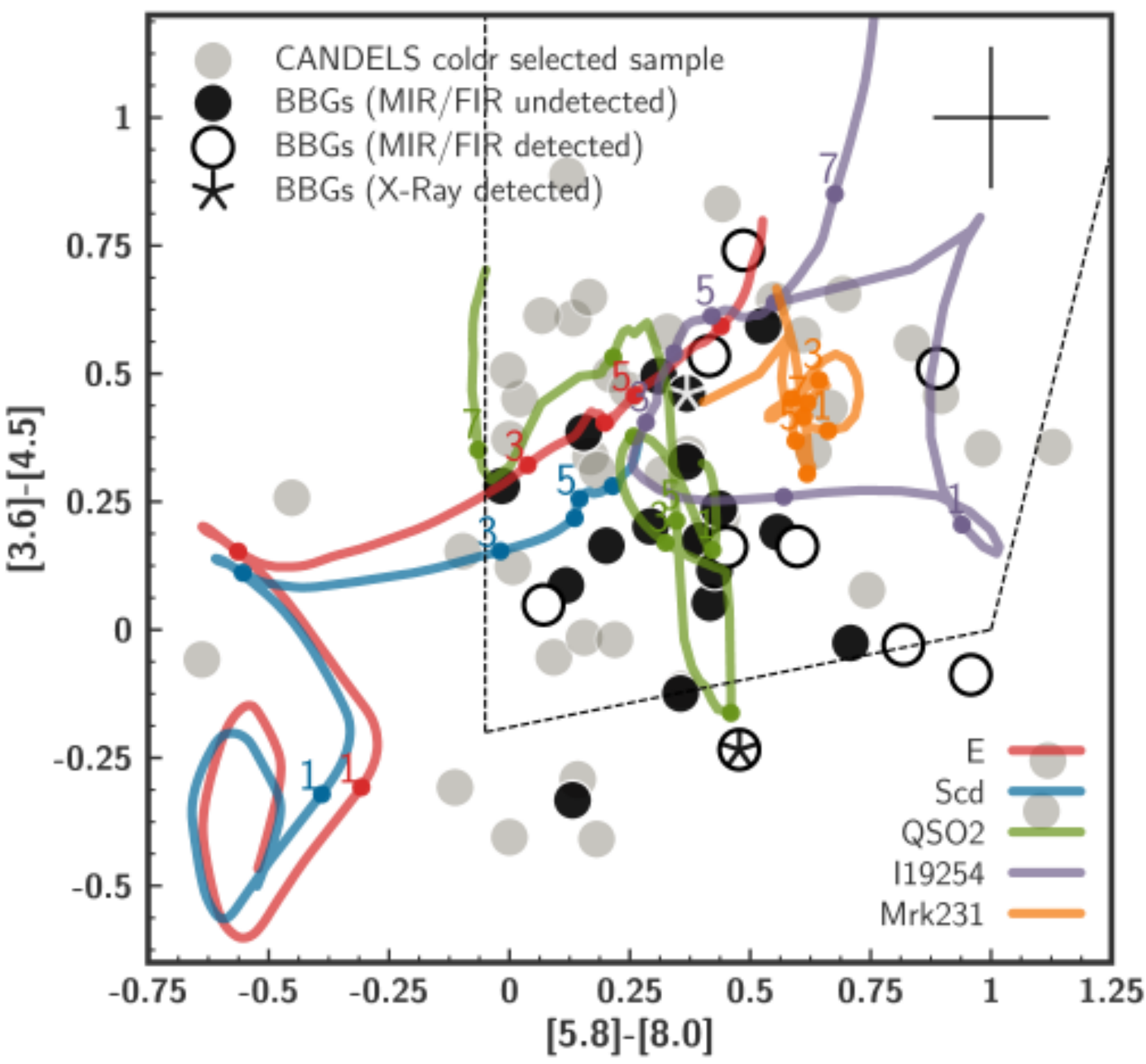}   
  \end{center}
  \caption{IRAC $[3.6]-[4.5]$ versus $[5.8]-[8.0]$ color-color diagram
    and the corresponding AGN selection wedge (dashed line) from
    \cite[][see also \citealt{2012ApJ...748..142D}]{stern2005mid}.
    Only 7 sources are not shown due to the non-detection in the IRAC
    5.8 and/or 8.0 $\mu$m channels.  We plot the expected colors for
    an elliptical and a late-type spiral galaxy (assuming the
    templates found in \cite{coleman1980colors} at different
    redshifts). We also depict the colors for a IR-bright Sy2 galaxy
    (I19254, \citealp{2011A&A...532A..49B}), a Type2 QSO
    \citep{polletta2007spectral}, and a Type1 Seyfert galaxy (Mrk231,
    \citealp{2011A&A...532A..49B}). These templates have been
    attenuated using a \cite{2000ApJ...533..682C} law and
    A(V)=1~mag. Error bars are not plotted for clarity but the average
    values are shown at the top-right corner.}
  \label{fig:agn}
\end{figure}

Here we study whether some of the BBGs could host an obscured AGN.  As
discussed in \S\ref{ssec:multiphot}, $\sim 25$\% (8/33) of the sample
is detected in MIPS24, and $\sim 15$\% (5/33) at longer wavelengths.
At the average redshift of the BBGs, $z\sim4$, these IR detections
probe the rest-frame mid-IR emission which can be linked to star
formation or obscured AGN activity. Figure\,\ref{fig:agn} shows the
distribution of the BBGs in the \cite{stern2005mid} IRAC color-color
diagram which is widely used to study the likelihood of AGN emission.
It is clear from the figure (see \citealt{2012ApJ...748..142D}) that
while this color-color plot is very efficient to identify strong
mid-IR emission in low and intermediate redshift galaxies, it is more
ambiguous at $z\gtrsim3-4$, given that the evolutionary tracks for all
the galaxy templates, from ULIRGs to elliptical galaxies, have colors
inside the AGN selection wedge (dashed line). Thus, it is not
surprising that most of the BBGs and the galaxies in the
color-selected sample are found in the AGN region.  We conclude that
this color-color diagram is not a reliable diagnostic to understand
the nature of the IR emission in our sample of high-z faint sources.
Therefore, we can only indicate that, based on the IR detections,
around 25-30\% of the BBGs may harbor (bright) obscured AGN, although
no conclusive proof can be presented at this stage. Many more galaxies
(up to 75\%) present colors which are consistent with Type 2 AGN
(similar to the I19254 or QSO2 templates) but also with
obscured/evolved star formation. In addition to the IR data, two of
the BBGs are detected in X-rays (stars in Figure~\ref{fig:agn}). Given
the high redshift of these galaxies, such detection implies a large
intrinsic luminosity ($L_X(2-10 keV)>10^{43}$~erg/s) typically
associated with the presence of a Type 2 or even Type 1 AGN. Note that
$\sim$25\% of the BBGs, that indeed correspond to the bluest sub-MS
and MS galaxies, are not shown in the diagram due to their undetection
beyond 4.5$\mu$m.

\setlength\tabcolsep{4pt}
\footnotesize
%\begin{landscape}
\begin{table*}
  \small
  \centering
  \begin{threeparttable}   	 
\vspace*{0.1cm}
\caption{\label{tab:stellarprop}Stellar properties of our sample of
  BBGs at $z>$3.}
    \hspace*{20pt}\begin{tabular}{>{\raggedright}p{0.5cm}cccccccccccccc}
    \toprule
 &       ID        &   $_{   }z_{   }$   & SFRsed& SFR$_{2800}$ &  SFR$_{\rm{IR}}$ &      SFR &    Mass               &     $\tau$                 & Age     & $\rm{t}_{m}$               & A(V) &UVJ \tnote{a} & SFR-M \tnote{b} \\ 
 &     &	       &[M$_{\odot}$yr$^{-1}$]&[$\mathrm{M}_\odot$\,yr$^{-1}$] & [M$_{\odot}$yr$^{-1}$] &[M$_{\odot}$yr$^{-1}$] & log[M/M$_{\odot}]$& [Myr]  & [Gyr] &  [Gyr] &[mag]&  &   \\ \hline \hline  \\   
  1  & GDN$\_$BBG01 &    3.8 &     17 &        6 &        1757 &   1763 &      $10.4\,_{10.4}^{10.5}$ &     $ 324\,_{286}^{379}$ &   $ 1.0\,_{0.9}^{1.1}$ &   0.5 &    $2.4\,_{2.4}^{2.5}$ &   dSFG        &  Starburst     \\[0.5ex]
  2  & GDN$\_$BBG02 &    4.8 &    292 &       14 &        2381 &   2395 &      $11.7\,_{11.7}^{11.8}$ &     $ 315\,_{280}^{353}$ &   $ 1.0\,_{0.9}^{1.1}$ &   0.5 &    $2.7\,_{2.7}^{2.8}$ &   dSFG        &  Starburst     \\[0.5ex]
  3  & GDN$\_$BBG03 &    5.2 &    318 &        1 &        1039 &   1040 &      $11.8\,_{11.7}^{11.8}$ &     $ 313\,_{281}^{358}$ &   $ 1.0\,_{0.9}^{1.1}$ &   0.5 &    $3.4\,_{3.3}^{3.4}$ &   dSFG        &  MS           \\[0.5ex]
  4  & GDN$\_$BBG04 &    5.1 &     26 &        3 &     \nodata &     26 &      $11.0\,_{11.0}^{11.1}$ &     $ 204\,_{177}^{230}$ &   $ 1.0\,_{0.9}^{1.1}$ &   0.6 &    $1.9\,_{1.8}^{2.0}$ &   dSFG        &  MS           \\[0.5ex]
  5  & GDN$\_$BBG05 &    4.4 &     23 &        3 &        1108 &   1111 &      $10.8\,_{10.8}^{10.9}$ &     $ 248\,_{222}^{279}$ &   $ 1.0\,_{0.9}^{1.1}$ &   0.6 &    $2.0\,_{2.0}^{2.0}$ &   dSFG        &  Starburst     \\[0.5ex]
  6  & GDN$\_$BBG06 &    4.2 &     10 &        1 &     \nodata &     10 &      $10.3\,_{10.3}^{10.4}$ &     $ 504\,_{444}^{569}$ &   $ 1.4\,_{1.3}^{1.6}$ &   0.7 &    $2.0\,_{1.9}^{2.1}$ &   dSFG        &  MS           \\[0.5ex]
  7  & GDN$\_$BBG07 &    4.8 &     12 &       10 &     \nodata &     12 &      $10.0\,_{ 9.9}^{10.1}$ &     $ 634\,_{430}^{872}$ &   $ 0.9\,_{0.7}^{1.1}$ &   0.4 &    $0.6\,_{0.5}^{0.8}$ &   bSFG        &  MS           \\[0.5ex]
  8  & GDN$\_$BBG08 &    3.8 &      6 &        0 &     \nodata &      6 &      $10.5\,_{10.4}^{10.5}$ &     $ 324\,_{284}^{380}$ &   $ 1.4\,_{1.2}^{1.6}$ &   0.9 &    $1.5\,_{1.4}^{1.7}$ &   dSFG        &  subMS        \\[0.5ex]
  9  & GDN$\_$BBG09 &    4.2 &      4 &        2 &     \nodata &      4 &      $10.4\,_{10.4}^{10.5}$ &     $ 316\,_{280}^{356}$ &   $ 1.4\,_{1.3}^{1.6}$ &   0.9 &    $1.5\,_{1.4}^{1.6}$ &   dSFG        &  subMS        \\[0.5ex]
  10 & GDN$\_$BBG10 &    5.1 &      4 &       37 &     \nodata &      4 &      $10.8\,_{10.7}^{10.8}$ &     $ 151\,_{105}^{180}$ &   $ 1.0\,_{0.9}^{1.1}$ &   0.7 &    $1.2\,_{1.0}^{1.4}$ &   Quiescent   &  subMS        \\[0.5ex]
  11 & GDN$\_$BBG11 &    6.6 &    154 &     1927 &     \nodata &    154 &      $11.1\,_{11.0}^{11.1}$ &     $ 338\,_{288}^{407}$ &   $ 0.7\,_{0.6}^{0.8}$ &   0.3 &    $2.0\,_{1.9}^{2.0}$ &   dSFG        &  MS           \\[0.5ex]
  12 & GDN$\_$BBG12 &    3.8 &      4 &        5 &     \nodata &      4 &      $10.4\,_{10.3}^{10.4}$ &     $ 317\,_{281}^{354}$ &   $ 1.4\,_{1.3}^{1.6}$ &   0.9 &    $2.0\,_{1.9}^{2.1}$ &   dSFG        &  subMS        \\[0.5ex]
  13 & GDN$\_$BBG13 &    4.4 &     20 &        5 &     \nodata &     20 &      $10.6\,_{10.5}^{10.6}$ &     $ 315\,_{281}^{354}$ &   $ 1.0\,_{0.9}^{1.1}$ &   0.5 &    $2.0\,_{1.9}^{2.0}$ &   dSFG        &  MS           \\[0.5ex]
  14 & GDN$\_$BBG14 &    5.9 &     31 &       38 &     \nodata &     31 &      $10.4\,_{10.3}^{10.5}$ &     $ 323\,_{286}^{385}$ &   $ 0.7\,_{0.6}^{0.8}$ &   0.3 &    $1.2\,_{1.1}^{1.3}$ &   bSFG        &  MS           \\[0.5ex]
  15 & GDN$\_$BBG15 &    5.0 &      3 &        3 &     \nodata &      3 &      $10.2\,_{10.1}^{10.3}$ &     $ 195\,_{169}^{222}$ &   $ 1.0\,_{0.9}^{1.1}$ &   0.6 &    $0.2\,_{0.1}^{0.2}$ &   Quiescent   &  subMS        \\[0.5ex]
  16 & GDN$\_$BBG16 &    4.6 &     48 &        0 &        9828 &   9828 &      $10.9\,_{10.9}^{11.0}$ &     $ 328\,_{286}^{401}$ &   $ 1.0\,_{0.9}^{1.1}$ &   0.5 &    $2.3\,_{2.3}^{2.5}$ &   dSFG        &  Starburst     \\[0.5ex]
  17 & GDN$\_$BBG17 &    4.2 &      7 &        3 &     \nodata &      7 &      $10.0\,_{10.0}^{10.1}$ &     $ 346\,_{294}^{415}$ &   $ 1.0\,_{0.9}^{1.1}$ &   0.5 &    $0.7\,_{0.6}^{0.9}$ &   dSFG        &  MS           \\[0.5ex]\hline \\[-0.5ex]
  18 & GDS$\_$BBG01 &    4.8 &     38 &        2 &     \nodata &     38 &      $10.8\,_{10.8}^{10.9}$ &     $ 318\,_{281}^{355}$ &   $ 1.0\,_{0.9}^{1.1}$ &   0.5 &    $1.7\,_{1.6}^{1.7}$ &   dSFG        &  MS           \\[0.5ex]
  19 & GDS$\_$BBG02 &    5.3 &    290 &        3 &        1839 &   1842 &      $11.7\,_{11.7}^{11.8}$ &     $ 317\,_{283}^{357}$ &   $ 1.0\,_{0.9}^{1.1}$ &   0.5 &    $2.7\,_{2.7}^{2.7}$ &   dSFG        &  Starburst     \\[0.5ex]
  20 & GDS$\_$BBG03 &    6.3 &    162 &        5 &     \nodata &    162 &      $11.1\,_{11.1}^{11.2}$ &     $ 315\,_{278}^{354}$ &   $ 0.7\,_{0.6}^{0.8}$ &   0.3 &    $1.8\,_{1.7}^{1.8}$ &   dSFG        &  MS           \\[0.5ex]
  21 & GDS$\_$BBG04 &    4.9 &      8 &        1 &     \nodata &      8 &      $10.3\,_{10.2}^{10.3}$ &     $ 324\,_{291}^{371}$ &   $ 1.1\,_{1.0}^{1.2}$ &   0.6 &    $1.2\,_{1.0}^{1.3}$ &   bSFG        &  MS           \\[0.5ex]
  22 & GDS$\_$BBG05 &    5.5 &     44 &        0 &     \nodata &     44 &      $10.9\,_{10.9}^{11.0}$ &     $ 317\,_{282}^{358}$ &   $ 1.0\,_{0.9}^{1.1}$ &   0.5 &    $2.0\,_{1.9}^{2.1}$ &   dSFG        &  MS           \\[0.5ex]
  23 & GDS$\_$BBG06 &    3.5 &      1 &        1 &     \nodata &      1 &      $10.2\,_{10.2}^{10.2}$ &     $ 805\,_{740}^{913}$ &   $ 1.5\,_{1.3}^{1.6}$ &   0.7 &    $2.0\,_{1.9}^{2.0}$ &   dSFG        &  subMS        \\[0.5ex]
  24 & GDS$\_$BBG07 &    4.6 &     37 &        2 &     \nodata &     37 &      $10.8\,_{10.8}^{10.9}$ &     $ 323\,_{285}^{375}$ &   $ 1.0\,_{0.9}^{1.1}$ &   0.5 &    $1.9\,_{1.8}^{2.0}$ &   dSFG        &  MS           \\[0.5ex]
  25 & GDS$\_$BBG08 &    5.5 &     19 &        2 &     \nodata &     19 &      $11.0\,_{10.9}^{11.0}$ &     $ 198\,_{177}^{225}$ &   $ 1.0\,_{0.9}^{1.1}$ &   0.6 &    $1.7\,_{1.7}^{1.8}$ &   dSFG        &  subMS        \\[0.5ex]
  26 & GDS$\_$BBG09 &    5.8 &    161 &        4 &         856 &    861 &      $11.1\,_{11.0}^{11.2}$ &     $ 329\,_{289}^{396}$ &   $ 0.7\,_{0.6}^{0.8}$ &   0.3 &    $2.5\,_{2.4}^{2.6}$ &   dSFG        &  Starburst     \\[0.5ex]
  27 & GDS$\_$BBG10 &    4.8 &     46 &        0 &     \nodata &     46 &      $10.9\,_{10.9}^{11.0}$ &     $ 316\,_{283}^{360}$ &   $ 1.0\,_{0.9}^{1.1}$ &   0.6 &    $2.0\,_{1.9}^{2.1}$ &   dSFG        &  MS           \\[0.5ex]
  28 & GDS$\_$BBG11 &    4.9 &     15 &        0 &     \nodata &     15 &      $11.2\,_{11.2}^{11.3}$ &     $ 158\,_{141}^{177}$ &   $ 1.0\,_{0.9}^{1.1}$ &   0.7 &    $2.0\,_{1.9}^{2.1}$ &   dSFG        &  subMS        \\[0.5ex]
  29 & GDS$\_$BBG12 &    3.4 &     26 &        0 &     \nodata &     26 &      $10.3\,_{10.2}^{10.3}$ &     $ 392\,_{349}^{441}$ &   $ 0.7\,_{0.6}^{0.8}$ &   0.3 &    $2.0\,_{1.9}^{2.1}$ &   bSFG        &  MS           \\[0.5ex]
  30 & GDS$\_$BBG13 &    5.1 &     19 &        0 &     \nodata &     19 &      $11.0\,_{10.9}^{11.1}$ &     $ 200\,_{178}^{224}$ &   $ 1.0\,_{0.9}^{1.1}$ &   0.6 &    $2.0\,_{1.9}^{2.0}$ &   dSFG        &  subMS        \\[0.5ex]
  31 & GDS$\_$BBG14 &    4.5 &      3 &        0 &     \nodata &      3 &      $11.1\,_{11.0}^{11.1}$ &     $ 125\,_{110}^{142}$ &   $ 1.0\,_{0.9}^{1.1}$ &   0.8 &    $0.9\,_{0.8}^{1.0}$ &   Quiescent   &  subMS        \\[0.5ex]
  32 & GDS$\_$BBG15 &    4.7 &      9 &        3 &     \nodata &      9 &      $11.0\,_{10.9}^{11.0}$ &     $ 157\,_{140}^{177}$ &   $ 1.0\,_{0.9}^{1.1}$ &   0.7 &    $2.0\,_{2.0}^{2.1}$ &   dSFG        &  subMS       \\[0.5ex]
  33 & GDS$\_$BBG16 &    4.5 &     31 &        1 &     \nodata &     31 &      $10.8\,_{10.7}^{10.8}$ &     $ 316\,_{280}^{353}$ &   $ 1.0\,_{0.9}^{1.1}$ &   0.5 &    $2.0\,_{1.9}^{2.0}$ &   dSFG        &  MS           \\[0.5ex] \hline      \bottomrule 
 	 \end{tabular}\hspace*{-5pt}%
    \begin{tablenotes}
    \item[a] Type of galaxy according to the $UVJ$ diagram: Blue (bSFG)
      or dusty (dSFG) star-forming galaxy.
    \item[b] Type of galaxy according to its position with respect to the
      MS in the SFR vs. stellar mass plot: starburst, MS or sub-MS
      galaxy.
      \end{tablenotes}
\end{threeparttable}   	 
\end{table*}

\normalsize

\subsection{Quantification of the role of BBGs in galaxy evolution}
\label{quant}

\begin{figure}[t]
  \begin{center}
     \includegraphics[width=1.\linewidth]{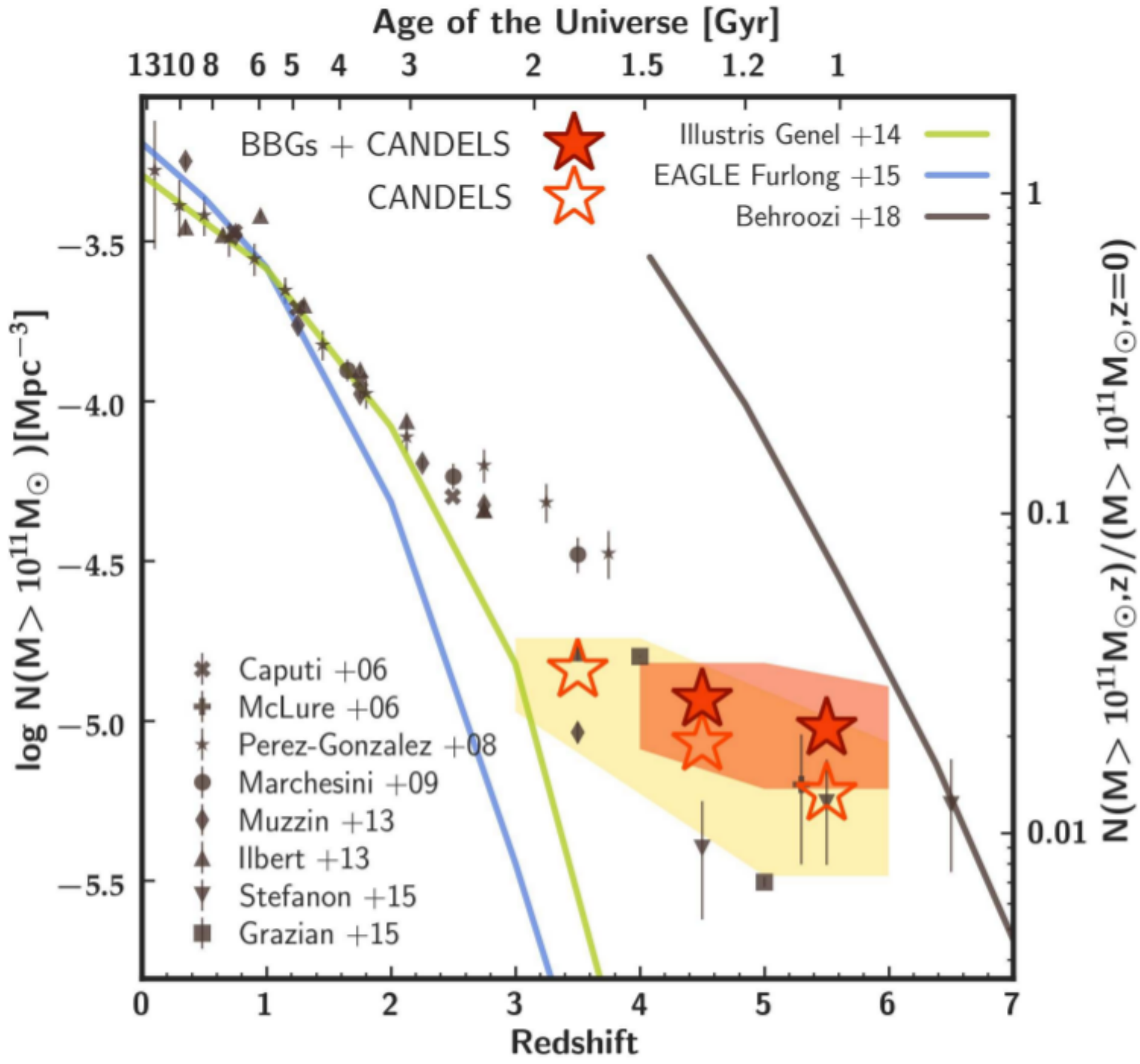}   
  \end{center}
  \caption{Number density of massive ($\rm{M}>10^{11} \rm{M}_{\odot}$)
    galaxies as a function of redshift (shown in the bottom horizontal
    axis and the corresponding age of the Universe in the top axis).
    The fraction of the local number density (computed as the average
    value of the numbers given in \citealt{Baldry2012} and
    \citealt{Bernardi2013}) is also shown in the secondary vertical
    axis.  Values from the literature are plotted with different grey
    symbols, as detailed in the legend. The values inferred from the
    CANDELS sample at redshifts $z=3-4$, $z=4-5$ and $z=5-6$ are shown
    with empty red stars, with errors depicted as a yellow shaded
    area.  The total number density, including the BBGs reported in
    this work, at the same redshift intervals are shown with filled
    red stars, with uncertainties plotted as a red shaded area.  The
    grey line corresponds to the number density threshold evolution
    for a stellar mass of $10^{11} \rm{M}_{\odot}$ presented in
    \citet{behroozi2018}.  The blue and green lines represent the
    number densities predicted by the EAGLE (\citealt{Furlong2015})
    and Illustris (\citealt{Genel2014}) simulations.}
  \label{fig:ndensity}
\end{figure}

Our understanding of the $z>3$ galaxy population relies largely on
samples of UV-selected galaxies typically characterized by blue colors
and prominent Lyman breaks \citep{2015ApJ...803...34B}.  However, it
is currently unknown if these galaxies are representative of the
massive $z>$3 galaxy population, or even if any of these
galaxies harbor evolved stellar populations outshined in the
UV/optical by recent bursts. In this sense, it is important to analyze
the contribution of our sample of BBGs to the known population of
$z>3$ galaxies. The most relevant statistical numbers are given in
Table~\ref{table:ndensity} and we discuss them in the following
paragraphs.

The CANDELS mass-limited sample presented at the beginning of
\S\,\ref{sec:Pysicalprop} comprises 414 galaxies (53 of them also
belong to the color-selected sample).  The 33 BBGs introduced in this
paper only represent a small fraction (8$\pm$1\%) of the
  general population (i.e., adding up the mass-limited sample and the
  BBGs) of massive ($\rm{M}>10^{10} \rm{M}_{\odot}$), $z>$3 galaxies
  in the GOODS fields. Nonetheless, they do represent a significant
  fraction (33$\pm$13\%) of the reddest sub-population of massive
galaxies at $z>$3.  Furthermore, our selection technique is specially
effective at selecting galaxies at $z=4-6$, recovering a fraction of
23$\pm$5\% and 43$\pm9$\% of the mass- and color-selected samples,
respectively.

The analysis of our sample of BBGs and the comparison samples also
points out that 80-100\% of the most massive ($\log
\rm{M}>10^{11}$~$\rm{M}_{\odot}$) galaxies at $4<z<6$ are red
($H-[3.6]>2.5$~mag). This percentage decreases to 25\%-30\% for $\log
\rm{M}=10^{10-11}\, \rm{M}_{\odot}$. With the sample of BBGs presented
in this paper, we have doubled the number of known red massive galaxy
candidates at $4<z<6$: the CANDELS catalog includes 32 $\log
\rm{M}>10^{10}\,\rm{M}_{\odot}$ galaxies and we detect 26 more. We
remark that our sample of BBGs is biased towards the massive end of
the stellar mass function: it accounts for $27\pm$17\% of the total
number density of galaxies at $4<z<5$ and 38$\pm19$\% at $5<z<6$, in
both cases for $\log \rm{M}>10^{11}\,\rm{M}_{\odot}$.  For lower mass
galaxies, our sample, and red galaxies in general, are a minor
contributor ($\sim10$\%) to the global population.

In absolute number density numbers, presented in
Figure~\ref{fig:ndensity}, and for $4<z<6$ and
$\rm{M}>10^{11}\,\rm{M}_{\odot}$, the mass-limited sample
extracted from the CANDELS catalog presents a number density
$7.3\pm0.2\times10^{-6}$~galaxies/$\rm{Mpc}^3$. This is consistent
with the estimations presented in \cite{Stefanon2015}, which range
between $5\times10^{-6}$ and $8.3\times10^{-6}$~galaxies/$\rm Mpc^3$
(depending on assumptions on the calculation of photometric redshifts)
and take into account very faint (or even undetected) NIR sources in
the ULTRAVista field detected in IRAC (down to $[3.6]$$\sim$23.4,
1~mag brighter than our analysis). These number densities are
  also consistent with the ones obtained by integrating the Schechter
  functions presented in \cite{Grazian2015}
  ($5.9\times10^{-6}$~galaxies/$\rm Mpc^3$) and \cite{Duncan2014}
  ($5.0\times10^{-6}$~galaxies/$\rm Mpc^3$) for stellar mass functions
  based on $H$-band selected samples.  The slight discrepancy can be
  in part attributed to the systematic differences between the
  Schechter function fits in those papers and the stellar mass
  functions datapoints at the massive end.  It is worth noticing that
  although models of galaxy evolution are capable of properly
  reproducing the number densities at low redshift ( $z\lesssim2$),
  current simulations underpredict the observed values of massive
  galaxies ($\log \rm{M}>10^{11}\,\rm{M}_{\odot}$) at higher
  redshifts. As shown in Figure~\ref{fig:ndensity}, EAGLE
  (\citealt{Furlong2015}) values plunge at $z\sim1.5$ while in
  Illustris (\citealt{Genel2014}) it occurs at $ z\sim2.5$. The reason
  for this mismatch between models and observations is unclear. But
  our results, which are consistent with other estimations at $z>2$
  shown in Figure~\ref{fig:ndensity}, clearly point out to a rapid
  early evolution of the star formation in some halos resulting in the
  appearance of very massive galaxies in the first Gyr of the history
  of the Universe, resembling more a quick monolithical collapse
  rather than a gentle hierarchical assembly. The observed values are,
  however, well below the threshold calculated by \citet{behroozi2018}
  as the limit of number densities for massive galaxies imposed by the
  current cosmological paradigm, which, according to these authors,
  could not be surpassed with our current knowledge of the Physics
  governing the evolution of the Universe. From the observational
  point of view, in other to place more robust constraints on modern
  theoretical models we need to better constrain the stellar masses
  and overcome the spatial resolution limitations that our current
  mid-IR data have.

To wrap up our results, taking into account our BBGs we have obtained
a more complete census of massive galaxies at $z>3$ by adding IRAC
sources undetected (or faint) in the $H$ and $K$ bands.  We conclude
that the total number density of $M>10^{11}$~$\rm{M}_{\odot}$ at
$4<z<6$ is $1.1\times10^{-5}$~galaxies/$\rm Mpc^3$. This corresponds
to $3$\% of the total number of local
$\rm{M}>10^{11}\,\rm{M}_{\odot}$ galaxies (considering the median
value of those obtained by integrating the local stellar mass
functions in \citealt{Baldry2012} and \citealt{Bernardi2013}), i.e.,
nearly 1 every 30 massive galaxies in the local Universe must have
assembled more than $10^{11}\,\rm{M}_{\odot}$ of their mass in the
first 1.5 Gyr of the Universe.

\begin{table}
\begin{threeparttable} 
	\caption{\label{table:ndensity} Number densities for the BBGs presented in this work 
	and the color-and mass-limited comparison samples, provided for different ranges of 
	redshift and mass. For each range, the total number of galaxies and the relative 
	percentages are also given in parenthesis.}
	\begin{center} {\scriptsize
	\begin{tabular}{llcccc}
	\hline\\
	\multicolumn{2}{c}{} & \multicolumn{4}{c}{$\Phi$ $[10^{-6} \rm{Mpc}^{-3}]$} \\[0.5ex] \cmidrule{3-6}
	\multicolumn{2}{l}{Redshift}& {Total}& {Mass-sel} & {Color-sel} & {BBGs} \\[-2.ex]
	\\\hline
	\hline\\
	$z>3$&$\rm M/\rm M_\sun>10^{11}$ & 12.6$\pm$2.0 & 10.4$\pm$1.8 & 4.9$\pm$1.2 & 2.5$\pm$0.9\\
	& & (42) & (81$\pm$15\%) & (38$\pm$11\%) & (19$\pm$7\%)\\
	&$\rm M/\rm M_\sun=10^{10-11}$ & 220.1$\pm$6.1 & 112.7$\pm$5.9 & 9.5$\pm$1.7 & 7.7$\pm$1.5 \\ 
	&	& (393) & (94$\pm$5\%) & (8$\pm$1\%) & (6$\pm$1\%)\\ 
	\hline\\
	{$4<z<6$}&$\rm M/\rm M_\sun>10^{11}$ & 10.6$\pm$2.0 & 7.3$\pm$2.0 & 6.1$\pm$1.9 & 3.4$\pm$1.4 \\
	& & (19) & (68$\pm$21\%) & (58$\pm$19\%) & (32$\pm$14\%)\\
	&$\rm M/\rm M_\sun=10^{10-11}$ & 90.0$\pm$7.1 & 79.4$\pm$6.7 & 11.7$\pm$2.6& 11.2$\pm$2.5 \\
	& & (162) & (88$\pm$8\%) & (13$\pm$3\%) & (12$\pm$3\%)\\
	\hline\\
	{$3<z<4$}&$\rm M/\rm M_\sun>10^{11}$ & 14.4$\pm$3.7 & 14.4$\pm$3.7 & \nodata &  \nodata \\
	& & (15) & (100$\pm$26\%) & (\nodata) & ( \nodata)\\
	&$\rm M/\rm M_\sun=10^{10-11}$ & 214.1$\pm$14.3 & 209.3$\pm$14.2 & 8.6$\pm$2.9& 4.8$\pm$2.1 \\
	& & (223) & (98$\pm$7\%) & (4$\pm$1\%) & (2$\pm$1\%)\\
	\hline\\
	{$4<z<5$}&$\rm M/\rm M_\sun>10^{11}$ & 11.7$\pm$3.5 & 8.5$\pm$3.0 & 6.4$\pm$2.6 & 3.2$\pm$1.8 \\
	& & (11) & (73$\pm$28\%) & (55$\pm$25\%) & (27$\pm$17\%)\\
	&$\rm M/\rm M_\sun=10^{10-11}$ & 120.8$\pm$11.3 & 107.0$\pm$10.6 & 17.0$\pm$4.2& 14.8$\pm$4.0 \\
	& & (115) & (88$\pm$9\%) & (14$\pm$4\%) & (12$\pm$3\%)\\
	\hline\\
	{$5<z<6$}&$\rm M/\rm M_\sun>10^{11}$ & 9.5$\pm$3.3 & 5.9$\pm$2.6 & 5.9$\pm$2.6 & 3.5$\pm$2.0 \\
	& & (8) & (62$\pm$31\%) & (62$\pm$31\%) & (38$\pm$24\%)\\
	&$\rm M/\rm M_\sun=10^{10-11}$ & 55.6$\pm$8.1 & 48.5$\pm$7.6 & 5.9$\pm$2.6& 7.1$\pm$2.9 \\
	& & (47) & (87$\pm$14\%) & (11$\pm$5\%) & (13$\pm$5\%)\\
	
	\\\hline
	\end{tabular} 
  \begin{tablenotes}
       \item Uncertainties in the densities and percentages have been calculated assuming Poisson statistics and taking into account the photometric redshift and stellar mass probability distributions functions.          
  \end{tablenotes}
  }
  \end{center}
\end{threeparttable}   
\end{table}

\section{Summary and conclusions}
\label{sec:results}

Combining ultra-deep data taken in the HST/WFC3 $F160W$ and {\it
  Spitzer}/IRAC 3.6 and 4.5~$\mu$m bands, we have identified a sample
of 33 IRAC bright/optically faint Balmer Break Galaxies (BBGs) at high
redshift within the 2 GOODS fields. Our sample is composed by
extremely red ($H-[3.6]\gtrsim2.5$~mag) and relatively bright mid-IR
($[3.6]<24.5$~mag) galaxies. This translates to the following physical
properties, according to our analysis of X-ray to radio spectral
energy distributions: the typical BBG is a massive galaxy with a
stellar mass $<log(M/M_sun)>=10.8$ lying at redshift $<z>=4.8$. BBGs
harbor relatively young stellar populations (mass-weighted age
$<t_{m}>=0.6$~Gyr) with significant amounts of dust ($<A(V)>=2$~mag),
although the range of stellar properties is wide.

We have analyzed the sample of BBGs by comparing them with a
mass-limited ($\rm{M}>10^{10} \rm{M}_{\odot}$ and $z>3$)
sample and a color-selected ($H-[3.6]\geqslant2.5$) sample extracted
from the CANDELS catalogs published for these fields. We have found
that our BBGs substantially differ from the galaxies in the
mass-limited sample, which are bluer in general, while their
physical properties are quite similar to those in the color-selected
sample. However, our BBGs are too faint in the rest-frame UV and
optical to be included in typical NIR selected samples in this
redshift range.

The $H-[3.6]$ red colors of most of our sources are compatible with
heavily extincted starbursts or relatively evolved
  populations. However, our sample includes a distinct population of
blue (both in their observed $H-[3.6]$ and their $UVJ$ rest-frame
colors) galaxies. This population has similar SEDs to galaxies from
the mass-limited sample.  They indeed present uncommon blue
[3.6]-[4.5] colors that might be caused by the presence of an emission
line in the [3.6] band (converting them in red sources in our
selection color $H-[3.6]$). We also note that these possible
line-emitters correspond to some of the less massive
($\rm{M}<10^{10.5} {\rm M}_{\odot}$) BBGs in our sample.  Therefore,
their detection might be a consequence of our improved photometric
technique to recover faint sources and reliable upper limits.

We have also demonstrated that an $H-[3.6]$ color and IRAC
magnitude cuts imply a redshift selection. The redshift distributions
of both the BBGs and the color-selected sample peak at $z=4-5$, while
the mass-limited sample presents an exponentially decreasing
redshift distribution (typical of flux limited samples).  We are also
more effective selecting galaxies at $z=4-5$ than any other sample of
$H$-band faint galaxies in the literature.  Our selection criterion is
also adequate to uncover the high mass end of the stellar
mass function ($\rm{M}\gtrsim10^{10} {\rm M}_{\odot}$). The BBG
stellar mass distribution peaks at $\rm{M}\sim10^{10.5} {\rm
  M}_{\odot}$.  The color-selected sample presents a comparable
histogram with a longer tail at higher masses due to their brighter
IRAC magnitudes.  The mass-limited sample, in contrast,
presents a distribution that decreases with increasing masses.

From the SED modeling, we find a strong evidence that massive red
galaxies at $z=3-6$ span a diverse range in stellar population
properties. In order to understand the nature of the heterogeneous
sample of BBGs, we have divided the sources in three star formation
regimes according their position with respect to the main sequence
(MS): starbursts, MS and, sub-MS galaxies.  Analyzing the average SEDs
of BBGs, we confirm that, in general, mass-limited galaxies
present bluer SEDs than those in the BBG and the color-selected
samples. However, as mentioned before, we identify a subsample of BBGs
in the MS which are blue and harder to separate from the general
population probed by a mass-limited sample. In
addition, we find a considerable number of sub-MS galaxies (33\% of
our sample), most of them with $\rm{M}<10^{10.5} {\rm M}_{\odot}$,
characterized by lower attenuations ($A(V)\sim1.5$~mag) and older
mass-weighted ages ($t_{m}\sim0.7$~Gyr).  On the other hand,
starbursts are found among the most massive ($\rm{M}>10^{10.5} {\rm
  M}_{\odot}$) galaxies in the BBG (20$\%$ of the total number of BBGs
are starburst) and color-selected (15$\%$) samples.  Starbursts are
characterized by high attenuations ($A(V)\sim2.5$~mag) and young ages
($t_{m}\sim0.5$~Gyr). We remark that the total IR luminosity
  for 5 out of the 6 starbursts has been calculated with 3-5
  datapoints (within the wavelength range that dominates the
  integrated IR luminosity), which translate to relatively small
  errors.  This suggests that a significant fraction of the BBGs
($\sim$25 and up to $\sim$75\%) might host an obscured AGN. MS
galaxies represent a constant proportion of BBGs ($\sim$50\%) and
color-selected ($\sim$60\%) up to the highest masses
$\rm{M}\sim10^{11.5} {\rm M}_{\odot}$. However, an important fraction
(25$\%$) of the MS galaxies from the color-selected sample have been
assigned with a SFR lower limit (given their detection by MIPS, but
their high redshift that prevent from obtaining a robust SFR
estimation) and may correspond to starburst galaxies. BBGs in the MS
present a larger scatter in their attenuations, mass-weighted ages and
$UVJ$ colors.

Subdividing BBGs by their rest-frame $UVJ$ colors, we find
that most of the BBGs correspond to dusty SFGs (80\% of the
sample), a smaller fraction to blue SFGs (10\%), and the rest to
quiescent galaxies.  Although several studies have reported the
existence of galaxies with suppressed star formation at that epoch
(e.g., \citealt{2014ApJ...783L..14S}), just 3 of our BBGs lie within
the quiescent wedge and 3 more BBGs have mass-weighted ages that are
old enough (t$_{m}\geqslant0.9$~Gyr) to be consistent with evolved
galaxies. However, 50\% of our sample (16 galaxies) is not detected
in the $H$-band down to magnitudes fainter than $\sim$27~mag and,
therefore, only count with $U-V$ lower limits. Out of those, 10
galaxies ($\sim$30\% of the entire sample), not detected by MIPS or
Herschel, could still be identified as quiescent galaxies in the
$UVJ$ diagram, although no conclusive proof of their nature can be
inferred given the high uncertainties.

We have found that the red BBGs presented in this work account
for 8\% of the total number density of $\log(\rm{M}/\rm{M}_\sun)
>10$ galaxies at $z>3$ found by public catalogs such as CANDELS' or
3D-HST's. Our BBGs are, however, a significant contributor (30\%) to
the general (adding cataloged galaxies and our BBGs) population of
$\log(\rm{M}/\rm{M}_\sun)>11$ galaxies at $4<z<6$.  Our analysis
also reveals that while 80-100\% of the most massive
($\rm{M}>10^{11}$~$\rm{M}_{\odot}$) galaxies at $4<z<6$ are red,
this percentage decreases for lower mass galaxies. We remark that
with the sample of BBGs presented in this paper, we have doubled the
number of known red massive galaxy candidates at $4<z<6$: the CANDELS
catalog includes 32 $\rm{M}>10^{10}\,\rm{M}_{\odot}$ galaxies and we
have detected 26 more. Hence, accounting for this kind of objects is
key to understand the population of massive galaxies at high redshift.
Adding the BBGs presented in this work to the known population of
$4<z<6$ and $\rm{M}>10^{11}\,\rm{M}_{\odot}$ we have found a total
number density of 1.1$\times10^{-5}$galaxies/$\rm Mpc^3$.
This represents 3\% of the the number density of local
  M$>10^{11}\,\rm{M}_{\odot}$, i.e., nearly 1 in 30 massive galaxies
in the local Universe must have assembled more than
$10^{11}\,\rm{M}_{\odot}$ in the first 1.5\,Gyr of the Universe.
We compare these numbers with state-of-the-art galaxy
  formation simulations, such as Illustris and EAGLE, finding that
  while the models do a reasonably good job up to $z\sim2$, they fail
  to predict very massive (M$>10^{11}\,\rm{M}_{\odot}$) galaxies at
  $z\gtrsim3.5$ such as those presented in this paper by orders of
  magnitude.

Spectroscopic follow-up observations in both the optical and NIR are
critical to confirm the redshifts and to better characterize the
properties of this heterogeneous population of red massive galaxies at
$z=3-6$ missed by deepest (mainly NIR selected) studies. Imaging in
near- and mid-infrared wavelengths together with spectroscopy from the
JWST will be essential to understand their nature and ALMA will be
crucial in constraining the amount of dust and gas in these systems,
as well as discriminating between dust-enshrouded star formation and
obscured AGN activity.

\acknowledgments

We acknowledge support from the Spanish Programa Nacional de
Astronomía y Astrofísica under grants AYA2015-63650-P and
BES-2013-065772. Nicolás Cardiel acknowledges financial support from
the Spanish Ministry of Economy and Competitiveness (MINECO) under
grant number AYA2016-75808-R, which is partly funded by the European
Regional Development Fund (ERDF). This work has made use of the
Rainbow Cosmological Surveys Database, which is operated by the
Universidad Complutense de Madrid (UCM) partnered with the University
of California Observatories at Santa Cruz (UCO/Lick,UCSC).  This
research has made use of the software packages SExtractor, IRAF
\citep{tody1993iraf} DAOPHOT routine and STILTS
(http://www.starlink.ac.uk/stilts/) software, provided by Mark Taylor
of Bristol University, England. This work also employed Astropy, a
community-developed core Python package for Astronomy
\citep{Astropy2013}; APLpy, an open-source plotting package for Python
\citep{APLpy2012}; Matplotlib \citep{MatplotlibHunter2007} and Numpy;
Photutils \citep{Photutils2016}.

\bibliography{ms}

%\printbibliography
%\appendix

\begin{appendices}

\setcounter{section}{0}
\setcounter{figure}{0}
\setcounter{table}{0}
\renewcommand{\thesection}{\Alph{section}}
\renewcommand{\thesubsection}{\Alph{section}.\arabic{subsection}}
\renewcommand{\thetable}{\Alph{section}.\arabic{table}}
\renewcommand{\thefigure}{\Alph{section}.\arabic{figure}}

\section{HST-based Photometric measurements for BBGs in optical and near-IR bands}
\label{A:PhotUncertainty}

\normalsize
%%MOTIVATION
In this Section, we present the method used to measure consistent and
reliable photometry in the optical and NIR HST bands for the BBGs
presented in this work. The intrinsically faint nature of these
galaxies in the optical and NIR makes the construction of robust SEDs
an extremely challenging task. Indeed, by definition, BBGs are very
faint and, in many cases, even undetected in the optical and NIR
bands. Only their mid-IR fluxes are strongly detected by IRAC and 
for 20\% of sources also by MIPS, and for 15\% also by Herschel in 
the far-IR.

As described in Section \ref{ssec:multiphot}, our BBGs cannot be
found, by definition of the selection, in the CANDELS (-and 3D-HST-)
photometric catalogs published by \citet{2013ApJS..207...24G} and
\citet{2014ApJS..214...24S}, respectively.  These galaxies were most
likely missed in the catalogs due to limitations (incompleteness) in
the source detection method. However, a forced photometric measurement
using small apertures does recover reliable fluxes in the optical and
NIR HST bands for some of our BBGs.

In order to measure those fluxes with the highest SNR possible (or at
least set upper limits based on the background noise), the choice of
an appropriate photometric aperture is critical. The CANDELS and
3D-HST photometric catalogs are based on isophotal magnitudes
corrected to \citet{Kron1980} magnitudes (see Section 3 in
\citealt{2013ApJS..207...24G}), with an imposed minimum aperture size
of 2.08 pixels (0.125\arcsec). Only a small fraction ($< 2.5\%$) of
the sources in the CANDELS catalog have isophotal radius smaller than
2 pixels, and we find that all our BBGs, that are marginally detected
in the HST stacked images, exhibit radius larger than $\sim$3-4 pixels
(see below).

%%% THE STACKS %%%%
We focus first in understanding and characterizing the brightness
profile of the BBGs to identify the most appropriate aperture size for
the flux measurement which maximizes the SNR. To do so, we created
several different stacks of the BBGs, and also of unresolved (stellar)
point-like sources and (faint) color-selected galaxies in the CANDELS
sample in order to compare their average properties.

\setlength\tabcolsep{4pt}
\footnotesize
%\begin{landscape}
\begin{table*}
  \small
  \centering
  \begin{threeparttable}   	 
\vspace*{0.1cm}
\caption{\label{tab:apertures} Properties of the photometric apertures used for the BBGs in this work.}
    \hspace*{20pt}\begin{tabular}{>{\raggedright}p{0.5cm}cccccccccccccc}
    \toprule
 & ID              & Fiducial $^{a}$  &   a   &   b    & angle& Other apertures considered  & Morphology in the HST stacks  $^{b}$         \\
 &                 &(arcsec)&(arcsec)&(arcsec)&degree& (arcsec) &                        \\ \hline \\[-2.5ex] 
 1 & GDN$\_$BBG01  & 0.4 &    \nodata &   \nodata   &  \nodata  &   0.65 ,  0.4$+$AC  &  HST ultra-faint  \\
 2 & GDN$\_$BBG02  & 0.4 &    \nodata &   \nodata   &  \nodata  &   0.65 ,  0.4$+$AC  &  HST ultra-faint  \\
 3 & GDN$\_$BBG03  & 0.4 &    \nodata &   \nodata   &  \nodata  &   0.65 ,  0.4$+$AC  &  HST ultra-faint  \\
 4 & GDN$\_$BBG04  & 0.4 &    \nodata &   \nodata   &  \nodata  &   0.65 ,  0.4$+$AC  &  HST ultra-faint  \\
 5 & GDN$\_$BBG05  & 0.4 &    \nodata &   \nodata   &  \nodata  &   0.65 ,  0.4$+$AC  &  HST ultra-faint  \\
 6 & GDN$\_$BBG06  & 0.4 &    \nodata &   \nodata   &  \nodata  &   0.65 ,  0.4$+$AC  &  HST ultra-faint  \\
 7 & GDN$\_$BBG07  & \nodata  &  0.75 &   0.53 &    0 &            \nodata      &  HST extended     \\
 8 & GDN$\_$BBG08  & 0.4 &    \nodata &   \nodata   &  \nodata  &   0.65 ,  0.4$+$AC  &  HST ultra-faint  \\
 9 & GDN$\_$BBG09  & 0.4 &    \nodata &   \nodata   &  \nodata  &   0.65 ,  0.4$+$AC  &  HST ultra-faint  \\
 10& GDN$\_$BBG10  & 0.4 &    \nodata &   \nodata   &  \nodata  &   0.65 ,  0.4$+$AC  &  HST ultra-faint  \\
 11& GDN$\_$BBG11  & 0.4 &    \nodata &   \nodata   &  \nodata  &   0.65 ,  0.4$+$AC  &  HST ultra-faint  \\
 12& GDN$\_$BBG12  & 0.4 &    \nodata &   \nodata   &  \nodata  &   0.65 ,  0.4$+$AC  &  HST ultra-faint  \\
 13& GDN$\_$BBG13  & 0.4 &    \nodata &   \nodata   &  \nodata  &   0.65 ,  0.4$+$AC  &  HST ultra-faint  \\
 14& GDN$\_$BBG14  & \nodata  &   0.8 &    0.5 &   90 &           \nodata       &  HST extended     \\
 15& GDN$\_$BBG15  & \nodata  &   0.6 &    0.5 &   60 &           \nodata       &  HST extended     \\
 16& GDN$\_$BBG16  & 0.4 &    \nodata &   \nodata   &  \nodata  &   0.65 ,  0.4$+$AC  &  HST ultra-faint  \\
 17& GDN$\_$BBG17  & \nodata  &   0.6 &    0.5 &   80 &           \nodata       &  HST extended     \\\hline \\[-2.5ex]
 18& GDS$\_$BBG01  & 0.4 &    \nodata &   \nodata   &  \nodata  &   0.65 ,  0.4$+$AC  &  HST ultra-faint  \\
 19& GDS$\_$BBG02  & 0.4 &    \nodata &   \nodata   &  \nodata  &   0.65 ,  0.4$+$AC  &  HST ultra-faint  \\
 20& GDS$\_$BBG03  & 0.4 &    \nodata &   \nodata   &  \nodata  &   0.65 ,  0.4$+$AC  &  HST ultra-faint  \\
 21& GDS$\_$BBG04  & 0.4 &    \nodata &   \nodata   &  \nodata  &   0.65 ,  0.4$+$AC  &  HST ultra-faint  \\
 22& GDS$\_$BBG05  & 0.4 &    \nodata &   \nodata   &  \nodata  &   0.65 ,  0.4$+$AC  &  HST ultra-faint  \\
 23& GDS$\_$BBG06  & 0.4 &    \nodata &   \nodata   &  \nodata  &   0.65 ,  0.4$+$AC  &  HST ultra-faint  \\
 24& GDS$\_$BBG07  & \nodata  &   0.7 &   0.36 &  115 &           \nodata       &  HST extended     \\
 25& GDS$\_$BBG08  & 0.4 &    \nodata &   \nodata   &  \nodata  &   0.65 ,  0.4$+$AC  &  HST ultra-faint  \\
 26& GDS$\_$BBG09  & 0.4 &    \nodata &   \nodata   &  \nodata  &   0.65 ,  0.4$+$AC  &  HST ultra-faint  \\
 27& GDS$\_$BBG10  & 0.4 &    \nodata &   \nodata   &  \nodata  &   0.65 ,  0.4$+$AC  &  HST ultra-faint  \\
 28& GDS$\_$BBG11  & 0.4 &    \nodata &   \nodata   &  \nodata  &   0.65 ,  0.4$+$AC  &  HST ultra-faint  \\
 29& GDS$\_$BBG12  & 0.4 &    \nodata &   \nodata   &  \nodata  &   0.65 ,  0.4$+$AC  &  HST ultra-faint  \\
 30& GDS$\_$BBG13  & 0.4 &    \nodata &   \nodata   &  \nodata  &   0.65 ,  0.4$+$AC  &  HST ultra-faint  \\
 31& GDS$\_$BBG14  & \nodata  &   0.9 &   0.34 &   55 &           \nodata       &  HST extended     \\
 32& GDS$\_$BBG15  & 0.4 &    \nodata &   \nodata   &  \nodata  &   0.65 ,  0.4$+$AC  &  HST ultra-faint \\
 33& GDS$\_$BBG16  & 0.4 &    \nodata &   \nodata   &  \nodata  &   0.65 ,  0.4$+$AC  &  HST ultra-faint  \\ 
 \hline   
    \bottomrule  \\[-5.ex]
 	 \end{tabular}\hspace*{-5pt}%
    \begin{tablenotes}
    \item[a] Fiducial photometric aperture radius (no aperture correction was applied).
    \item[b] Type of galaxy according to the HST stacks.
      \end{tablenotes}
\end{threeparttable}   	 
\end{table*}

\normalsize

%% Unify Near-IR vs. NIR notation in the text
\begin{figure}
  \begin{center}
    \includegraphics[width=1.0\textwidth]{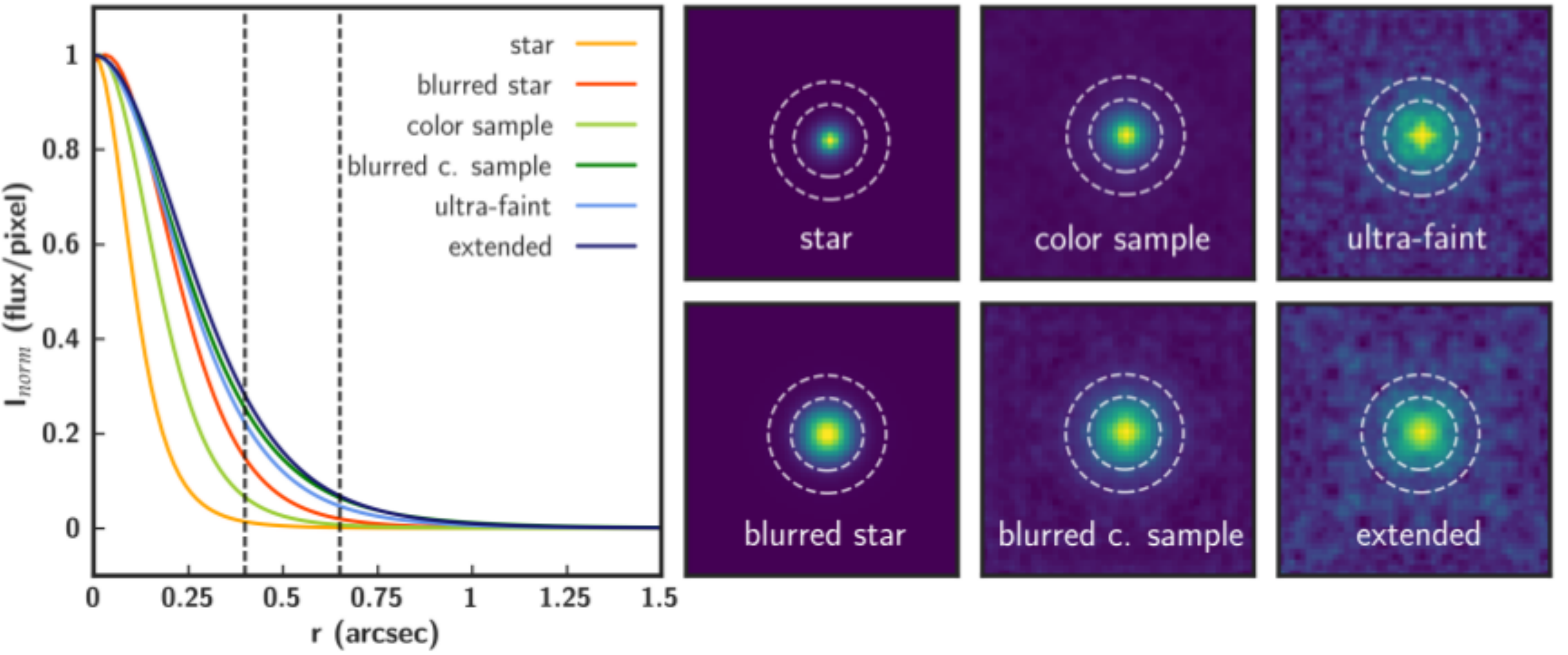}
  \end{center}
  \caption{\textit{Left:} Intensity profiles for the different stacked
    HST images described in the text. These stacks include: an average
    stellar profile, a stellar "blurred profile" taking into account
    random centering errors with rms $0.3\arcsec$, the average and
    blurred profiles for the CANDELS color-selected sample, the
    average profile for the 6 extended BBGs, and the average profile
    for the rest (27) of BBGs.  {\it Right}: stacked images for the
    previously mentioned samples of sources. Vertical lines on the
    left and circles on the postage stamps on the right show the
    photometric apertures considered in the paper (as discussed in
    Appendix \ref{A:PhotUncertainty}).}
    \label{fig:Profiles_A}
\end{figure}

%%PHOTOMETRIC STACK
First we created individual stacks for each one of the 33 BBGs by
combining all the deep HST optical and NIR imaging. These stacks are
shown in Appendix \ref{A:SEDs} (postage stamp in the lower left
corner). 20 out the 33 BBGs are clearly ($>5\sigma$) detected in these
stacks, and 6 out of those 20 (e.g., GDN$\_$BBG07 or GDN$\_$BBG15)
are marginally resolved or exhibit multiple-knots in their morphology.
For these 6 resolved galaxies, the photometric apertures are
determined based on their isophotal sizes measured in the HST stack
(i.e., as in the CANDELS catalog). Their radii range from $r =
0.6\arcsec$ to $0.9\arcsec$ (see Table~\ref{tab:apertures}).  For the
other 14 galaxies detected in the stacks, we analyze their growth
curves, and we find that the SNR of the recovered flux is maximum for
apertures of $r\sim0.4\arcsec$. For comparison, the median and
quartiles for the Kron-based apertures for faint ($H>26$~mag) galaxies
present in the CANDELS catalogs is $0.4^{0.5}_{0.3}\arcsec$.
Similarly, this value is also consistent with the aperture size of the
photometric measurements in the 3D-HST catalog is $r=0.35\arcsec$
\citep{2014ApJS..214...24S}.

%%GALAXY STACKS
We further study the average brightness profile of the BBGs by
creating a stack of all the galaxies. We first stack all the HST bands
for each galaxy, and then we stack the galaxies together. For clarity,
we create two of such stacks, one for the 6 extended galaxies and
another one for the remaning 27, as we expect their profiles to be
slightly different. The galaxy stacks are computed following the
method of \citet{Dole2006}. Briefly, the procedure consists of three
steps.  First, we create a WFC3 stack in each field. Second, we
extract ($3\arcsec\times3\arcsec$ as shown in Figure
\ref{fig:Profiles_A}) square images centered around each source, and
mask all known sources (i.e., those in the CANDELS catalog). Finally,
we sum up all postages after applying different rotations to them.
More precisely, we sum each image and its horizontally-flipped
analogous, and we also add each one of these images rotated by $90^o$,
$180^o$, and $270^o$, i.e., we use the image for each source 8 times
in total. During the last stacking step, outlier pixels were rejected.
The last steps increase the SNR of the stack and provides more
accurate average light profiles. The final average light profile of
the BBGs is shown in Figure~\ref{fig:Profiles_A}.

%%% STELLAR AND COLOR-SELECTED STACKS
For comparison purposes, we create also two additional stacks, one for
clean, well-detected, point-like sources (stars) in the field, and
another one for faint galaxies in the CANDELS color-selected sample
described in Section 5. The sample of point-like sources is selected
based on photometric and morphological criteria (see \citealp{
  perez2008stellar,2011ApJS..193...13A,2011ApJS..193...30B}).
Specifically, we selected stars in the fields with a FWHM value
smaller than 4.2 pixels, i.e., we rejected sources with FWHM larger
than $0.25\arcsec$ (the nominal FWHM in F160W is $0.17-0.19\arcsec$
-\citealp{koekemoer2011candels, 2013ApJS..207...24G}-). Then, we
create three different stellar stacks by combining the individual
postages in different ways, namely: 1) a direct stack of all of them,
2) a rotated \& flipped stack such as the one described above for
BBGs, and 3) a stack in which the center of each individual postage is
shifted randomly within the typical rms error in the position of BBGs
without HST detection ($0.3\arcsec$, see section
\ref{ssec:multiphot}). Figure~\ref{fig:Profiles_A} shows the stellar
stacks computed with methods 1 and 3 as well as two stacks for the
color-selected sample computed using either the rotation+flip method
or the random variation of the centering ({\it blurring}) within the
typical astrometric precision for the BBGs. The stacks based on the
first two methods yield very similar FWHMs of 0.17 and $0.20\arcsec$,
almost identical to the nominal FWHM in F160W. The third stack,
however, exhibits a noticeable broadening of the light profile
($0.43\arcsec$), very similar to that for the BBGs, although the
latter present brighter wings.

Based on the comparison of all the stacks we conclude that the bulk of
the BBG sample (27 galaxies, $\sim80\%$) consist of unresolved or
marginally resolved (at the resolution of F160W) galaxies, while the
other 6 BBGs, which are detected individually, might present some
extended emission (e.g., a faint disk). The FWHM obtained from the
stack of unresolved BBGs ($r=0.42\arcsec$) is fully consistent with
the value obtained by fitting the profile of the color-selected stack
($r=0.46\arcsec$), and it is also similar to the typical sizes of the
faint galaxies in the CANDELS catalog, whose aperture sizes are given
above.

%%% THE PHOTOMETRY BASED ON THE ANALYSIS OF THE APERTURES
Based on the results from the analysis of the light profiles, we
decided to use three different methods to measure the HST-based
photometry of the BBGs. This means that for each galaxy we analyzed 3
different SEDs with the same IRAC fluxes, but varying optical/NIR
photometry. We use these 3 different SED types to quantify the impact
of the intrinsic faintness of the objects and the uncertainties in
size measurements on the SED-derived properties.

The first photometric method is based on fixed circular apertures with
a radius of $r=0.4\arcsec$, and we do not apply any aperture
correction, i.e., we assume that the BBGs are small faint sources
similar to the color-selected sample in the CANDELS catalog (which do
not apply any aperture correction to the Kron apertures of similar
size to ours). As mentioned above, this aperture size provides the
highest SNR in the individual HST stacks and therefore we chose it as
our fiducial value for analysis purposes thorough the paper. The
second method is based on the same aperture photometry with
$r=0.4\arcsec$, but in this case we do apply an aperture correction to
account for possible missing flux. We compute this correction from the
average BBG profile (building a growth curve with it) measured in the
galaxy stack of undetected BBGs (light green curve in
Figure~\ref{fig:Profiles_A}). For this measurement, we also assume
that the full size of the average profile is given by the radius where
the flux is equal to 1$\sigma$ of the background in the stacked image,
$r=0.65\arcsec$. The aperture correction implied by the average light
profile for BBGs from $0.4\arcsec$ to $r=0.65\arcsec$ is 0.25~mag.
Note that for a point-like source (a star) accounting for an
uncertainty in the center determination, that aperture correction
would be 0.18~mag. For reference, with this aperture we recover
reliable ($>5\sigma$) fluxes in the F160W image for 17 galaxies, and
the typical SNR for these measurements was 7.5. Lastly, the third
method is based on circular apertures with a radius of
$r=0.65\arcsec$.  As mentioned above, this radius roughly corresponds
to the full size (i.e., 100\% of the flux) of the BBGs as determined
from the galaxy stack. Therefore, no aperture correction was applied
in this case. For reference, using this aperture we recovered reliable
($>5\sigma$) fluxes in the F160W band for 14 galaxies, and the typical
SNR for these measurements is 6.

%%% SUMMARY

The comparison between the integrated F160W magnitudes obtained with
the small (r$=0.4\arcsec$ with aperture correction) and large
($r=0.65\arcsec$) apertures for the 11 unresolved galaxies with
reliable ($>5\sigma$ ) $F160W$ detections yields an average difference
and rms of $\Delta m=0.2\pm0.2$, i.e., both types of photometry are
consistent within errors.

In the main text, the $r=0.4\arcsec$ aperture photometry without
aperture correction is used as the fiducial photometry (unless the
source was clearly detected in the HST stack, see
Table~\ref{fig:Profiles_A}). Nonetheless, throughout the paper we use
the 2 other photometric measurements to study the impact of varying
the HST-based fluxes on the overall SED of the BBGs and on other
SED-based properties such as redshifts and stellar masses. For
example, we use the redshift probability density functions (zPDFs)
derived from the fitting of the SEDs obtained with the different
methods to estimate the uncertainties in the photometric redshift
distribution of the BBGs.

% % % % % % % % % % % % % % % % % % % % % % % % % % % % % % % % % % % % % % % % % % % % % % % % % % % % % % % %%\setcounter{section}{0}
\setcounter{figure}{0}
\setcounter{table}{0}

% % % % % % % % % % % % % % % % % % % % % % % % % % % % % % % % % % % % % % % % % % % % % % % % % % % % % % % %
\section{Comparison samples from the public CANDELS catalogs}
\label{A:CANDELSample}

\subsection{Construction of the CANDELS samples}

\begin{figure*}
  \begin{center}
    \includegraphics[width=1.\textwidth]{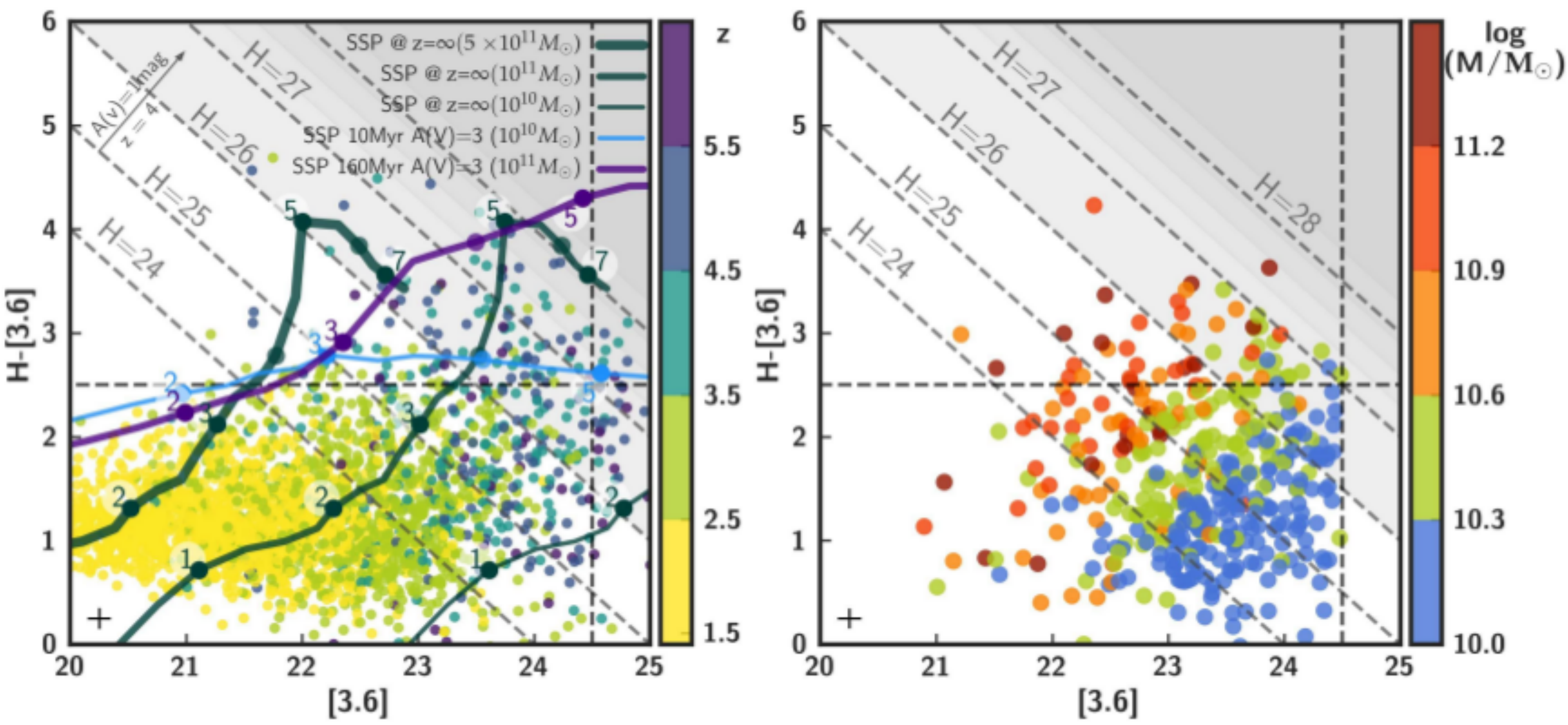}
  \end{center}
  \caption{\textit{Left panel:} $H-$[3.6] color plotted versus the
    observed [3.6] magnitude for $\rm{M}>10^{10}\,\rm{M}_{\odot}$
    CANDELS sources at $z>$1.4 color-coded by their redshift. Error
    bars are not plotted for clarity but the average values are shown
    in the bottom-left corner. We show \cite{bruzual2003stellar}
    evolutionary tracks of a maximally-old single stellar population
    (SSP) for different redshifts (marked with filled circles) with
    masses $10^{10}$, $10^{11}$ and $5\times10^{11}\,\rm{M}_{\odot}$.
    We also plot evolutionary tracks for extincted
      (A(V)=3~mag) SSPs with age 10 ($\rm{M}=10^{10}\,\rm{M}_{\odot}$)
      and 160~Myr ($\rm{M}=10^{11}\,\rm{M}_{\odot}$) in cyan and
      purple, respectively. The 1~mag attenuation vector at $\rm{z}=4$
      is shown in the upper-left corner. Dashed transverse lines
    depict constant $H$-band magnitudes. Dashed horizontal and
    vertical lines show the color and magnitude limits imposed in our
    selection of the BBG candidates presented in this work and the
    color-selected sample. \textit{Right panel}: Observed-frame
    $H-[3.6]$ vs.  [3.6] for the mass-limited comparison
    sample color-coded by stellar mass.  Error bars are not plotted
    for clarity but the average values are shown in the bottom-left
    corner. These plots can be directly compared to
    Figure~\ref{fig:col_mag_ios} in the main text, where we plot the
    results for our sample of BBGs.}
  \label{fig:col_mag_CANDELS}
\end{figure*}

The sample of red BBGs presented in this work is built up by extremely
faint galaxies at optical wavelengths. In order to understand their
nature, we constructed two comparison samples of IRAC bright ([3.6]
and [4.5]$<$ 24.5\,mag), $H$-band detected ($H\lesssim27$~mag) objects
based on the GOODS-S and GOODS-N CANDELS catalogs.

We first built a sample of optically faint ([$H>25$~mag]), extremely
red ($H-[3.6]>2.5$~mag) sources (hereafter, the color-selected
sample).  This is directly comparable to our selection biases, since
we imposed an IRAC magnitude cut of 24.5 and our sources were (in
principle) $H$-band non-detections, which means that they are fainter
than $\sim$27~mag.  This translates to a color $H-[3.6]>2.5$~mag. In
addition, we constructed a complementary stellar mass limited
(M$>10^{10} M_{\odot}$) sample cut at $z>$3 (mass-limited
sample, hereafter). As we have shown in \S\ref{ssec:SEDphotozprop},
these cuts roughly characterize our sample of BBGs.

In the construction of comparison samples, we discarded sources with
uncertain photometry in the 3.6~$\mu$m, 4.5~$\mu$m and $H$-bands.
Specifically, those sources for which the synthetic (inferred from SED
fitting) and observed photometry did not match
($m_\textrm{synth}-m_\textrm{obs}>0.6~\textrm{mag}$) were removed.
This magnitude difference was identified with contamination from
nearby objects (e.g., spikes from stars). We note that the synthetic
magnitudes could be biased due to the presence of strong emission
lines.  However, for IRAC, strong emission lines in high-redshift
sources typically affect the photometry by less than 0.5~mag (see
\citealp{2013ApJ...777...67S,2013ApJ...763..129S}). We also excluded
sources located in the edges or regions where the F160W exposure time
($<$ 1.5~ks) and limiting magnitude (5$\sigma \sim$27\,mag) are lower.
That low accuracy in the fitting could lead to false photo-z values
and consequently to erroneous predicted properties.  Those sources
with high uncertainties in F160W band ($>0.3$\,mag) were also
discarded. In summary, the mass-limited sample comprises 414
galaxies (i.e, 1.4 sources/arcmin$^{-2}$; 193 in GOODS-N and 221 in
GOODS-S) with $\rm{M}>10^{10}~{\rm M}_{\odot}$, $[3.6]$,
$[4.5]<24.5$~mag and $z>$3, while the color-selected sample comprises
53 galaxies (i.e, 0.18 sources/arcmin$^{-2}$; 20 in GOODS-N and 33 in
GOODS-S) with IRAC $[3.6]$,$[4.5]<24.5$~mag, $H>25$~mag and
$H$-[3.6]$\geqslant$2.5~mag. The median values, 1$^{\rm{st}}$ and
3$^{\rm{rd}}$ quartile of their main properties are summarize in
Table\,\ref{table:CANDELSprop}.

\begin{table}
\setlength\tabcolsep{2pt}
 \normalsize
 \begin{center} 
   \caption{Statistical properties of the CANDELS samples. Median
     values, 1$^\mathrm{st}$ and 3$^\mathrm{rd}$ quartiles of their
     redshift, magnitudes, colors and masses are shown.}
 \label{table:CANDELSprop}
 \begin{tabular}{lcccccc}
 \hline\\
 &{z} &{H} & {[3.6]}& {[4.5]} & {H$-$[3.6]} &{M}\\[-3ex]
 \raisebox{3ex}{Sample} 
 &{ } &{mag} & {mag}& {mag} & {mag}&{M$_{\odot}$}\\[0ex]
 \\ [-2ex] \hline
 \hline\\
 & $3.8^{4.7}_{3.3}$ & $24.9^{25.7}_{24.2}$&$23.3^{23.8}_{22.8}$&$23.2^{23.8}_{22.6}$ & $1.6_{2.1}^{1.1}$ & $10.4_{10.1}^{10.6}$\\[-2ex]
 \raisebox{3ex}{mass-limited}
 & $4.7^{5.3}_{4.1}$ &$ 26.5^{26.8}_{25.9}$&$23.6^{23.9}_{23.1} $&$23.4^{23.9}_{22.7}$&$ 2.8_{3.1}^{2.6}$ & $10.8_{10.4}^{11.1}$\\[-2ex]
 \raisebox{3ex}{Color-selected}\\[-2ex]
 \hline
 \end{tabular} 
 \end{center} 
\end{table}

\subsection{Colors, redshifts and masses of the comparison samples}
\label{Assec:CANDELS_colorMz}

The left panel of Figure~\ref{fig:col_mag_CANDELS} shows massive
($\rm{M}>10^{10}~{\rm M}_{\odot}$) galaxies at $z>$1.4 galaxies from
the CANDELS GOODS-N and GOODS-S catalogs.
Figure~\ref{fig:col_mag_CANDELS} reveals a rough relationship between
the position in a color-magnitude diagram ($H-[3.6]$ vs [3.6]) and the
redshifts and masses of the galaxies: increasing redshifts lead to
redder colors and fainter magnitudes. These red colors can be caused
by either the Balmer Break / D4000 spectral feature
(redshifted beyond the $F160W$ filter at $z>$3) typical of
intermediate an evolved stellar populations, or by a dusty starburst
with significant UV attenuation.  The right panel of
Figure\,\ref{fig:col_mag_CANDELS} shows only massive
($\rm{M}>10^{10}$) galaxies at $z>$3. This mass-limited
sample unveils how masses (roughly traced by the IRAC magnitudes)
increase parallel to the $H$-band magnitude constant lines. For
galaxies with similar stellar mass, fainter ones present redder
colors, while for a given $[3.6]$ magnitude, redder colors will
correspond to more massive galaxies and higher redshifts. As indicated
by the evolutionary tracks for maximally-old quiescent galaxies
($\rm{M}=10^{10}$, $5\times10^{10}$, $10^{11}$ and
$5\times10^{11}\rm{M}_{\odot}$), a red $H-[3.6]~\gtrsim2.5$~mag color
is a good proxy to identify massive evolved galaxies or massive dusty
galaxies at $z>3$ (BBGs). Young star-forming galaxies would
  enter the $H-[3.6]~\gtrsim2.5$~mag region at $z\sim2$ for
  attenuations of $A(V)\gtrsim3$~mag. We note that those $z<3$ galaxies
  with very high attenuations would most probably be detected by MIPS.
  We also remark that at $z<3$, this kind of very dusty starburst
  would be characterized by relatively bright IRAC magnitudes
  ($[3.6]\lesssim22$~mag), and very few sources are found in that
  region of the color-magnitude diagram for the mass-selected sample.
  With all this in mind, we conclude that the color and magnitude cuts
  for both the BBGs and color-selected sample (shown with the dashed
  vertical and horizontal lines) are effective at selecting massive
  ($\rm{M} \gtrsim 10^{10}$) galaxies at high redshift ($z\gtrsim3$),
  with little contamination from lower redshift galaxies or low mass
  objects. It is also important to notice that most of the
  color-selected sources are indeed included in the mass-selected
  sample.

\begin{figure}
  \begin{center}
    \includegraphics[width=.57\textwidth]{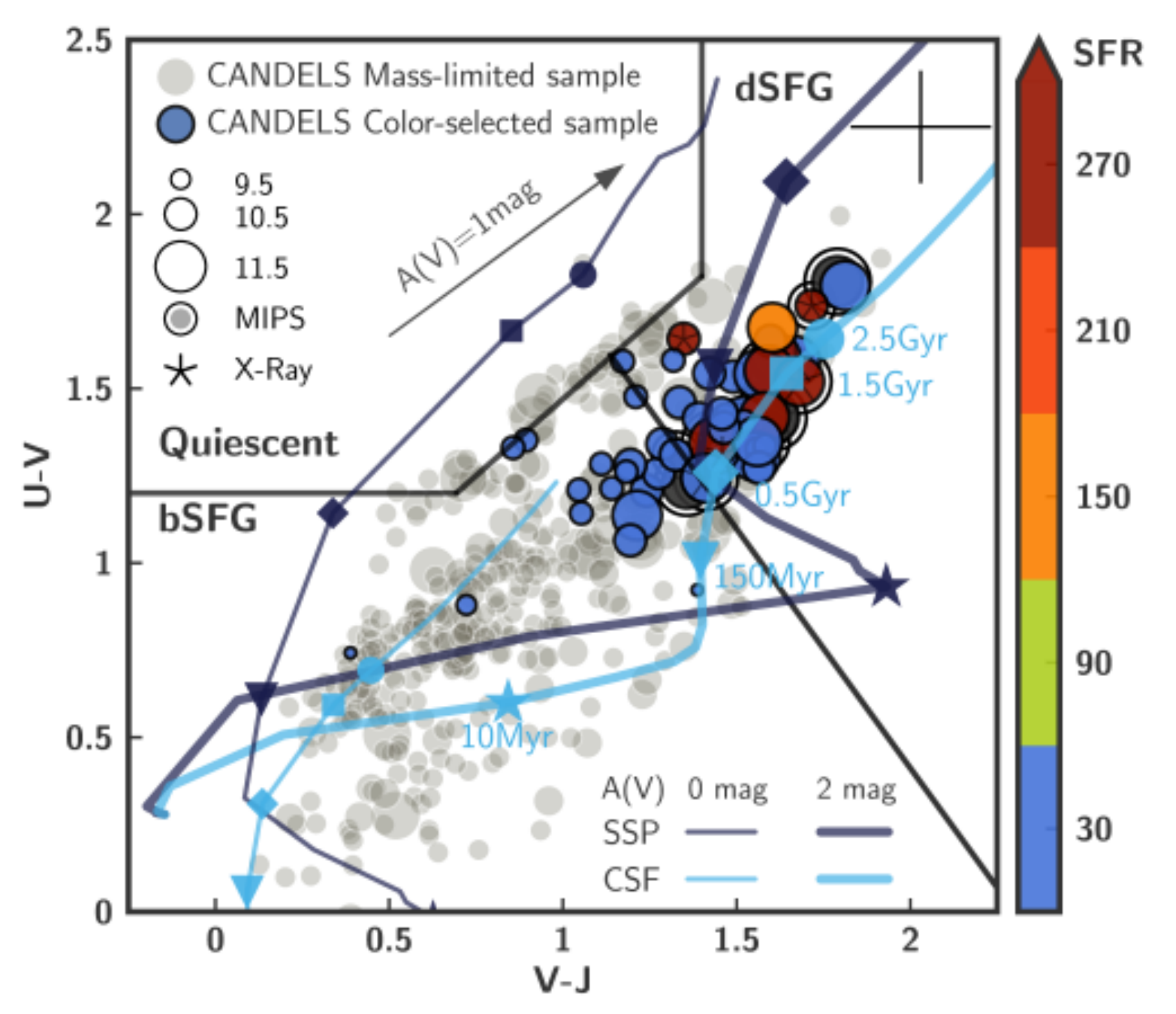}
  \end{center}
  \caption{Rest-frame $U-V$ vs. $V-J$ colors for the CANDELS
    color-selected sample color-coded by SFR and sized by stellar
    mass. SFR lower limits are shown in dark grey. The
    mass-limited sample is plotted in grey with the size also
    scaled according their masses.  MIPS detected galaxies are
    surrounded with another circle and X-ray detected galaxies show a
    star inside the symbol.  The \citet{Whitaker2011} upper boundary
    (black wedge) separates quiescent galaxies (top left) from SFGs
    (bottom). The black diagonal line denotes an additional criterion
    to separate blue (bSFGs) from dusty (dSFGs) star-forming galaxies.
    Error bars are not plotted for clarity but the average values are
    shown next to the right boundary of the quiescent region.  The 0
    and 2\,magnitude extincted single stellar population (SSP) and
    constant star formation (CSF) models are shown with thin and thick
    solid lines, respectively.  The 1~mag attenuation vector computed
    assuming a \cite{2000ApJ...533..682C} reddening law is also
    shown.}
    \label{fig:UVJ_CANDELS}
\end{figure}

We also assessed the effect of mass and SFR in the $U-V$ and $V-J$
rest-frame synthetic colors (Figure\,\ref{fig:UVJ_CANDELS}). Both
colors tend to increase parallel to the attenuation vector.  The $UVJ$
diagram has been proven to be very efficient to isolate the red
sequence of galaxies at $z<3$. We use the definition found in
\citet{Whitaker2011} to differentiate quiescent galaxies from those
actively forming stars:

\begin{eqnarray}
	(U-V) > 1.2  \nonumber \\
	(V-J) < 1.4  \nonumber \\
	(U-V) > 0.88 \times (V-J) + 0.6 
\end{eqnarray}

Within the $UVJ$ star-forming region, there exists a wide range
of colors that could be caused by either differences in the amount
of reddening, and/or differences in the SFHs. Several criteria have
been proposed to distinguish between red an blue SFGs using an
additional $U-V$ and/or $V-J$ color criteria (i.e.
\citealp{2007ApJ...655...51W, 2014ApJ...787L..36S,
2014ApJ...783L..14S,2014ApJ...796...35F}). The reddening vector,
computed by assuming a \cite{2000ApJ...533..682C} reddening law,
suggests that the large range of colors observed in galaxies can be
explained by different amounts of extinction. But there also exists
of a degeneracy between age, attenuation, and SFR as suggested by
earlier shallower surveys (e.g., \citealp{2010ApJ...725.1277M}). In
order to understand the nature of our sample of BBGs, we have tested
in Figure\,\ref{fig:UVJ_CANDELS} the behavior in the $UVJ$ diagram
of single stellar population (SSP) and constant SFR (CSF) models
with different amounts of extinction.  Figure\,\ref{fig:UVJ_CANDELS}
also shows the implications of our definition of bSFGs and dSFGs.
Dusty SFGs present $U-V$ and $V-J$ colors which can be reproduced by
models with $A(V)=2$~mag of attenuation and ages older than
$\sim$10~Myr and $\sim500$~Myr for an SSP or CSF SFH, respectively.
Younger or less extincted systems would be qualified as bSFGs.
Figure\,\ref{fig:UVJ_CANDELS} shows that in order for an object to
enter the quiescent galaxy wedge, it must harbor SSPs (or short
star-forming burst, i.e., SFHs which differ significantly from a CSF
model) with ages older than 0.5~Gyr and no dust. If some dust is
present, younger galaxies might be found in the quiescent locus, and
old ($>2$~Gyr) galaxies might leave that region and be found in the
upper-right corner of the $UVJ$ diagram.

Not surprisingly, analyzing the comparison samples in terms of the
$UVJ$ colors, we find that the mass-limited sample presents
significantly bluer colors and lower masses than the color-selected
one. If we only consider the color-selected sample, the most massive
sources are those with the reddest colors (as also reported by
\citealt{2010ApJ...713..738W}) and mainly correspond to dSFGs. We find
22 bSFGs, out of which 1 ($5\%$) is detected by MIPS 24~$\mu$m. We
have 31 dSFGs in the color-selected sample, 11 of them ($35\%$)
showing IR emission. The upper left sector that delimits the quiescent
region contains a few galaxies from the mass-limited sample
but none from the color-selected sample.  However, their closeness to
the boundary together with the high uncertainties makes its
classification quite uncertain.  Moreover, there is a small number of
galaxies in the color-selected sample that lie very close to the
boundary and are characterized by very low SFRs and high mass-weighted
ages. Therefore, they might be either quiescent or post-starburst
galaxies.
 
Not surprisingly, most of the FIR detected galaxies (26\%) in the
color-selected sample are found among the most massive sources.  They
are located within the dSFG locus and present very red $UVJ$ colors.
Considering the possible presence of an obscured AGN, we note the
existence of 6 X-ray emitters also in the same area. Among them, 3
($50\%$) present MIPS 24~$\mu$m emission and 2 are also detected by
PACS and SPIRE, or even at 850 or 1200~$\mu$m.  Compared to the
mass-limited comparison sample, it is evident and expectable
that the $H-[3.6]>2.5$~mag color cut biases the selection against
bSFGs.

% % % % % % % % % % % % % % % % % % % % % % % % % % % % % % % % % % % % % % % % % % % % % % % % % % % % %

%\setcounter{section}{0}
\setcounter{figure}{0}
\setcounter{table}{0}

\section{SEDs and postage stamps of BBG candidates}
\label{A:SEDs}

\begin{figure*}[h]
  \caption{\label{fig:SEDs_A} Postage stamps
    ($12\arcsec\times12\arcsec$) and SEDs for the 33 BBGs presented in
    this work. For each galaxy, on the left we plot images taken by
    HST in the $H$-band, by {\it Spitzer} in the $[3.6]$ and 24~$\mu$m
    bands, the stack with all HST data (including ACS and WFC3
    images), and the stack for SHARDS data (in GOODS-N). Green circles
    in all stamps show sources included in the public catalogs
    released by the CANDELS team (Guo et al. 2013). The red circle
    marks the BBG. On the right, we show SEDs including all measured
    fluxes and upper limits. Bands (most probably) dominated by
    stellar emission are plotted with black dots, and bands probing
    the dust emission are plotted with grey dots. These SEDs have been
    fitted to stellar population models (using 2 codes,
    \texttt{synthesizer} and \texttt{FAST}) whose main physical
    properties are written in the plot.  Dust emission is also fitted
    with Rieke et al.  (2009) templates when the galaxy is detected by
    MIPS and/or Herschel. The inset in these SED plots shows the zPDF
    obtained for the 3 photometric methods (based on the different
    apertures written in the legend) explained in Appendix~A, except
    for extended sources (see Table~A.1), which only shows one zPDF.}
	\begin{minipage}[b]{0.44\linewidth}
	 \centering
			\begin{minipage}[b]{0.315\linewidth}
				\includegraphics[width=1.\linewidth]{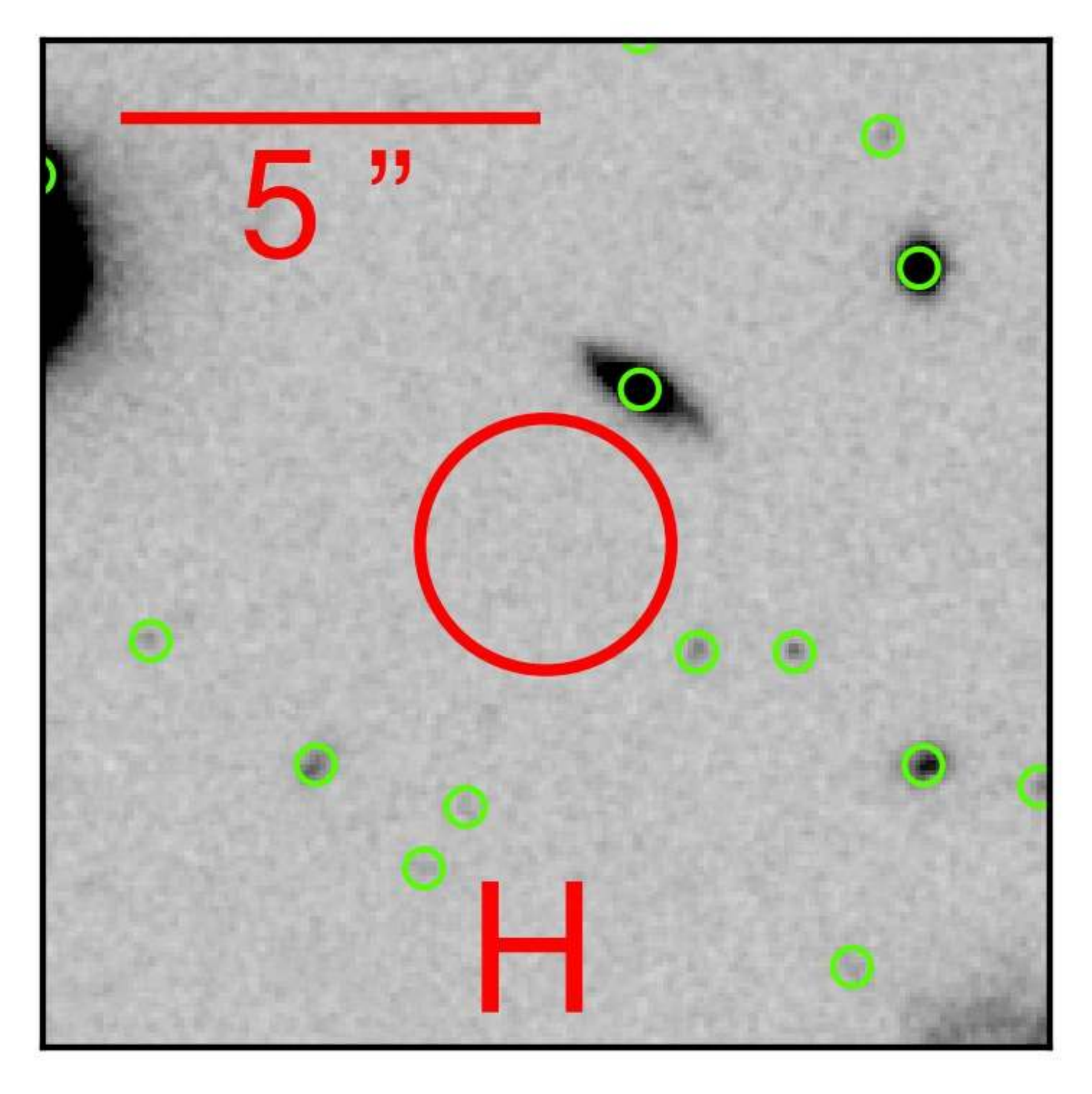}
			\end{minipage}\setlength\tabcolsep{1pt}
			\begin{minipage}[b]{0.315\linewidth}
				\includegraphics[width=1.\linewidth]{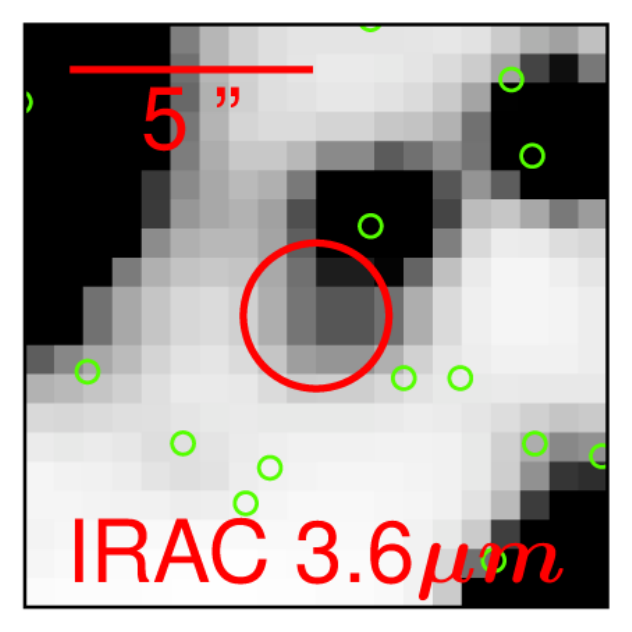}		
			\end{minipage}	
			\begin{minipage}[b]{0.315\linewidth}
				\includegraphics[width=1.\linewidth]{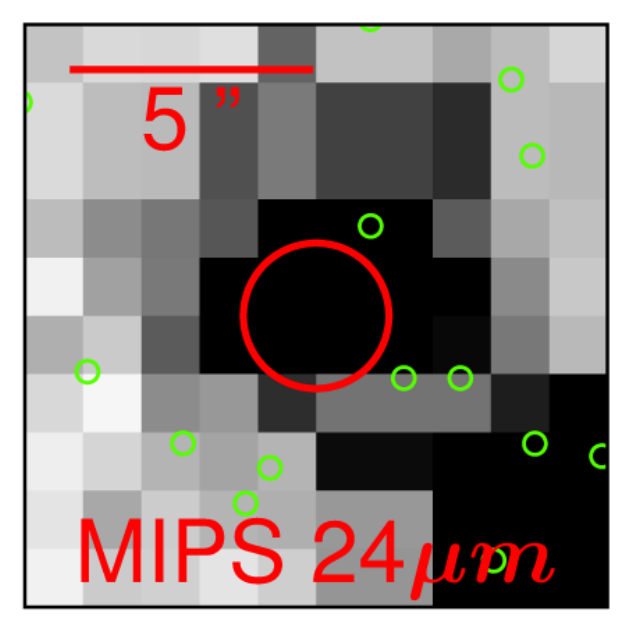}		
			\end{minipage}			
			\includegraphics[width=.49\linewidth]{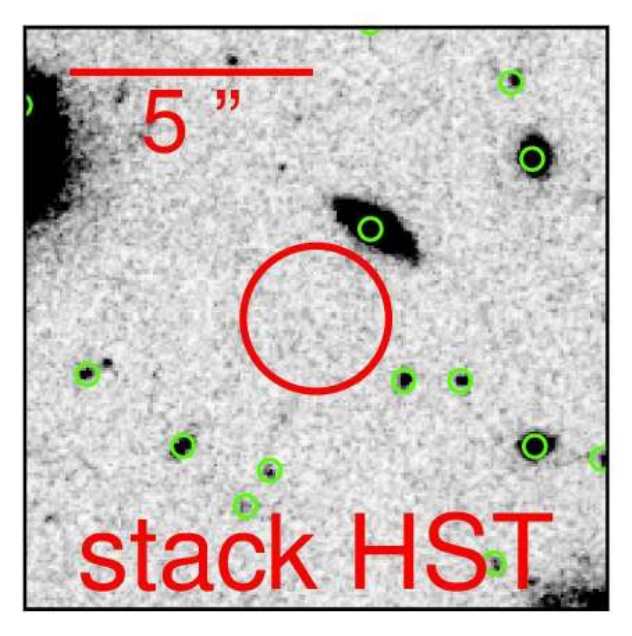}
			\includegraphics[width=.49\linewidth]{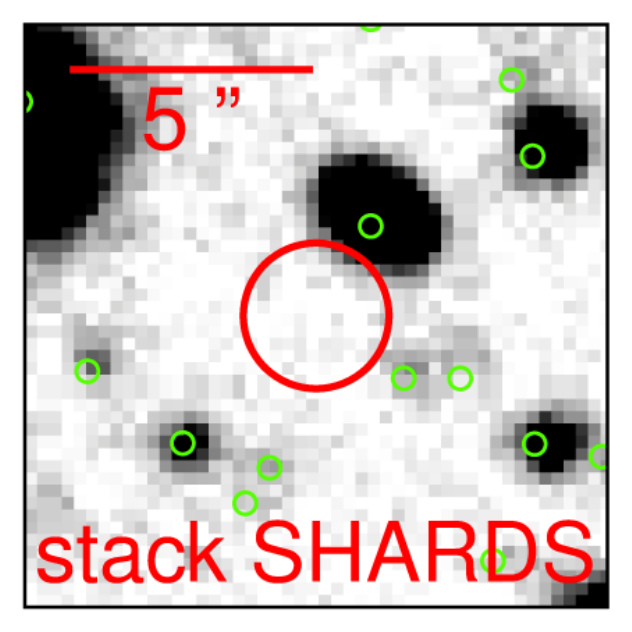}
			\centering
	\end{minipage}
	\quad
	\begin{minipage}[b]{0.52\linewidth}
	\begin{center}
		\includegraphics[width=1.\linewidth]{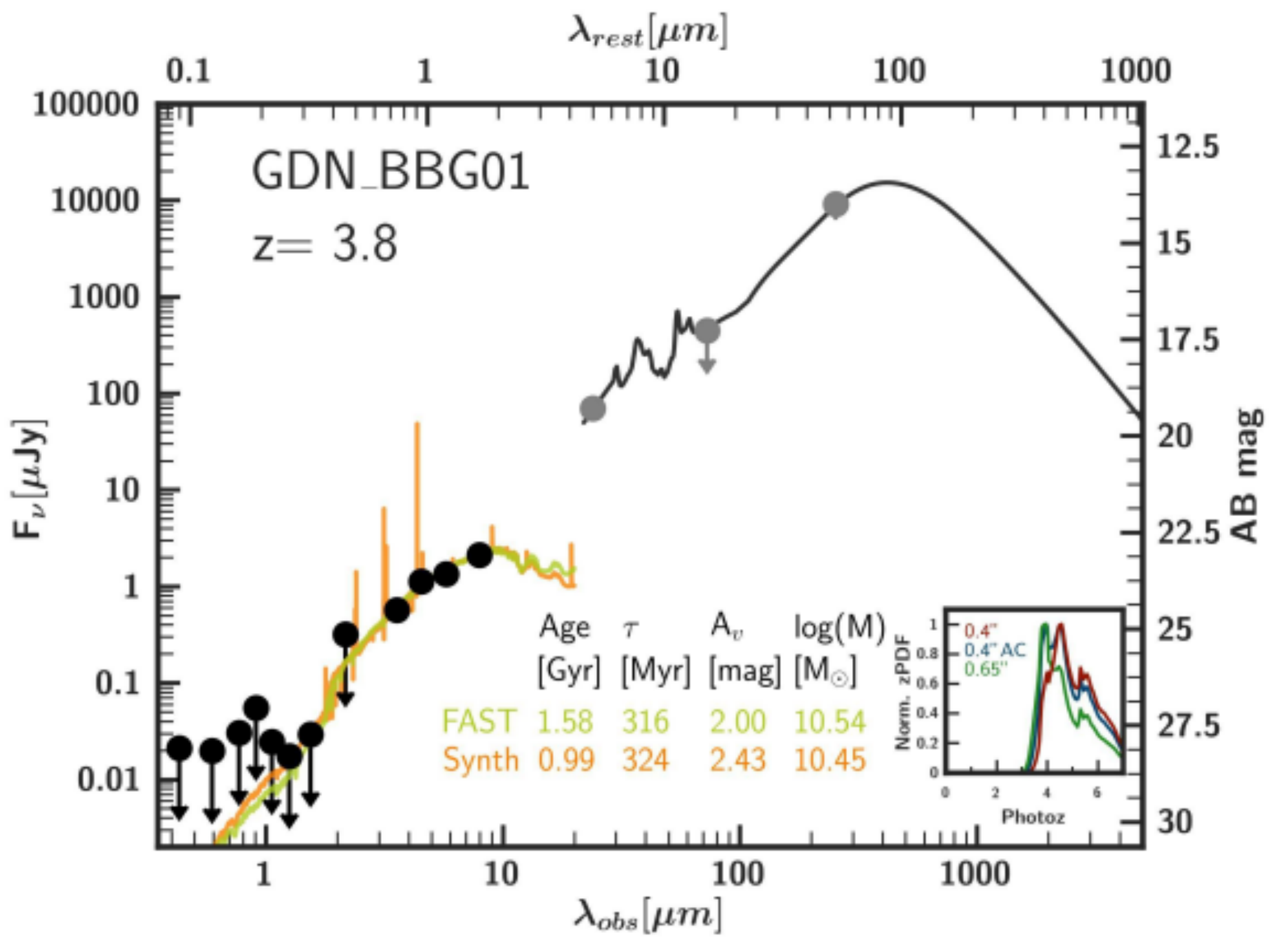}
	\end{center}
	\end{minipage}
\end{figure*}

In Figure~\ref{fig:SEDs_A}, we present SEDs, stellar and dust emission
models, and postage stamps in several bands (including stacked images)
for all the BBGs presented in this paper.

% source 2 gdn
\begin{figure*}
	\begin{minipage}[b]{0.44\linewidth}
		\begin{minipage}[b]{0.315\linewidth}
			\includegraphics[width=1.\linewidth]{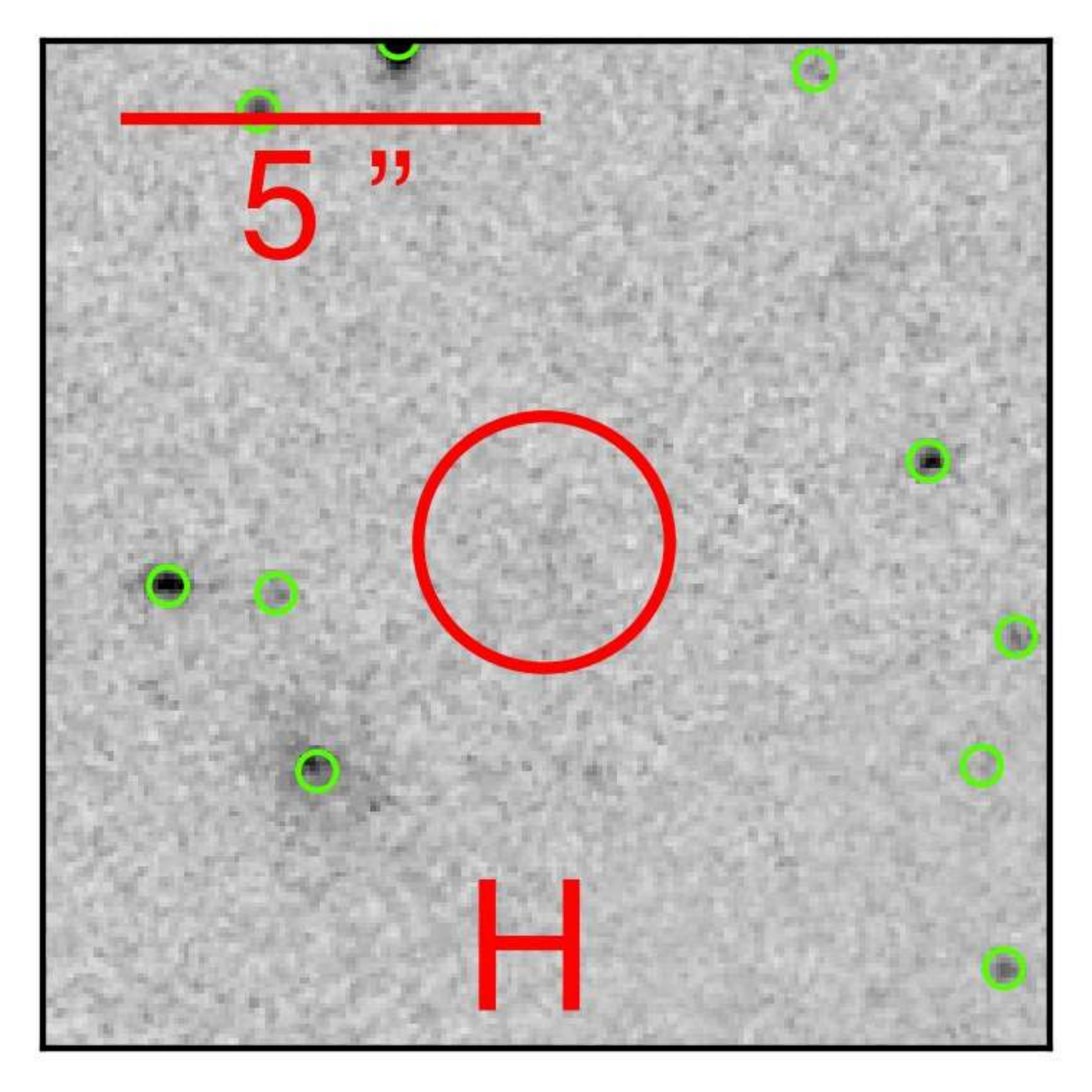}
		\end{minipage}
		\begin{minipage}[b]{0.315\linewidth}
			\includegraphics[width=1.\linewidth]{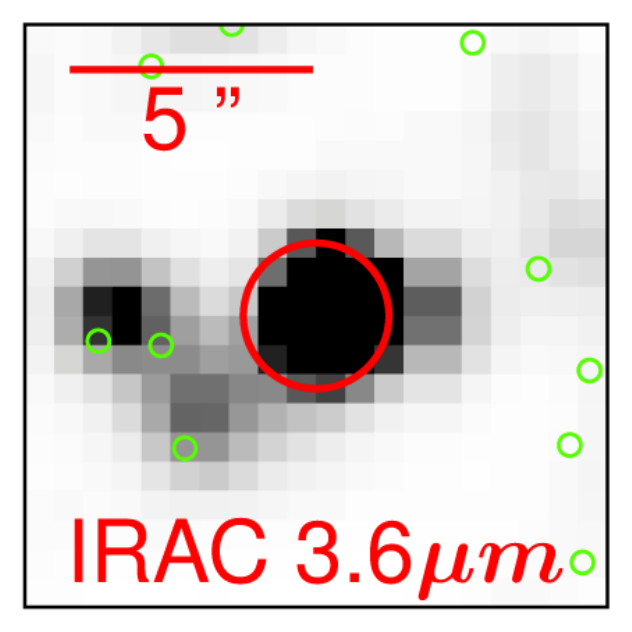}		
		\end{minipage}	
		\begin{minipage}[b]{0.315\linewidth}
			\includegraphics[width=1.\linewidth]{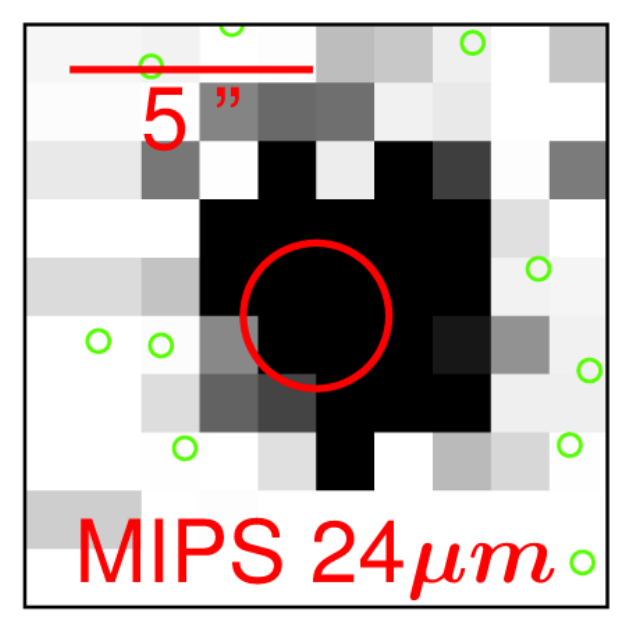}		
		\end{minipage}			
		\includegraphics[width=.49\linewidth]{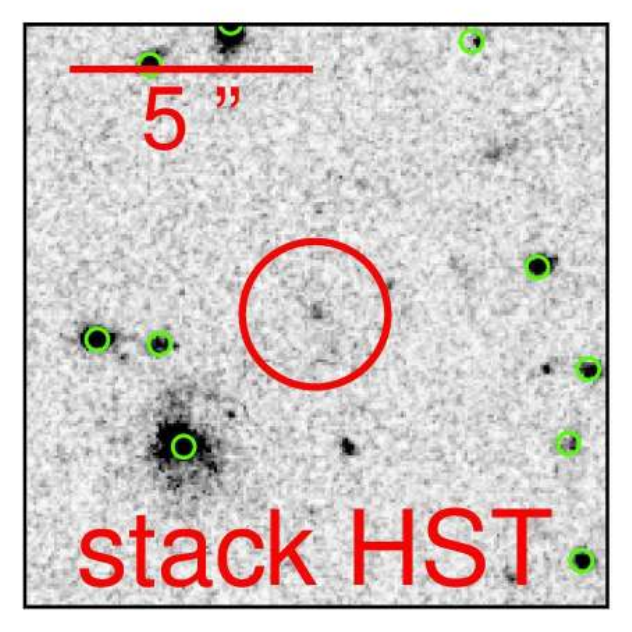}
		\includegraphics[width=.49\linewidth]{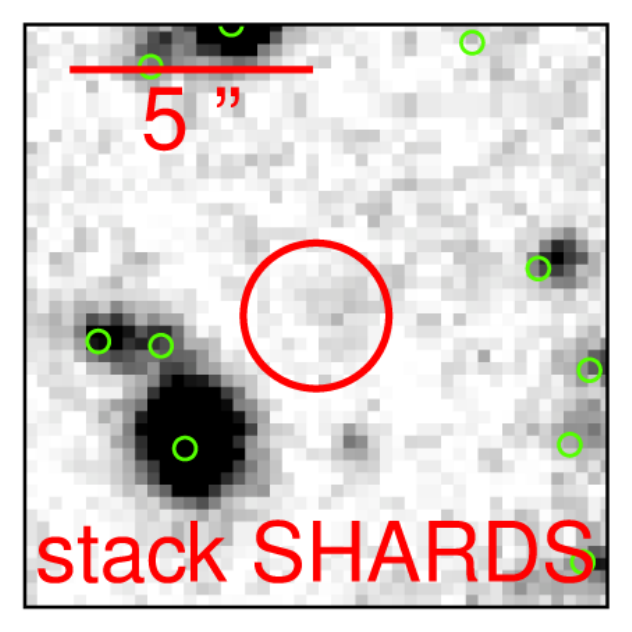}
		\centering
	\end{minipage}
	\quad
	\begin{minipage}[b]{0.52\linewidth}
		\includegraphics[width=1.\linewidth]{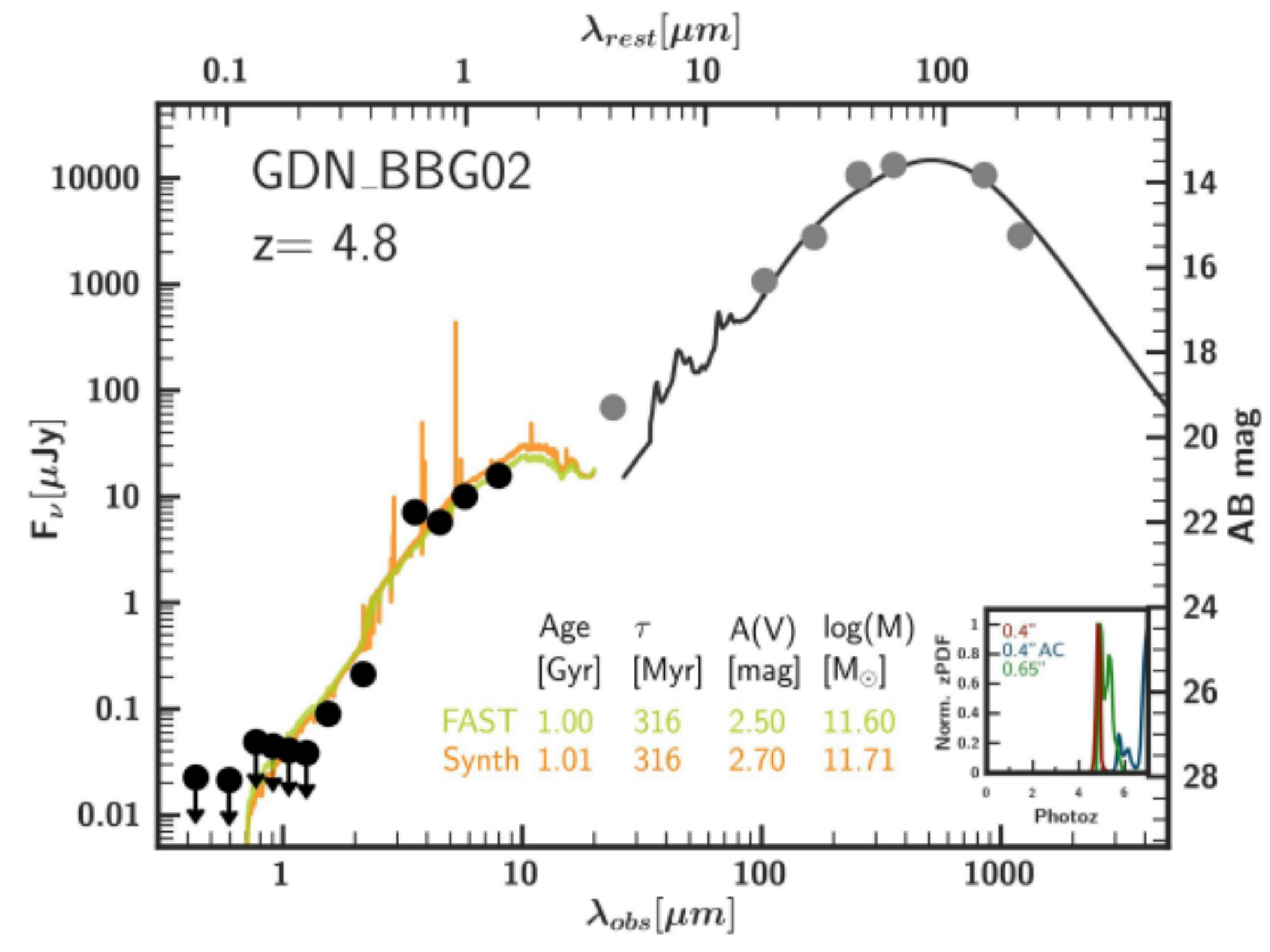}
		\centering
	\end{minipage}
% Source 3}
	\begin{minipage}[b]{0.44\linewidth}
		\begin{minipage}[b]{0.315\linewidth}
			\includegraphics[width=1.\linewidth]{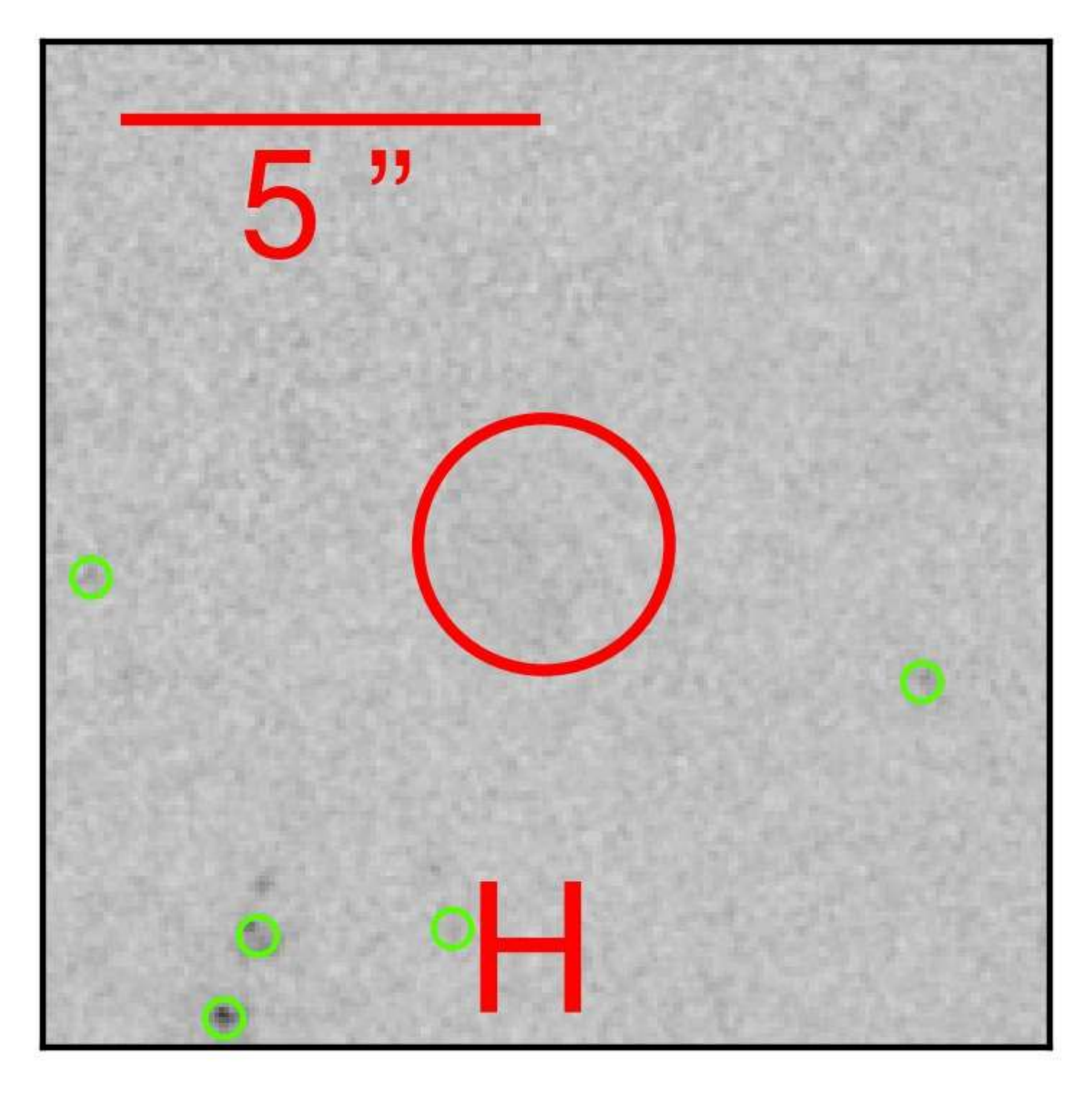}
		\end{minipage}
		\begin{minipage}[b]{0.315\linewidth}
			\includegraphics[width=1.\linewidth]{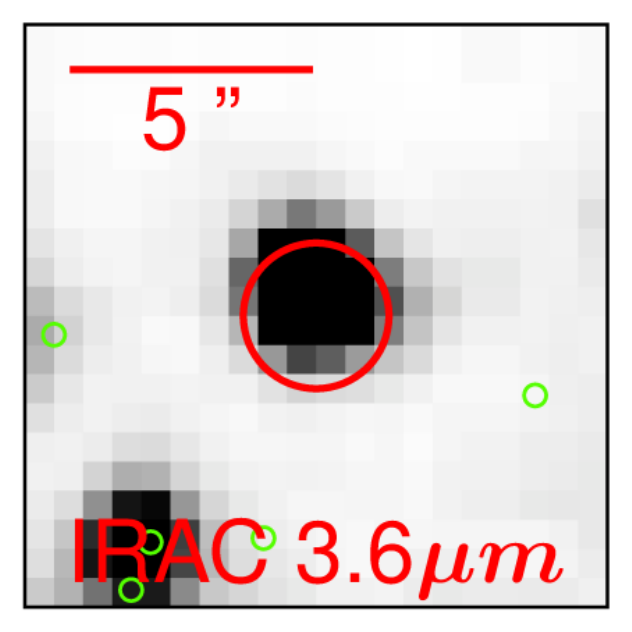}		
		\end{minipage}	
		\begin{minipage}[b]{0.315\linewidth}
			\includegraphics[width=1.\linewidth]{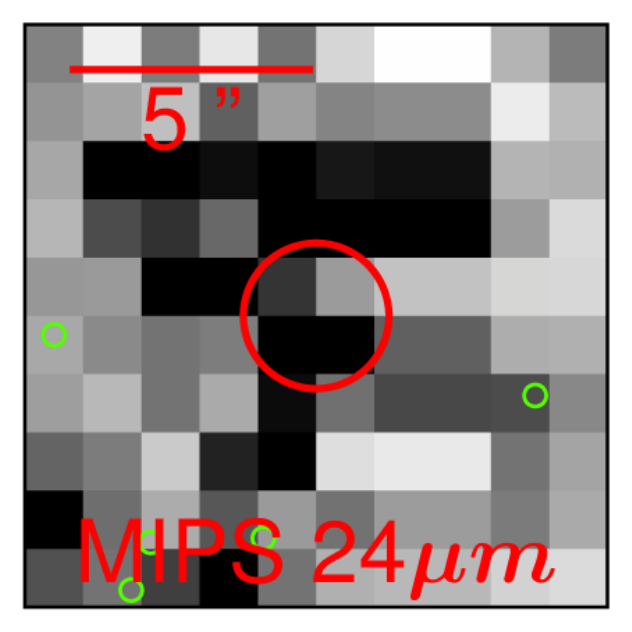}		
		\end{minipage}			
		\includegraphics[width=.49\linewidth]{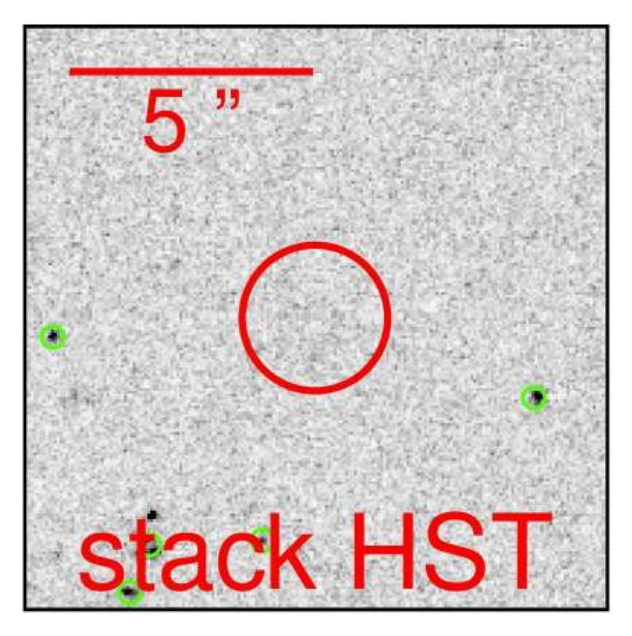}
		\includegraphics[width=.49\linewidth]{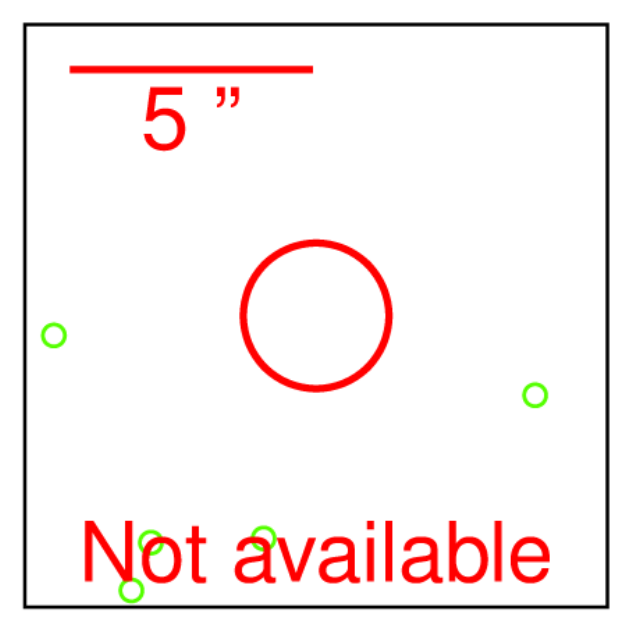}
		\centering
	\end{minipage}
	\quad
	\begin{minipage}[b]{0.52\linewidth}
		\includegraphics[width=1.\linewidth]{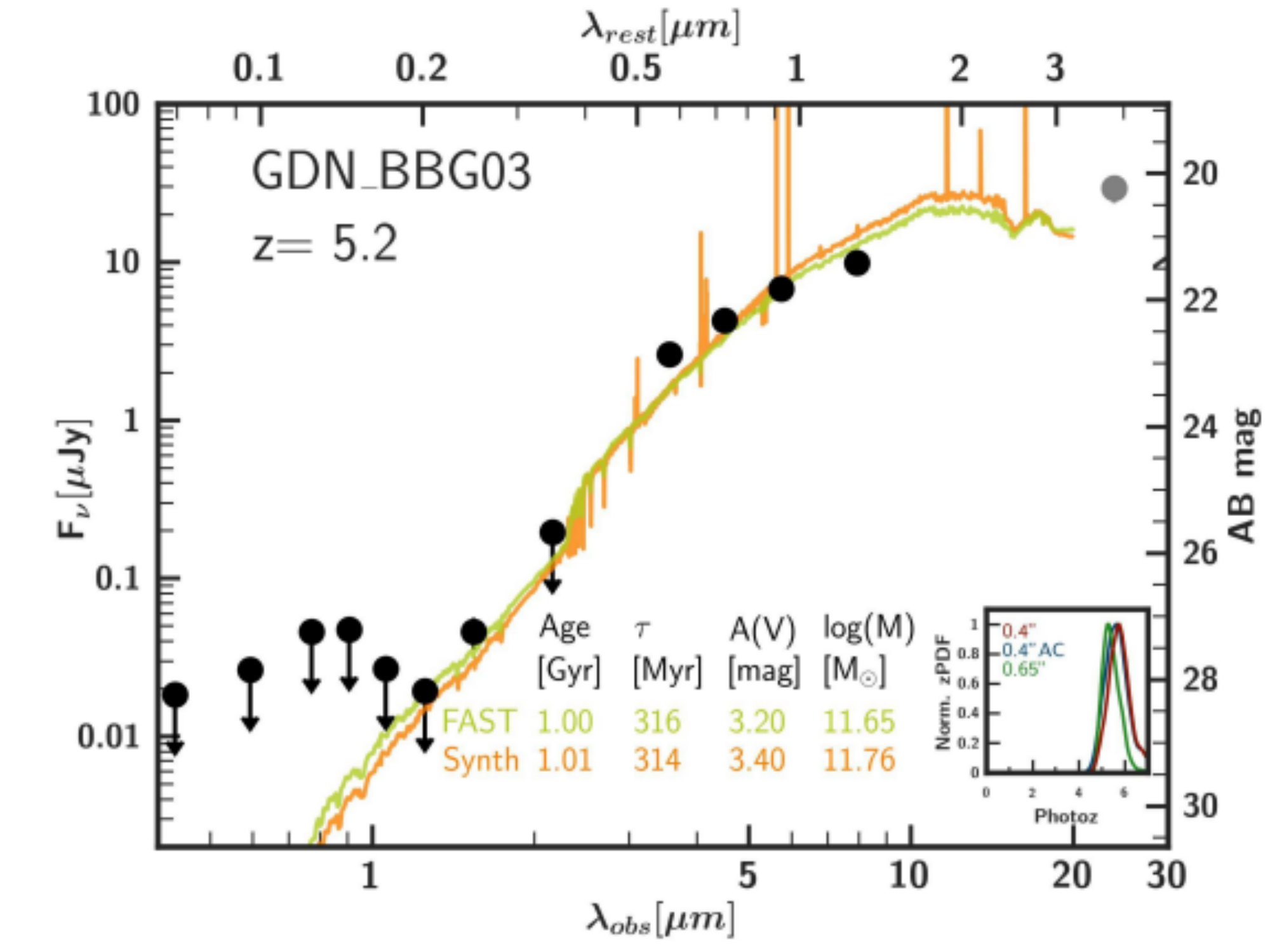}
		\centering
	\end{minipage}

% Source 4}

	\begin{minipage}[b]{0.44\linewidth}
	 \centering
			\begin{minipage}[b]{0.315\linewidth}
				\includegraphics[width=1.\linewidth]{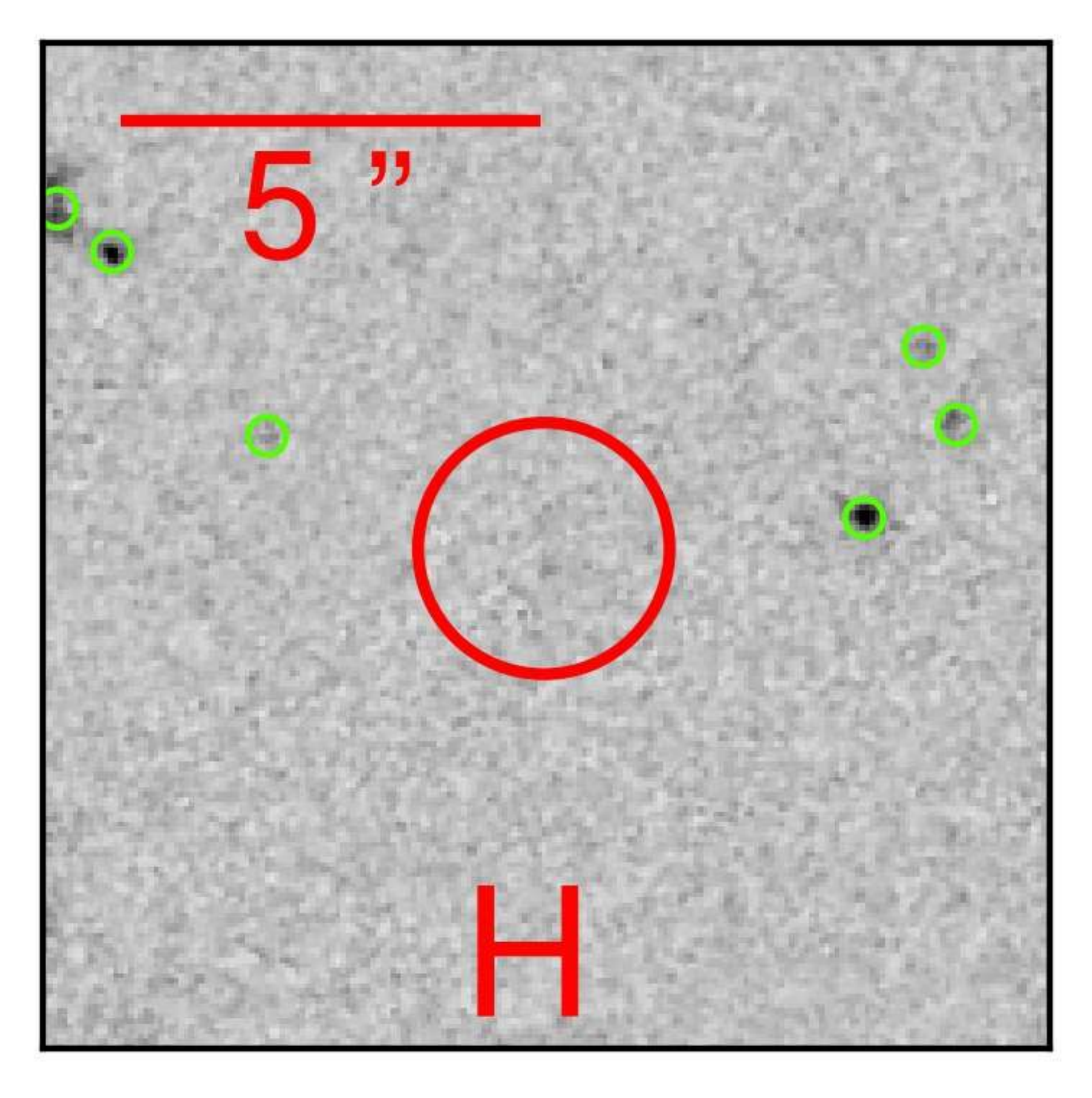}
			\end{minipage}
			\begin{minipage}[b]{0.315\linewidth}
				\includegraphics[width=1.\linewidth]{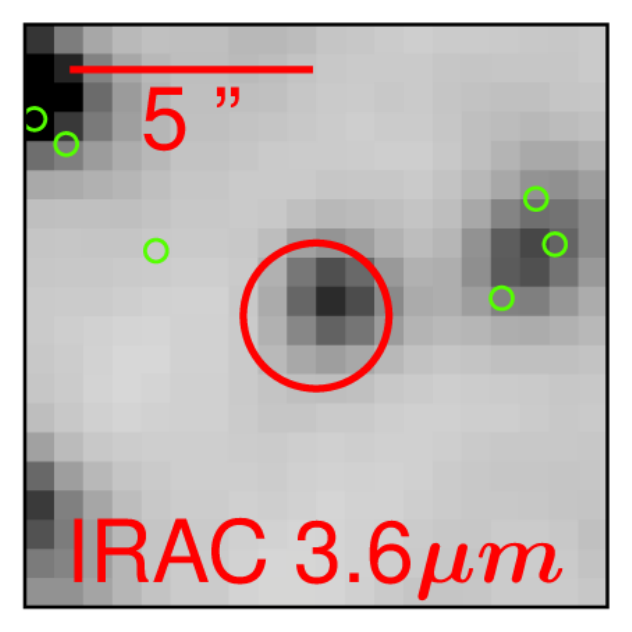}		
			\end{minipage}	
			\begin{minipage}[b]{0.315\linewidth}
				\includegraphics[width=1.\linewidth]{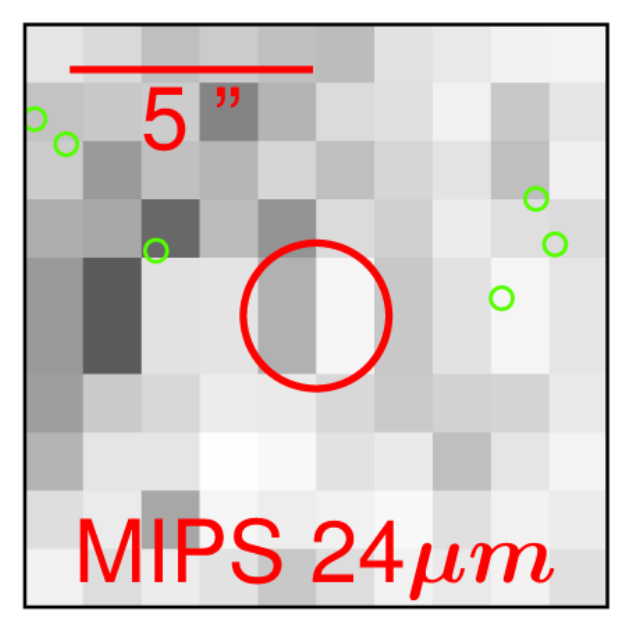}		
			\end{minipage}			
			\includegraphics[width=.49\linewidth]{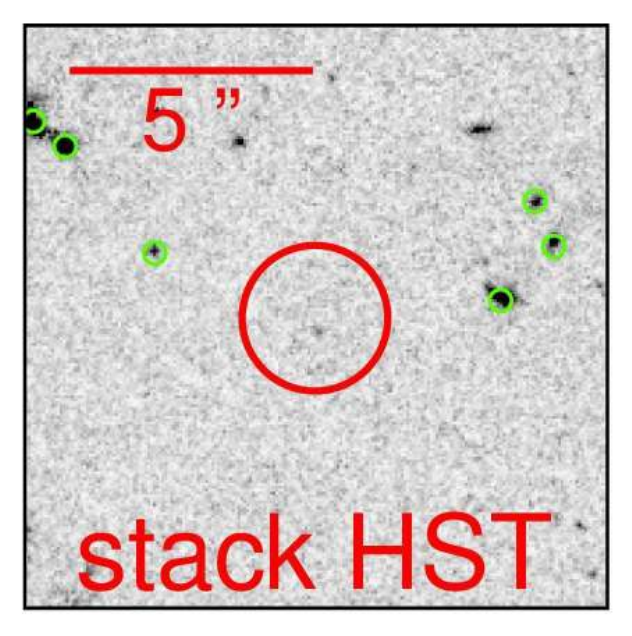}
			\includegraphics[width=.49\linewidth]{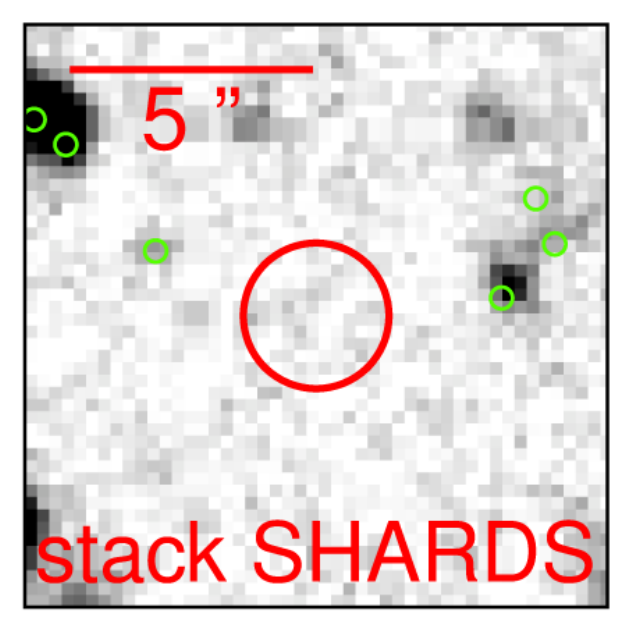}
			\centering
	\end{minipage}
	\quad
	\begin{minipage}[b]{0.52\linewidth}
		\includegraphics[width=1.\linewidth]{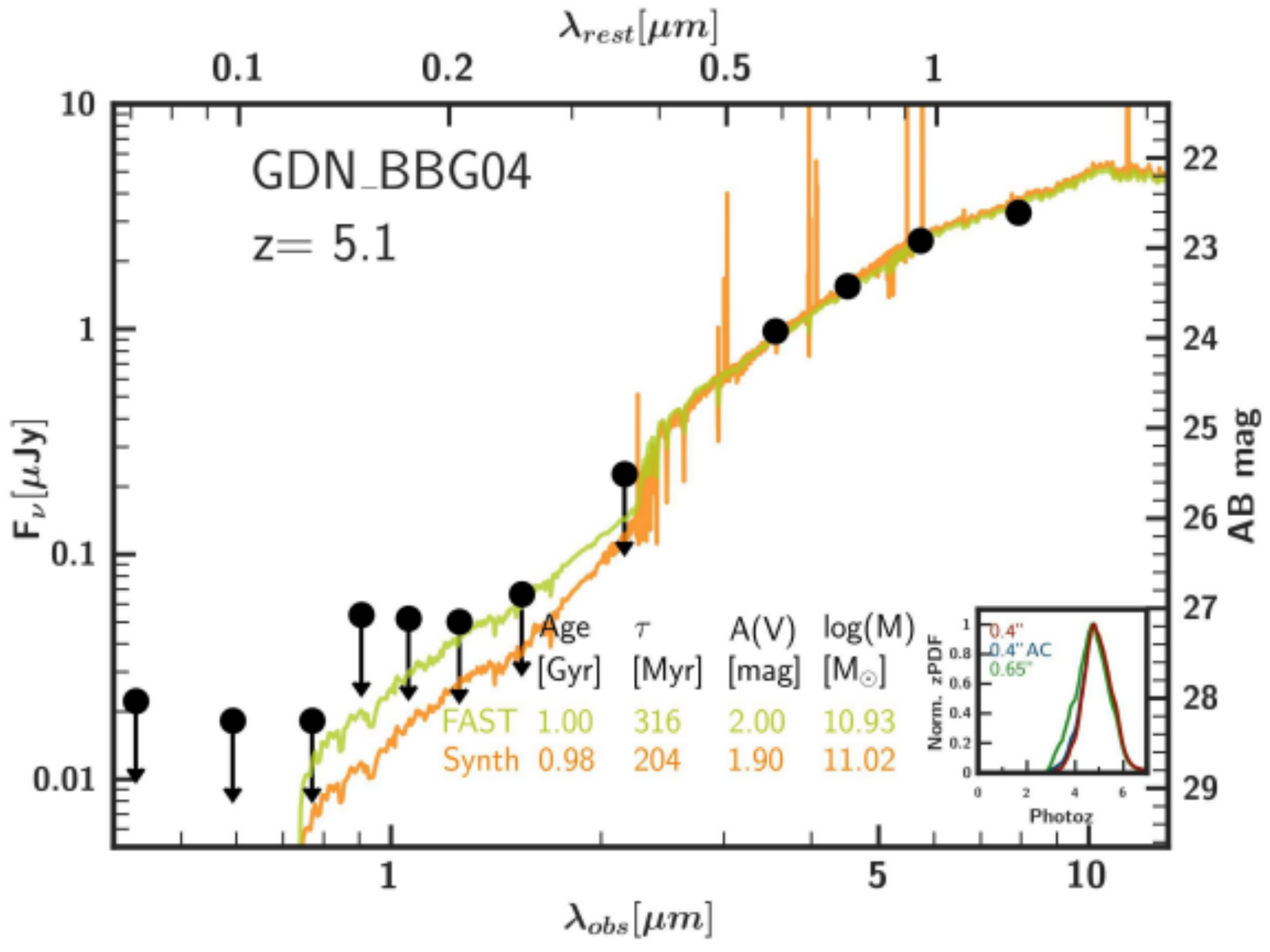}
		\centering
	\end{minipage}
\end{figure*}

% Source 5}
\begin{figure*}
	\begin{minipage}[b]{0.44\linewidth}
	 \centering
			\begin{minipage}[b]{0.315\linewidth}
				\includegraphics[width=1.\linewidth]{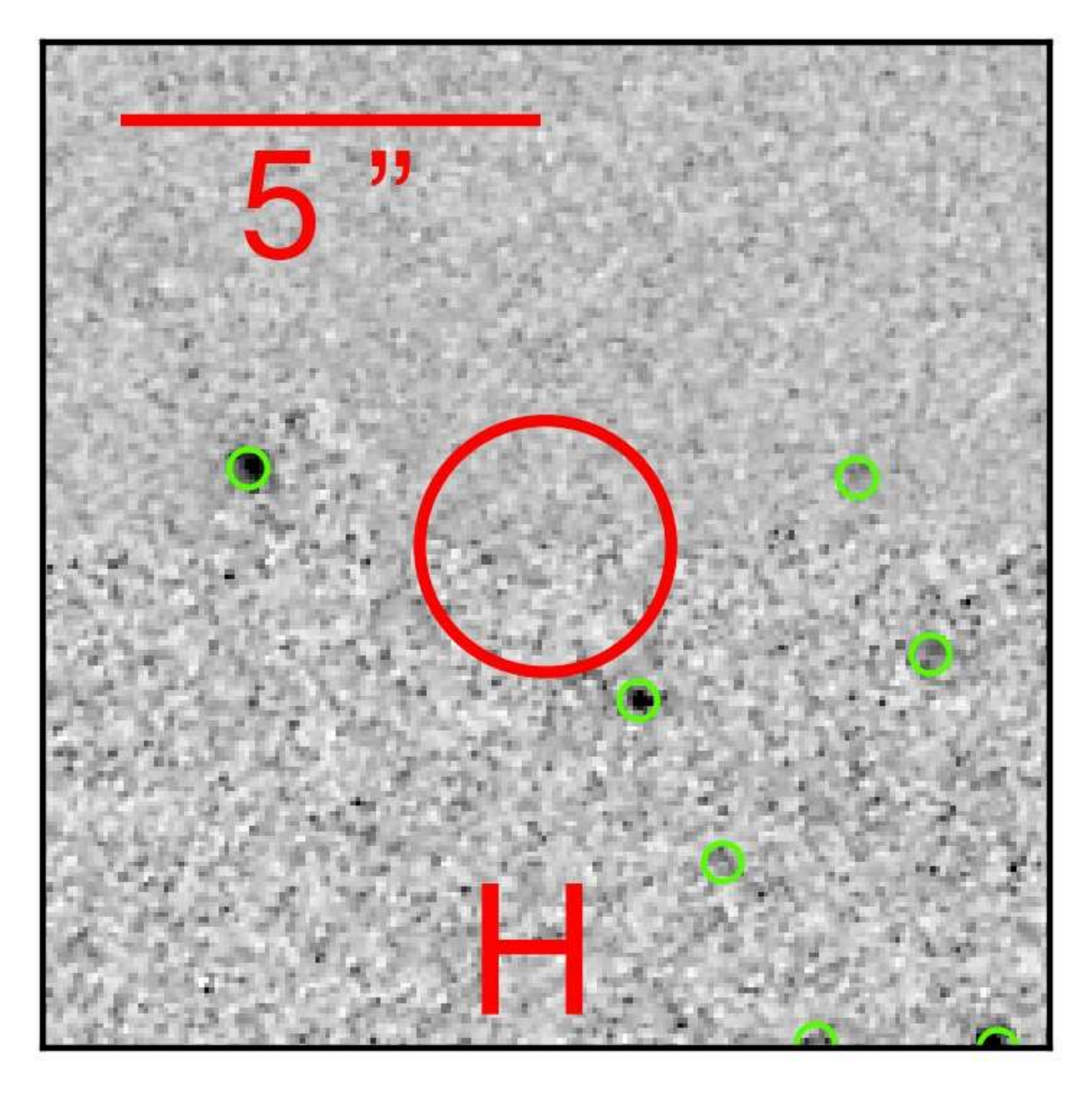}
			\end{minipage}
			\begin{minipage}[b]{0.315\linewidth}
				\includegraphics[width=1.\linewidth]{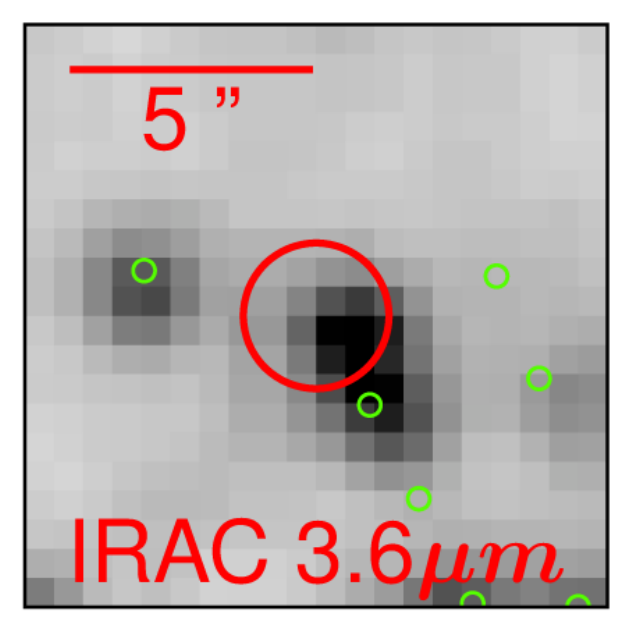}		
			\end{minipage}	
			\begin{minipage}[b]{0.315\linewidth}
				\includegraphics[width=1.\linewidth]{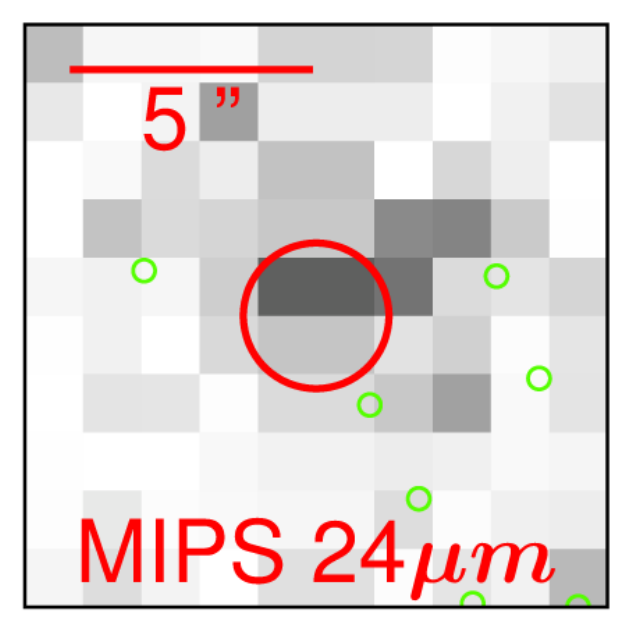}		
			\end{minipage}			
			\includegraphics[width=.49\linewidth]{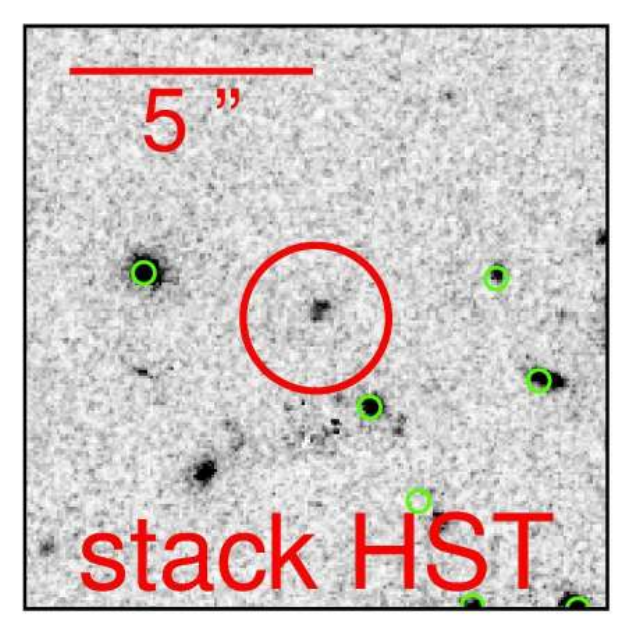}
			\includegraphics[width=.49\linewidth]{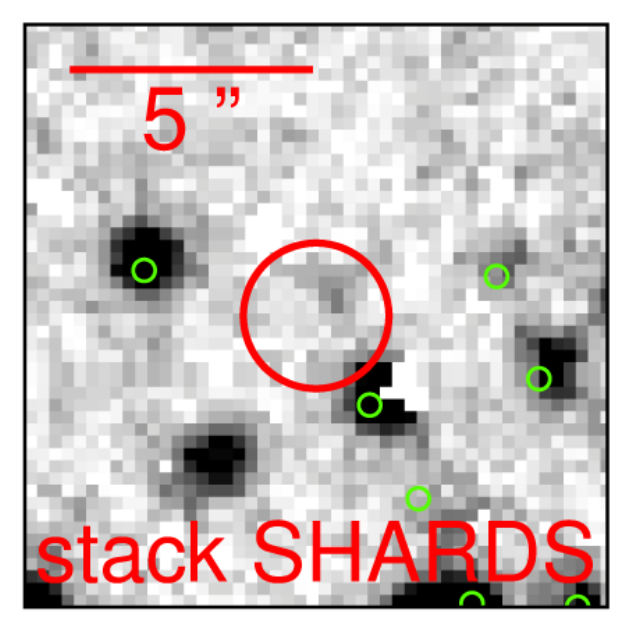}
			\centering
	\end{minipage}
	\quad
	\begin{minipage}[b]{0.52\linewidth}
		\includegraphics[width=1.\linewidth]{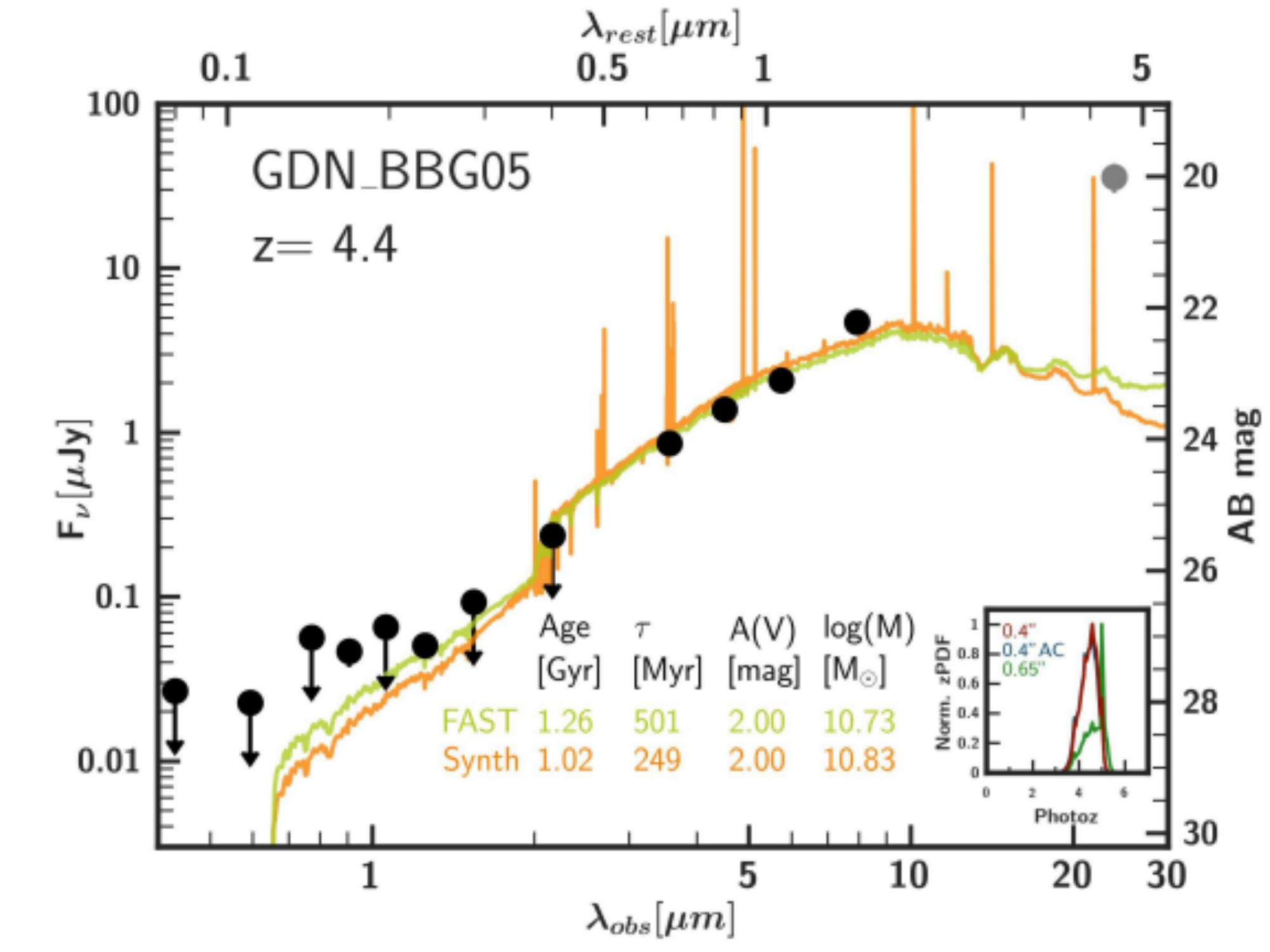}
		\centering
	\end{minipage}

% Source 6}

	\begin{minipage}[b]{0.44\linewidth}
	 \centering
			\begin{minipage}[b]{0.315\linewidth}
				\includegraphics[width=1.\linewidth]{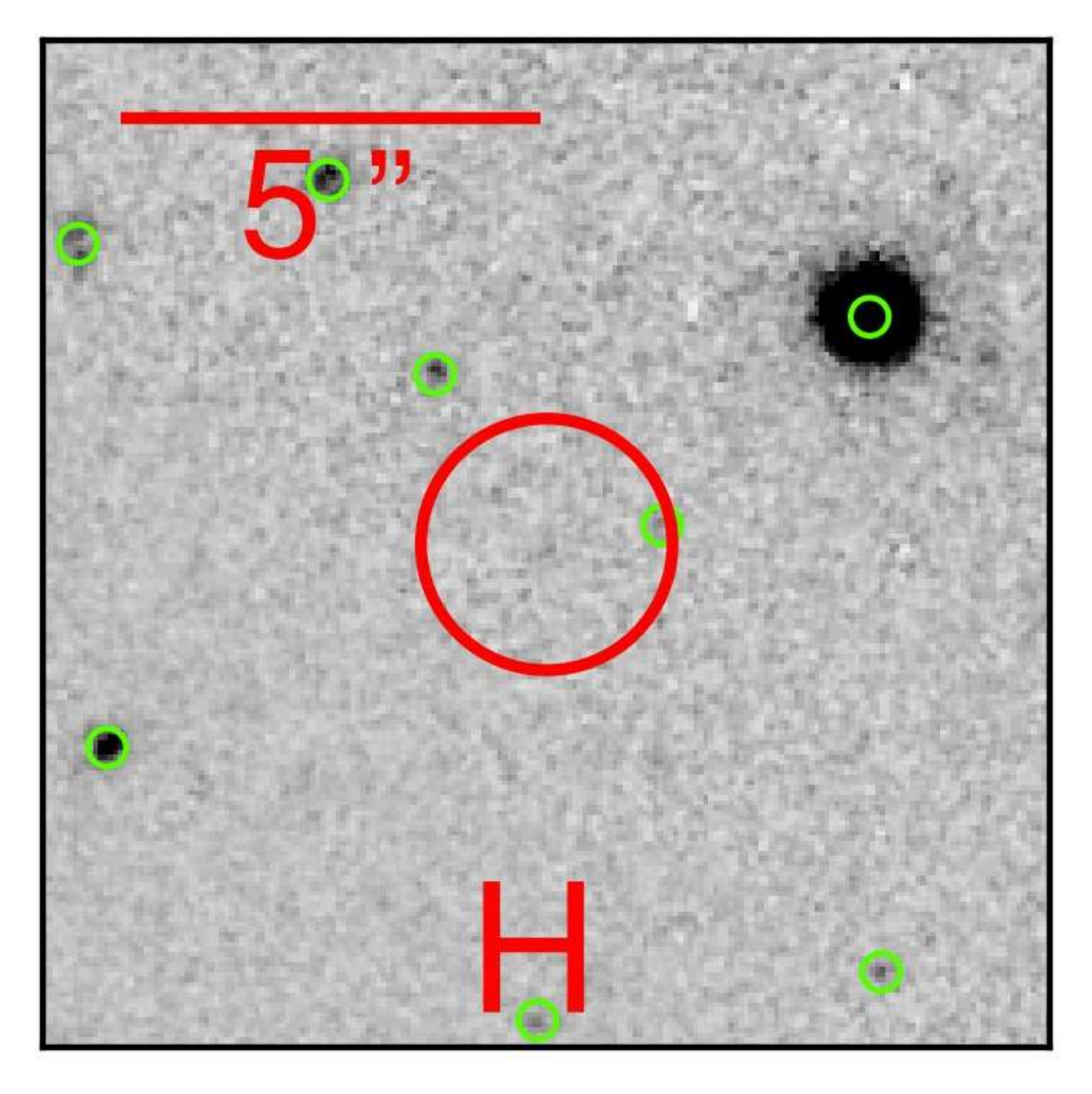}
			\end{minipage}
			\begin{minipage}[b]{0.315\linewidth}
				\includegraphics[width=1.\linewidth]{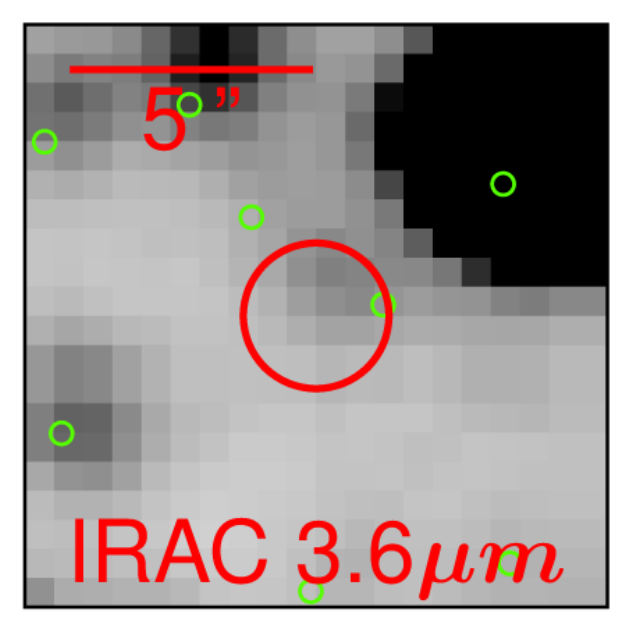}		
			\end{minipage}	
			\begin{minipage}[b]{0.315\linewidth}
				\includegraphics[width=1.\linewidth]{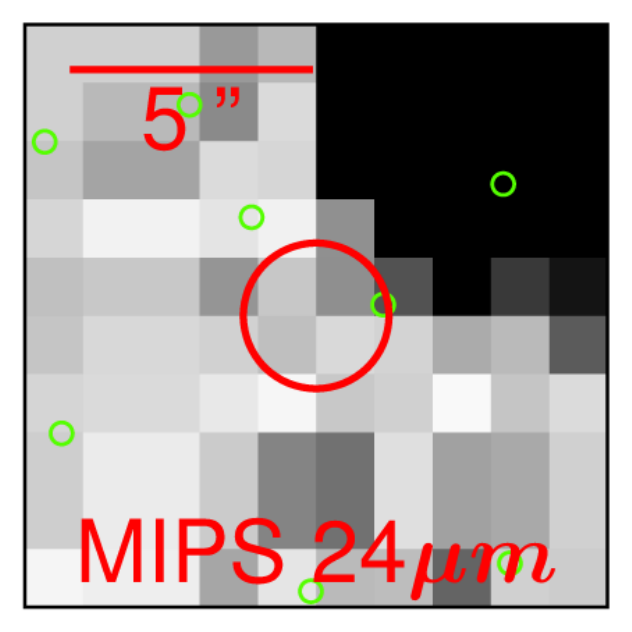}		
			\end{minipage}			
			\includegraphics[width=.49\linewidth]{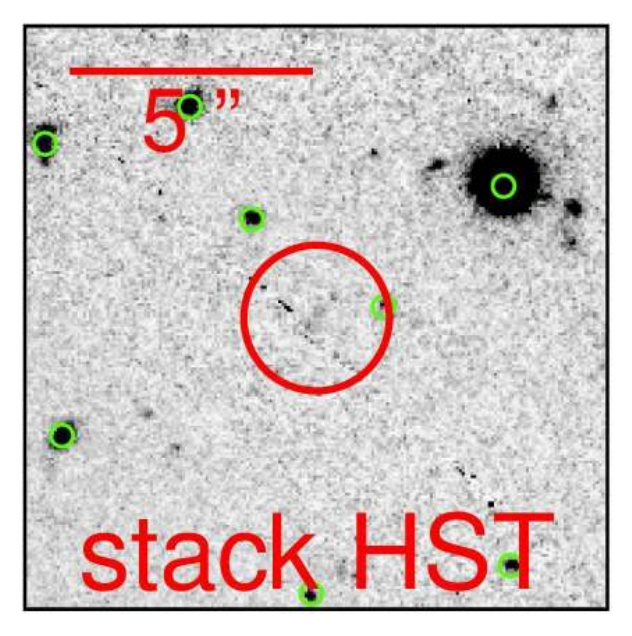}
			\includegraphics[width=.49\linewidth]{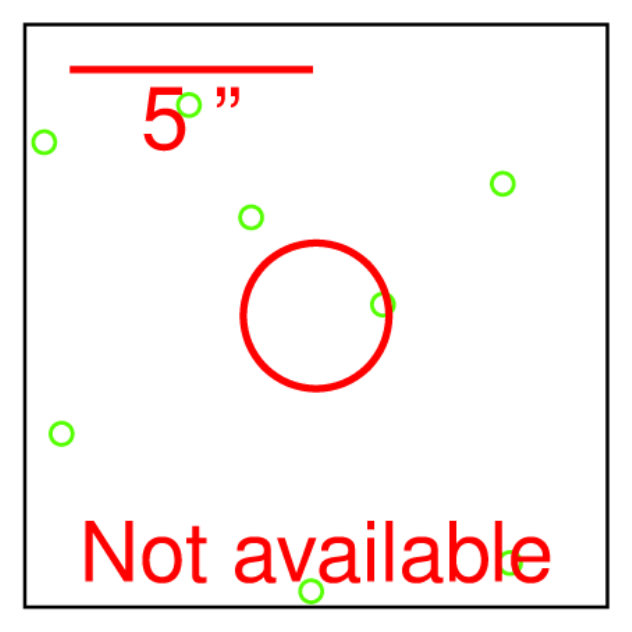}
			\centering
	\end{minipage}
	\quad
	\begin{minipage}[b]{0.52\linewidth}
		\includegraphics[width=1.\linewidth]{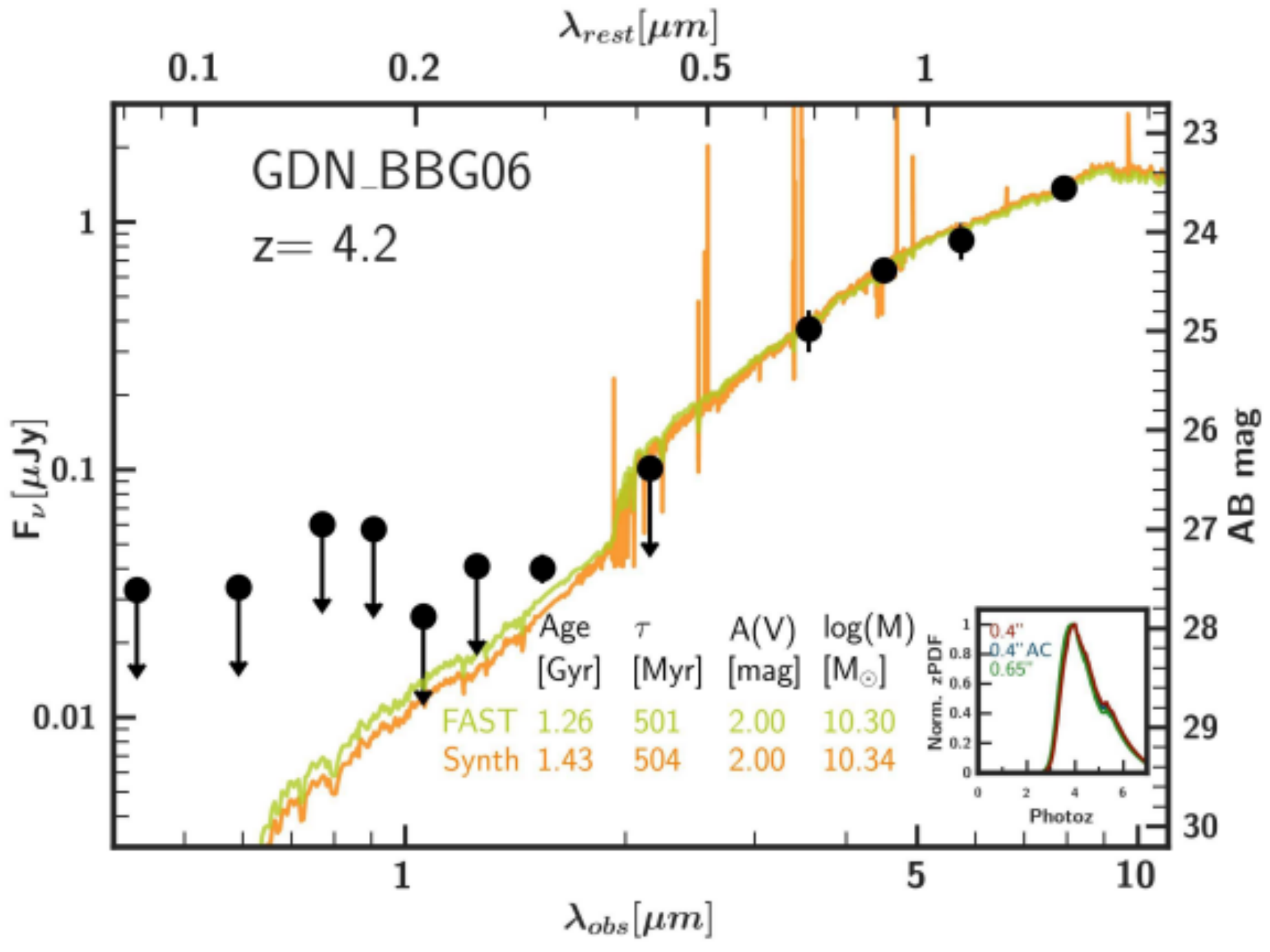}
		\centering
	\end{minipage}

% Source 7}

	\begin{minipage}[b]{0.44\linewidth}
	 \centering
			\begin{minipage}[b]{0.315\linewidth}
				\includegraphics[width=1.\linewidth]{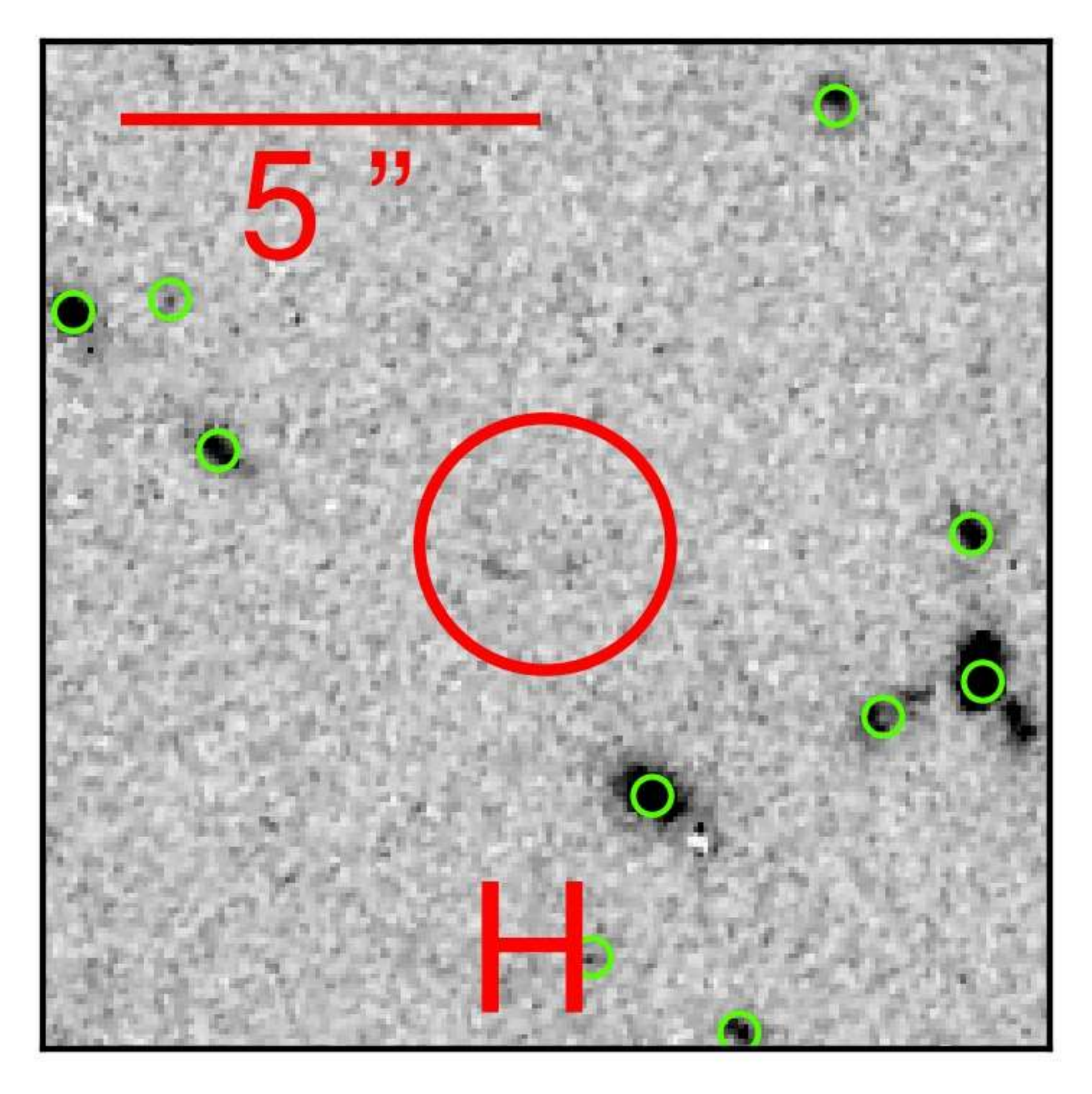}
			\end{minipage}
			\begin{minipage}[b]{0.315\linewidth}
				\includegraphics[width=1.\linewidth]{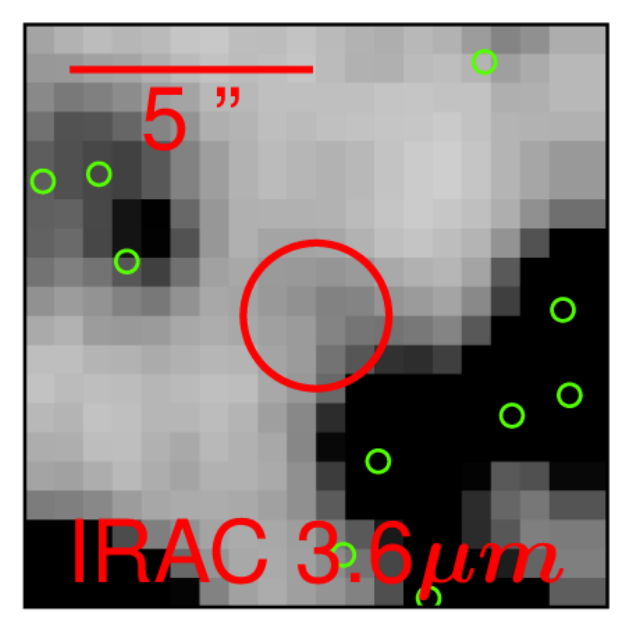}		
			\end{minipage}	
			\begin{minipage}[b]{0.315\linewidth}
				\includegraphics[width=1.\linewidth]{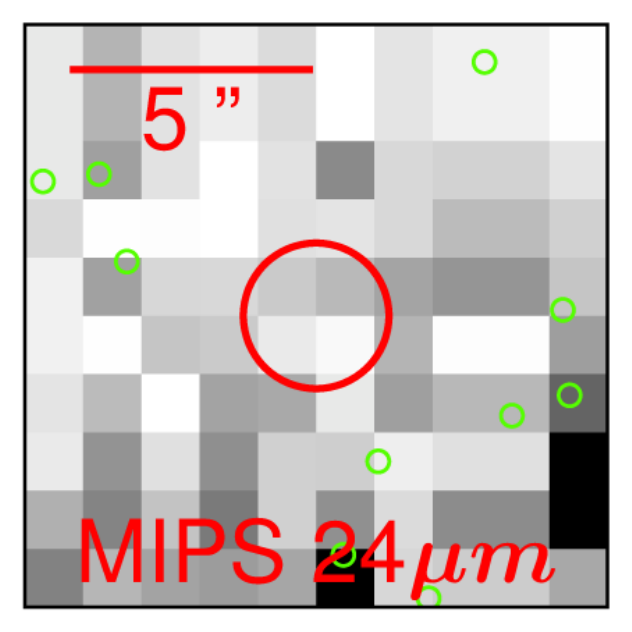}		
			\end{minipage}			
			\includegraphics[width=.49\linewidth]{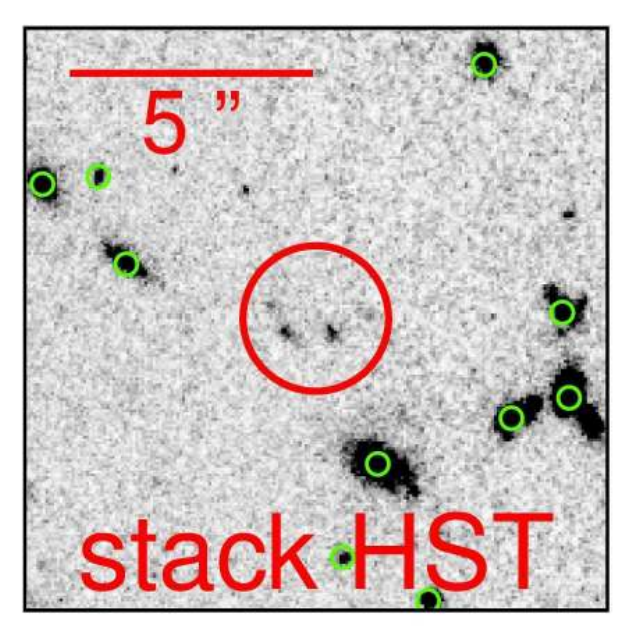}
			\includegraphics[width=.49\linewidth]{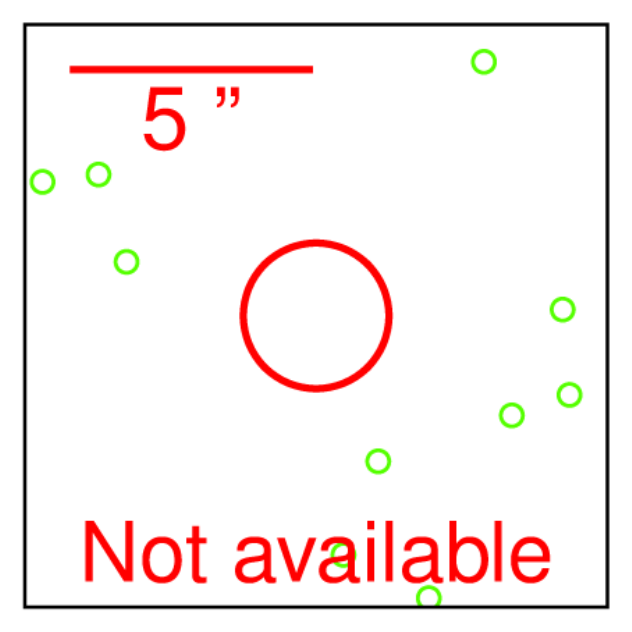}
			\centering
	\end{minipage}
	\quad
	\begin{minipage}[b]{0.52\linewidth}
		\includegraphics[width=1.\linewidth]{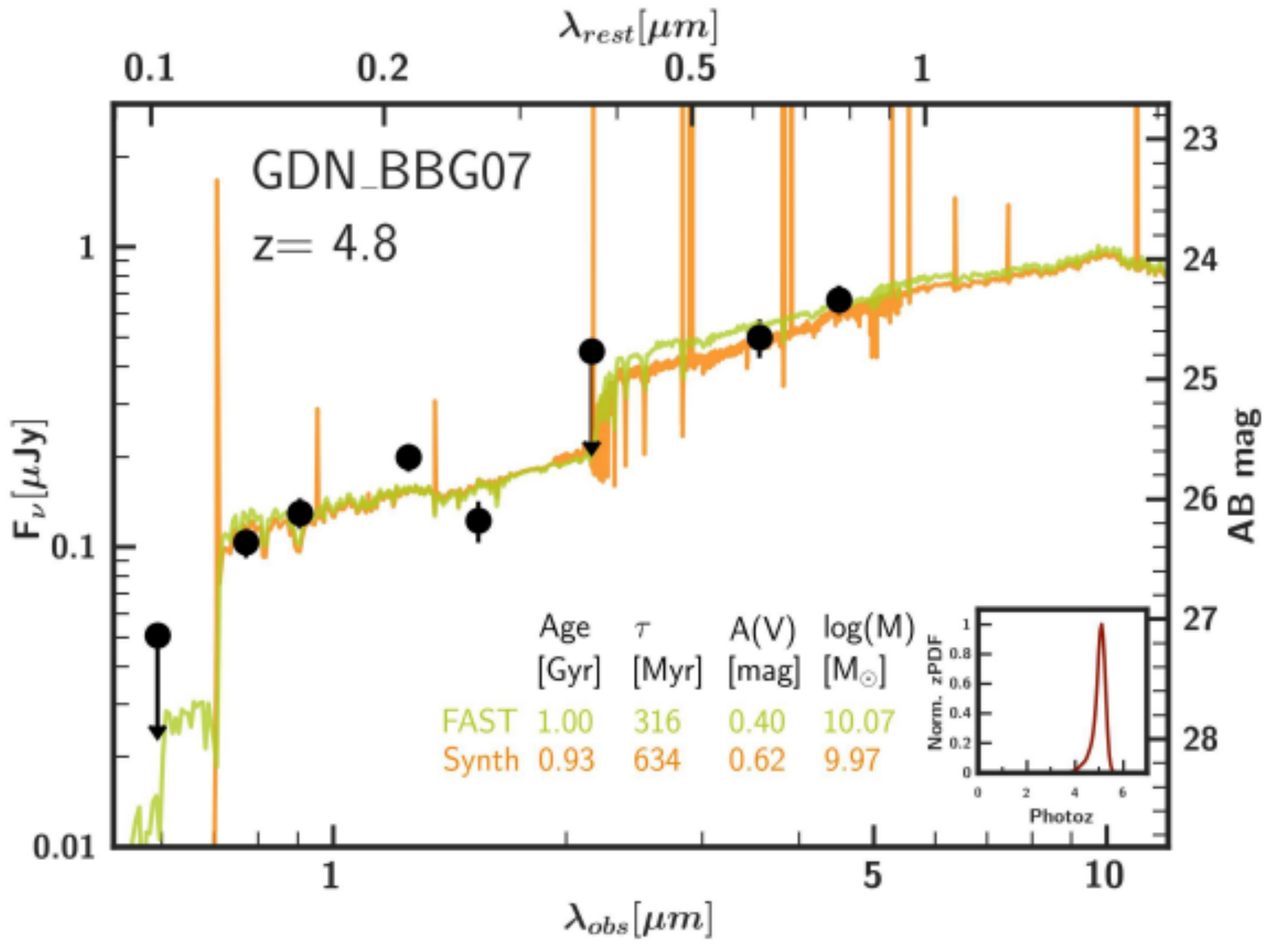}
		\centering
	\end{minipage}
\end{figure*}

% Source 8}
\begin{figure*}
	\begin{minipage}[b]{0.44\linewidth}
		\begin{minipage}[b]{0.315\linewidth}
			\includegraphics[width=1.\linewidth]{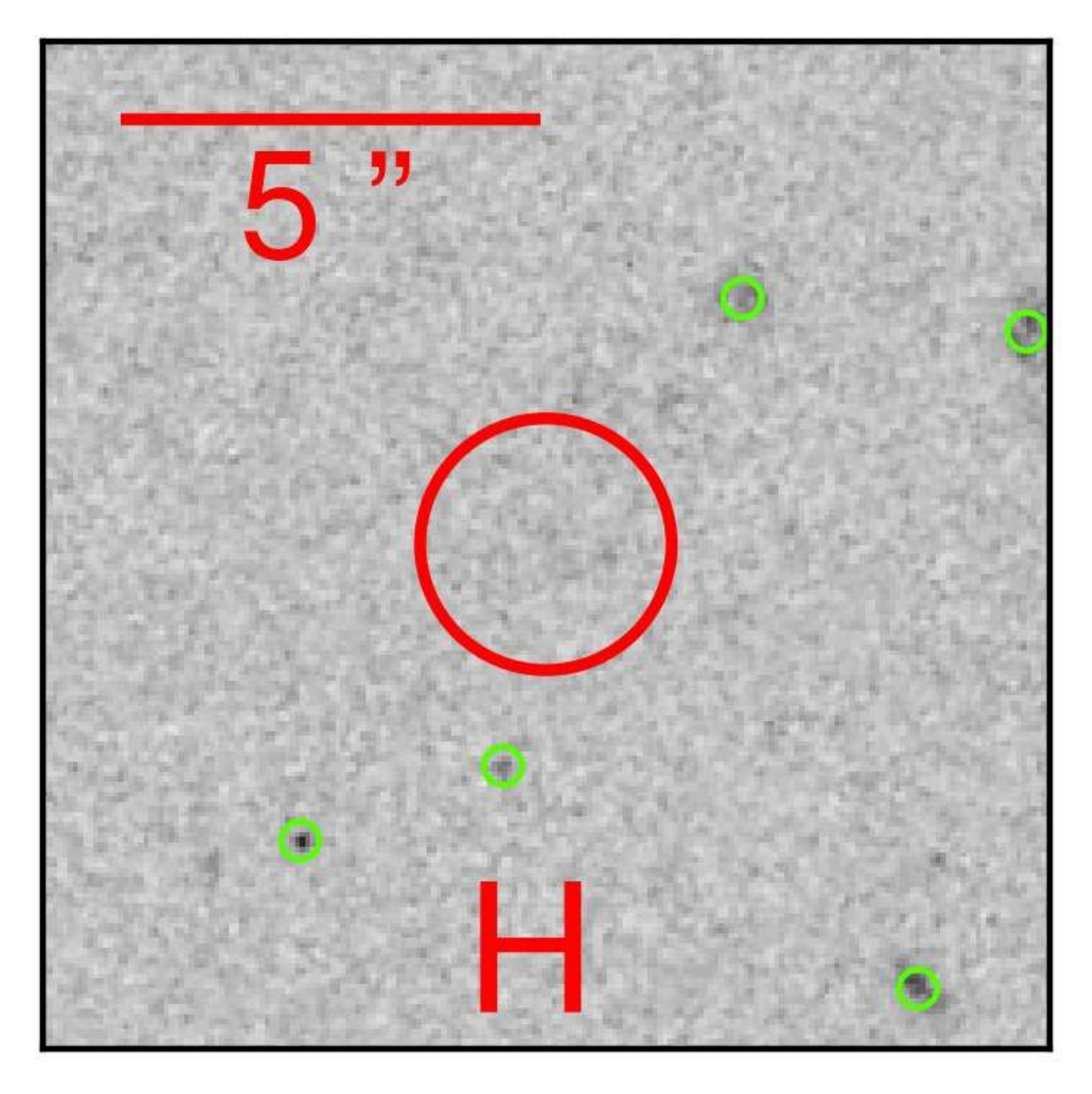}
		\end{minipage}
		\begin{minipage}[b]{0.315\linewidth}
			\includegraphics[width=1.\linewidth]{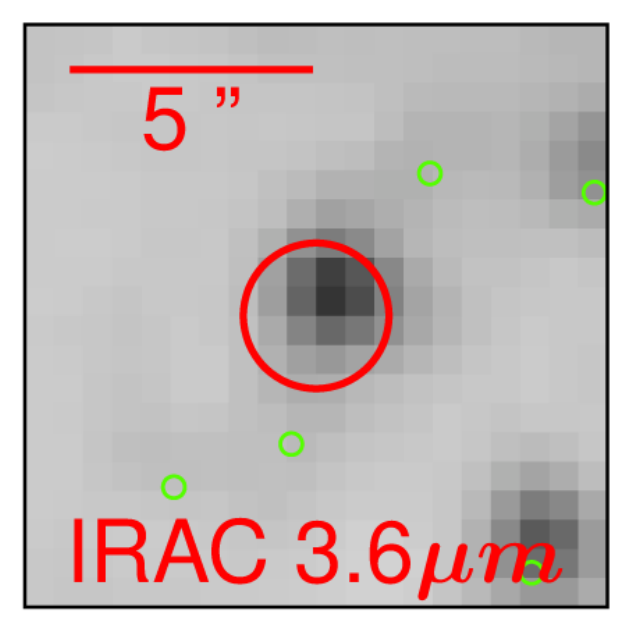}		
		\end{minipage}	
		\begin{minipage}[b]{0.315\linewidth}
			\includegraphics[width=1.\linewidth]{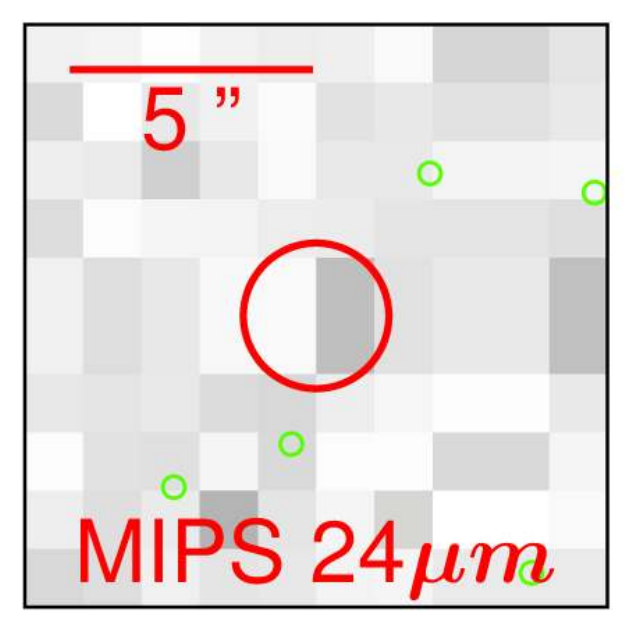}		
		\end{minipage}			
		\includegraphics[width=.49\linewidth]{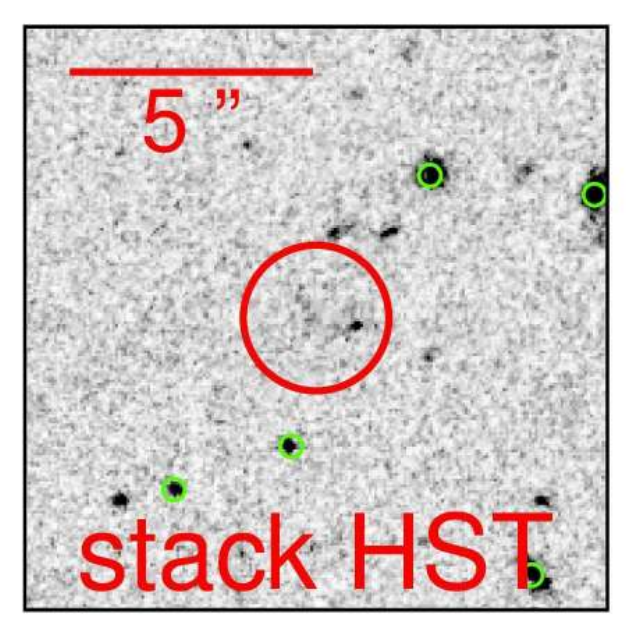}
		\includegraphics[width=.49\linewidth]{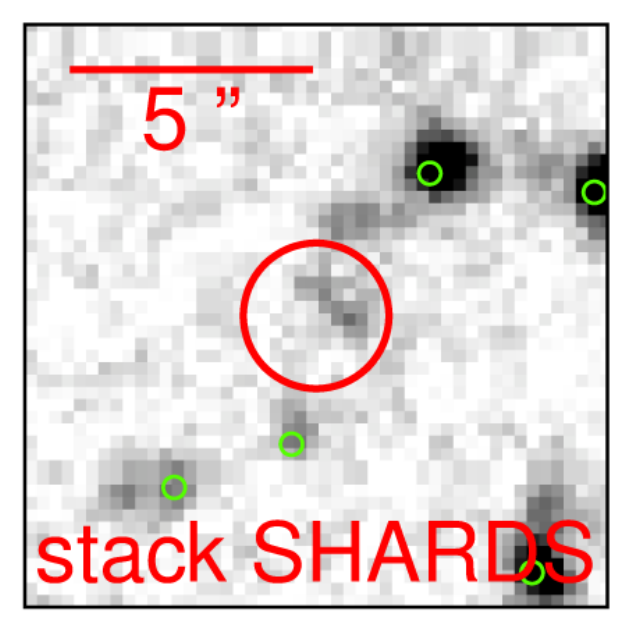}
		\centering
	\end{minipage}
	\quad
	\begin{minipage}[b]{0.52\linewidth}
		\includegraphics[width=1.\linewidth]{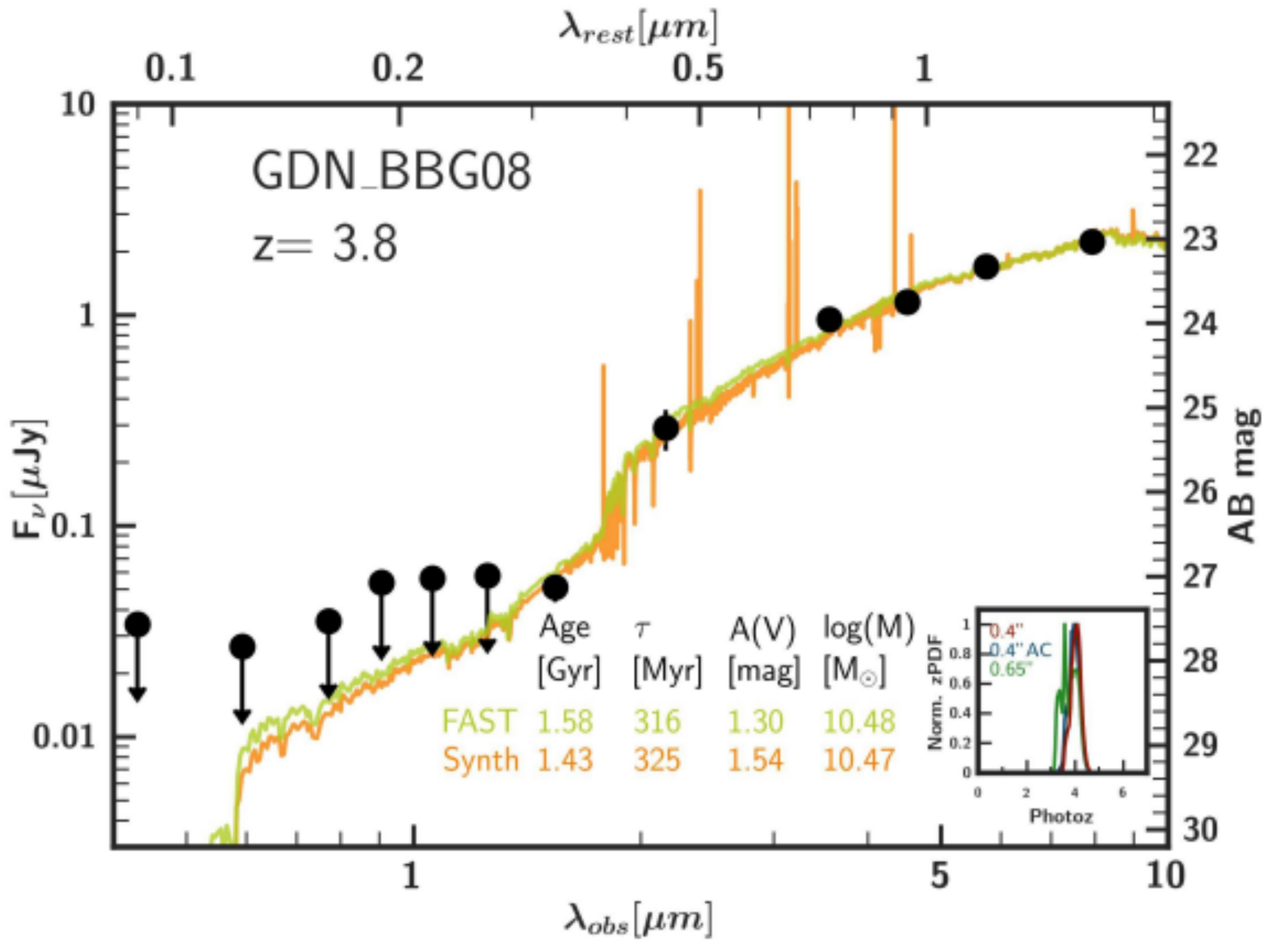}
		\centering
	\end{minipage}
	
% Source 9}
	\begin{minipage}[b]{0.44\linewidth}
		\begin{minipage}[b]{0.315\linewidth}
			\includegraphics[width=1.\linewidth]{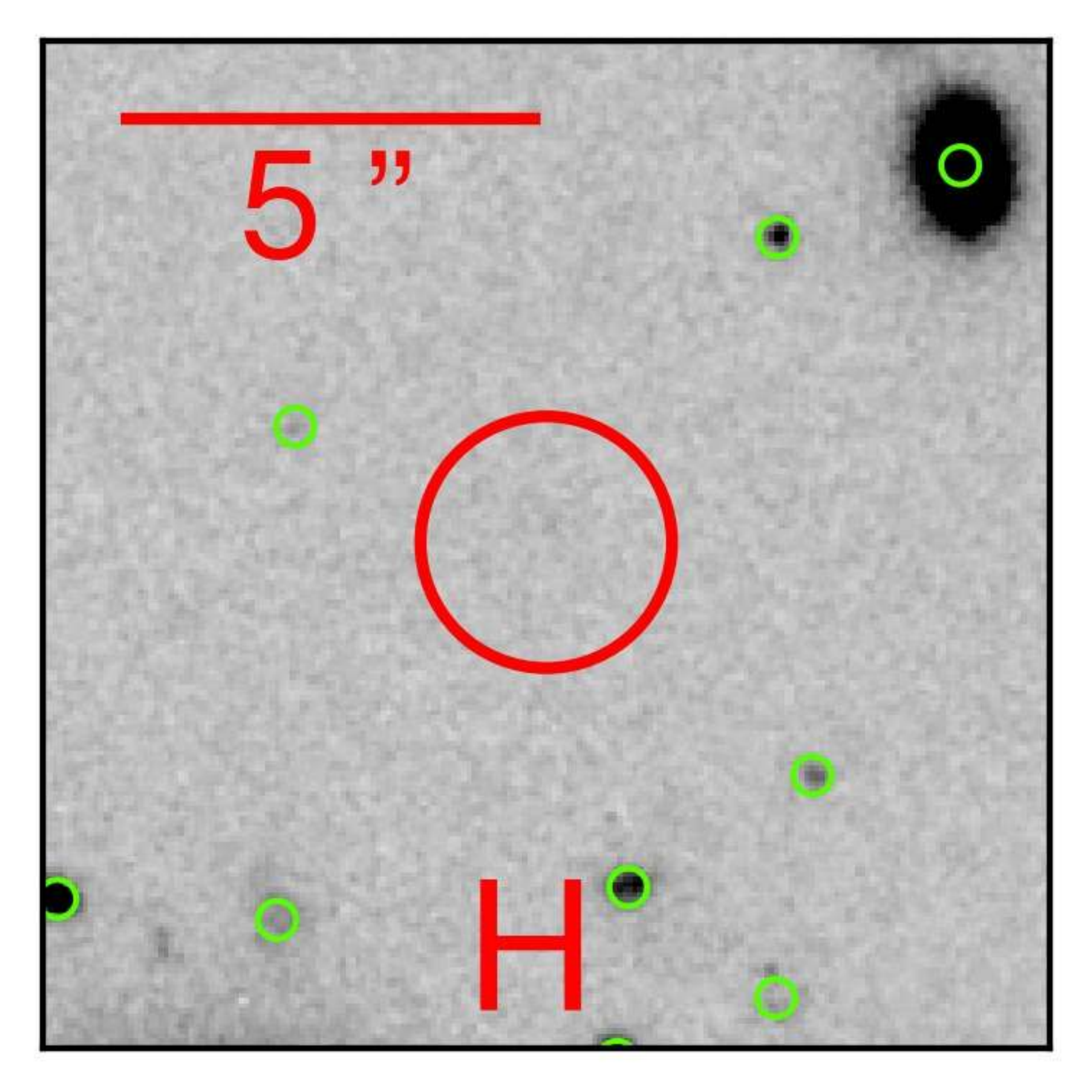}
		\end{minipage}
		\begin{minipage}[b]{0.315\linewidth}
			\includegraphics[width=1.\linewidth]{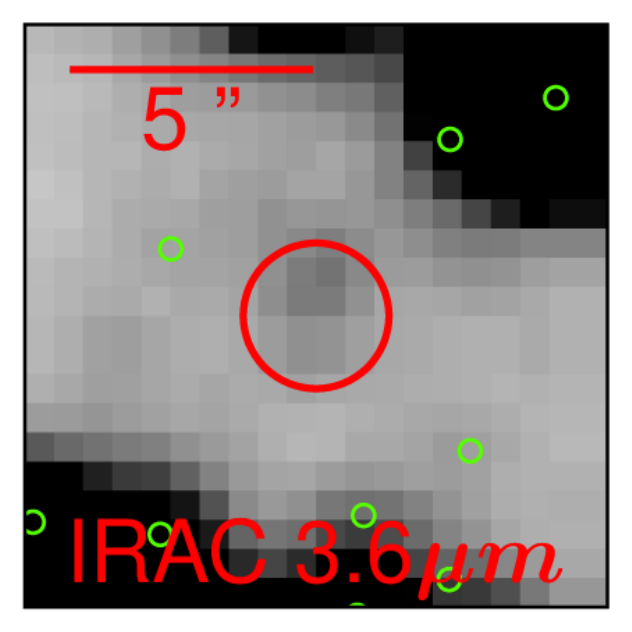}		
		\end{minipage}	
		\begin{minipage}[b]{0.315\linewidth}
			\includegraphics[width=1.\linewidth]{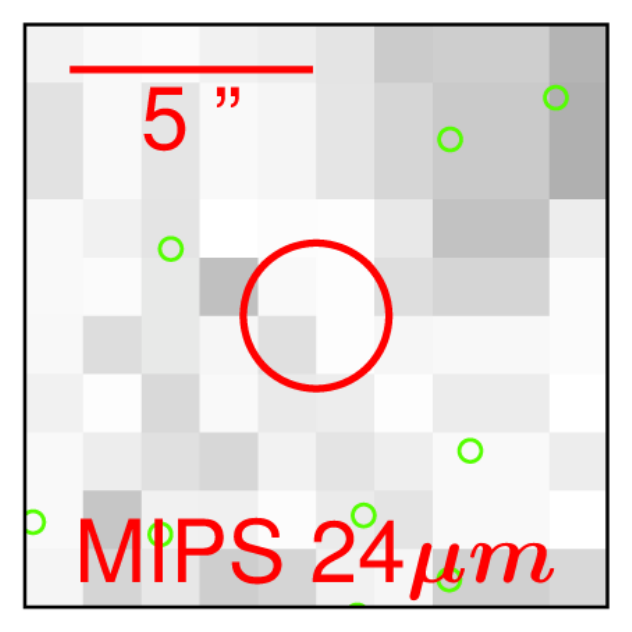}		
		\end{minipage}			
		\includegraphics[width=.49\linewidth]{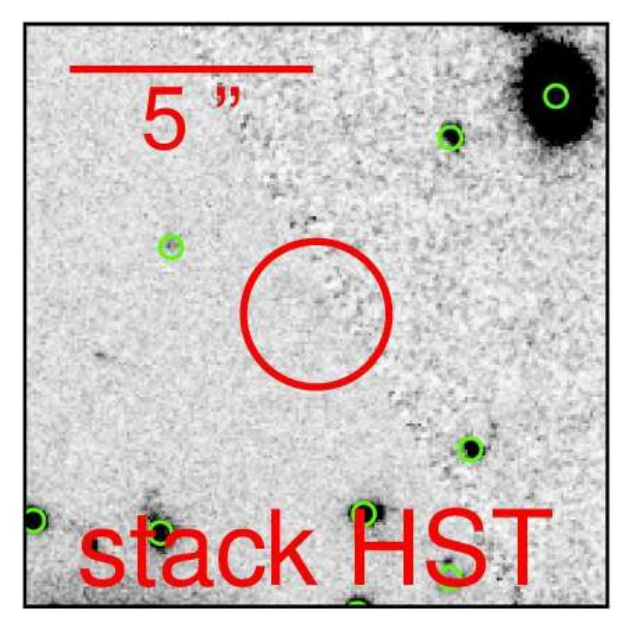}
		\includegraphics[width=.49\linewidth]{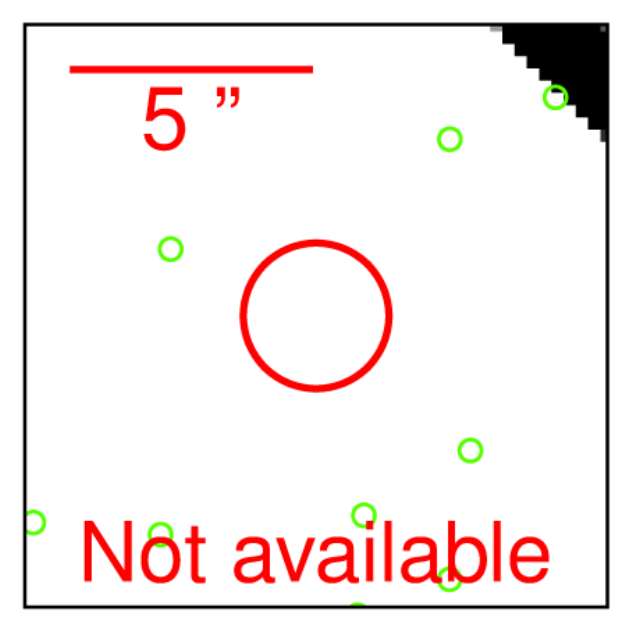}
		\centering
	\end{minipage}
	\quad
	\begin{minipage}[b]{0.52\linewidth}
		\includegraphics[width=1.\linewidth]{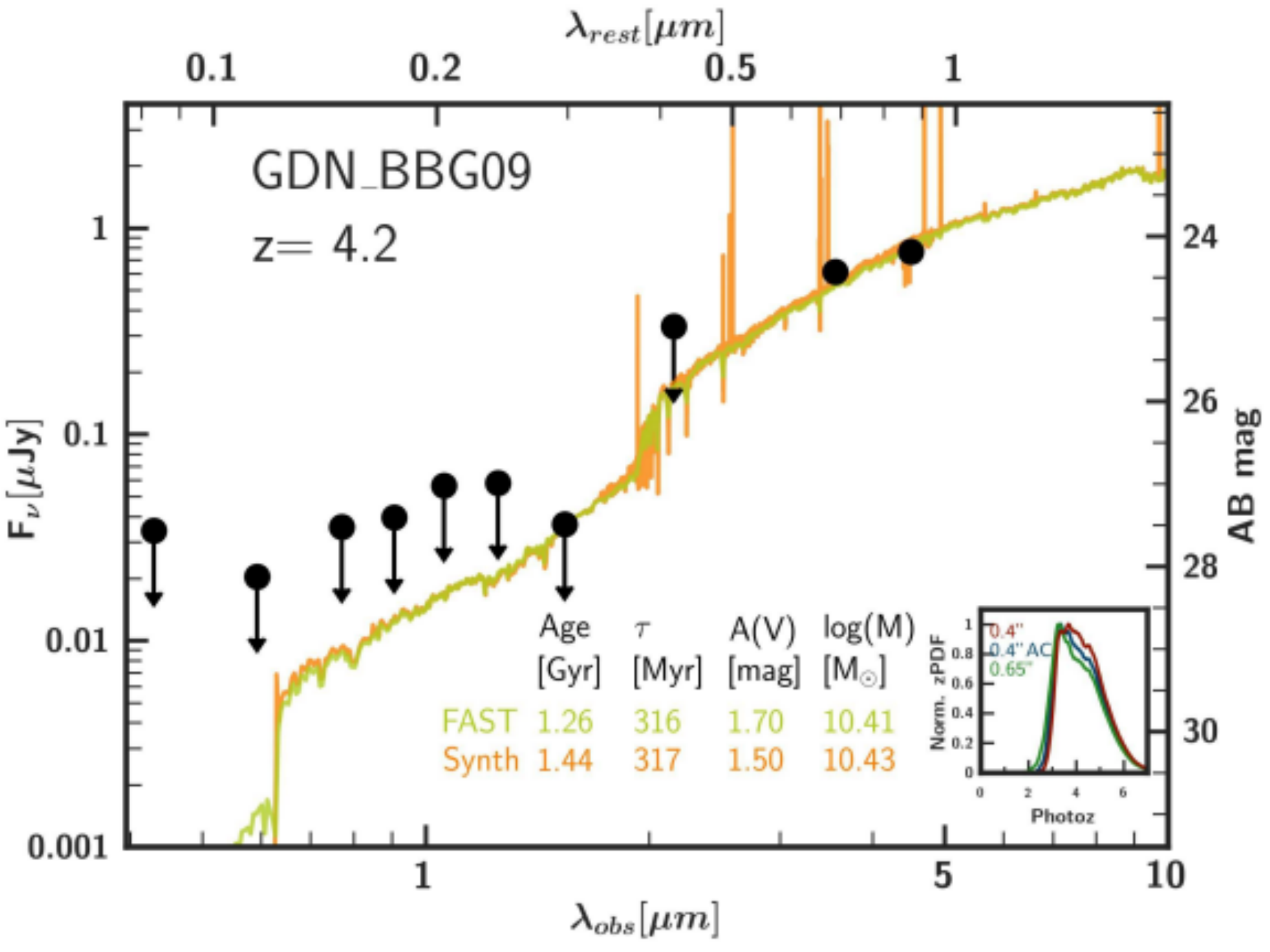}
		\centering
	\end{minipage}

% source 10  GDN
	\begin{minipage}[b]{0.44\linewidth}
		\begin{minipage}[b]{0.315\linewidth}
			\includegraphics[width=1.\linewidth]{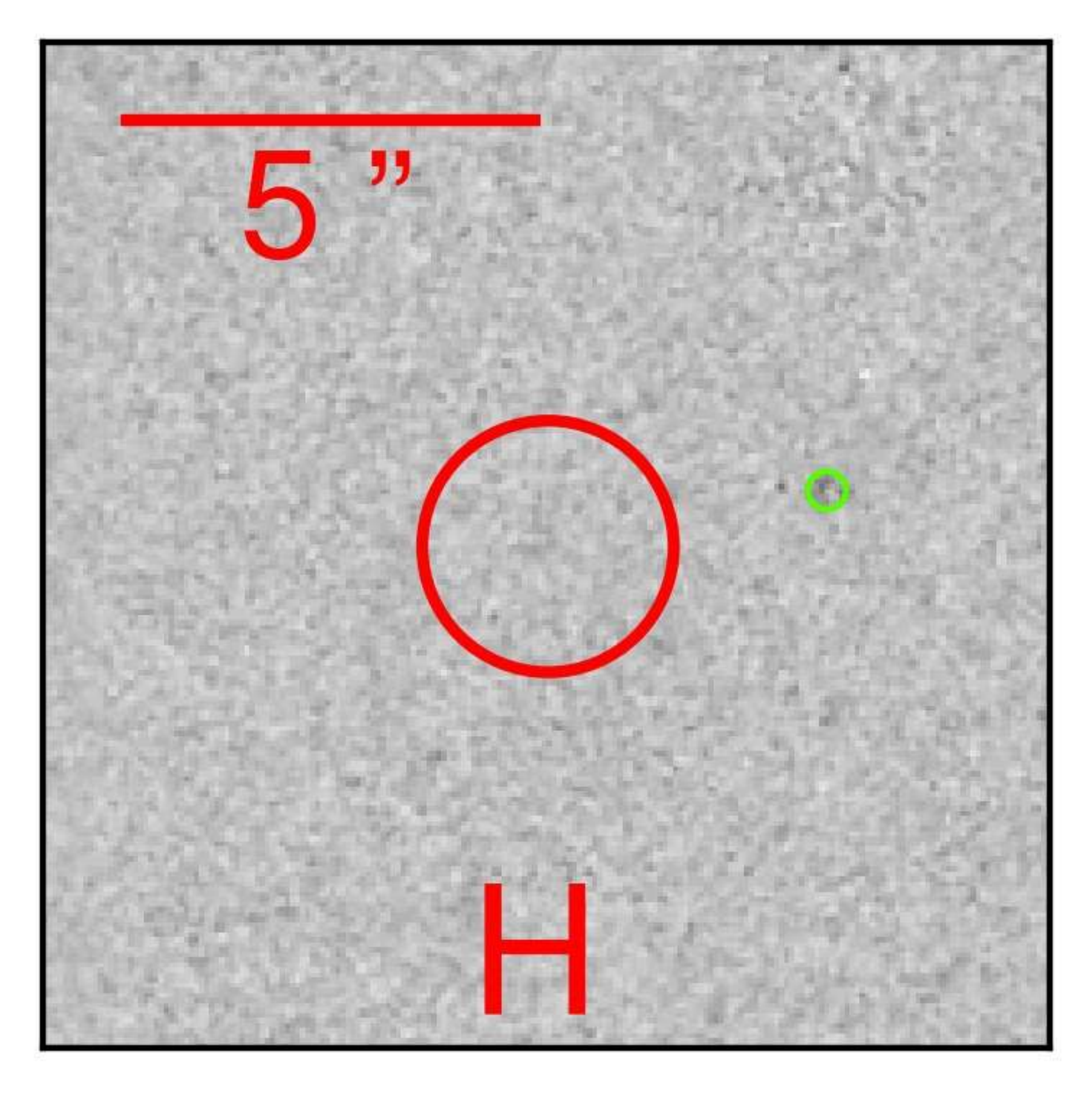}
		\end{minipage}
		\begin{minipage}[b]{0.315\linewidth}
			\includegraphics[width=1.\linewidth]{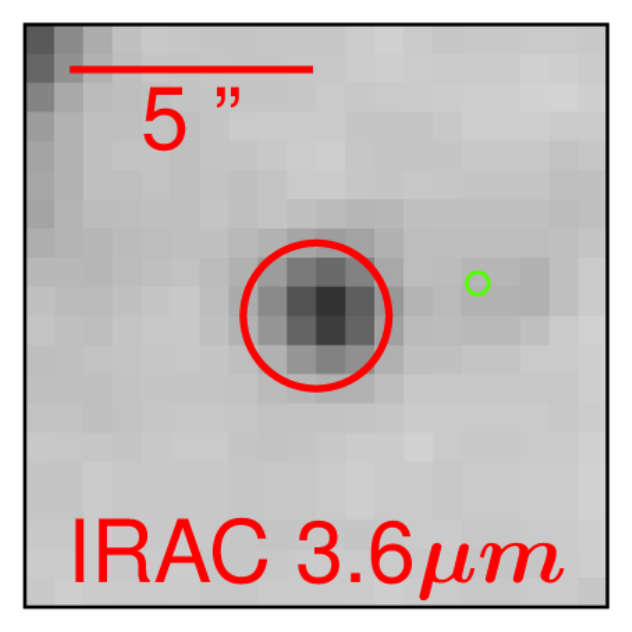}		
		\end{minipage}	
		\begin{minipage}[b]{0.315\linewidth}
			\includegraphics[width=1.\linewidth]{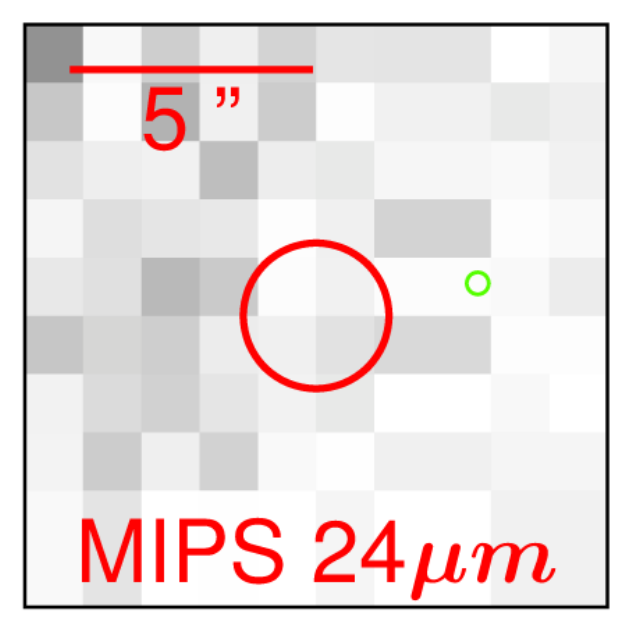}		
		\end{minipage}			
		\includegraphics[width=.49\linewidth]{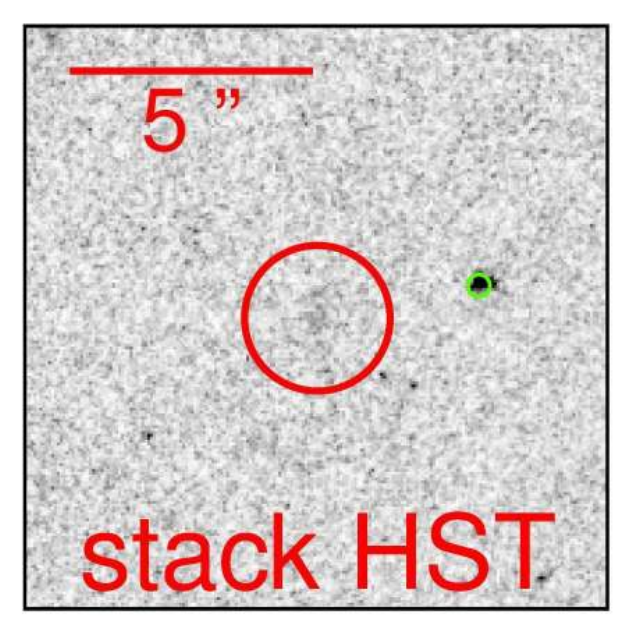}
		\includegraphics[width=.49\linewidth]{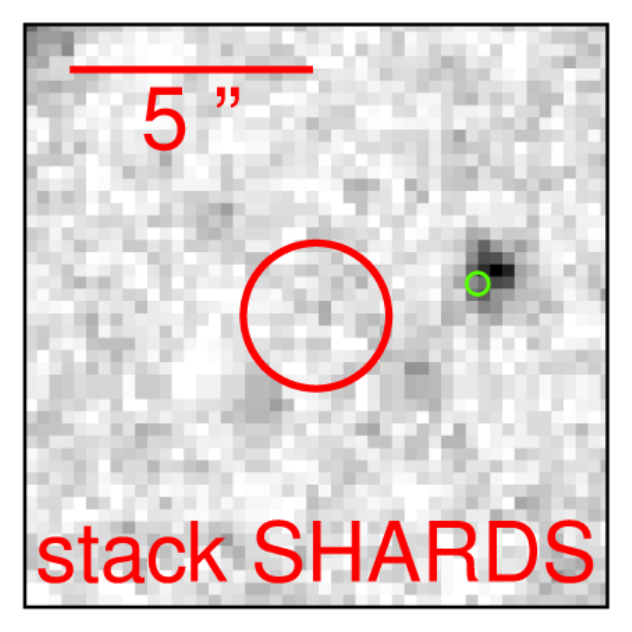}
		\centering
	\end{minipage}
	\quad
	\begin{minipage}[b]{0.52\linewidth}
		\includegraphics[width=1.\linewidth]{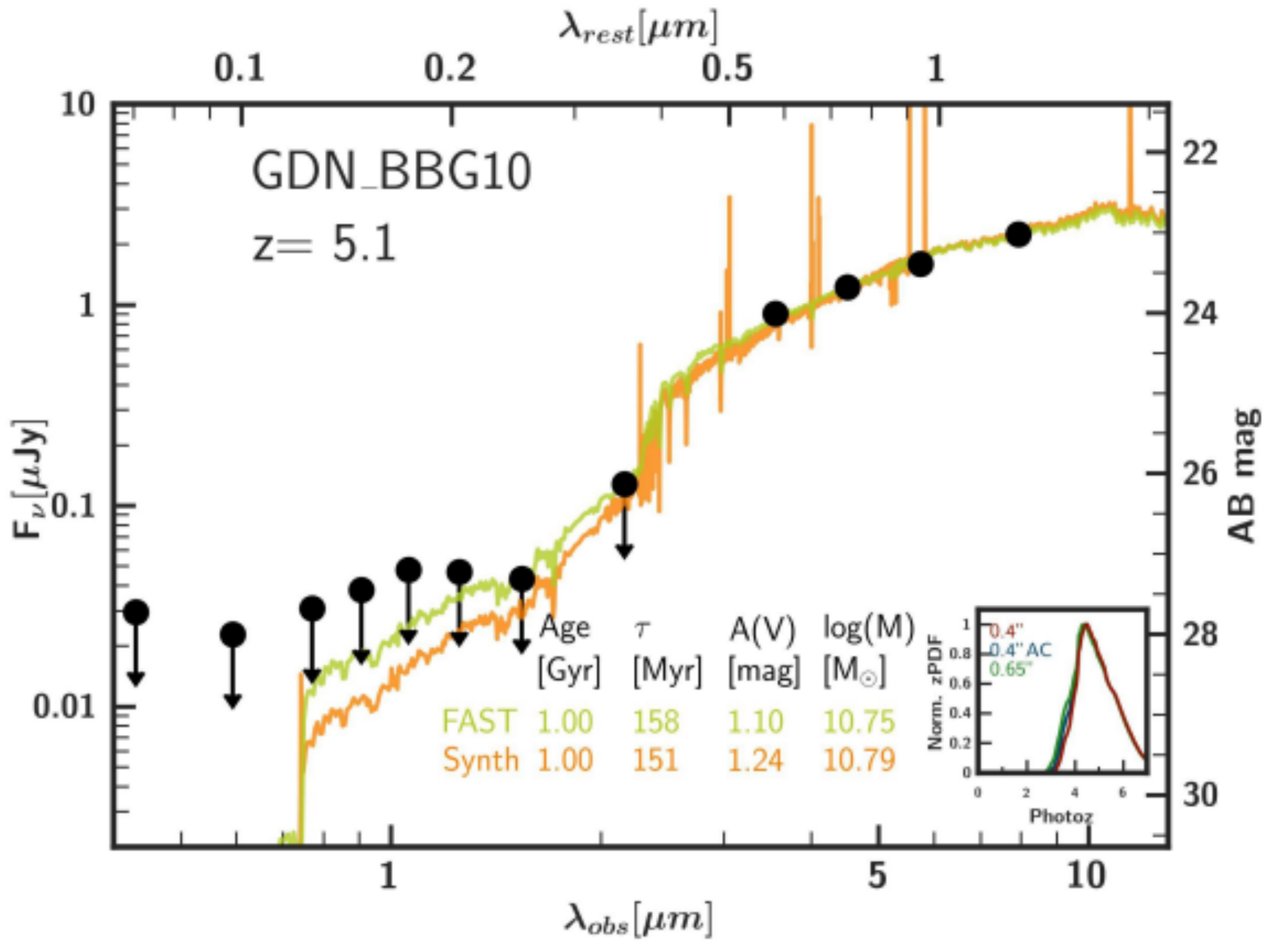}
		\centering
	\end{minipage}
\end{figure*}

% Source 11}
\begin{figure*}
	\begin{minipage}[b]{0.44\linewidth}
	 \centering
			\begin{minipage}[b]{0.315\linewidth}
				\includegraphics[width=1.\linewidth]{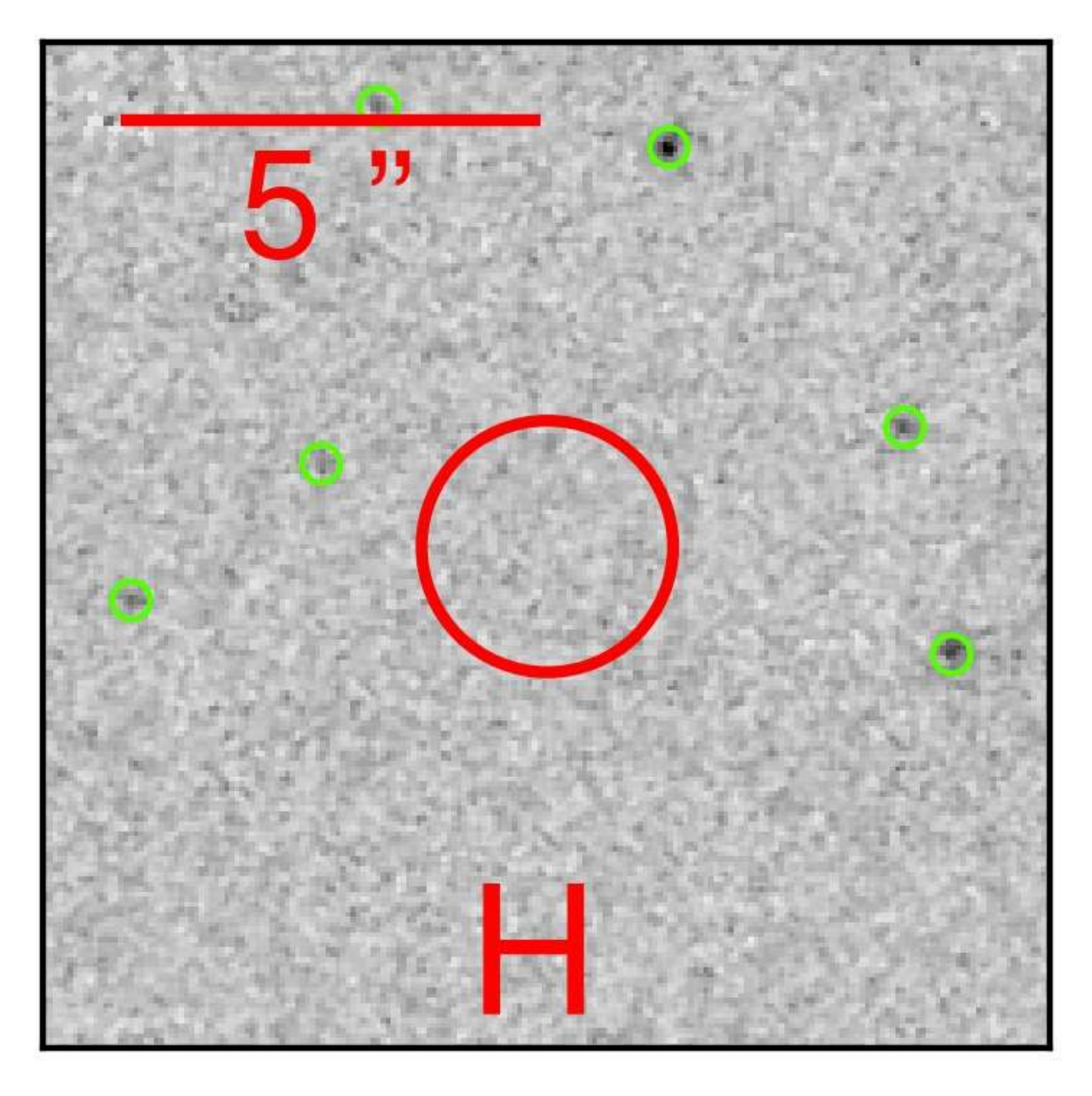}
			\end{minipage}
			\begin{minipage}[b]{0.315\linewidth}
				\includegraphics[width=1.\linewidth]{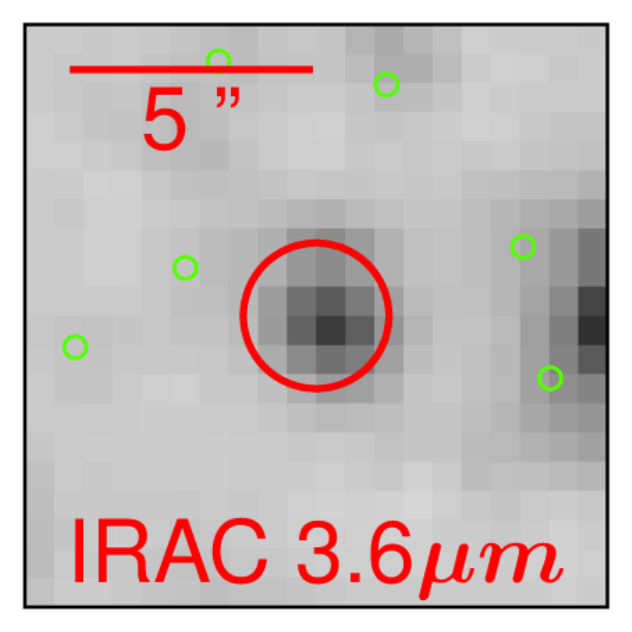}		
			\end{minipage}	
			\begin{minipage}[b]{0.315\linewidth}
				\includegraphics[width=1.\linewidth]{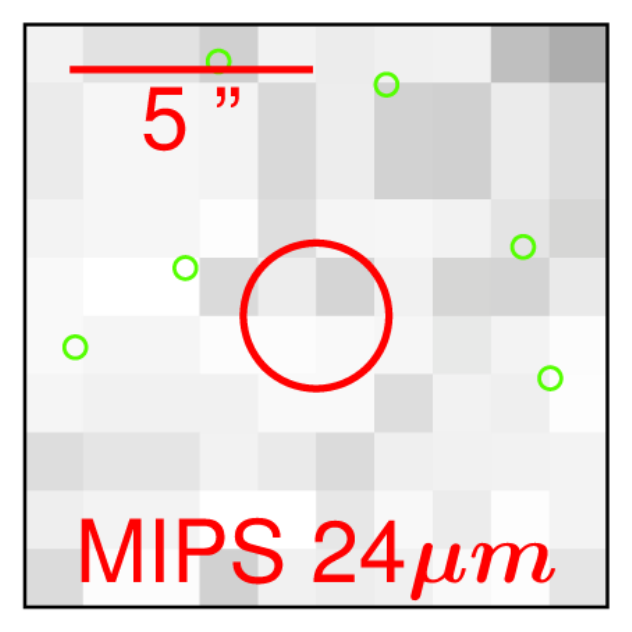}		
			\end{minipage}			
			\includegraphics[width=.49\linewidth]{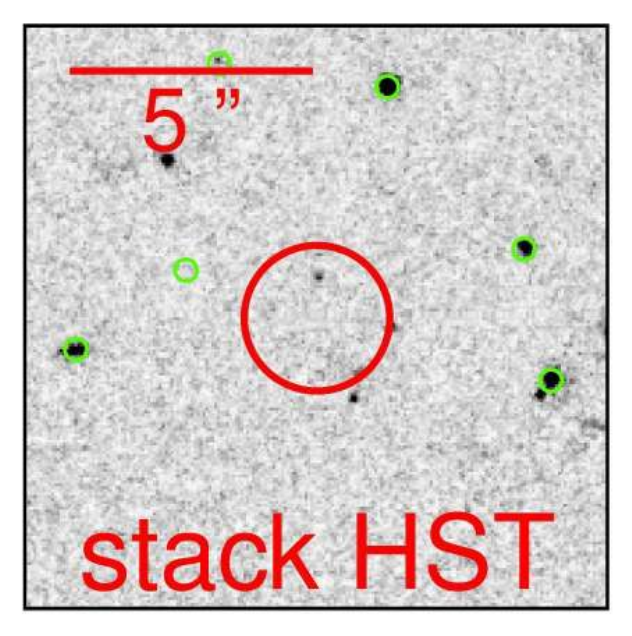}
			\includegraphics[width=.49\linewidth]{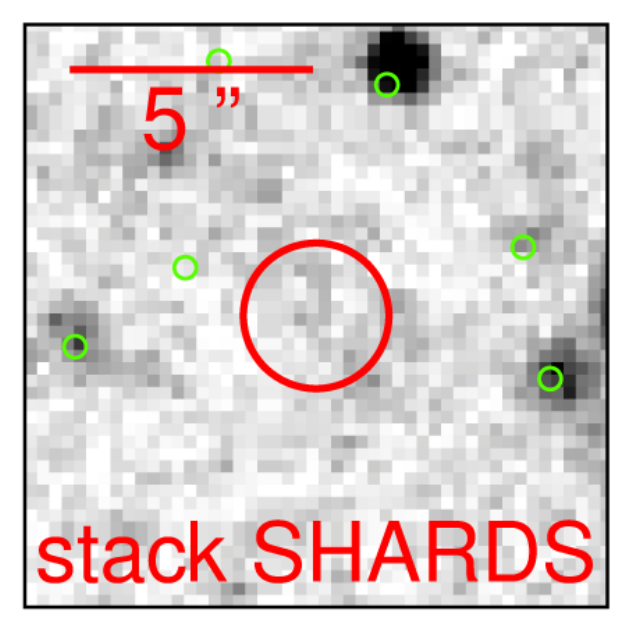}
			\centering
	\end{minipage}
	\quad
	\begin{minipage}[b]{0.52\linewidth}
	\begin{center}
		\includegraphics[width=1.\linewidth]{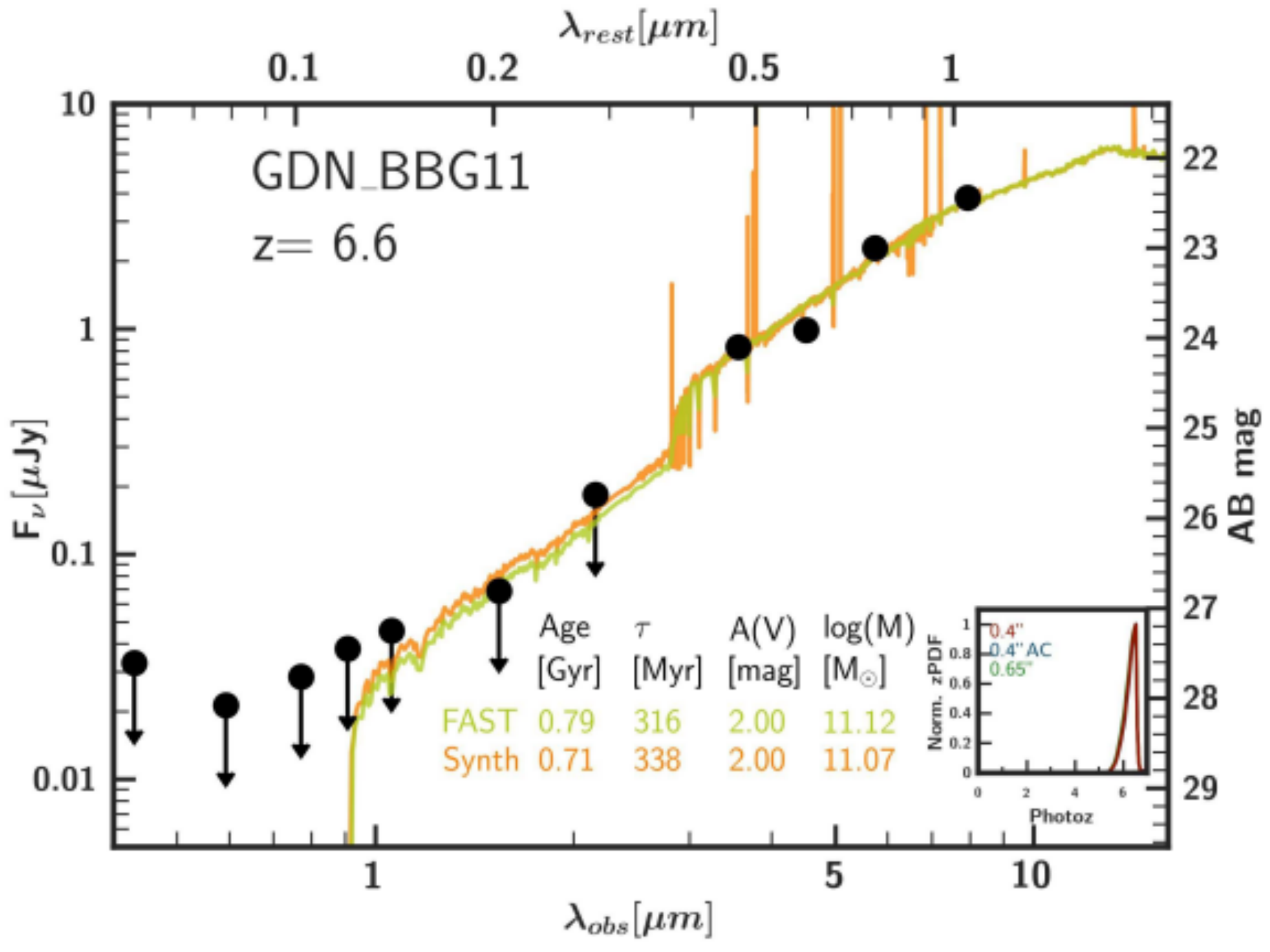}
	\end{center}
	\end{minipage}

% Source 12}

	\begin{minipage}[b]{0.44\linewidth}
		\begin{minipage}[b]{0.315\linewidth}
			\includegraphics[width=1.\linewidth]{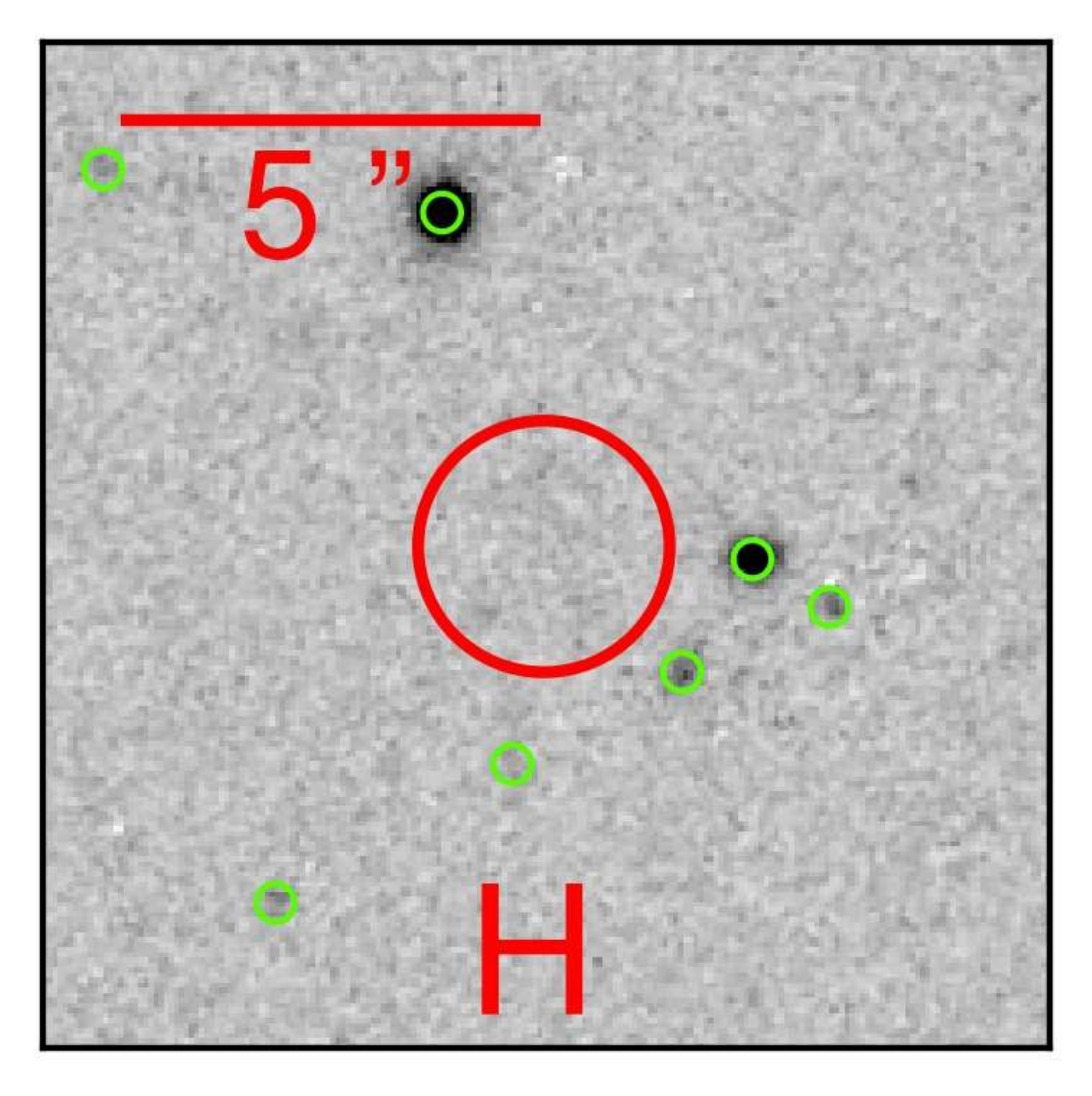}
		\end{minipage}
		\begin{minipage}[b]{0.315\linewidth}
			\includegraphics[width=1.\linewidth]{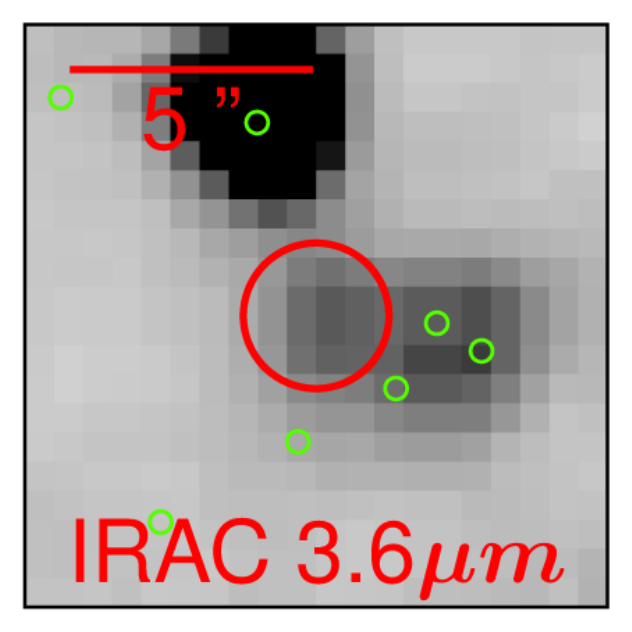}		
		\end{minipage}	
		\begin{minipage}[b]{0.315\linewidth}
			\includegraphics[width=1.\linewidth]{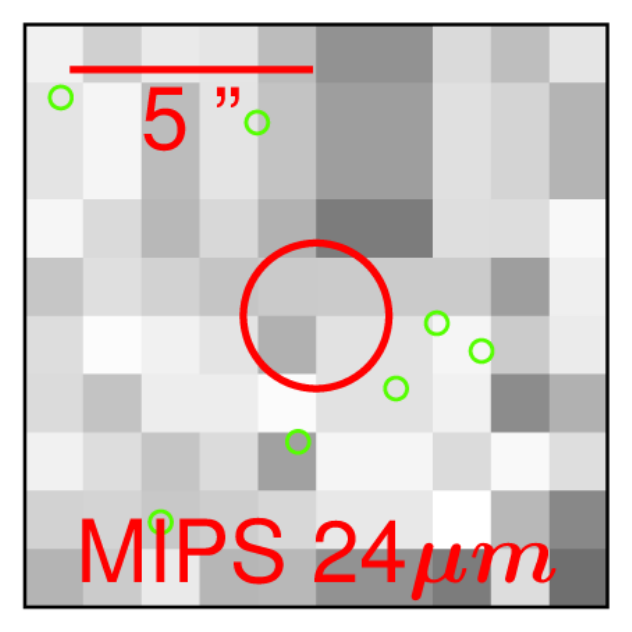}		
		\end{minipage}			
		\includegraphics[width=.49\linewidth]{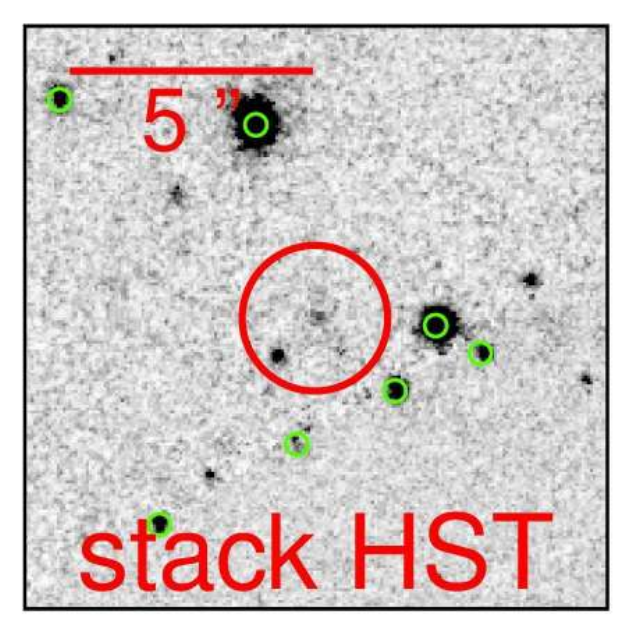}
		\includegraphics[width=.49\linewidth]{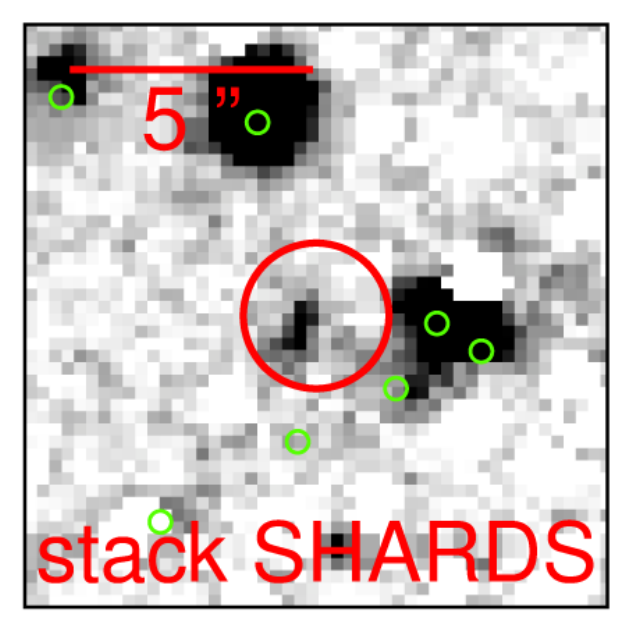}
		\centering
	\end{minipage}
	\quad
	\begin{minipage}[b]{0.52\linewidth}
		\includegraphics[width=1.\linewidth]{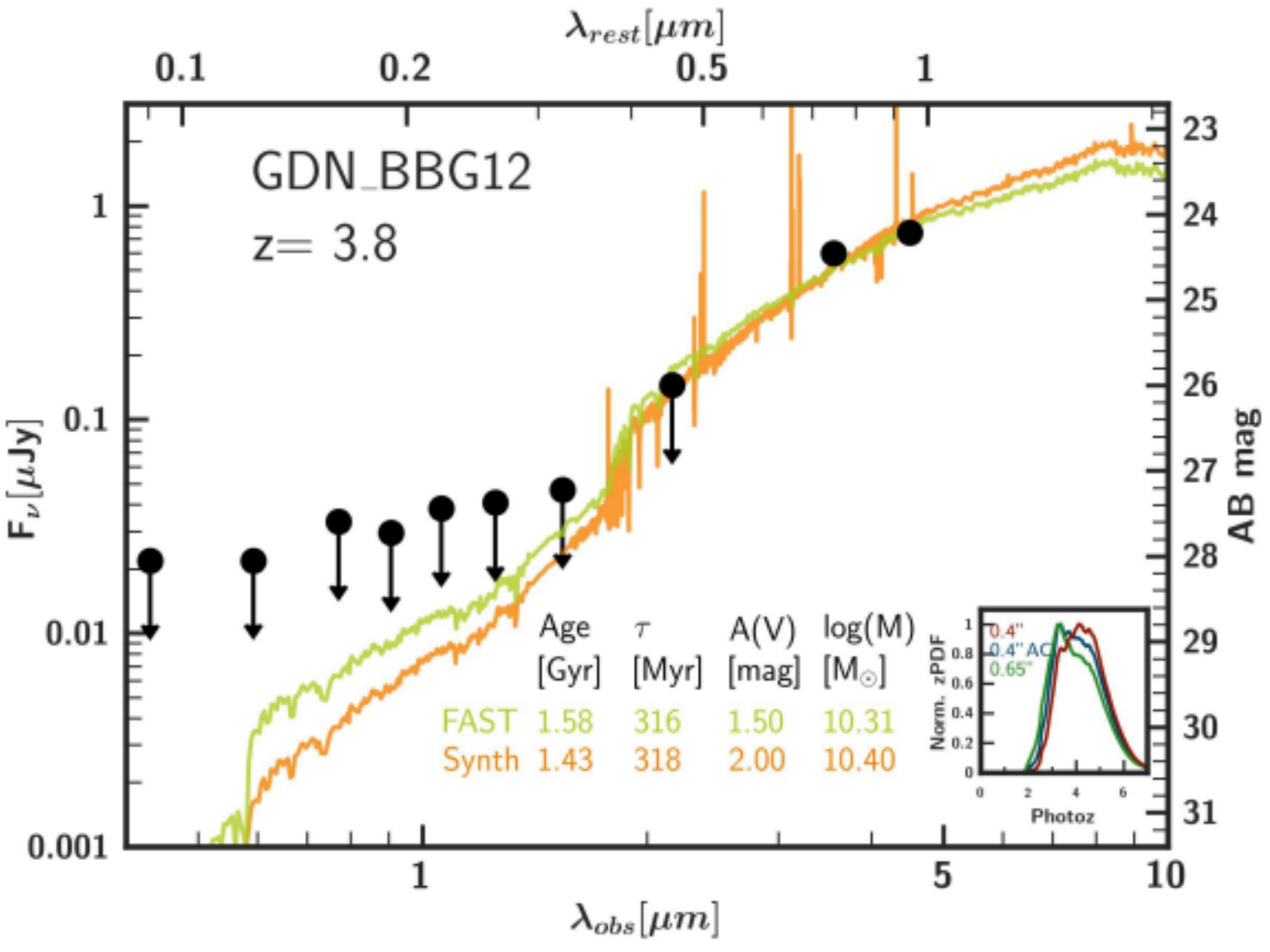}
		\centering
	\end{minipage}

% Source 13}
	\begin{minipage}[b]{0.44\linewidth}
		\begin{minipage}[b]{0.315\linewidth}
			\includegraphics[width=1.\linewidth]{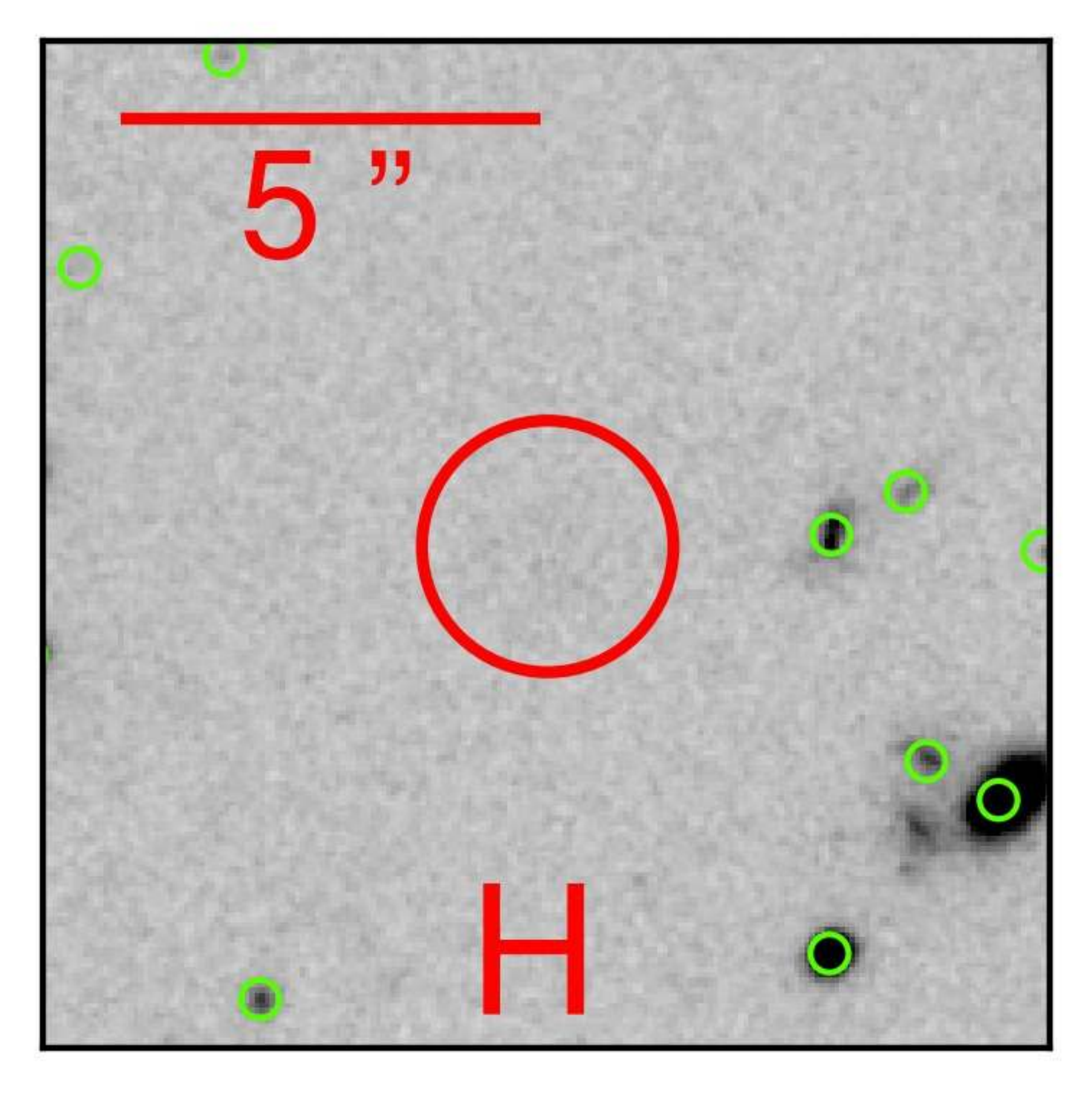}
		\end{minipage}
		\begin{minipage}[b]{0.315\linewidth}
			\includegraphics[width=1.\linewidth]{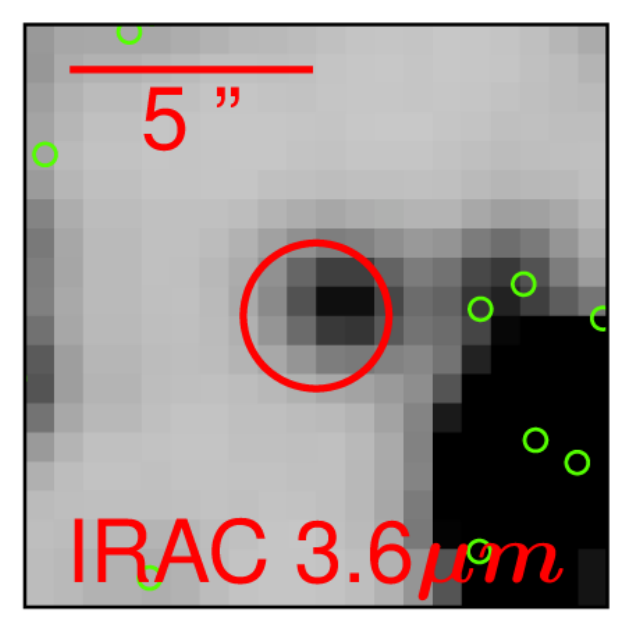}		
		\end{minipage}	
		\begin{minipage}[b]{0.315\linewidth}
			\includegraphics[width=1.\linewidth]{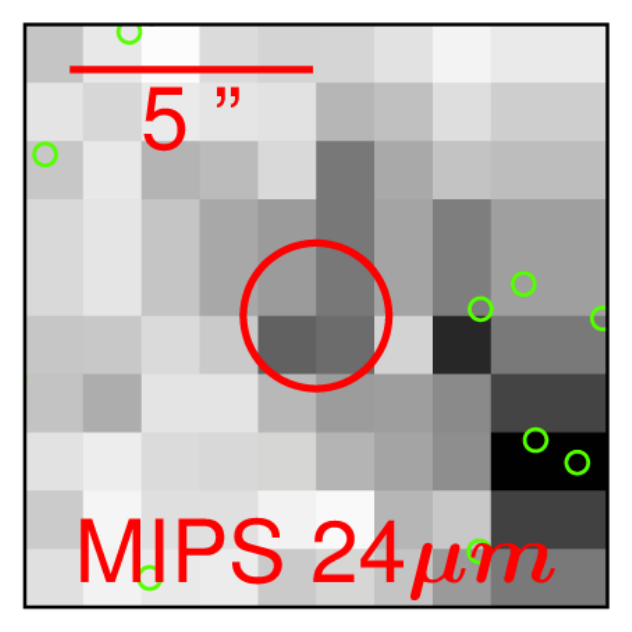}		
		\end{minipage}			
		\includegraphics[width=.49\linewidth]{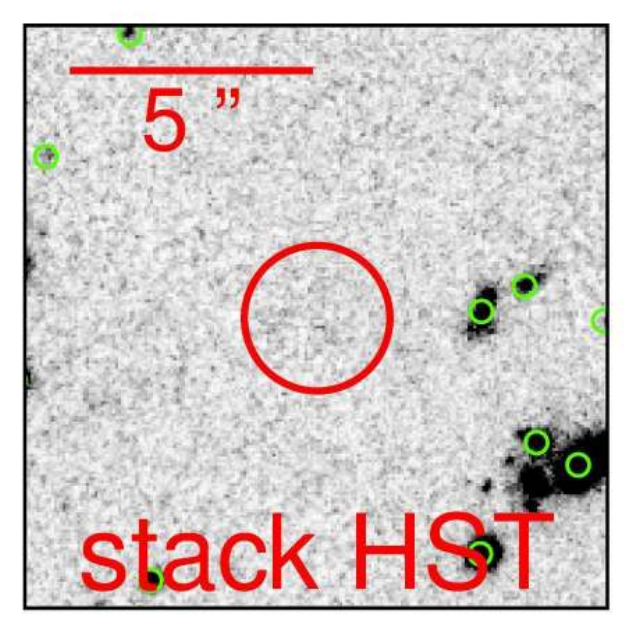}
		\includegraphics[width=.49\linewidth]{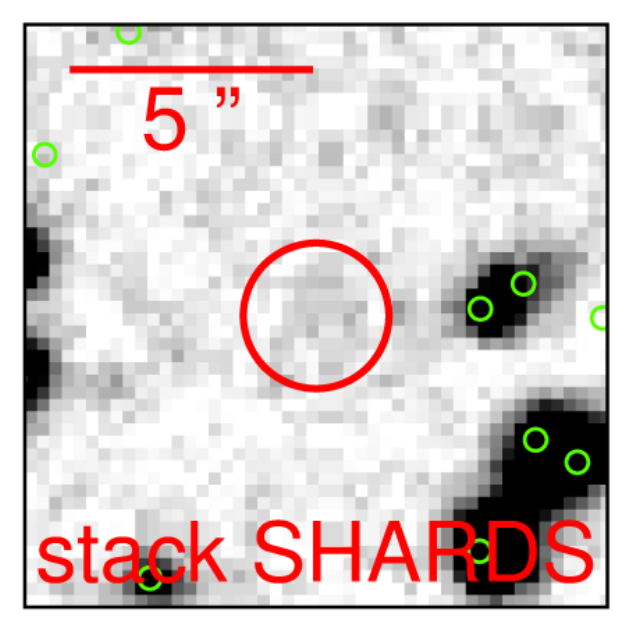}
		\centering
	\end{minipage}
	\quad
	\begin{minipage}[b]{0.52\linewidth}
		\includegraphics[width=1.\linewidth]{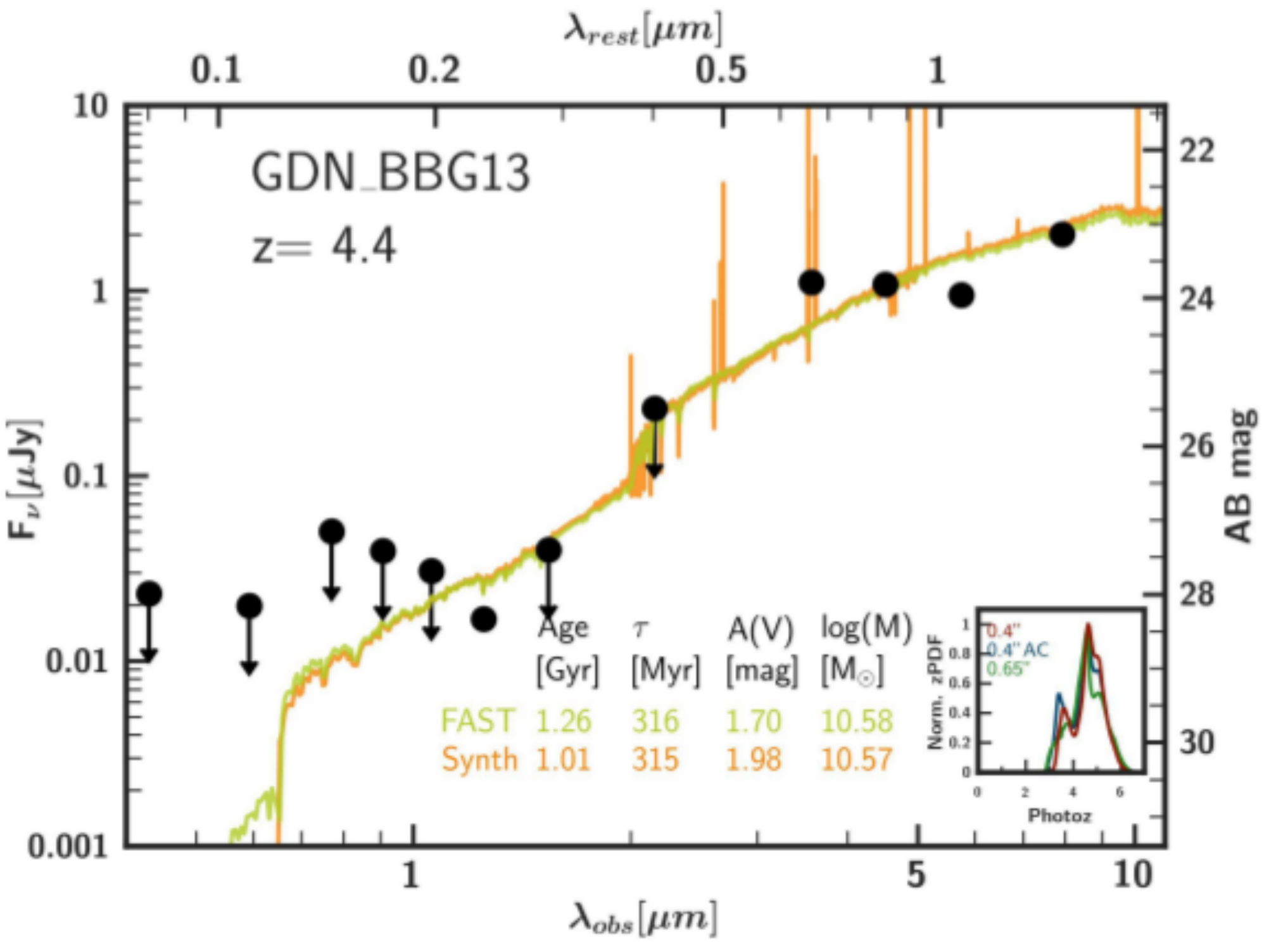}
		\centering
	\end{minipage}
\end{figure*}

% Source 14}
\begin{figure*}
	\begin{minipage}[b]{0.44\linewidth}
		\begin{minipage}[b]{0.315\linewidth}
			\includegraphics[width=1.\linewidth]{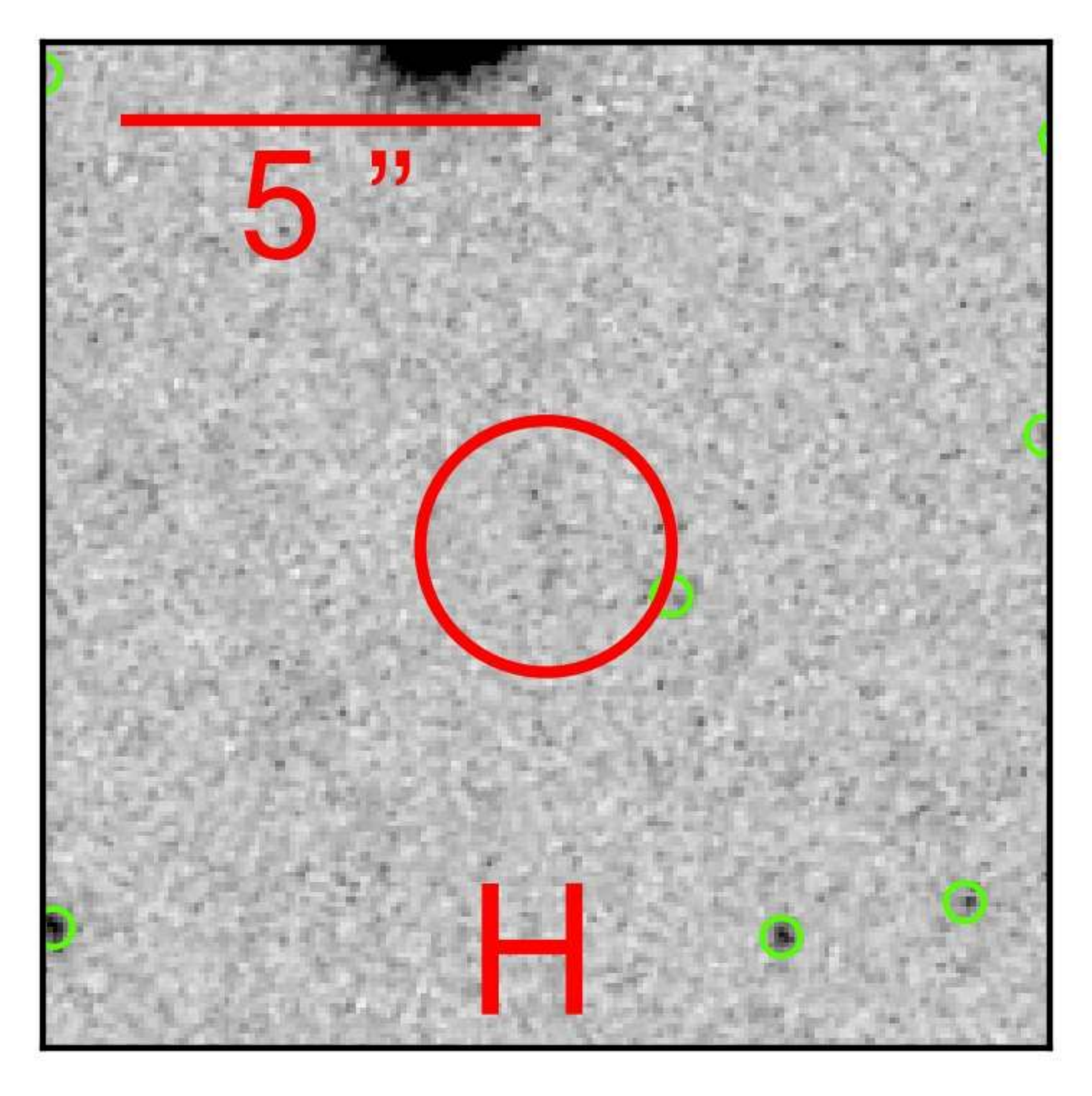}
		\end{minipage}
		\begin{minipage}[b]{0.315\linewidth}
			\includegraphics[width=1.\linewidth]{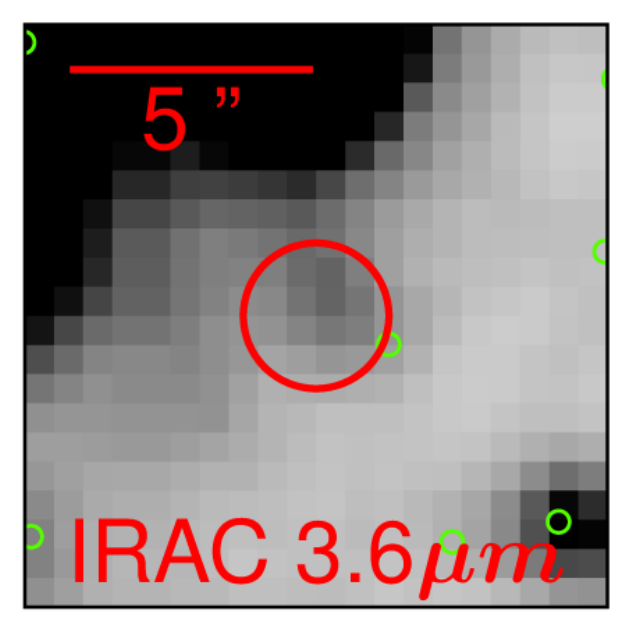}		
		\end{minipage}	
		\begin{minipage}[b]{0.315\linewidth}
			\includegraphics[width=1.\linewidth]{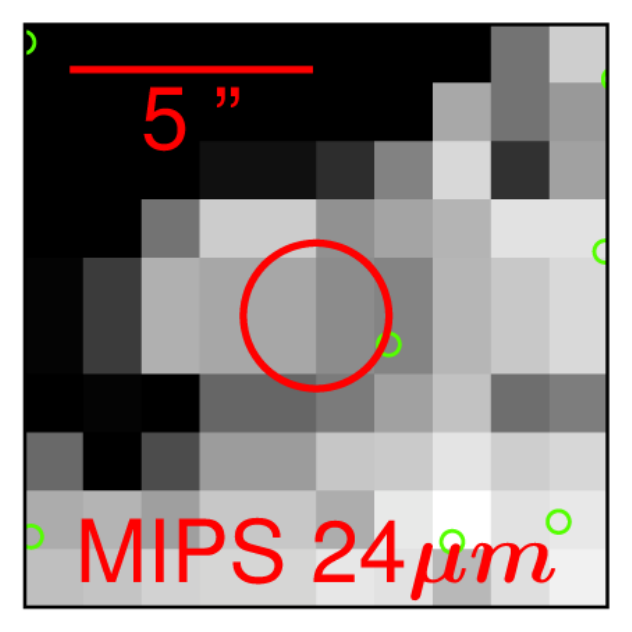}		
		\end{minipage}			
		\includegraphics[width=.49\linewidth]{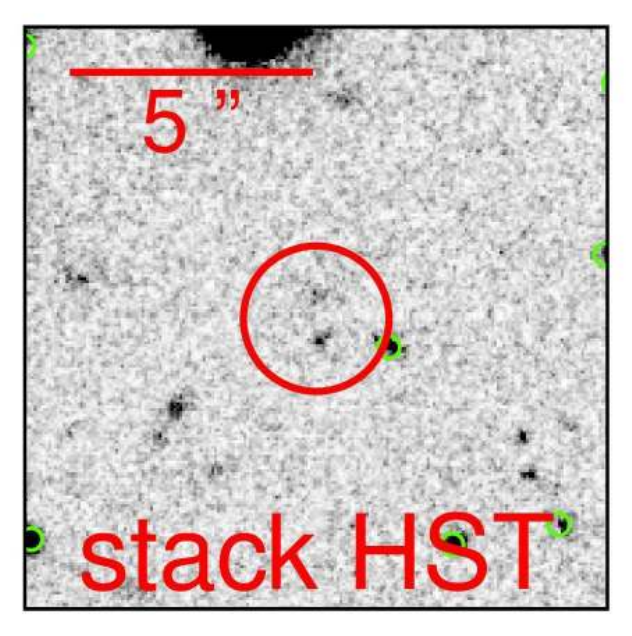}
		\includegraphics[width=.49\linewidth]{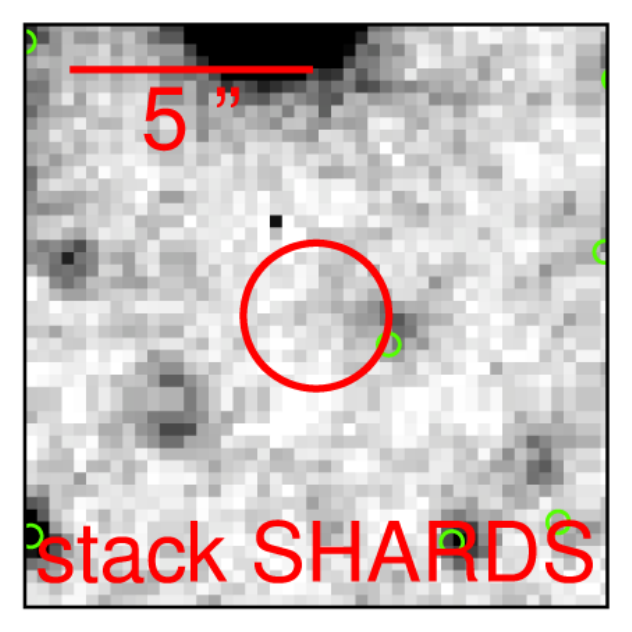}
		\centering
	\end{minipage}
	\quad
	\begin{minipage}[b]{0.52\linewidth}
		\includegraphics[width=1.\linewidth]{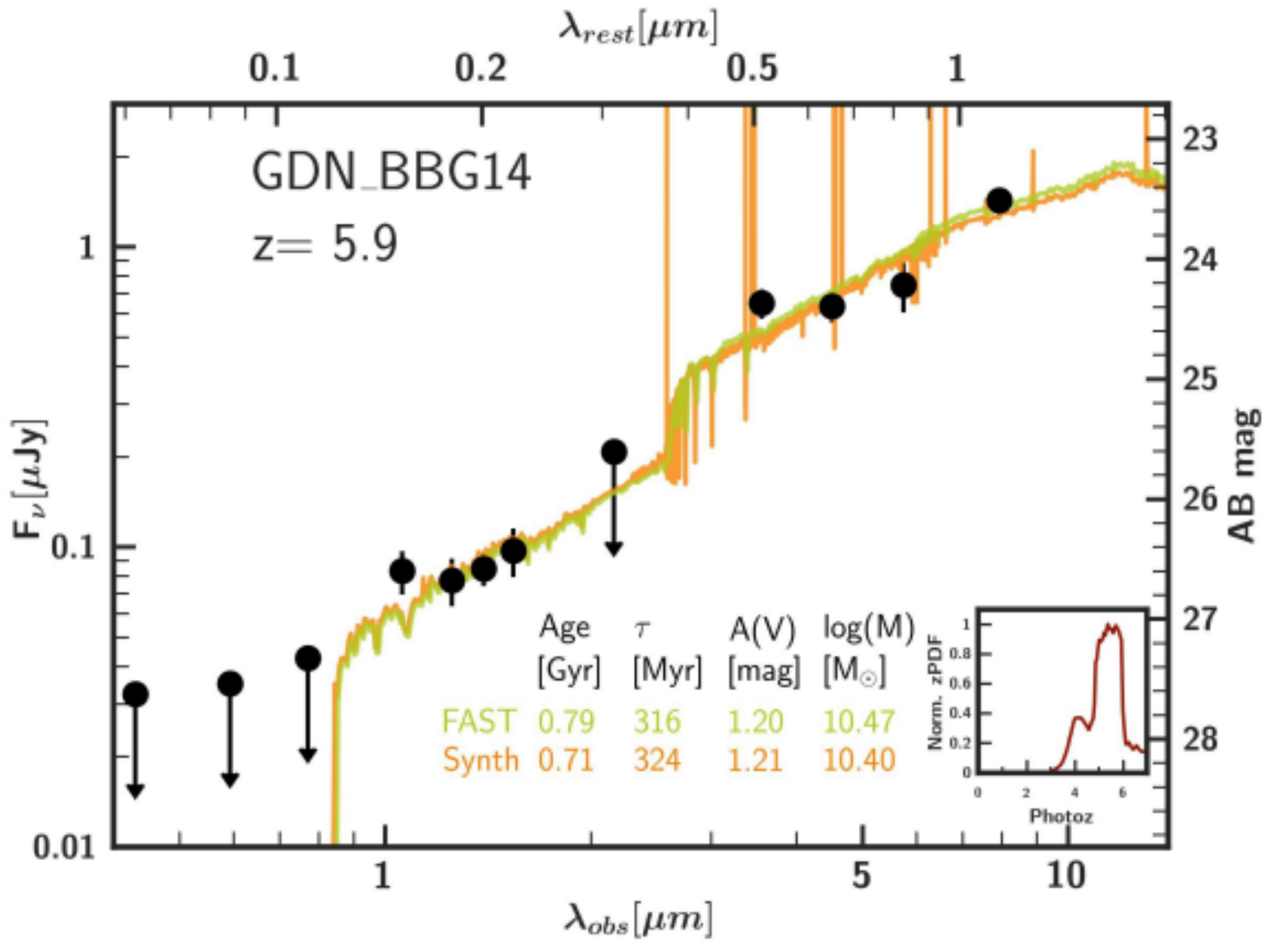}
		\centering
	\end{minipage}

% Source 15}

	\begin{minipage}[b]{0.44\linewidth}
	 \centering
			\begin{minipage}[b]{0.315\linewidth}
				\includegraphics[width=1.\linewidth]{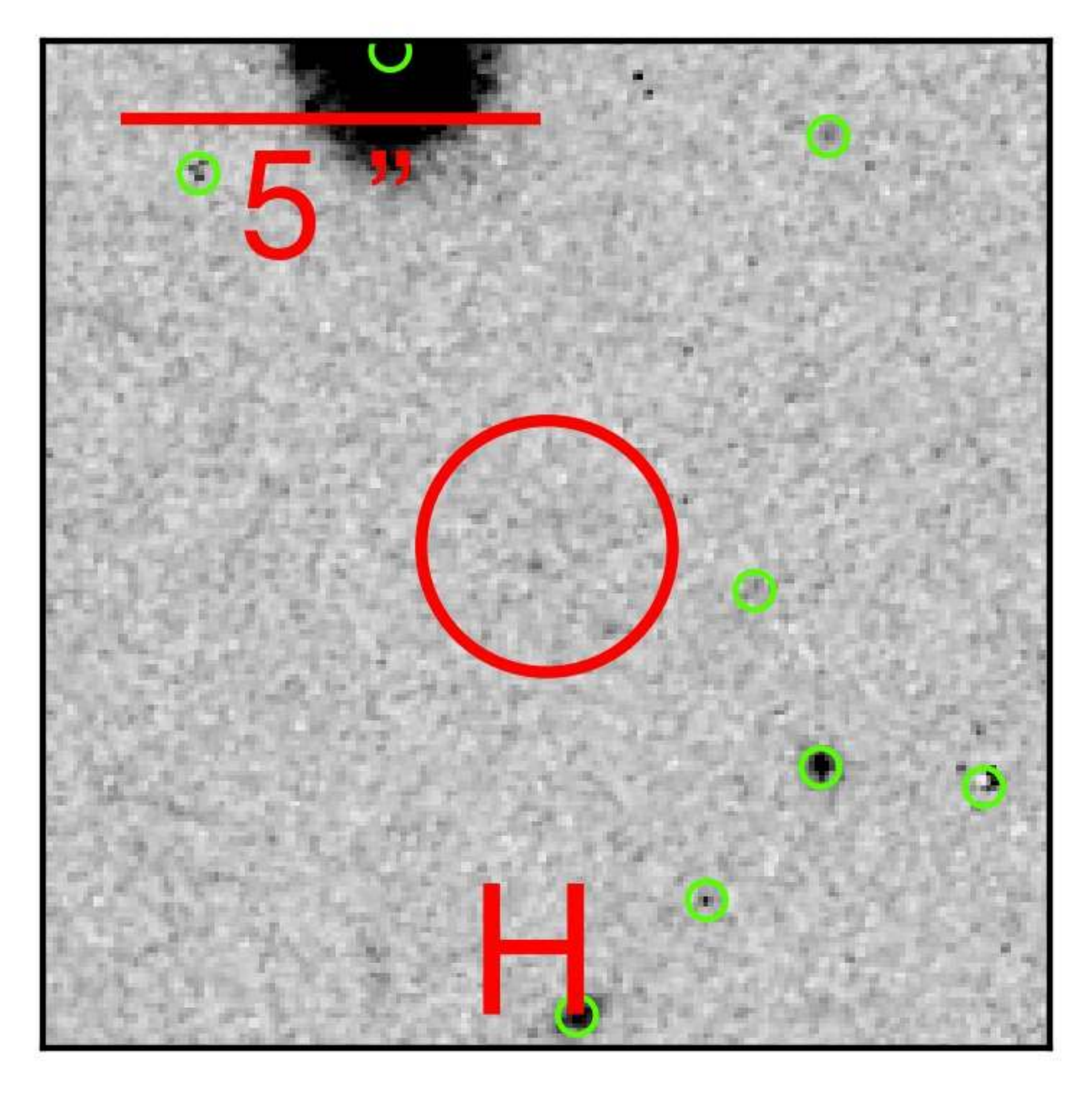}
			\end{minipage}
			\begin{minipage}[b]{0.315\linewidth}
				\includegraphics[width=1.\linewidth]{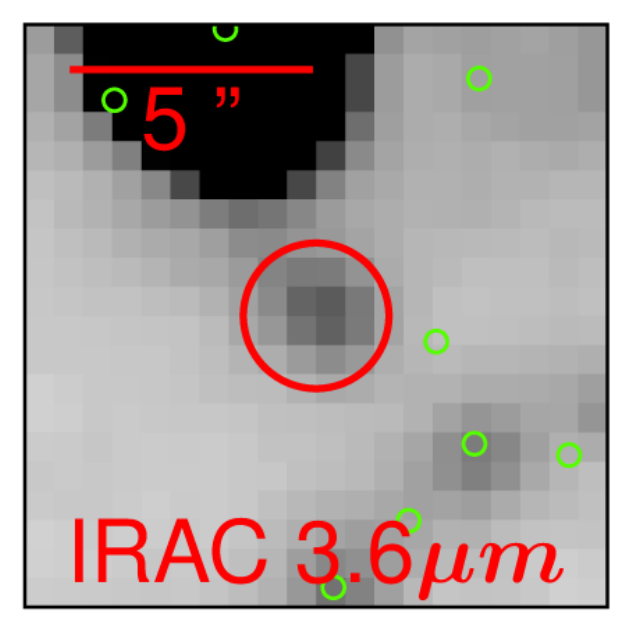}		
			\end{minipage}	
			\begin{minipage}[b]{0.315\linewidth}
				\includegraphics[width=1.\linewidth]{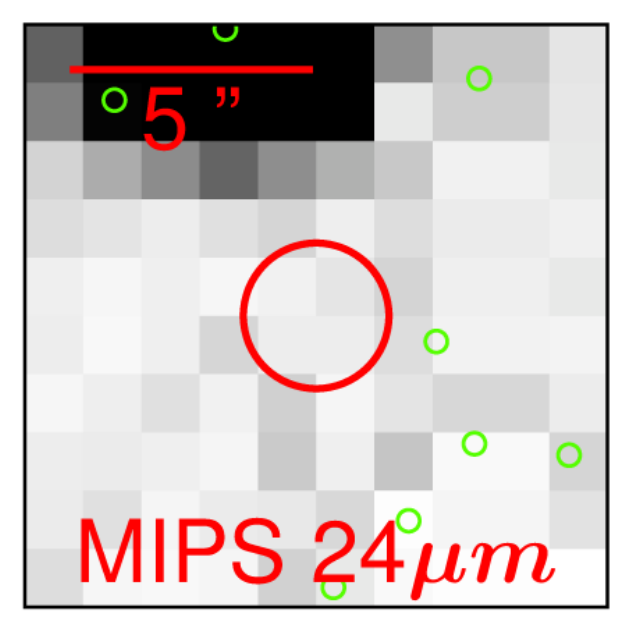}		
			\end{minipage}			
			\includegraphics[width=.49\linewidth]{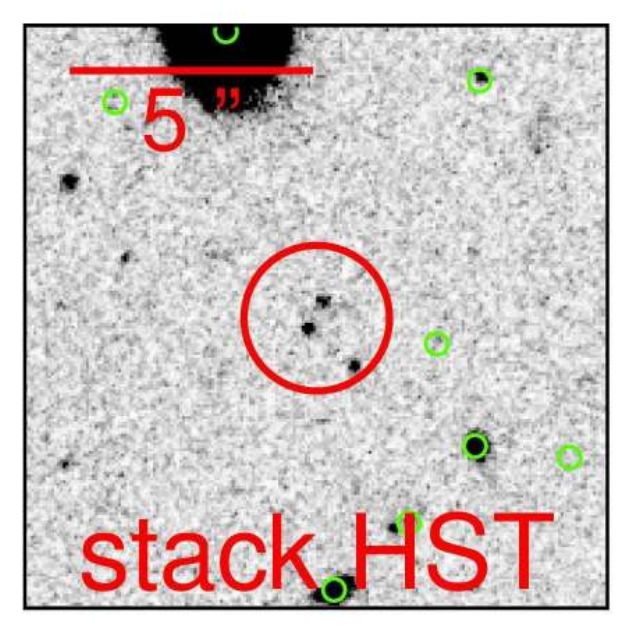}
			\includegraphics[width=.49\linewidth]{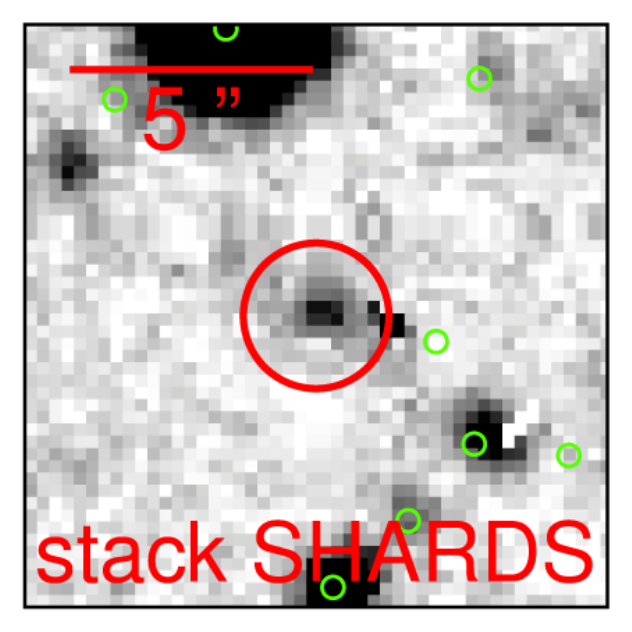}
			\centering

	\end{minipage}
	\quad
	\begin{minipage}[b]{0.52\linewidth}
	\begin{center}
		\includegraphics[width=1.\linewidth]{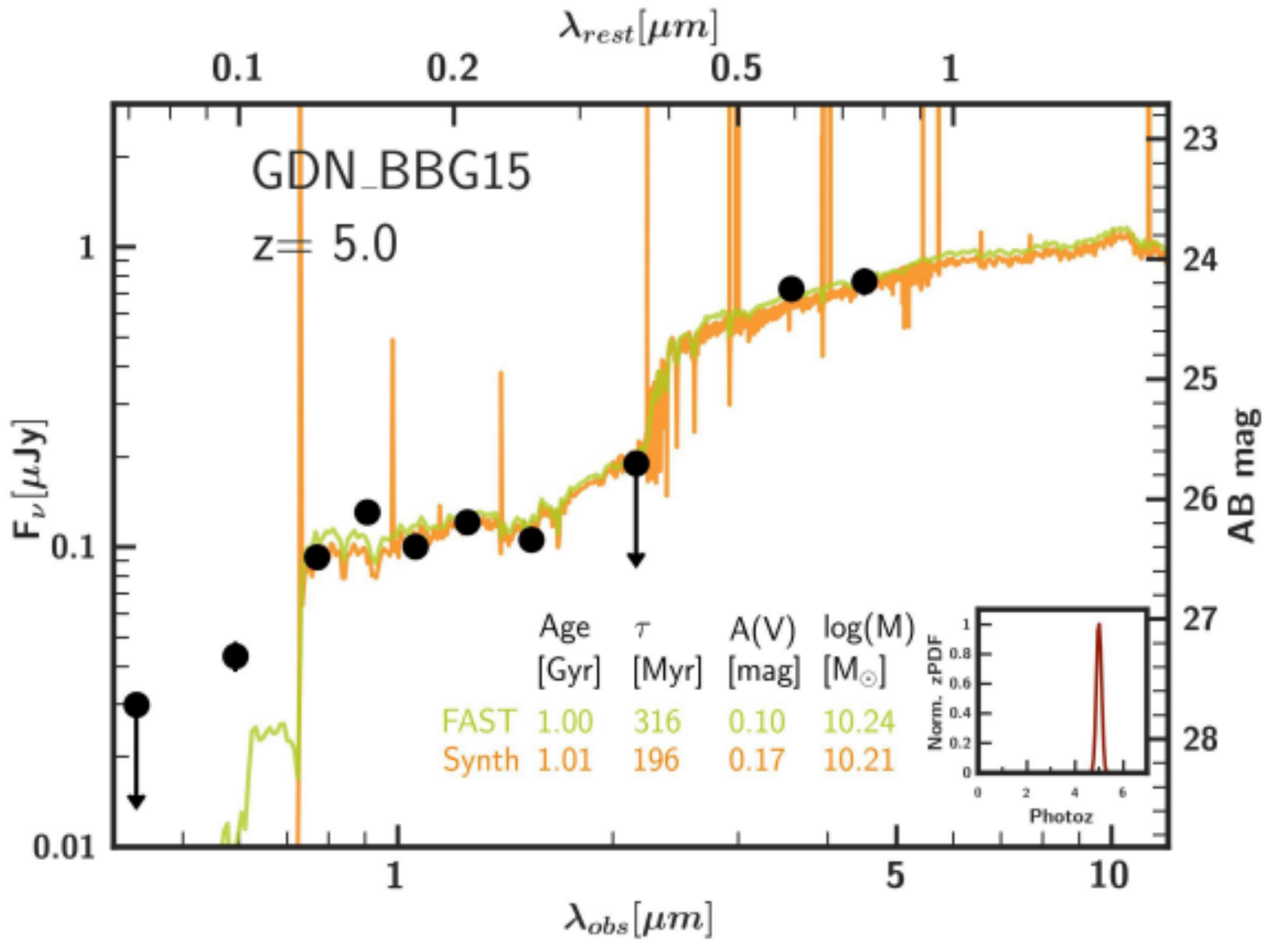}
	\end{center}
	\end{minipage}

% Source 16}

	\begin{minipage}[b]{0.44\linewidth}
		\begin{minipage}[b]{0.315\linewidth}
			\includegraphics[width=1.\linewidth]{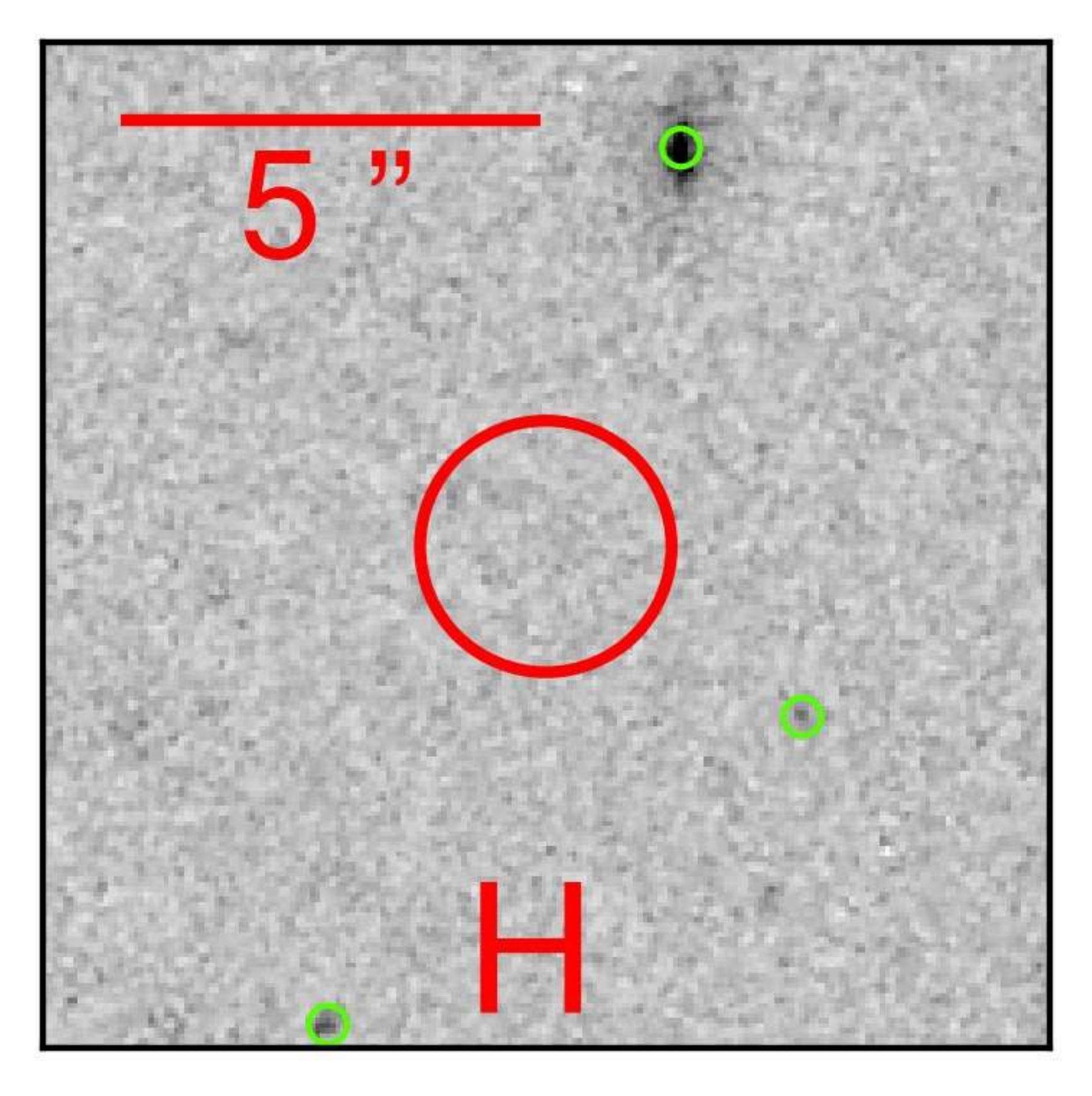}
		\end{minipage}
		\begin{minipage}[b]{0.315\linewidth}
			\includegraphics[width=1.\linewidth]{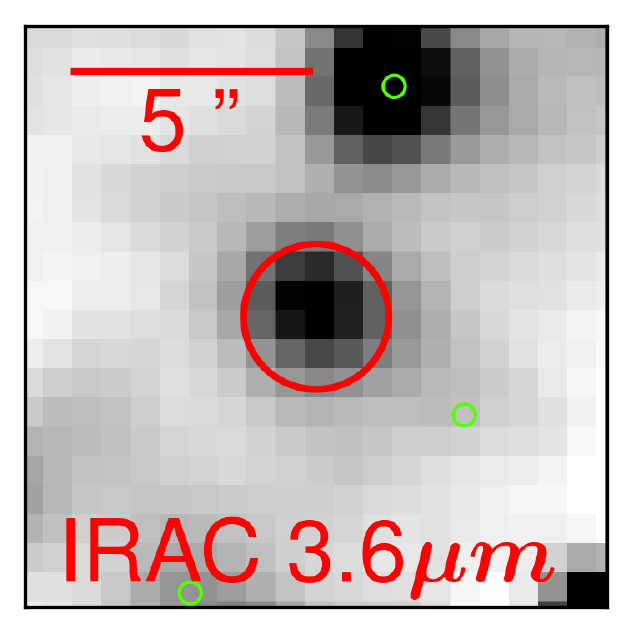}		
		\end{minipage}	
		\begin{minipage}[b]{0.315\linewidth}
			\includegraphics[width=1.\linewidth]{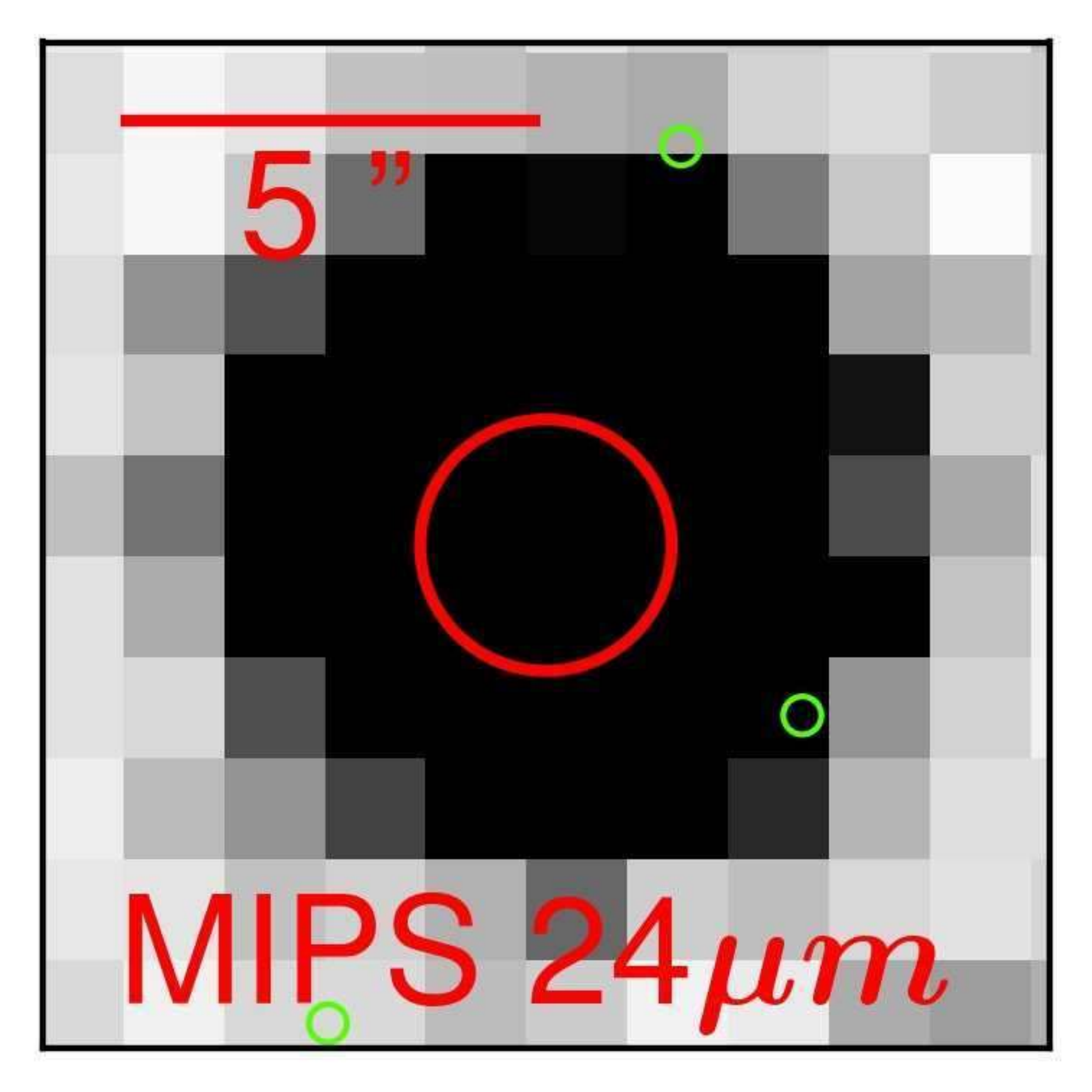}		
		\end{minipage}			
		\includegraphics[width=.49\linewidth]{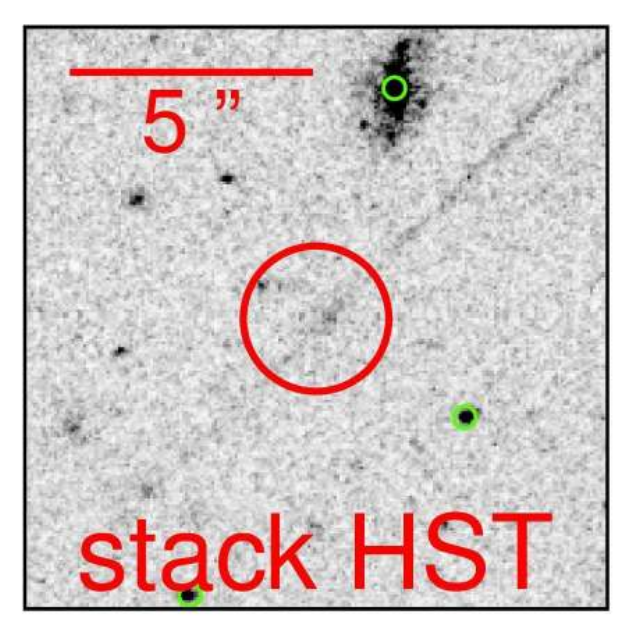}
		\includegraphics[width=.49\linewidth]{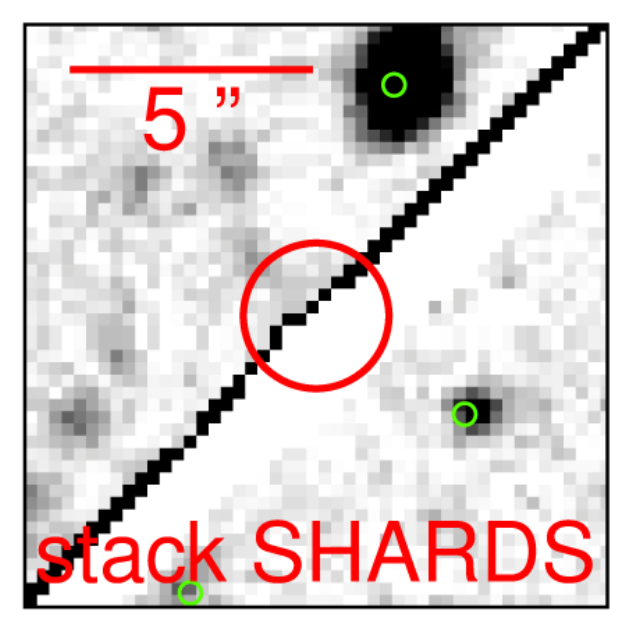}
		\centering
	\end{minipage}
	\quad
	\begin{minipage}[b]{0.52\linewidth}
		\includegraphics[width=1.\linewidth]{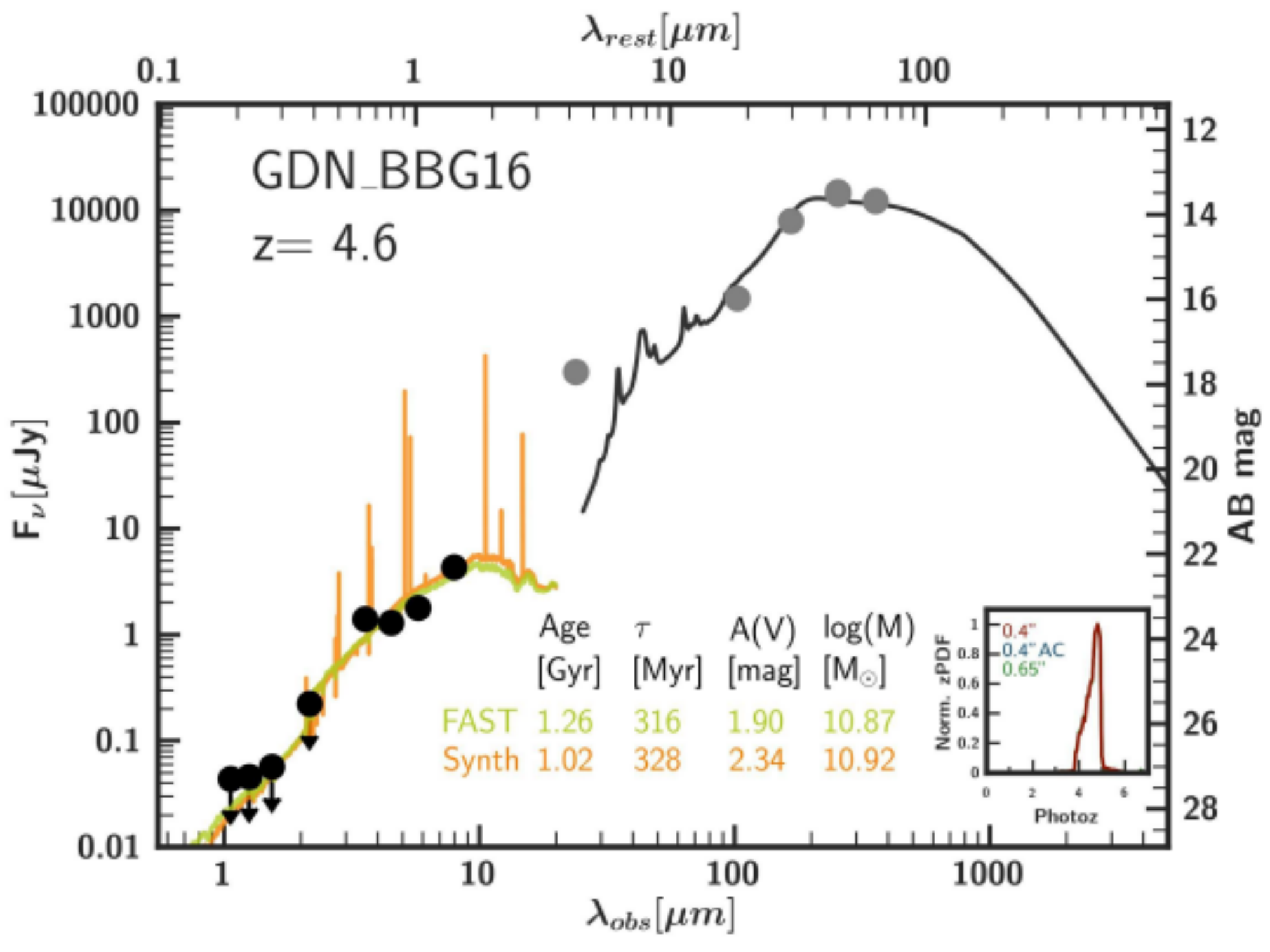}
		\centering
	\end{minipage}
\end{figure*}

% Source 17}
\begin{figure*}
	\begin{minipage}[b]{0.44\linewidth}
		\begin{minipage}[b]{0.315\linewidth}
			\includegraphics[width=1.\linewidth]{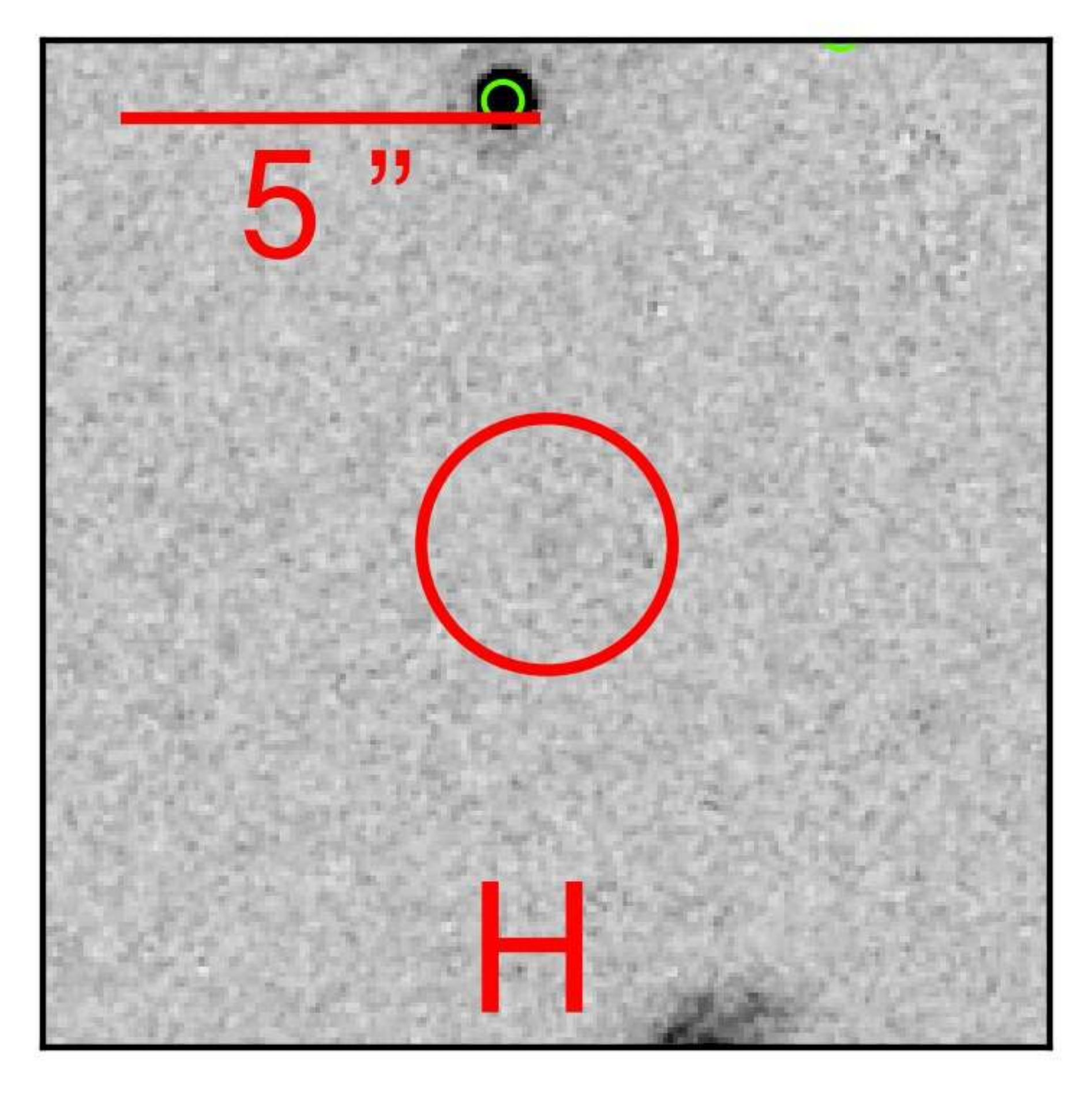}
		\end{minipage}
		\begin{minipage}[b]{0.315\linewidth}
			\includegraphics[width=1.\linewidth]{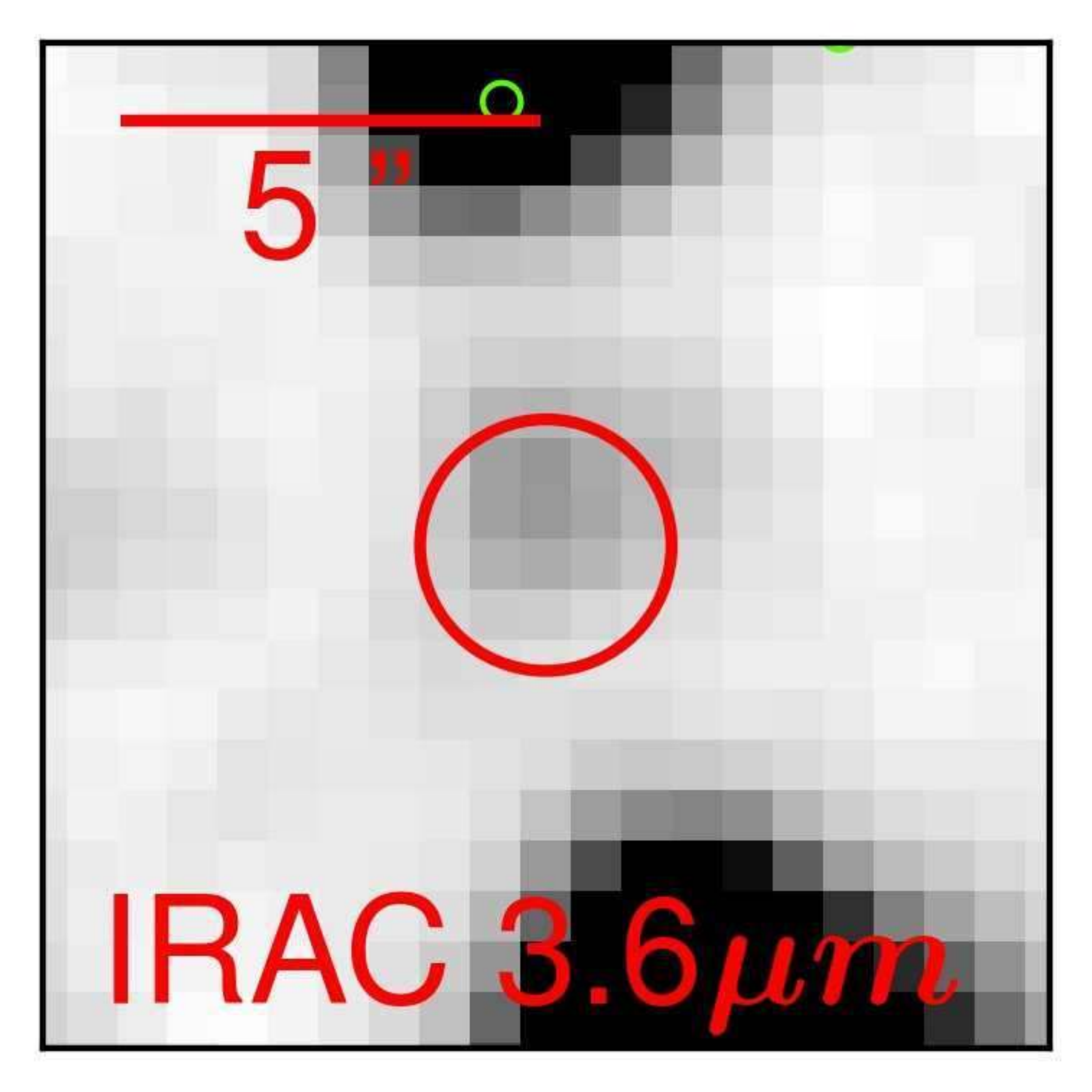}		
		\end{minipage}	
		\begin{minipage}[b]{0.315\linewidth}
			\includegraphics[width=1.\linewidth]{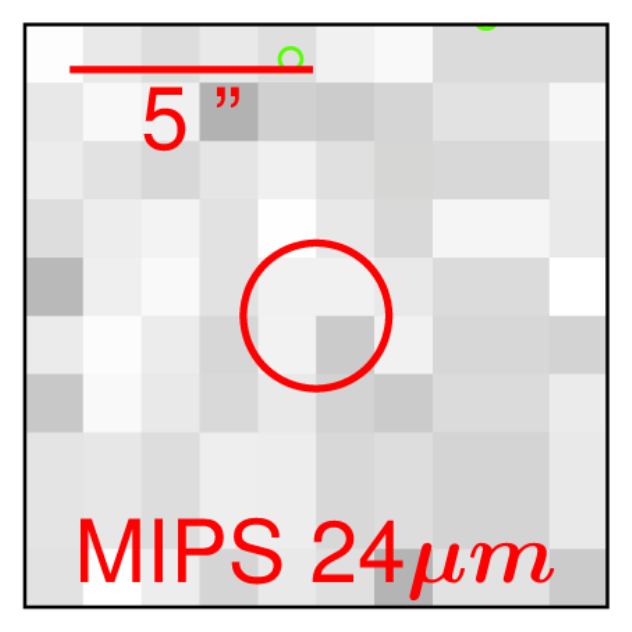}		
		\end{minipage}			
		\includegraphics[width=.49\linewidth]{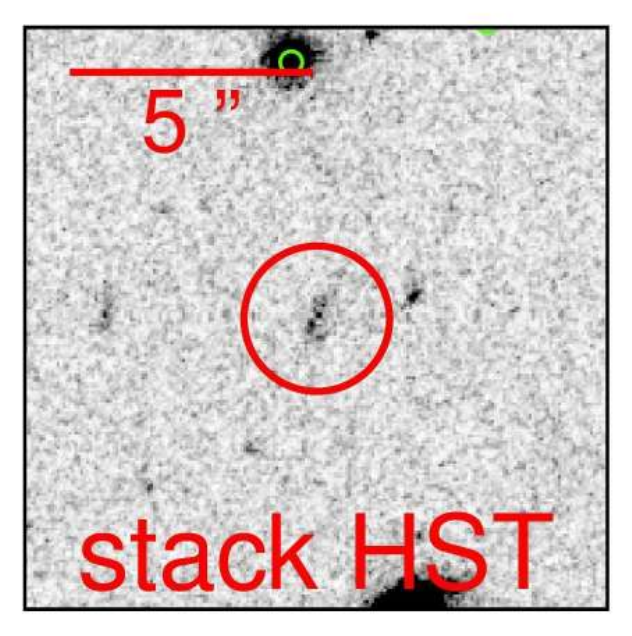}
		\includegraphics[width=.49\linewidth]{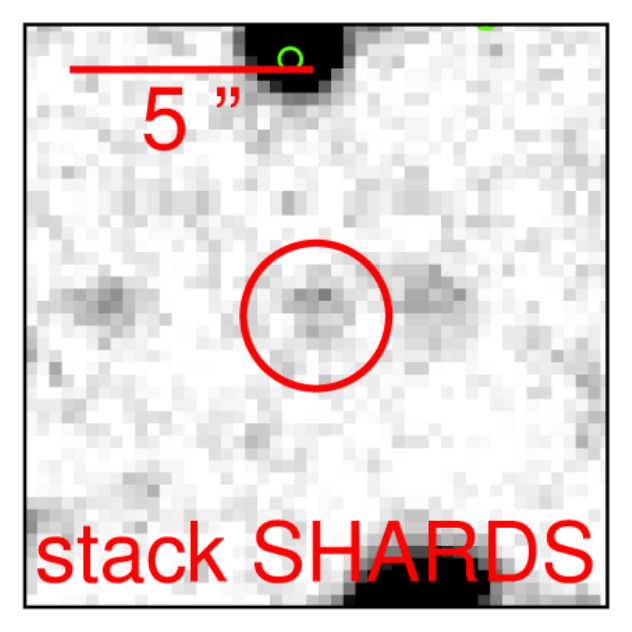}
		\centering
	\end{minipage}
	\quad
	\begin{minipage}[b]{0.52\linewidth}
		\includegraphics[width=1.\linewidth]{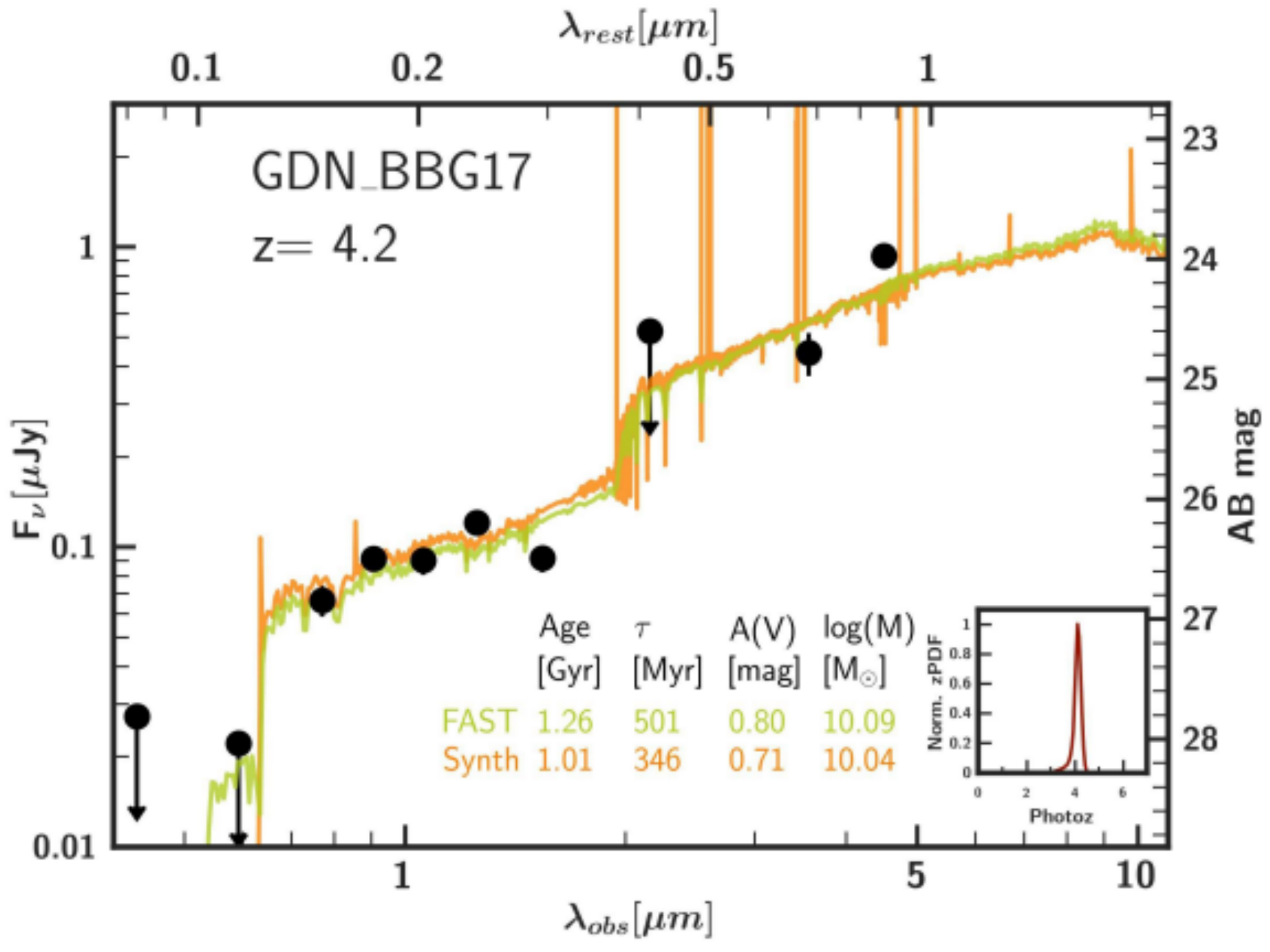}
		\centering
	\end{minipage}
\end{figure*}

\clearpage

% Source 1}
\begin{figure*}
	\begin{minipage}[b]{0.44\linewidth}
		\centering
		\begin{minipage}[b]{0.315\linewidth}
			\includegraphics[width=1.\linewidth]{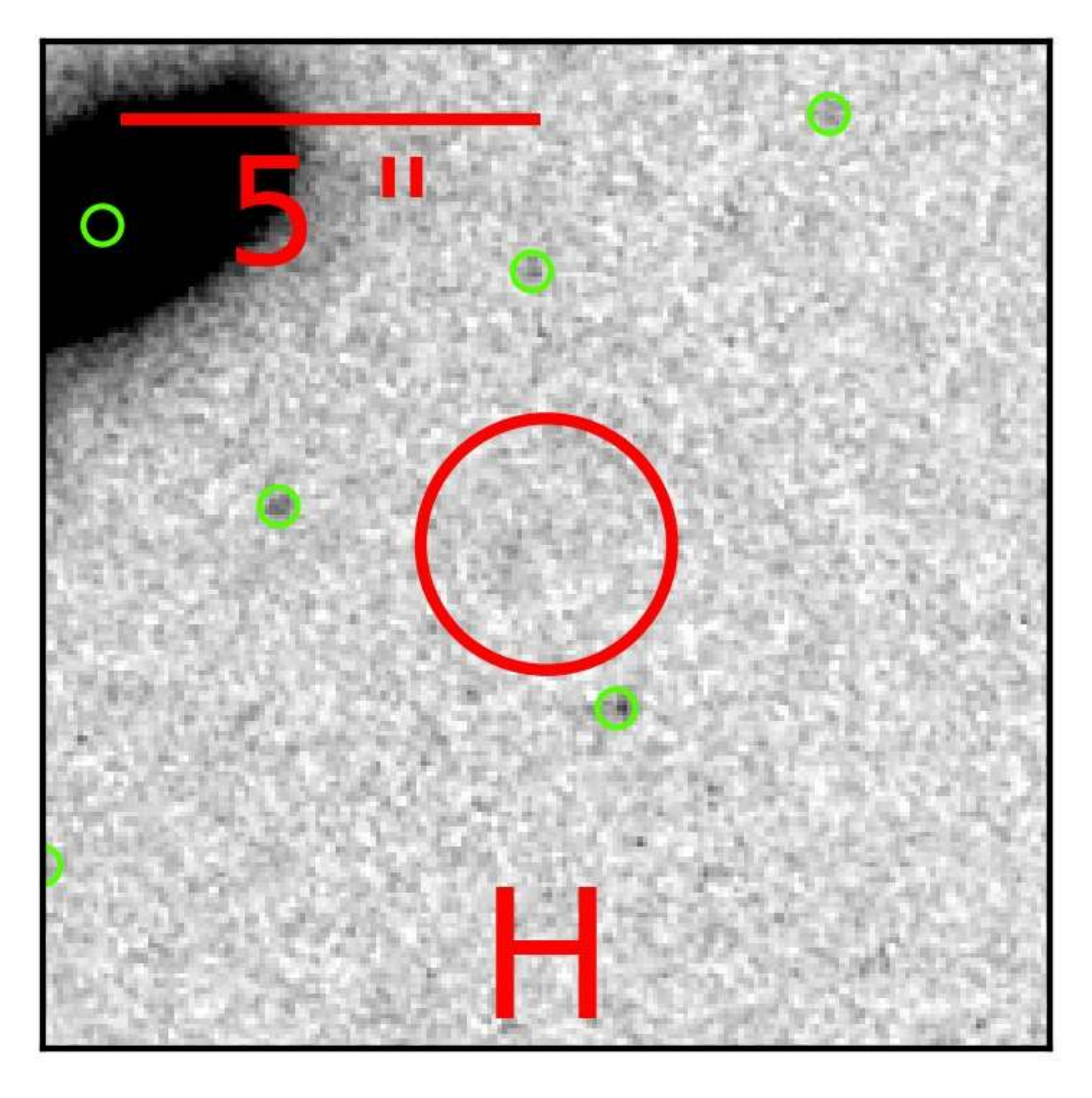}
		\end{minipage}
		\begin{minipage}[b]{0.315\linewidth}
			\includegraphics[width=1.\linewidth]{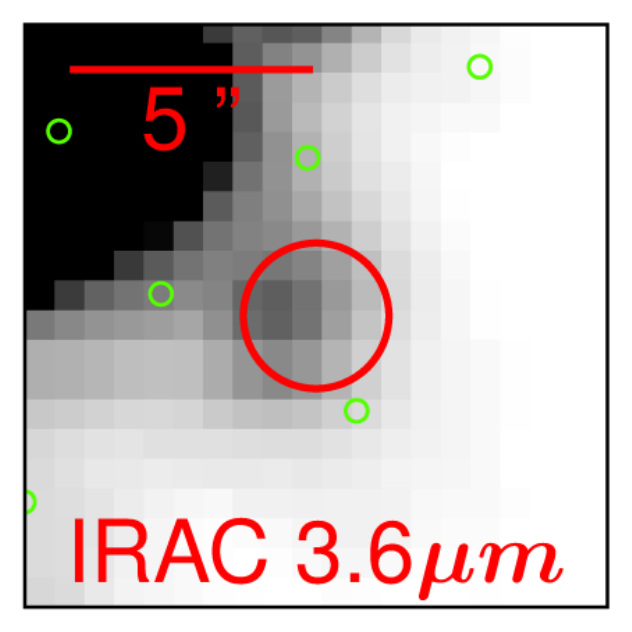}		
		\end{minipage}	
		\begin{minipage}[b]{0.315\linewidth}
			\includegraphics[width=1.\linewidth]{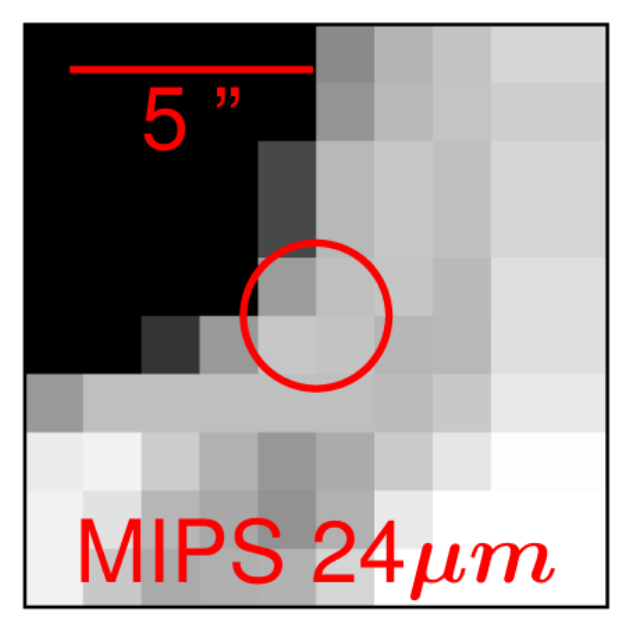}		
		\end{minipage}			
		\includegraphics[width=.49\linewidth]{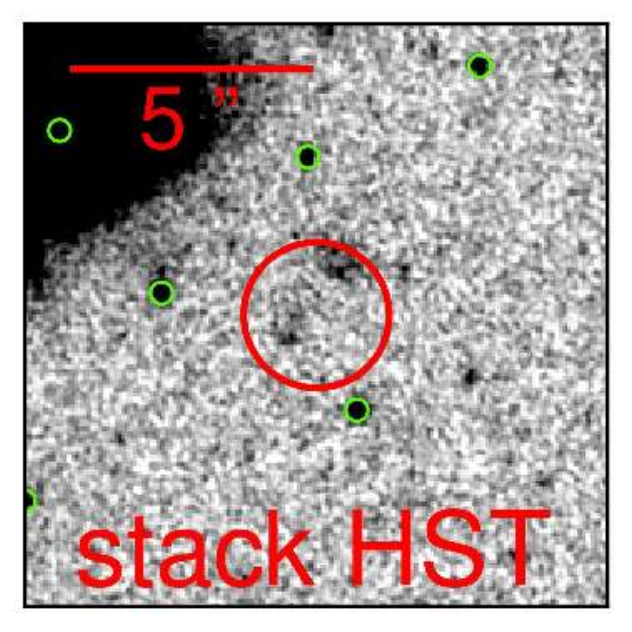}
		\includegraphics[width=.49\linewidth]{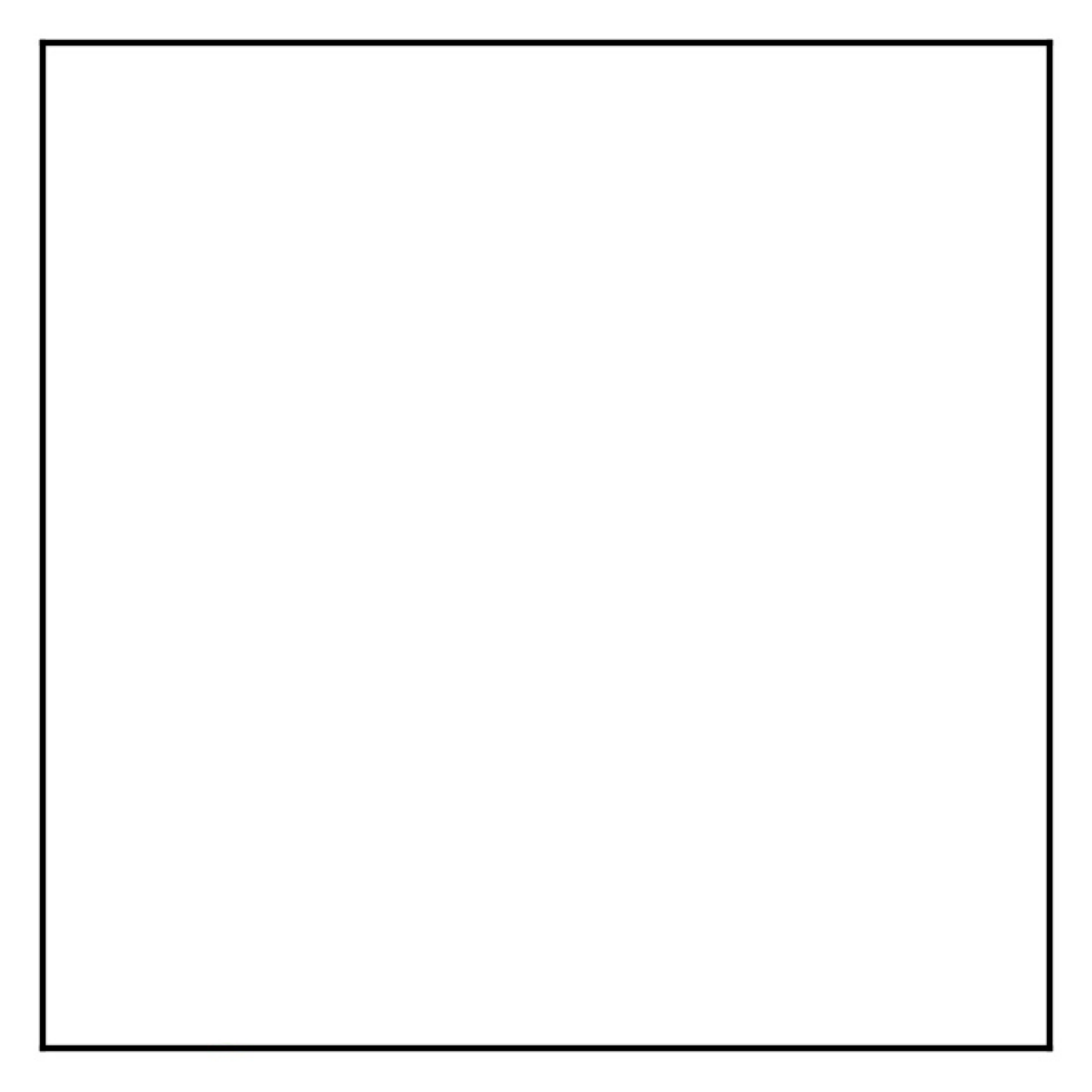}
		\centering
	\end{minipage}
	\quad
	\begin{minipage}[b]{0.52\linewidth}
		\begin{center}
			\includegraphics[width=1.\linewidth]{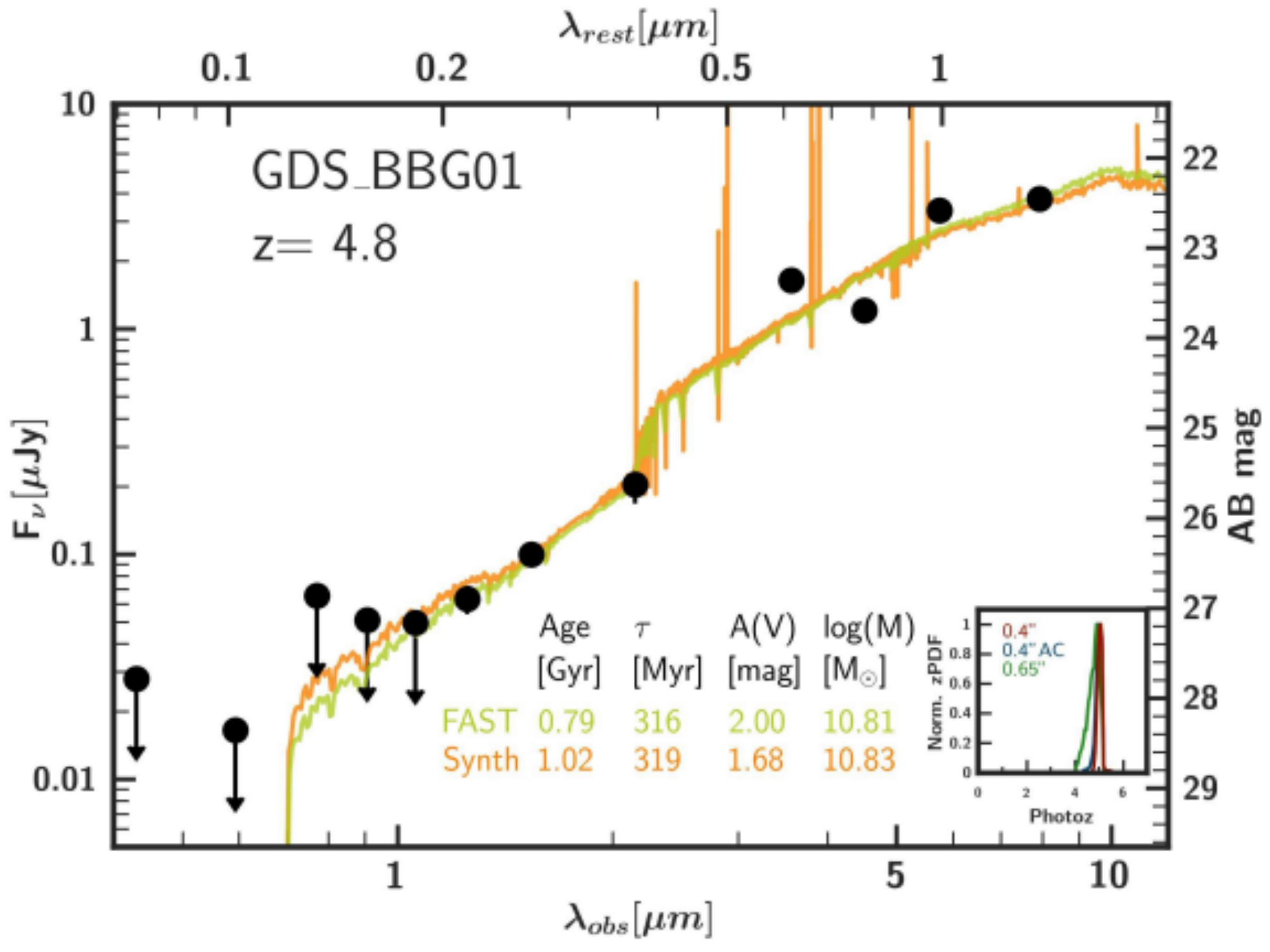}
		\end{center}
	\end{minipage}

% Source 2}
	\begin{minipage}[b]{0.44\linewidth}
		\centering
		\begin{minipage}[b]{0.315\linewidth}
			\includegraphics[width=1.\linewidth]{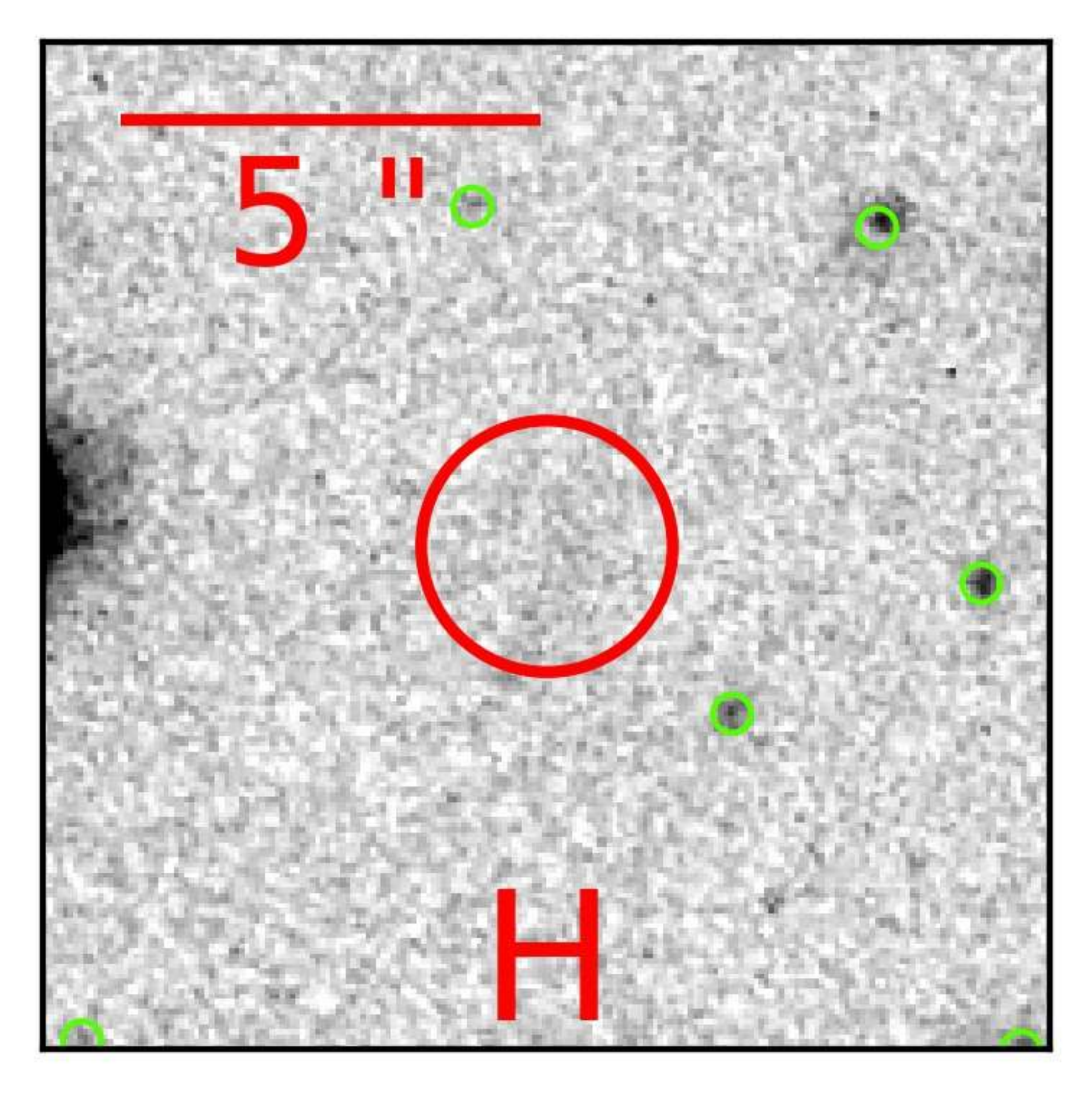}
		\end{minipage}
		\begin{minipage}[b]{0.315\linewidth}
			\includegraphics[width=1.\linewidth]{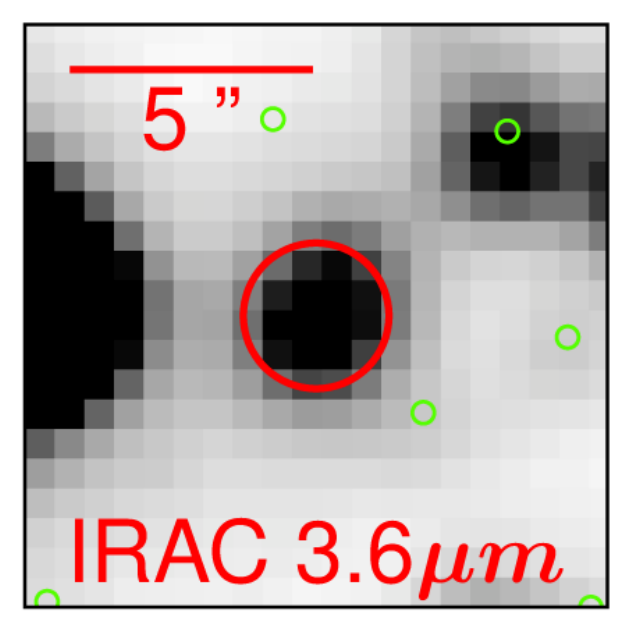}		
		\end{minipage}	
		\begin{minipage}[b]{0.315\linewidth}
			\includegraphics[width=1.\linewidth]{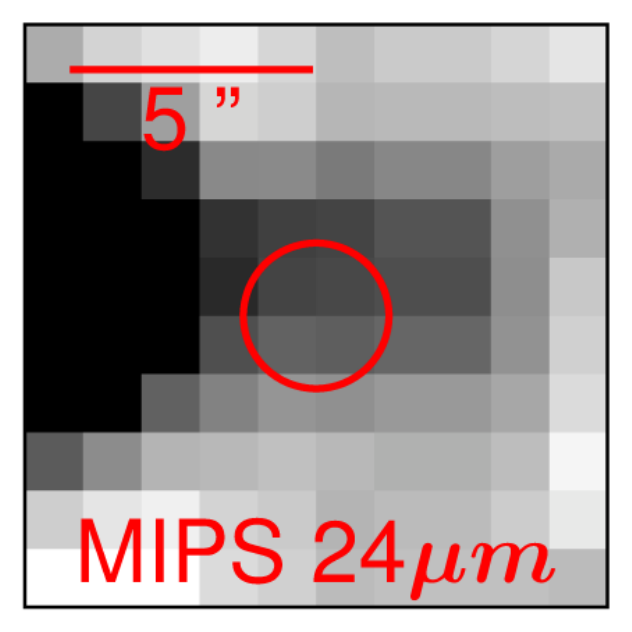}		
		\end{minipage}			
		\includegraphics[width=.49\linewidth]{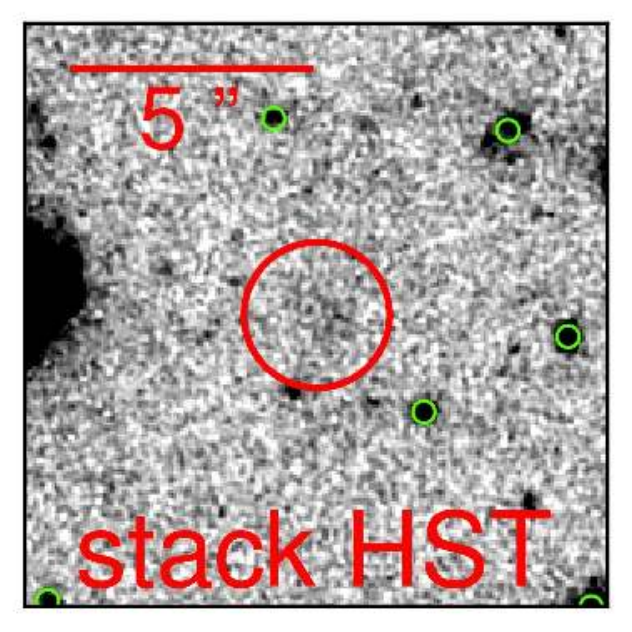}
		\includegraphics[width=.49\linewidth]{SH_ios_gs.pdf}
		\centering
	\end{minipage}
	\quad
	\begin{minipage}[b]{0.52\linewidth}
		\begin{center}
			\includegraphics[width=1.\linewidth]{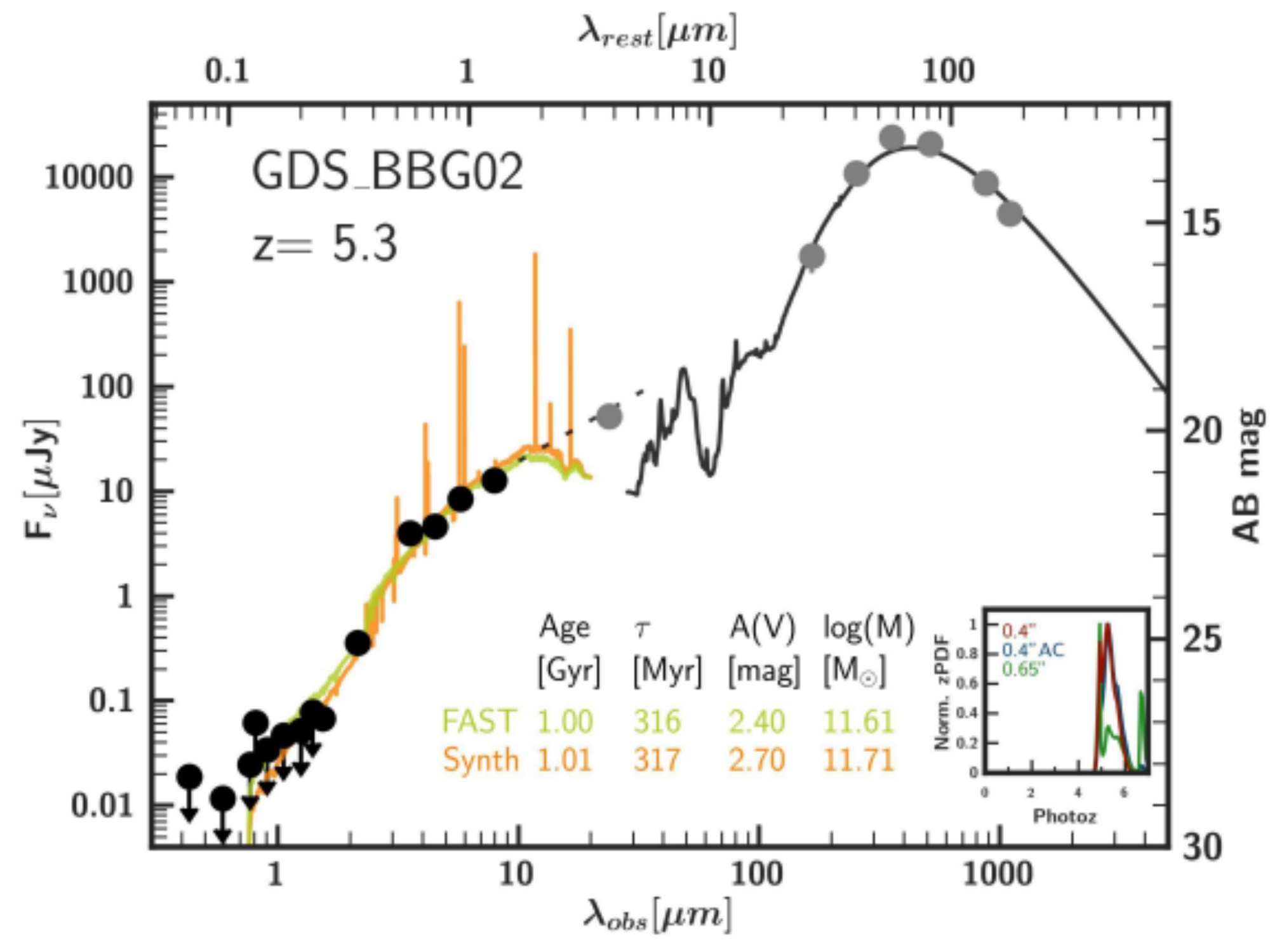}
		\end{center}
	\end{minipage}

% Source 3}
	\begin{minipage}[b]{0.44\linewidth}
		\begin{minipage}[b]{0.315\linewidth}
			\includegraphics[width=1.\linewidth]{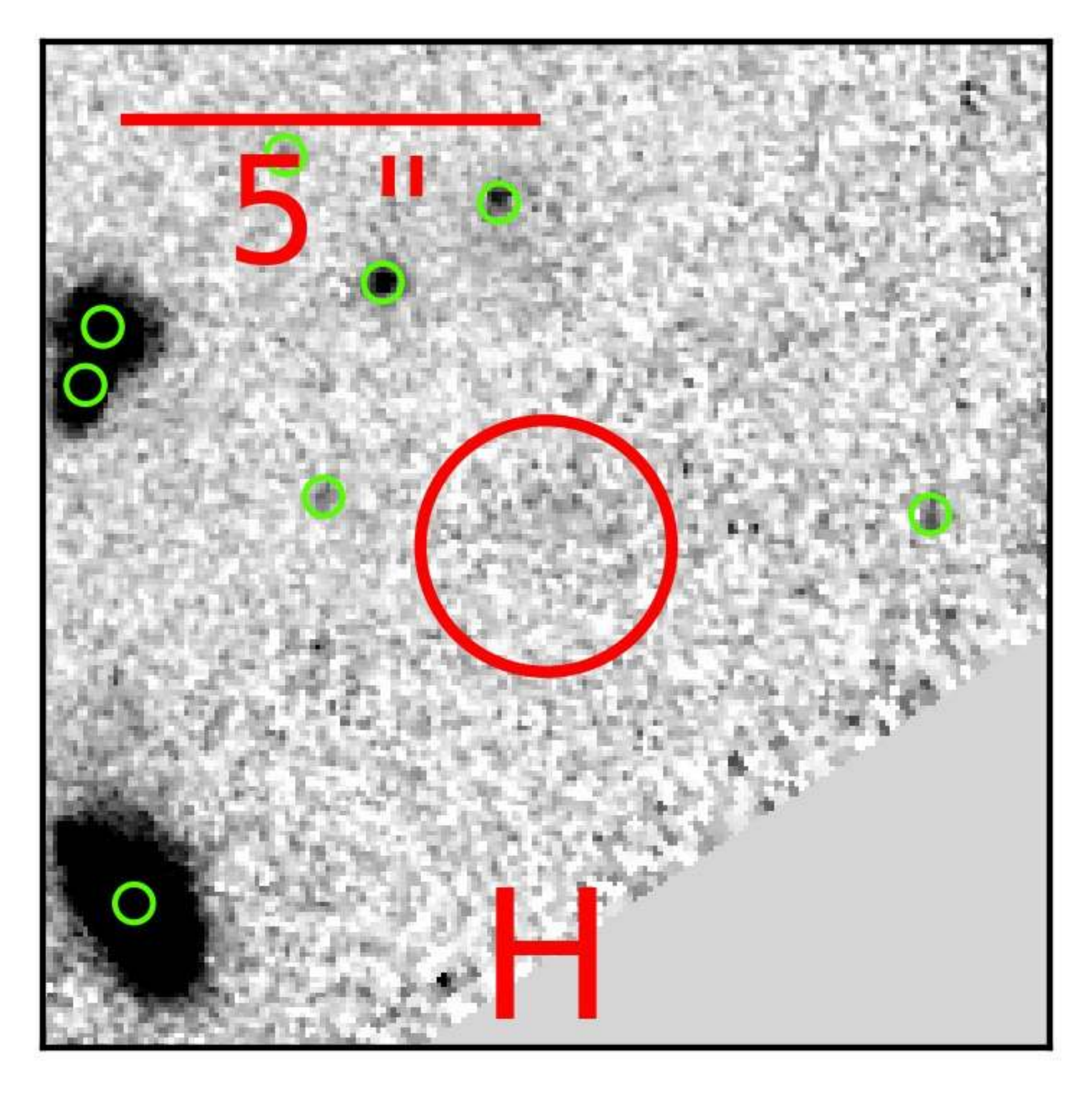}
		\end{minipage}
		\begin{minipage}[b]{0.315\linewidth}
			\includegraphics[width=1.\linewidth]{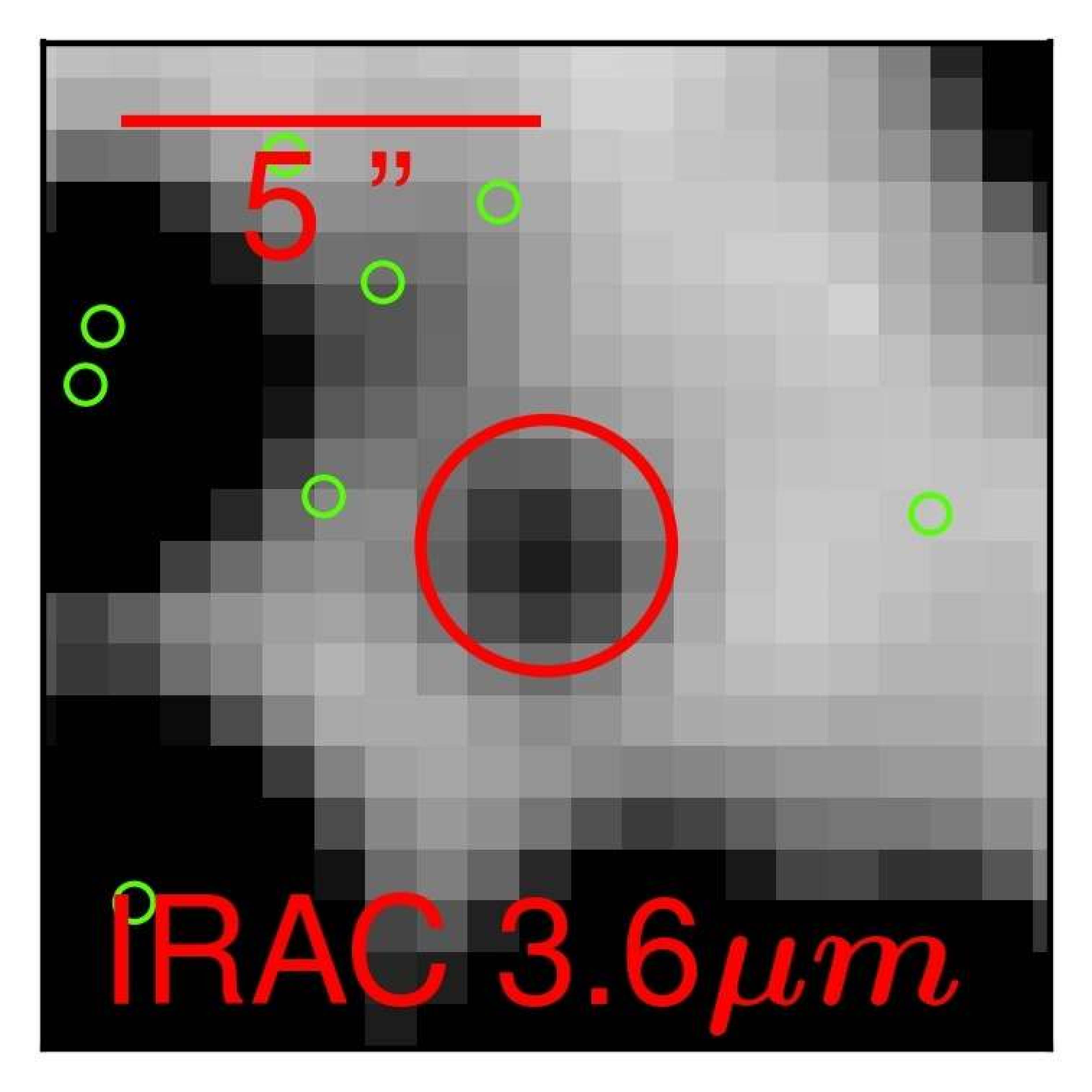}		
		\end{minipage}	
		\begin{minipage}[b]{0.315\linewidth}
			\includegraphics[width=1.\linewidth]{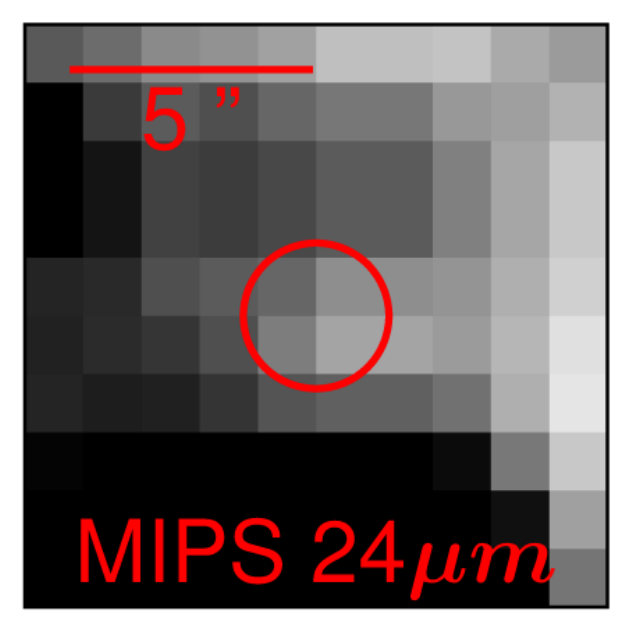}		
		\end{minipage}			
		\includegraphics[width=.49\linewidth]{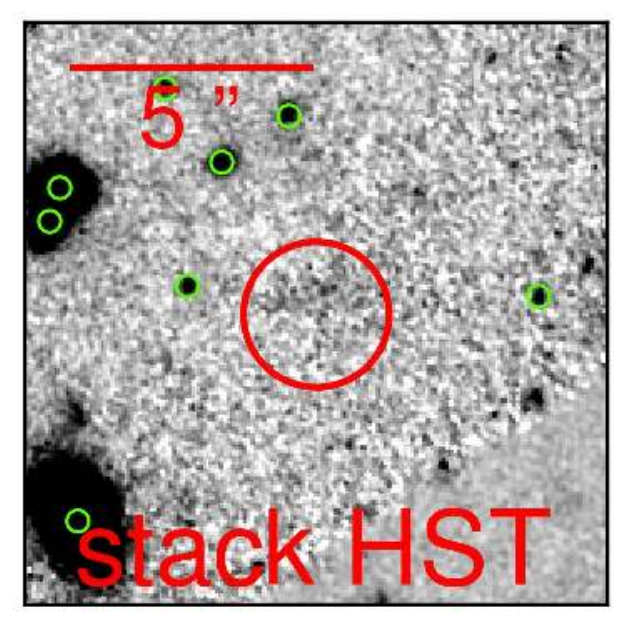}
		\includegraphics[width=.49\linewidth]{SH_ios_gs.pdf}
		\centering
	\end{minipage}
	\quad
	\begin{minipage}[b]{0.52\linewidth}
		\includegraphics[width=1.\linewidth]{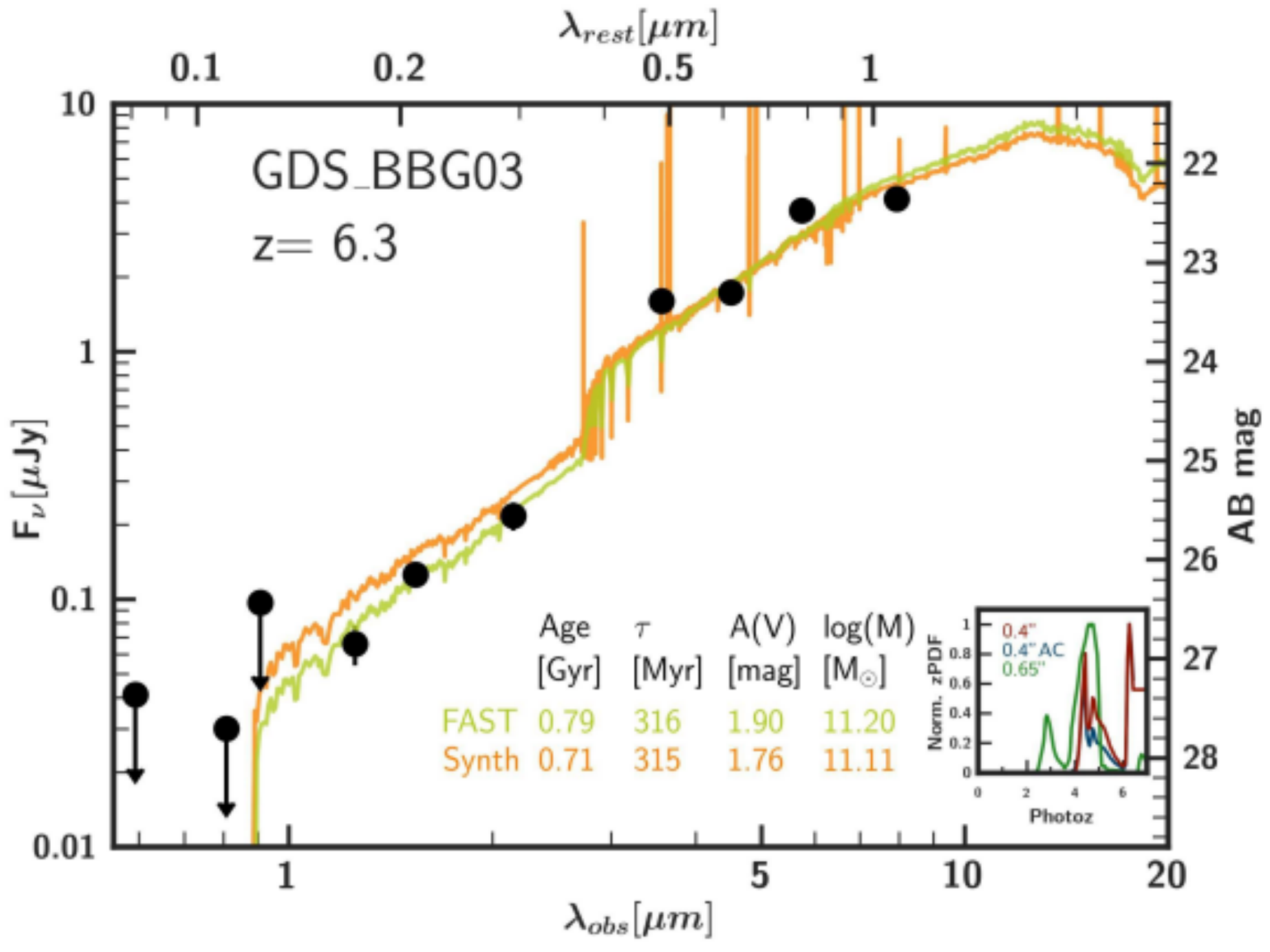}
		\centering
	\end{minipage}
\end{figure*}

% Source 4}
\begin{figure*}
	\begin{minipage}[b]{0.44\linewidth}
		\centering
		\begin{minipage}[b]{0.315\linewidth}
			\includegraphics[width=1.\linewidth]{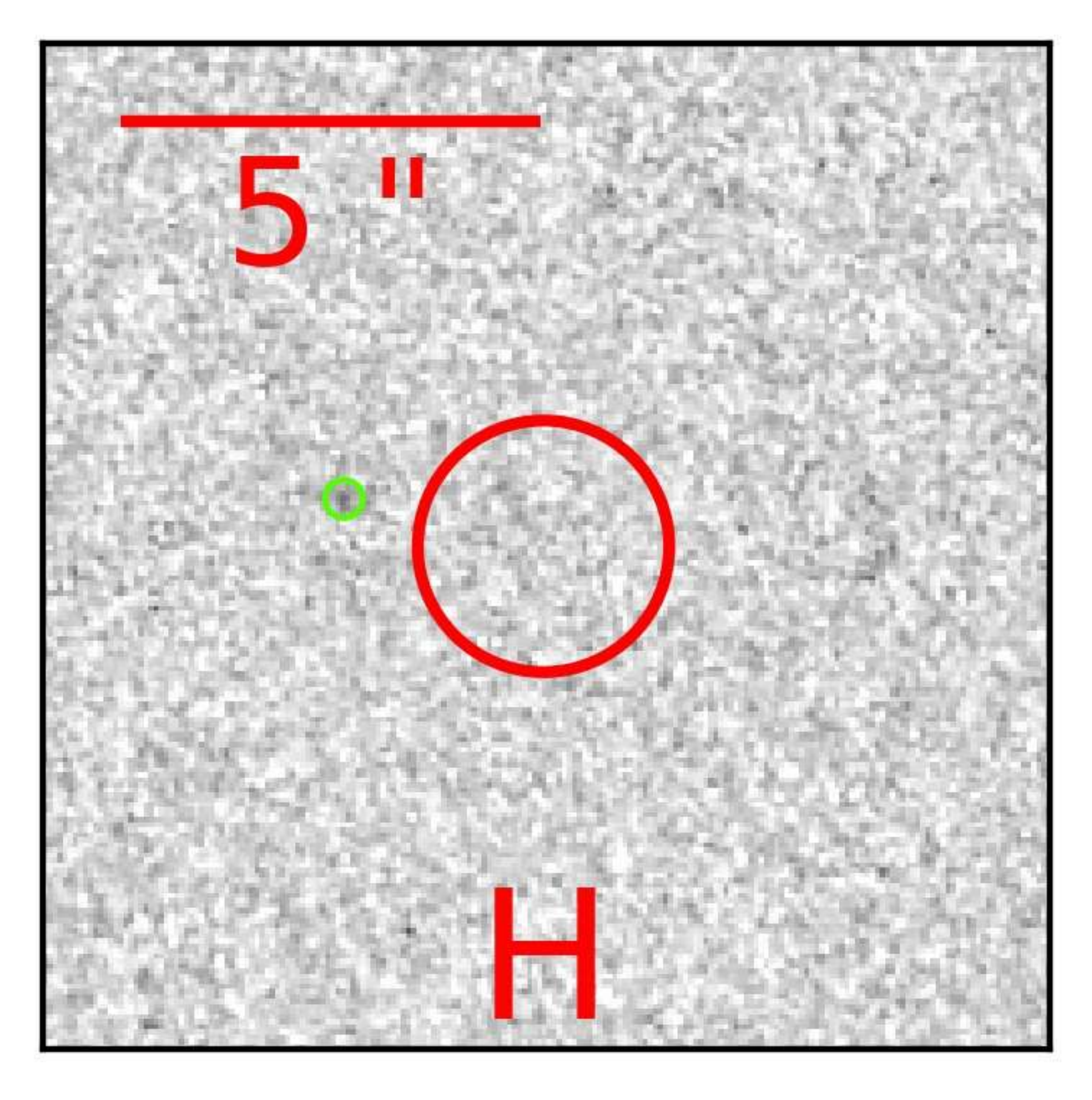}
		\end{minipage}
		\begin{minipage}[b]{0.315\linewidth}
			\includegraphics[width=1.\linewidth]{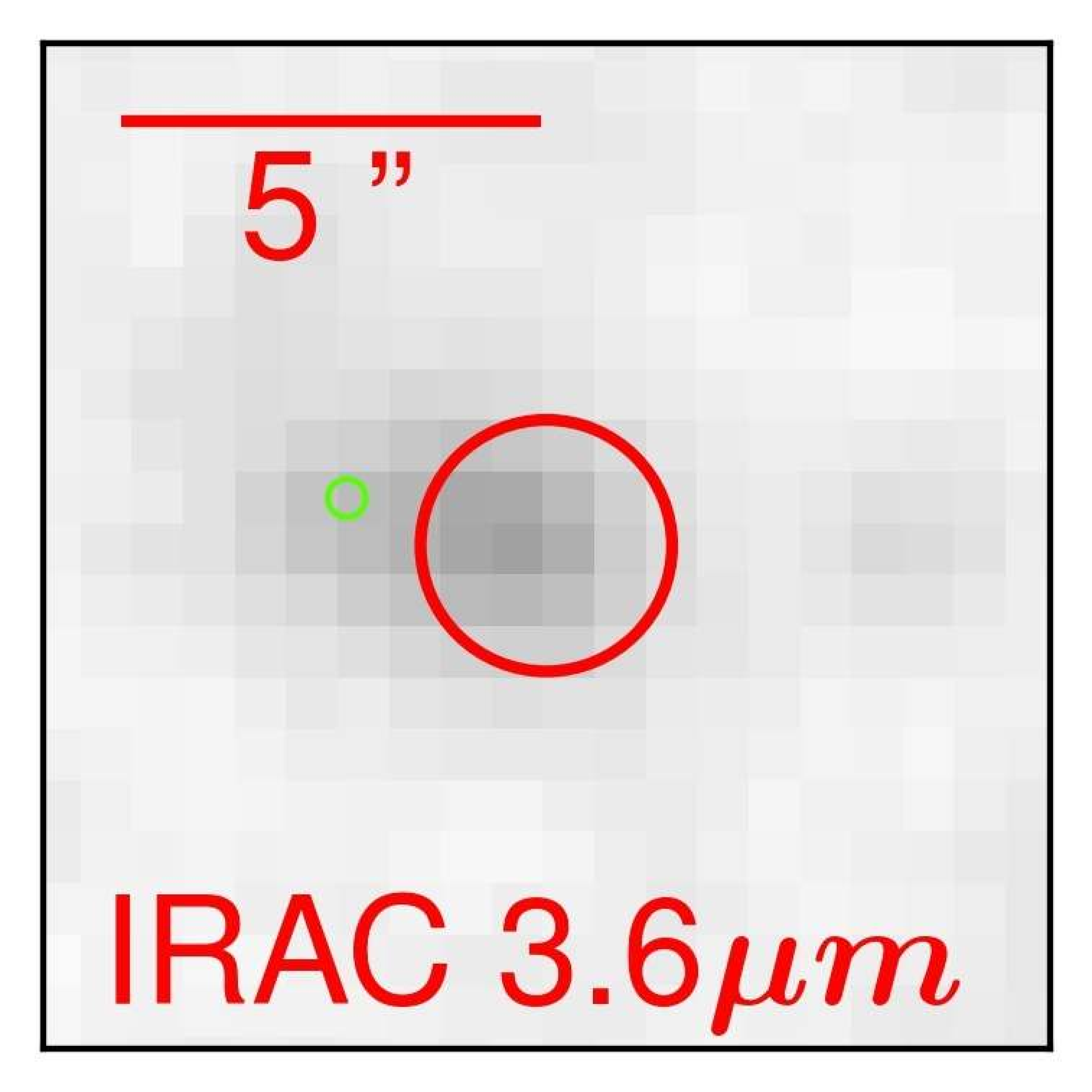}		
		\end{minipage}	
		\begin{minipage}[b]{0.315\linewidth}
			\includegraphics[width=1.\linewidth]{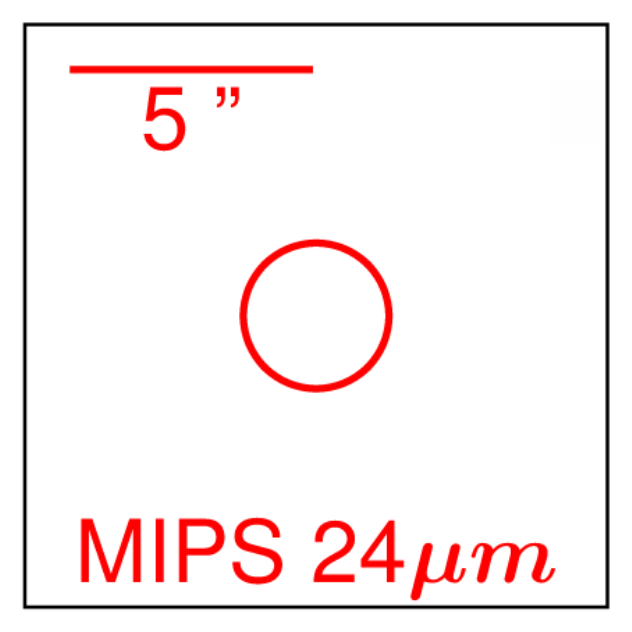}		
		\end{minipage}			
		\includegraphics[width=.49\linewidth]{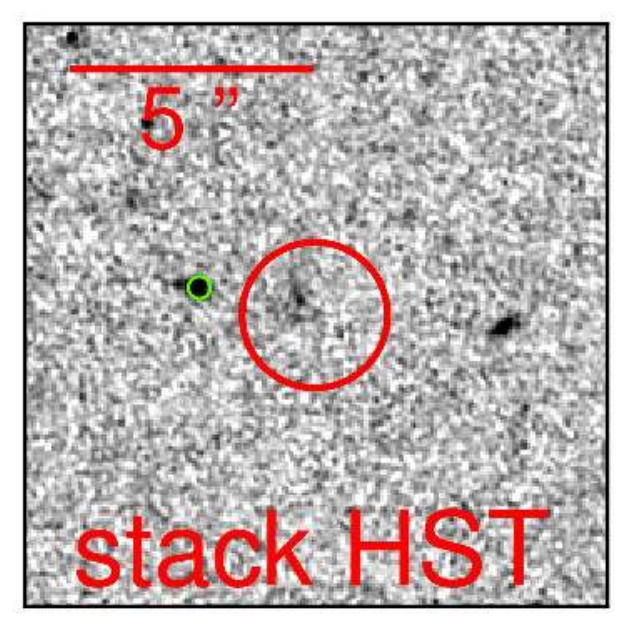}
		\includegraphics[width=.49\linewidth]{SH_ios_gs.pdf}
		\centering
	\end{minipage}
	\quad
	\begin{minipage}[b]{0.52\linewidth}
		\begin{center}
			\includegraphics[width=1.\linewidth]{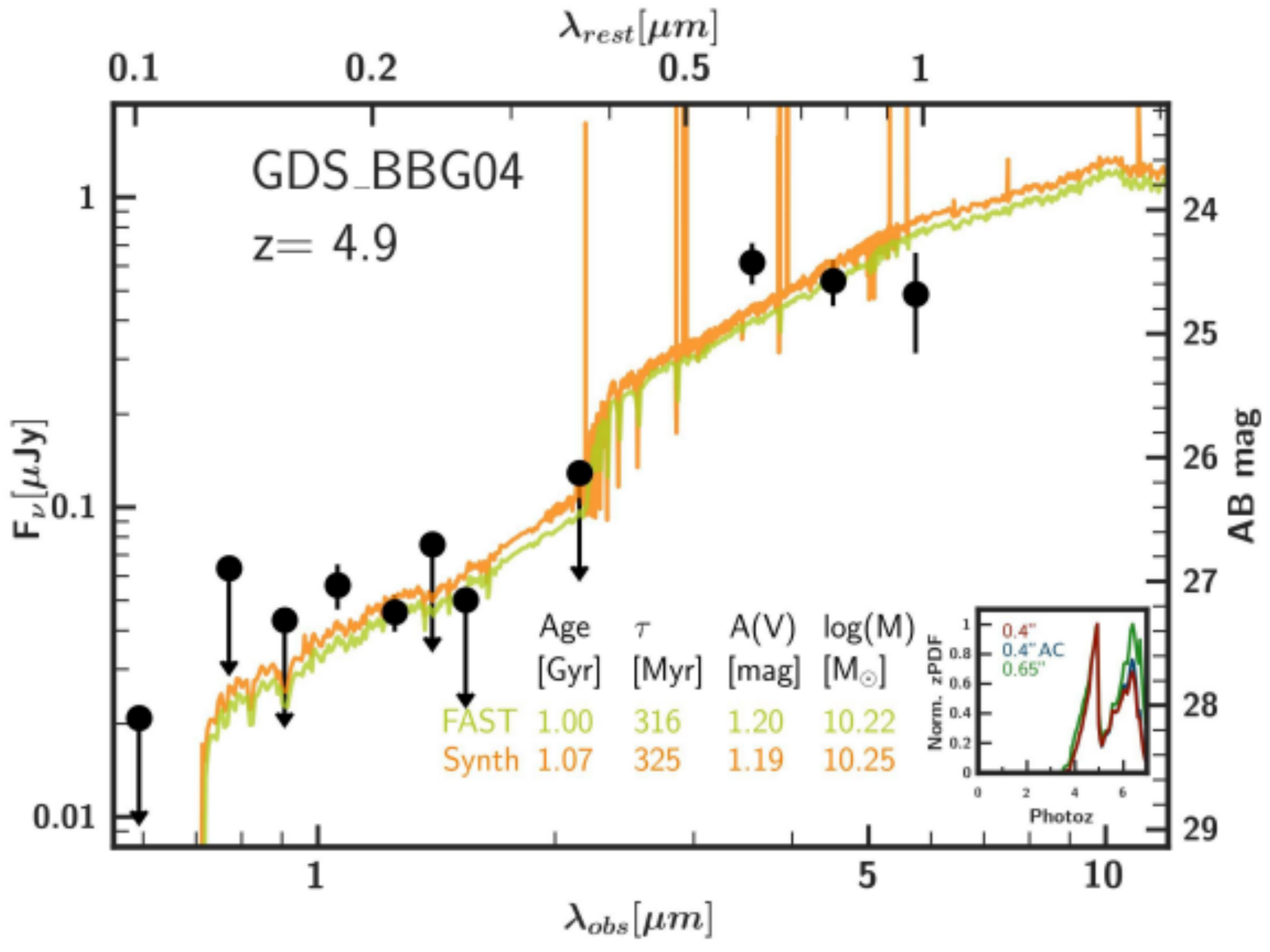}
		\end{center}
	\end{minipage}

% Source 5}
	\begin{minipage}[b]{0.44\linewidth}
		\centering
		\begin{minipage}[b]{0.315\linewidth}
			\includegraphics[width=1.\linewidth]{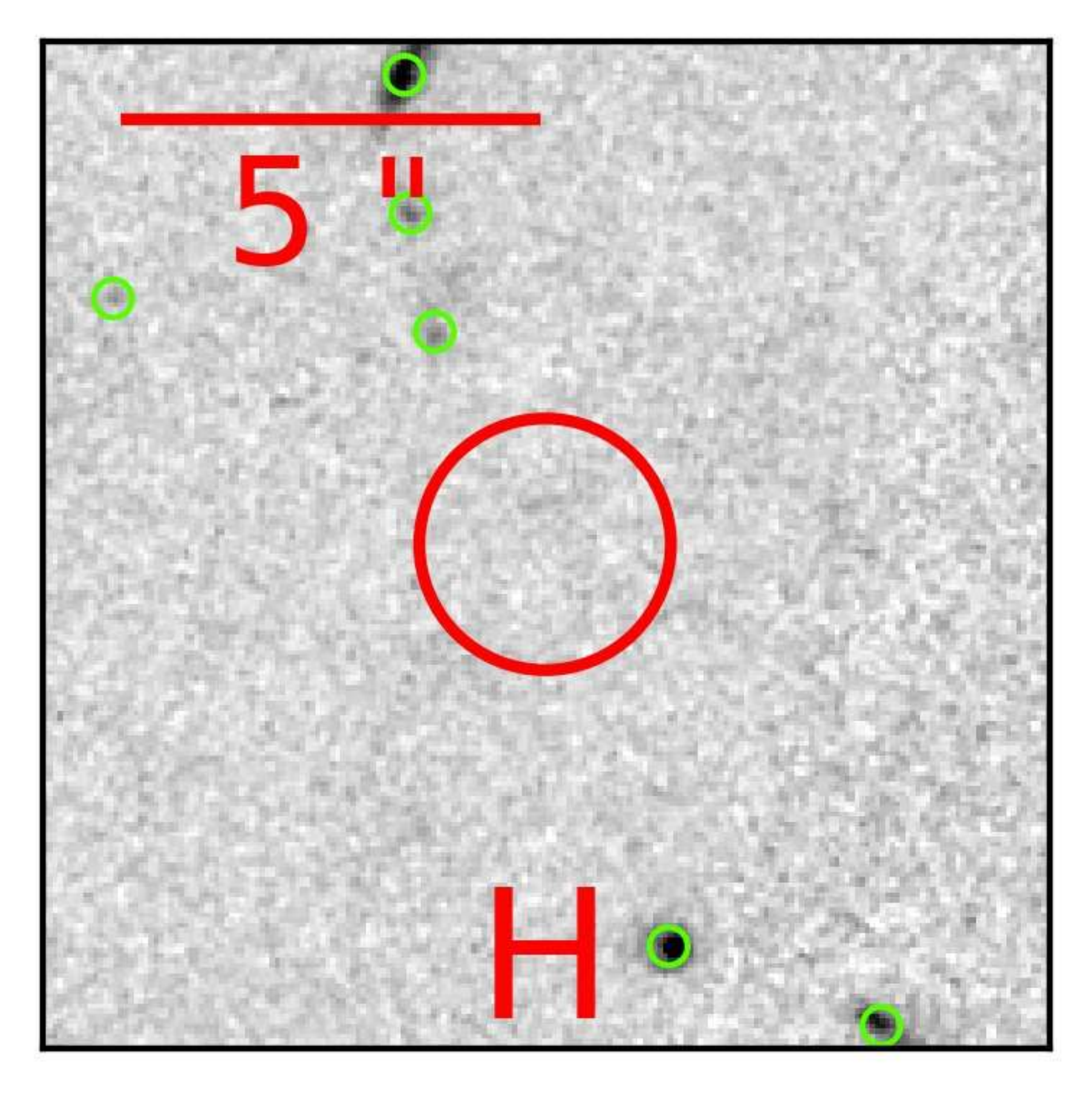}
		\end{minipage}
		\begin{minipage}[b]{0.315\linewidth}
			\includegraphics[width=1.\linewidth]{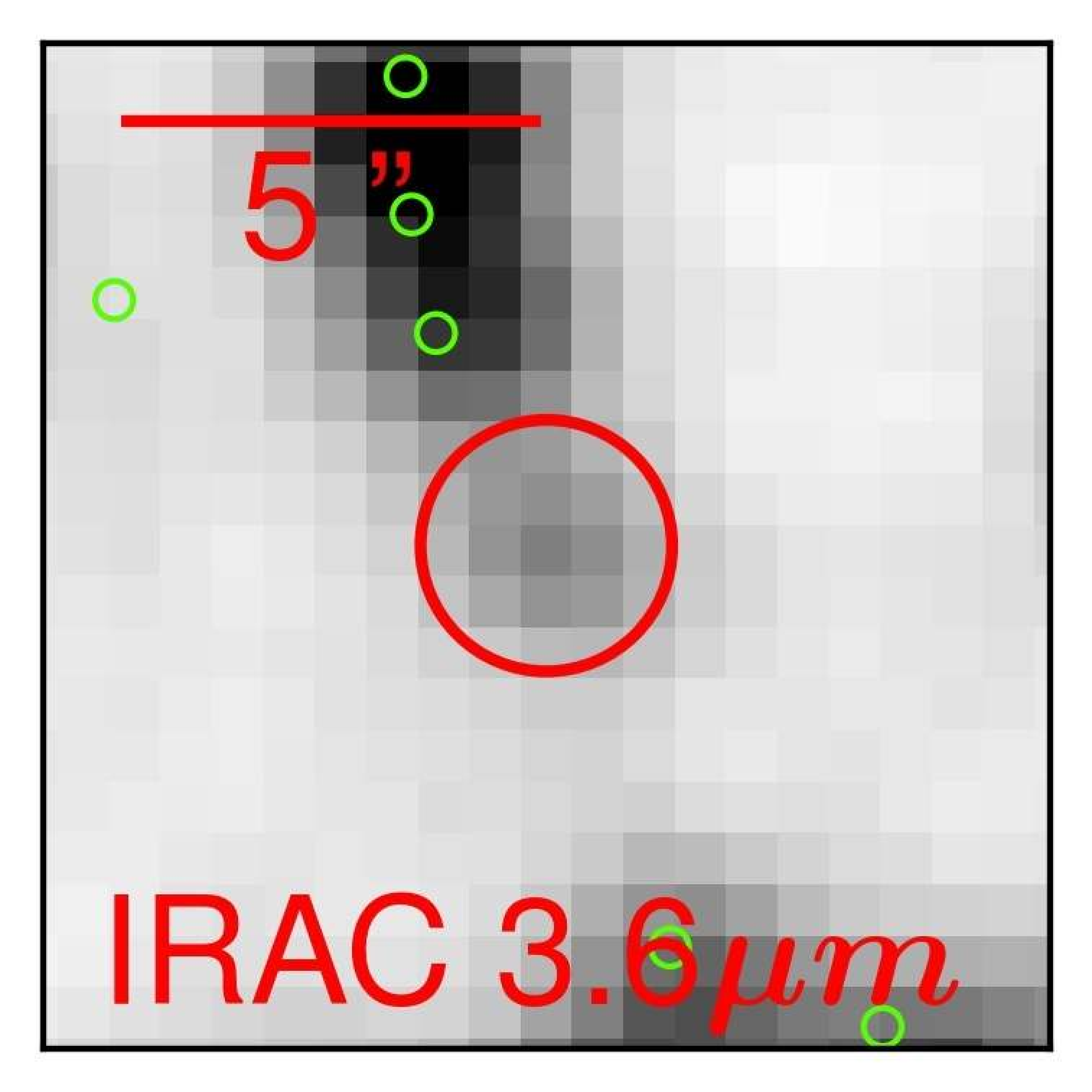}		
		\end{minipage}	
		\begin{minipage}[b]{0.315\linewidth}
			\includegraphics[width=1.\linewidth]{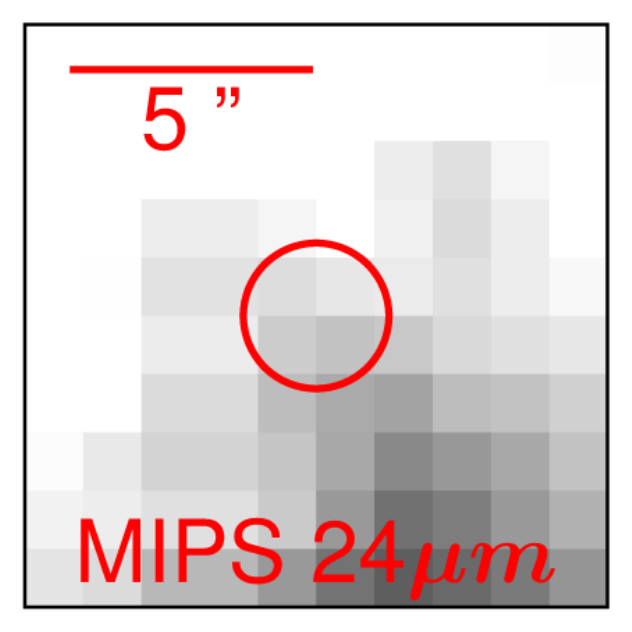}		
		\end{minipage}			
		\includegraphics[width=.49\linewidth]{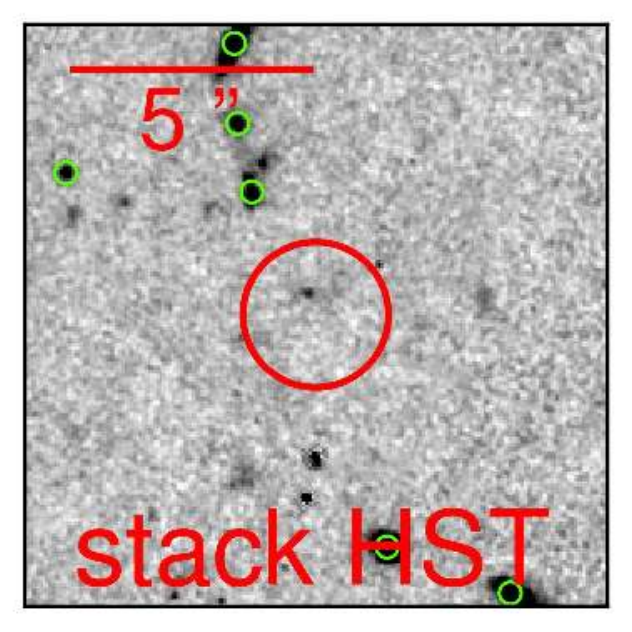}
		\includegraphics[width=.49\linewidth]{SH_ios_gs.pdf}
		\centering
	\end{minipage}
	\quad
	\begin{minipage}[b]{0.52\linewidth}
		\includegraphics[width=1.\linewidth]{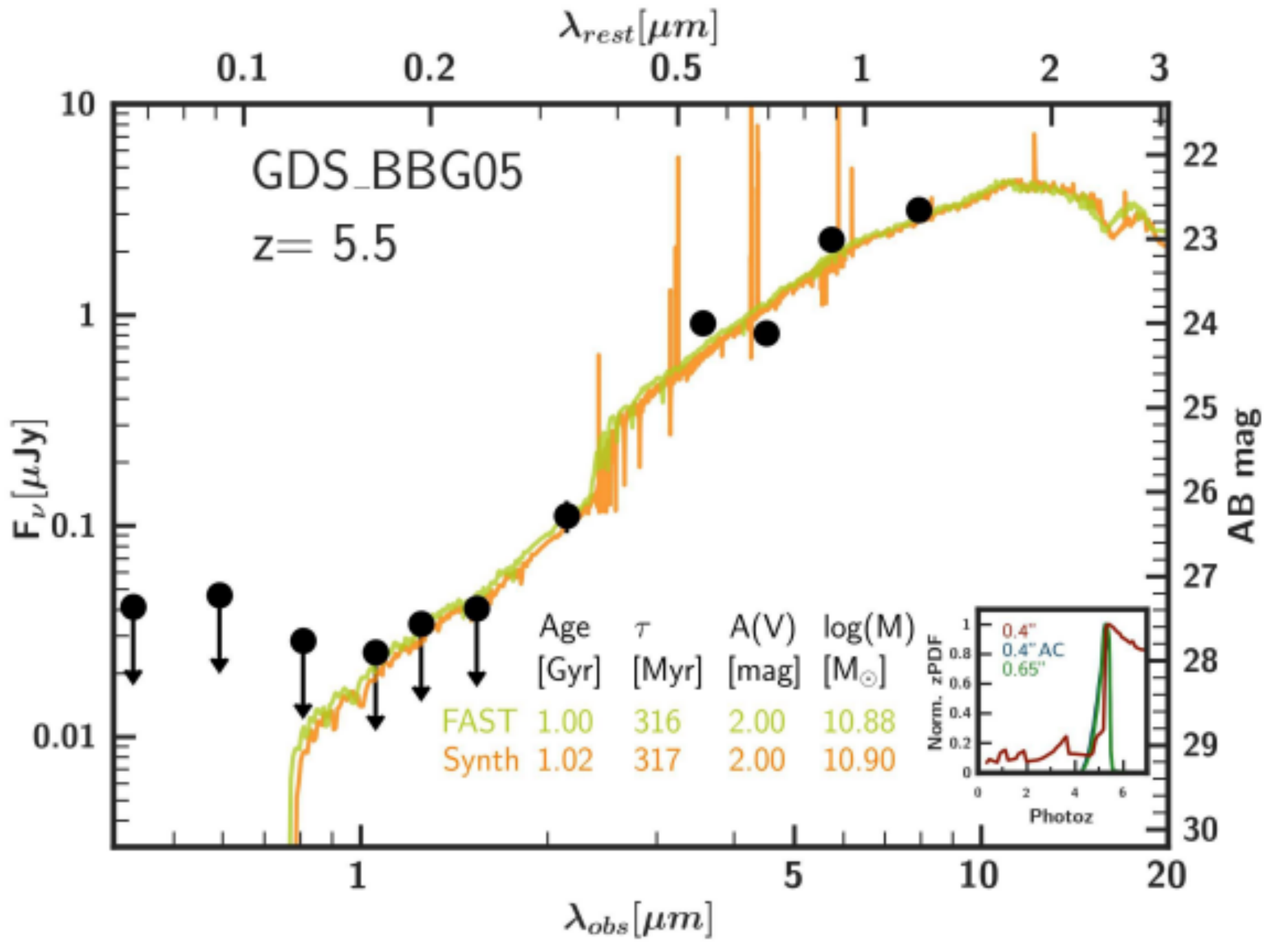}
		\centering
	\end{minipage}

% Source 6}
	\begin{minipage}[b]{0.44\linewidth}
		\centering
		\begin{minipage}[b]{0.315\linewidth}
			\includegraphics[width=1.\linewidth]{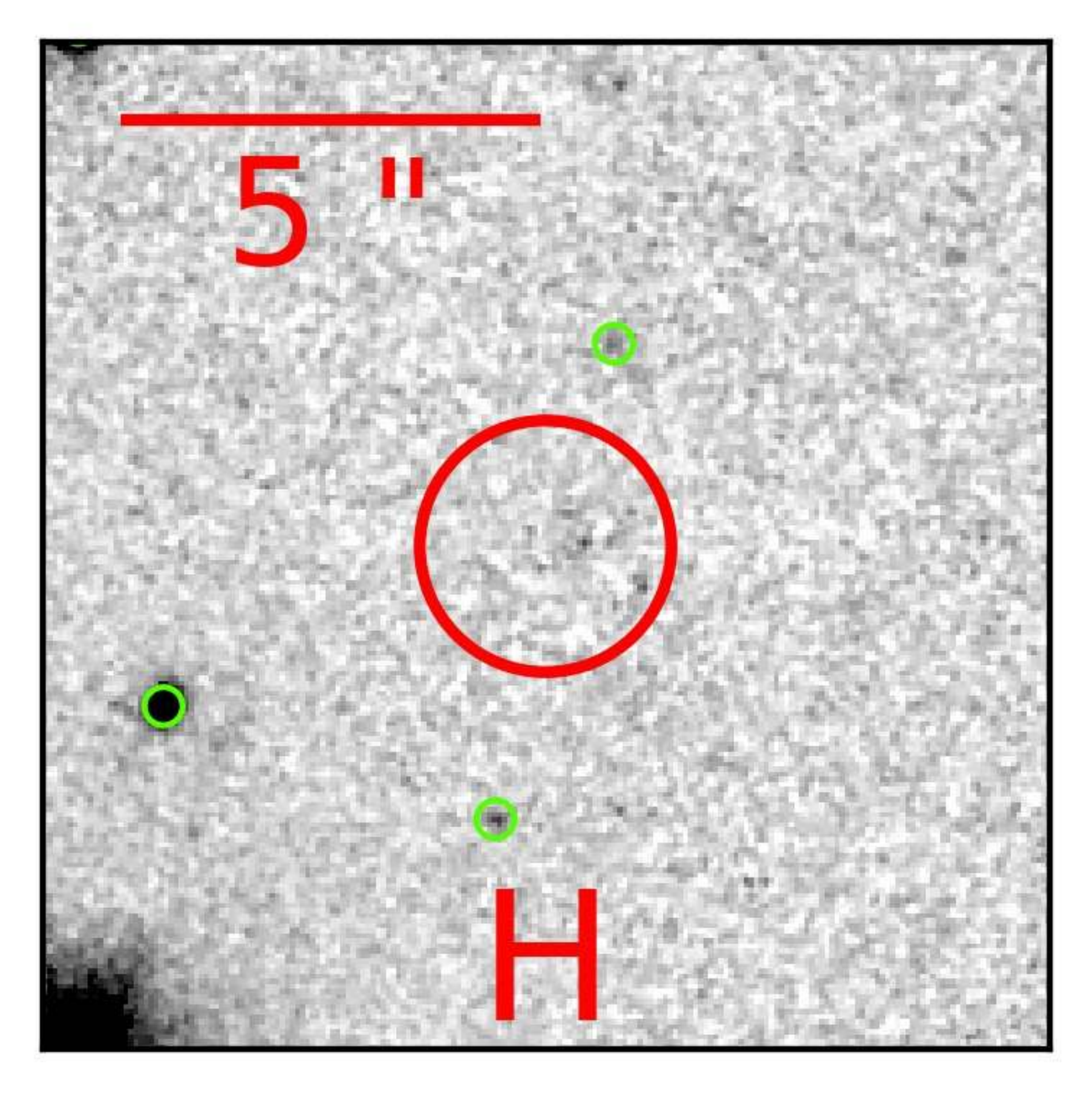}
		\end{minipage}
		\begin{minipage}[b]{0.315\linewidth}
			\includegraphics[width=1.\linewidth]{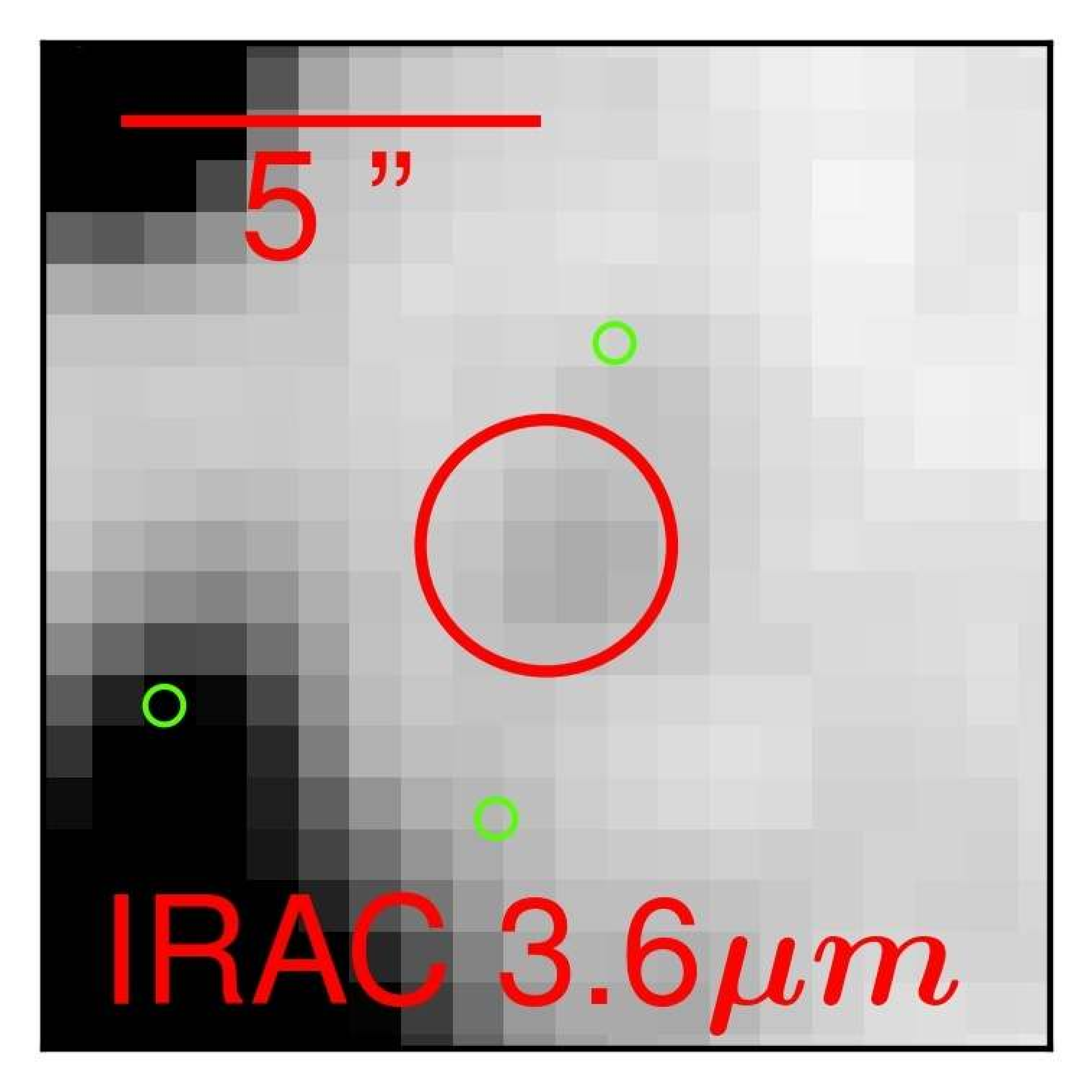}		
		\end{minipage}	
		\begin{minipage}[b]{0.315\linewidth}
			\includegraphics[width=1.\linewidth]{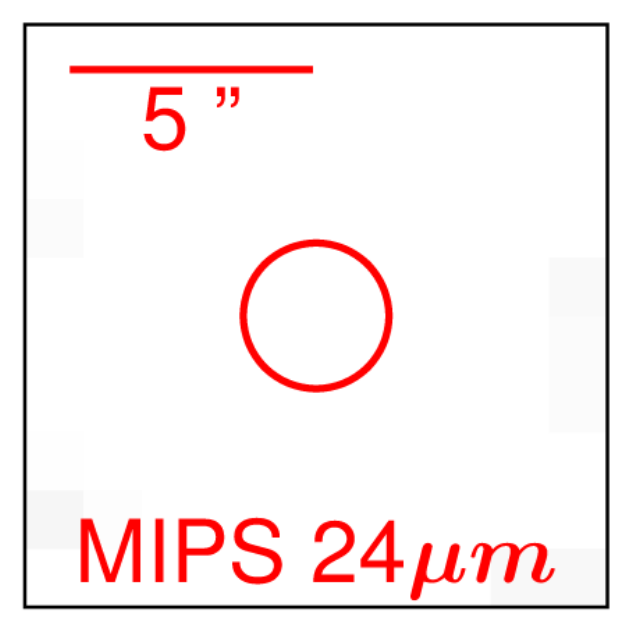}		
		\end{minipage}			
		\includegraphics[width=.49\linewidth]{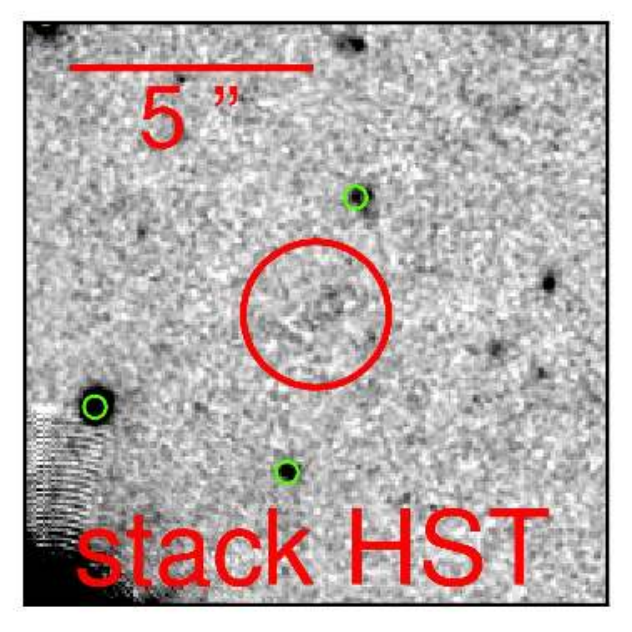}
		\includegraphics[width=.49\linewidth]{SH_ios_gs.pdf}
		\centering
	\end{minipage}
	\quad
	\begin{minipage}[b]{0.52\linewidth}
		\begin{center}
			\includegraphics[width=1.\linewidth]{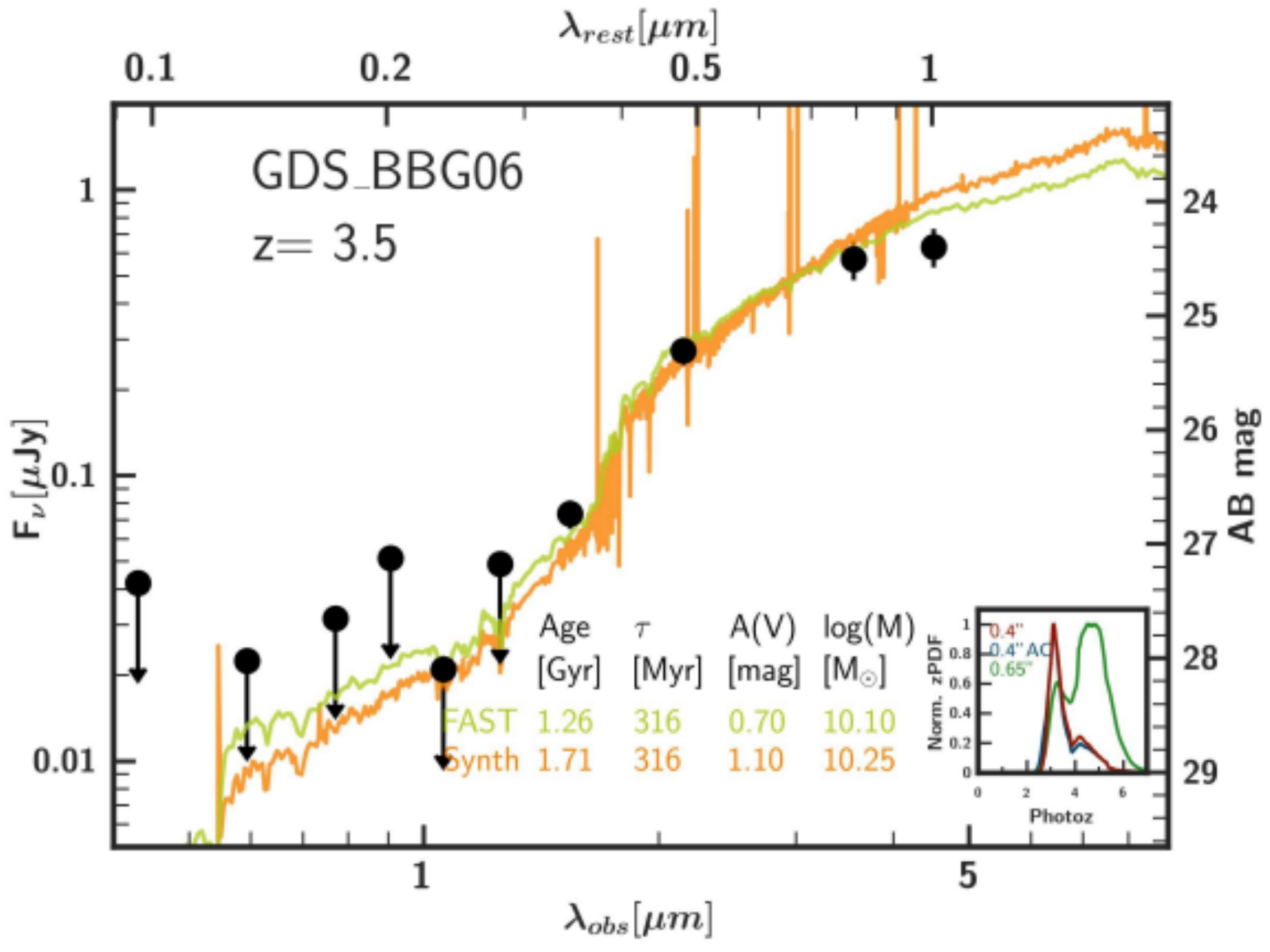}
		\end{center}
	\end{minipage}
\end{figure*}
% Source 7}
\begin{figure*}
	\begin{minipage}[b]{0.44\linewidth}
		\centering
		\begin{minipage}[b]{0.315\linewidth}
			\includegraphics[width=1.\linewidth]{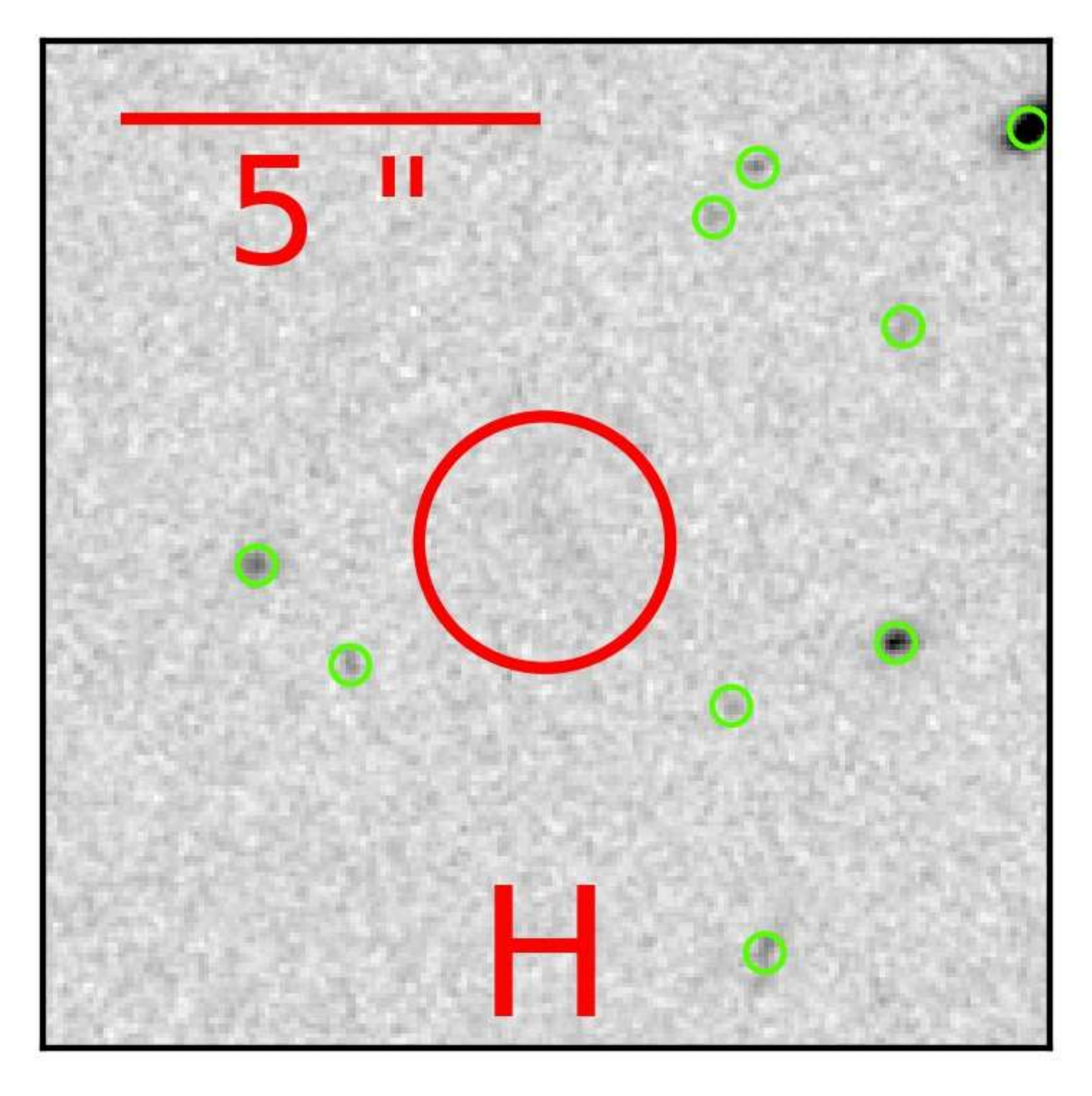}
		\end{minipage}
		\begin{minipage}[b]{0.315\linewidth}
			\includegraphics[width=1.\linewidth]{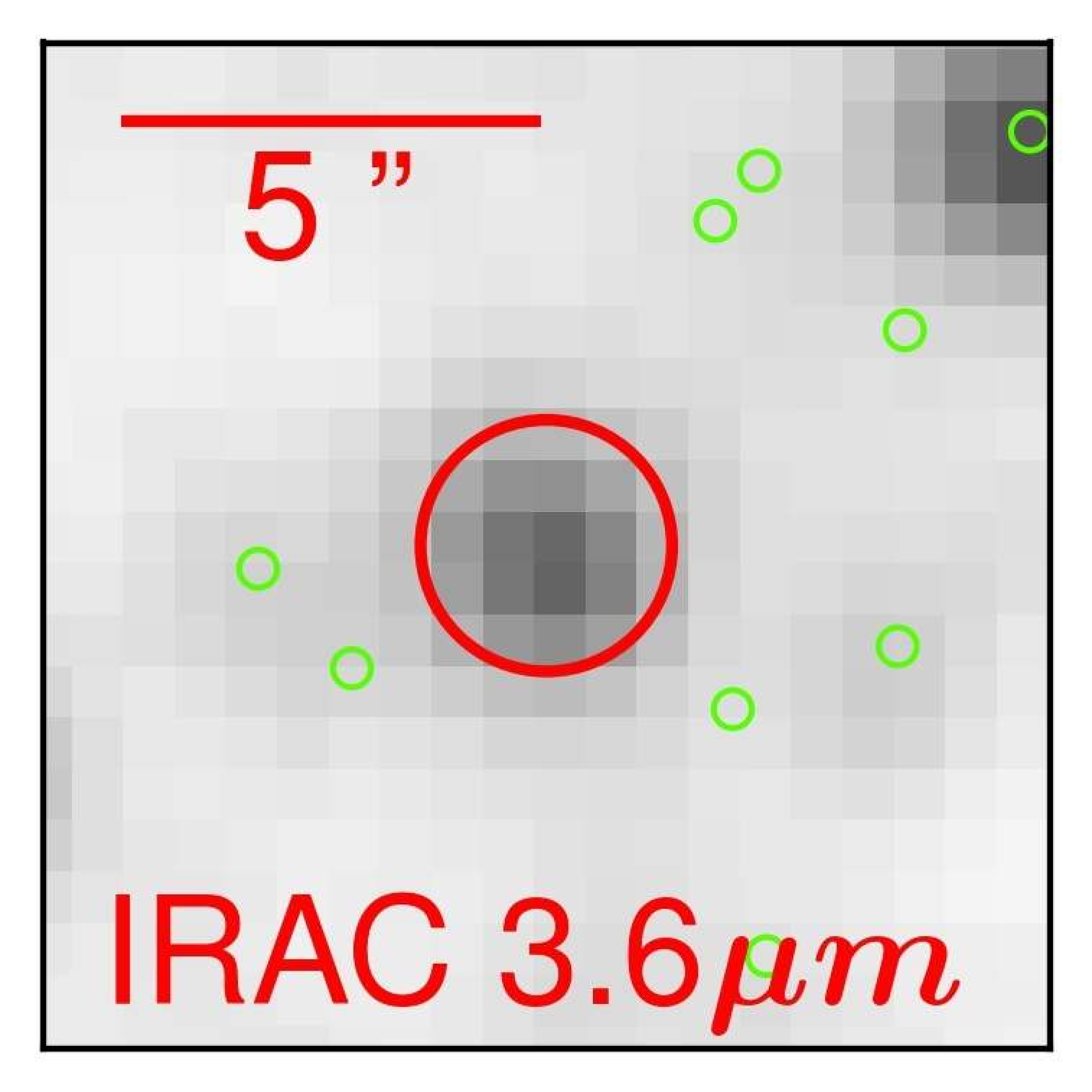}		
		\end{minipage}	
		\begin{minipage}[b]{0.315\linewidth}
			\includegraphics[width=1.\linewidth]{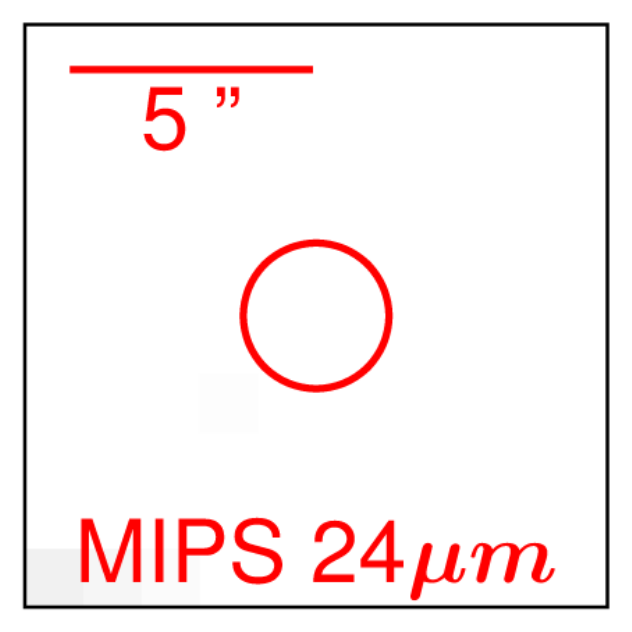}		
		\end{minipage}			
		\includegraphics[width=.49\linewidth]{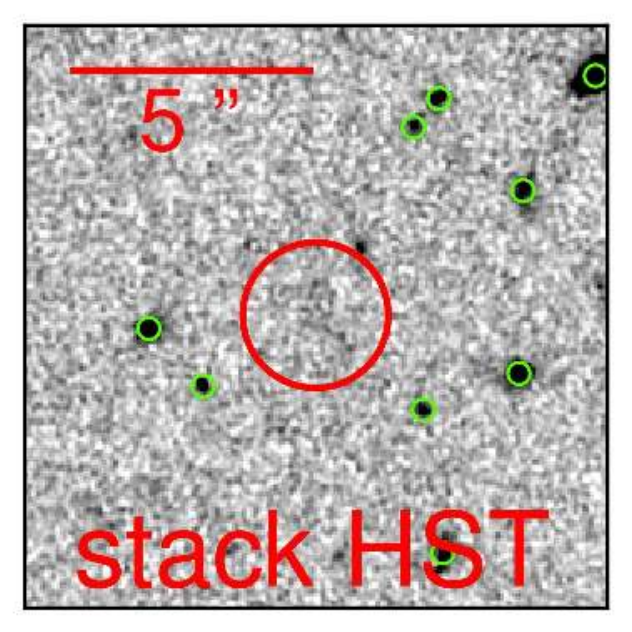}
		\includegraphics[width=.49\linewidth]{SH_ios_gs.pdf}
		\centering
	\end{minipage}
	\quad
	\begin{minipage}[b]{0.52\linewidth}
		\begin{center}
			\includegraphics[width=1.\linewidth]{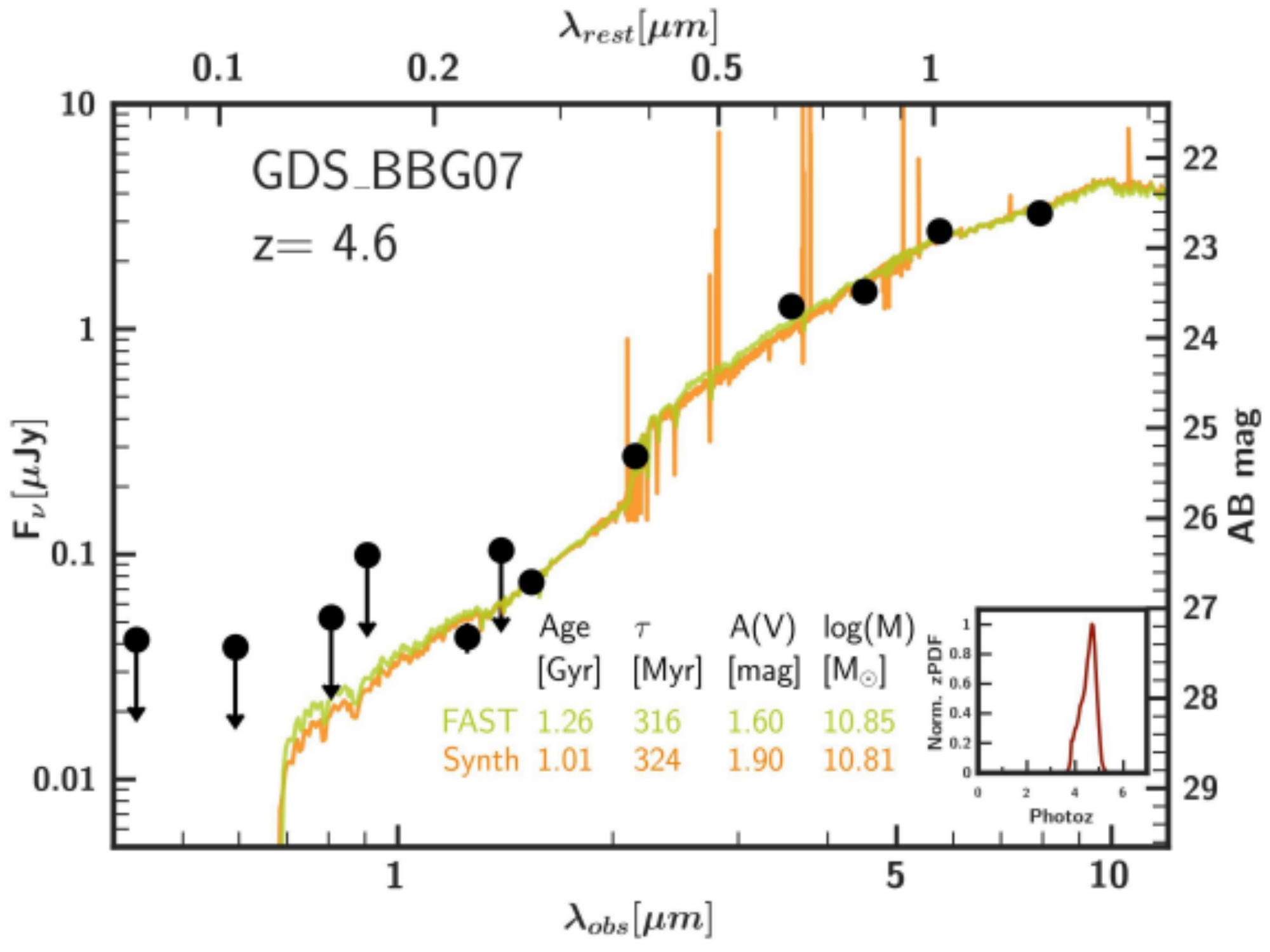}
		\end{center}
	\end{minipage}

% Source 8}

	\begin{minipage}[b]{0.44\linewidth}
		\centering
		\begin{minipage}[b]{0.315\linewidth}
			\includegraphics[width=1.\linewidth]{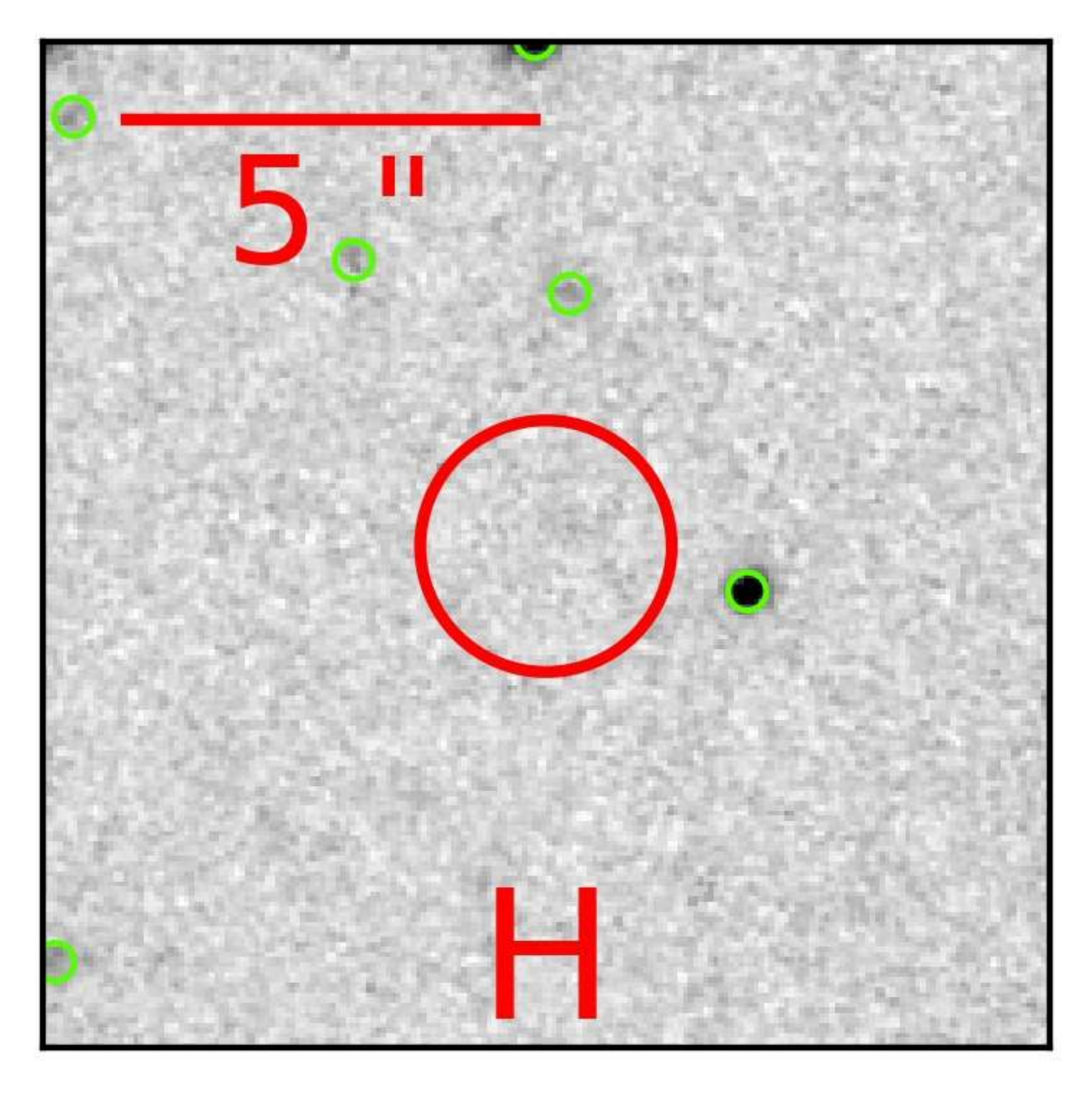}
		\end{minipage}
		\begin{minipage}[b]{0.315\linewidth}
			\includegraphics[width=1.\linewidth]{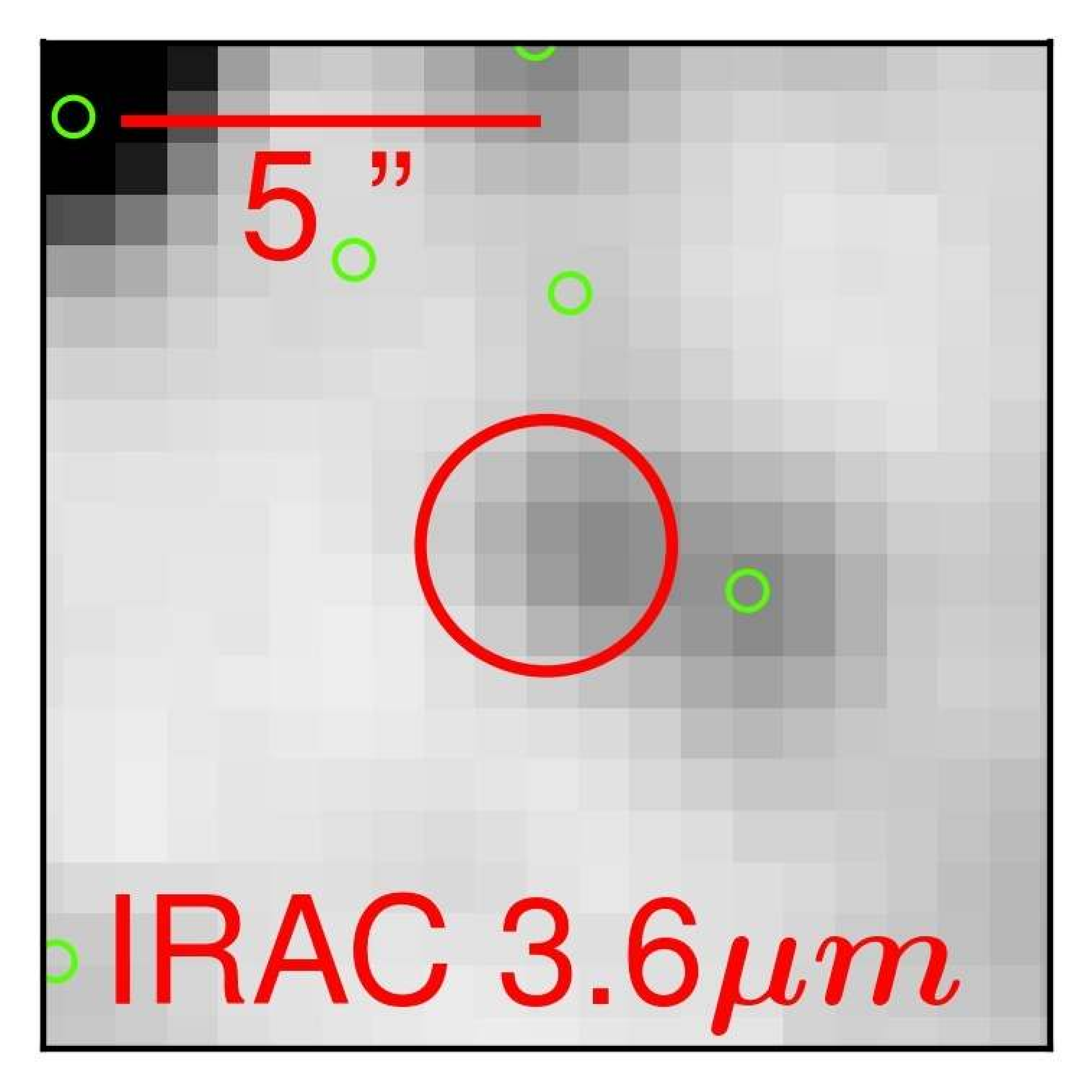}		
		\end{minipage}	
		\begin{minipage}[b]{0.315\linewidth}
			\includegraphics[width=1.\linewidth]{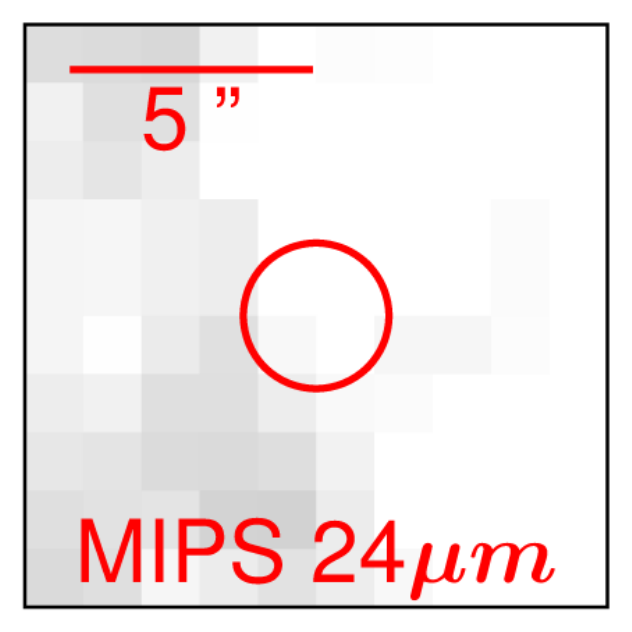}		
		\end{minipage}			
		\includegraphics[width=.49\linewidth]{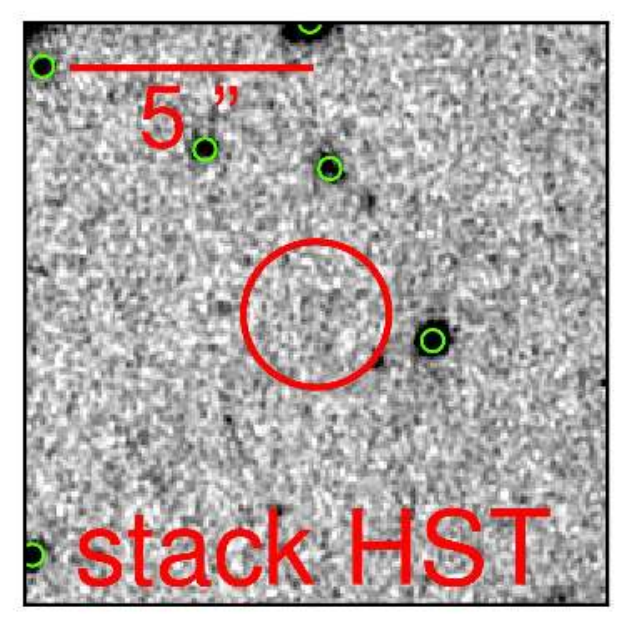}
		\includegraphics[width=.49\linewidth]{SH_ios_gs.pdf}
		\centering
	\end{minipage}
	\quad
	\begin{minipage}[b]{0.52\linewidth}
		\begin{center}
			\includegraphics[width=1.\linewidth]{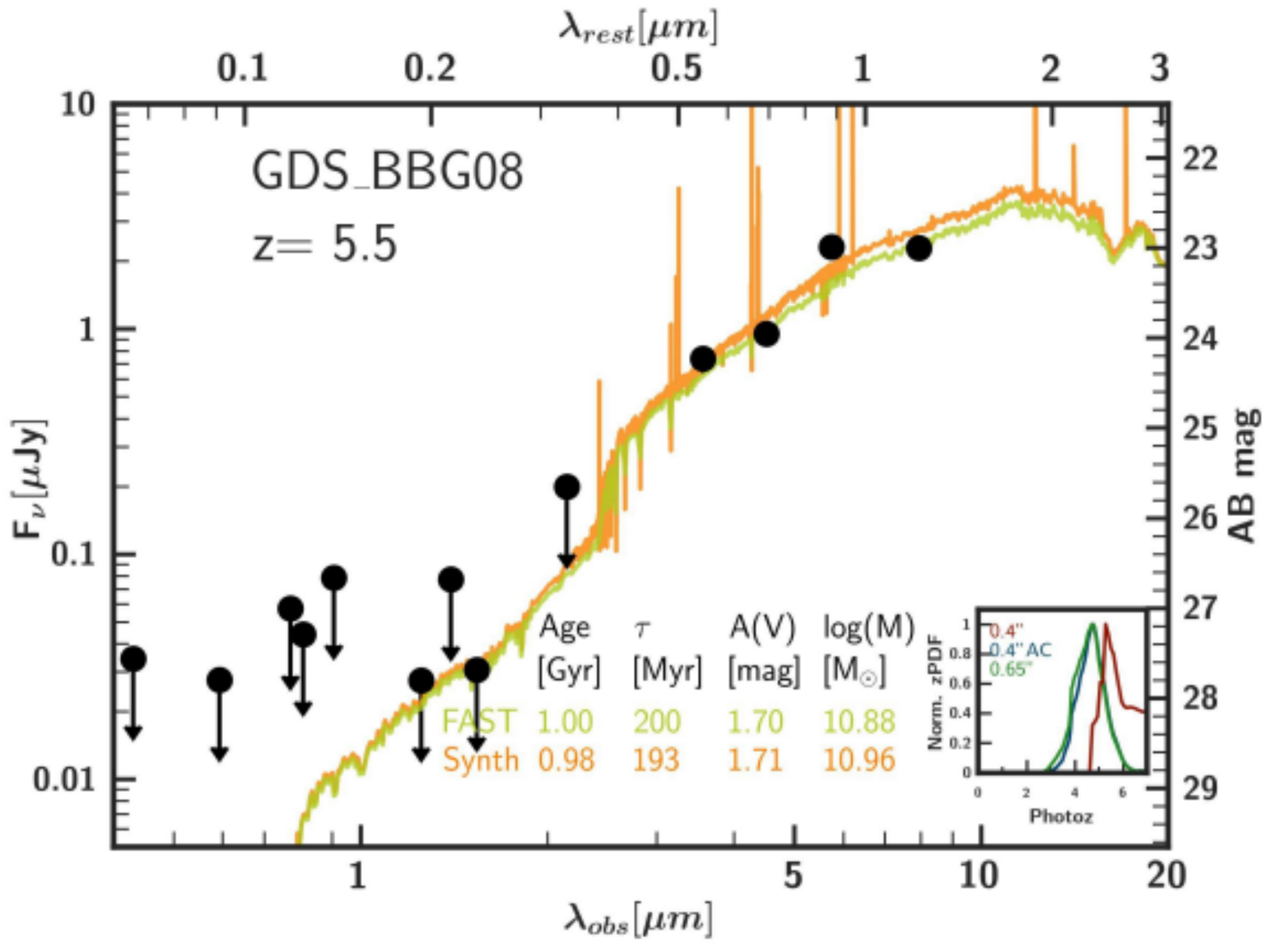}
		\end{center}
	\end{minipage}

% Source 9}

	\begin{minipage}[b]{0.44\linewidth}
		\begin{minipage}[b]{0.315\linewidth}
			\includegraphics[width=1.\linewidth]{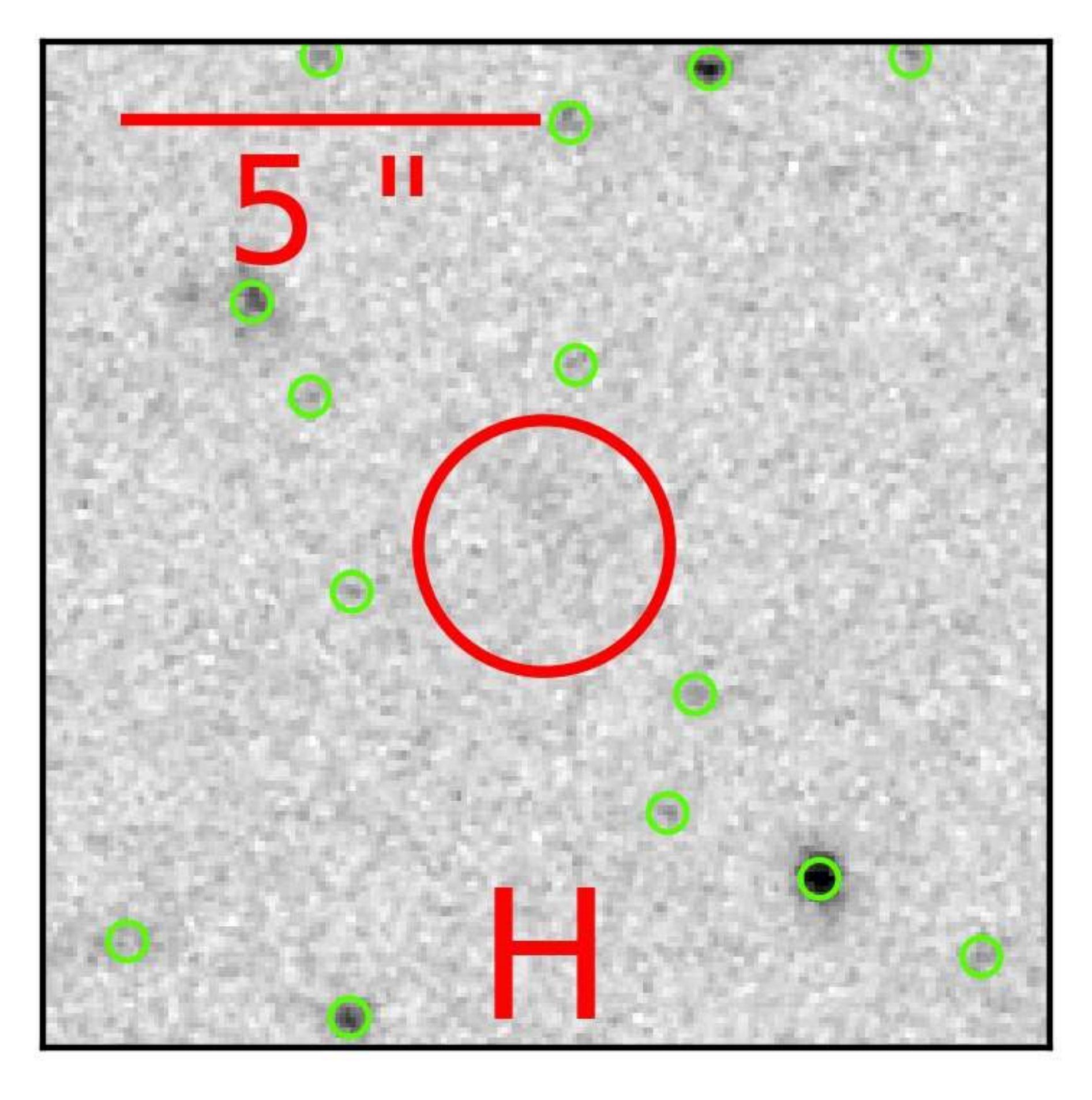}
		\end{minipage}
		\begin{minipage}[b]{0.315\linewidth}
			\includegraphics[width=1.\linewidth]{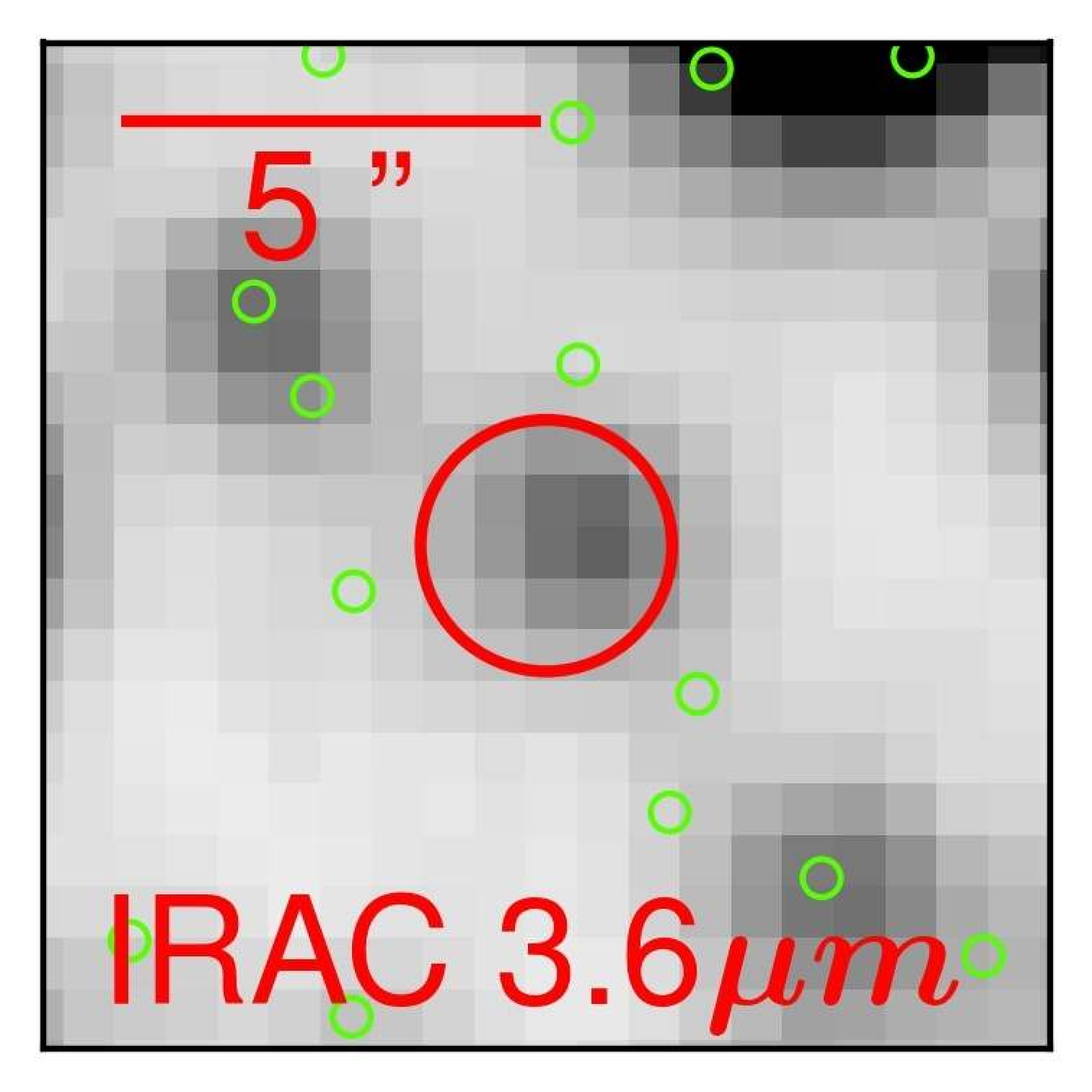}		
		\end{minipage}	
		\begin{minipage}[b]{0.315\linewidth}
			\includegraphics[width=1.\linewidth]{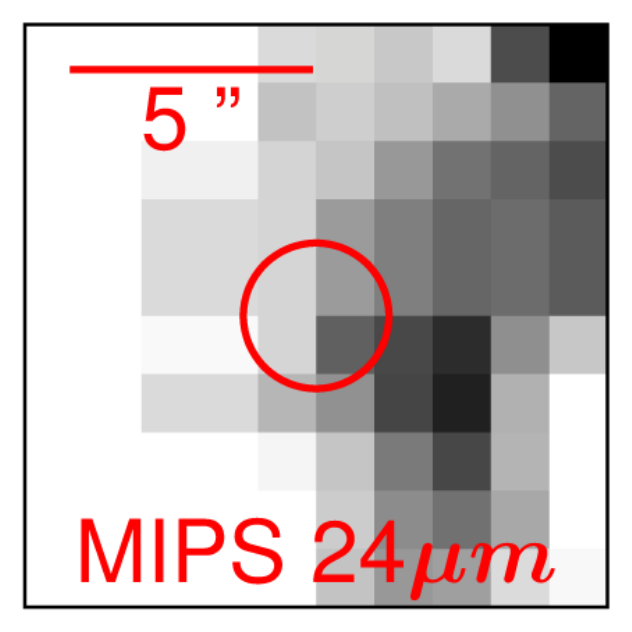}		
		\end{minipage}			
		\includegraphics[width=.49\linewidth]{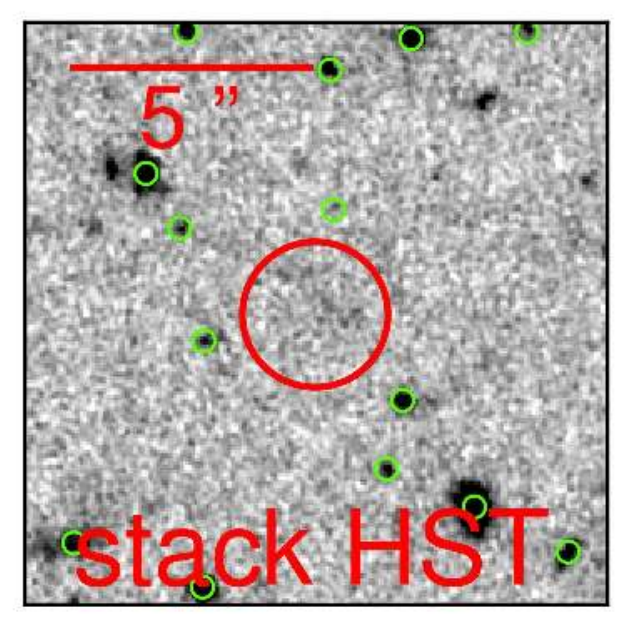}
		\includegraphics[width=.49\linewidth]{SH_ios_gs.pdf}
		\centering
	\end{minipage}
	\quad
	\begin{minipage}[b]{0.52\linewidth}
		\includegraphics[width=1.\linewidth]{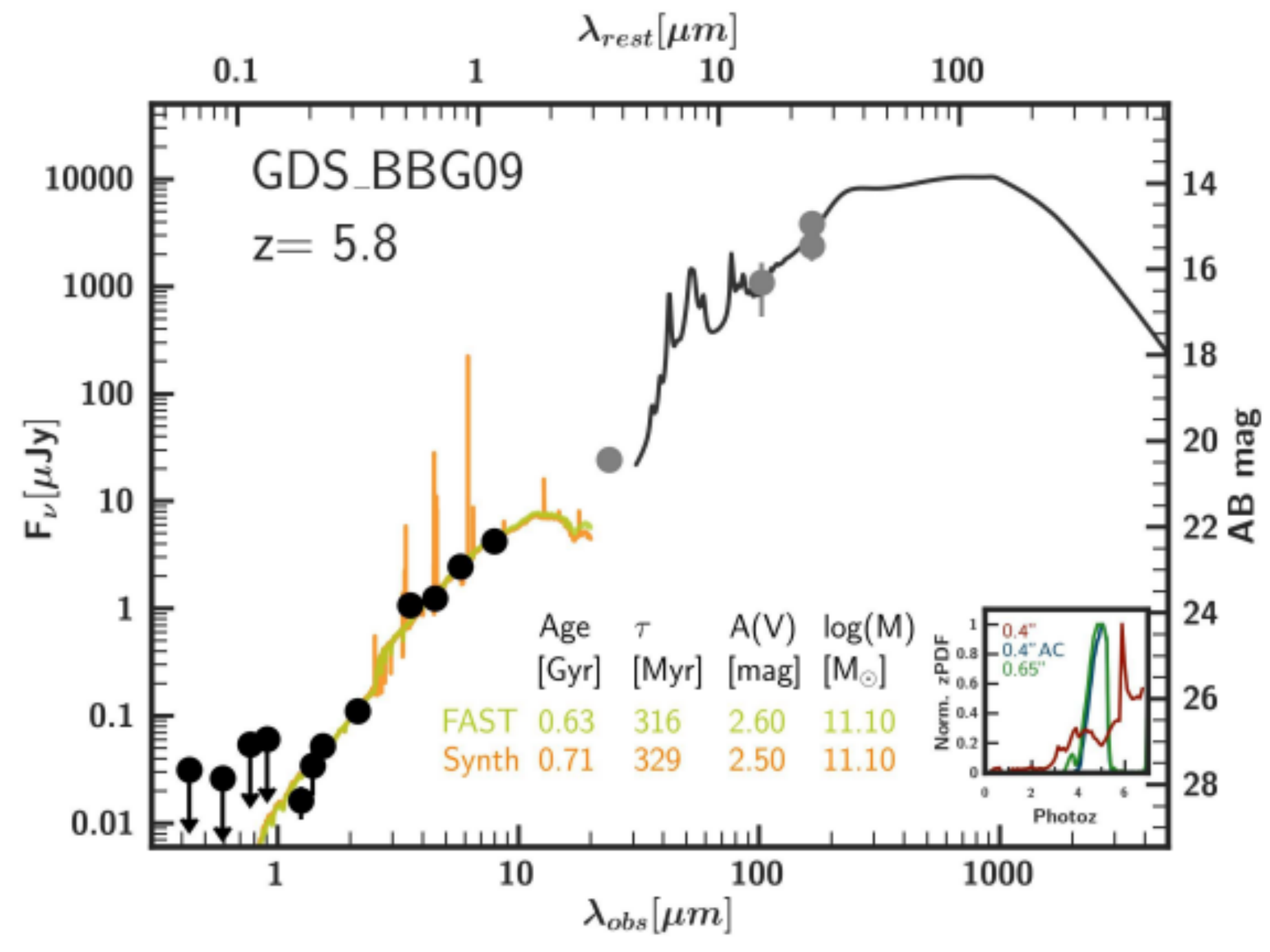}
		\centering
	\end{minipage}
\end{figure*}

% Source 10}
\begin{figure*}
	\begin{minipage}[b]{0.44\linewidth}
		\centering
		\begin{minipage}[b]{0.315\linewidth}
			\includegraphics[width=1.\linewidth]{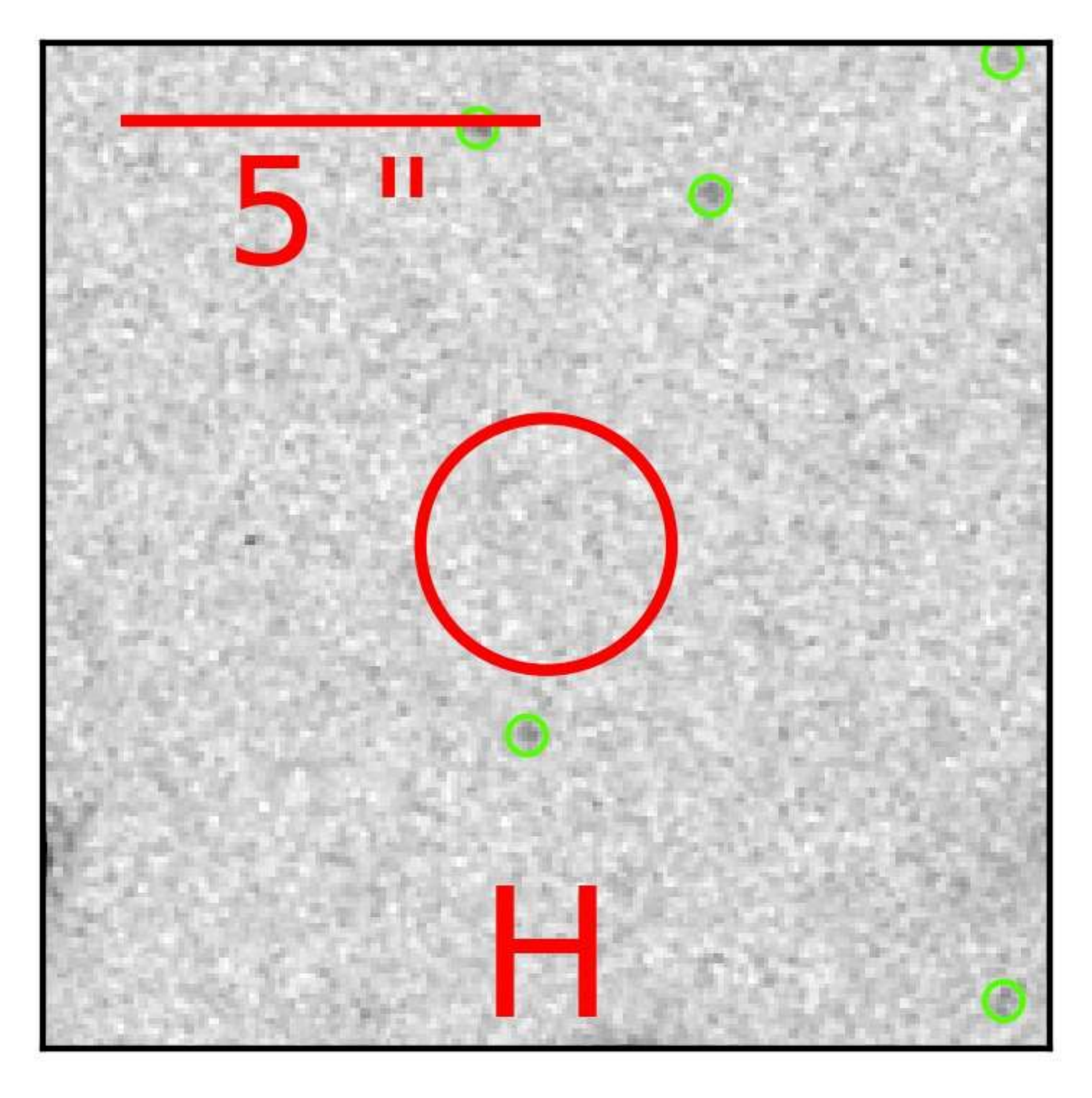}
		\end{minipage}
		\begin{minipage}[b]{0.315\linewidth}
			\includegraphics[width=1.\linewidth]{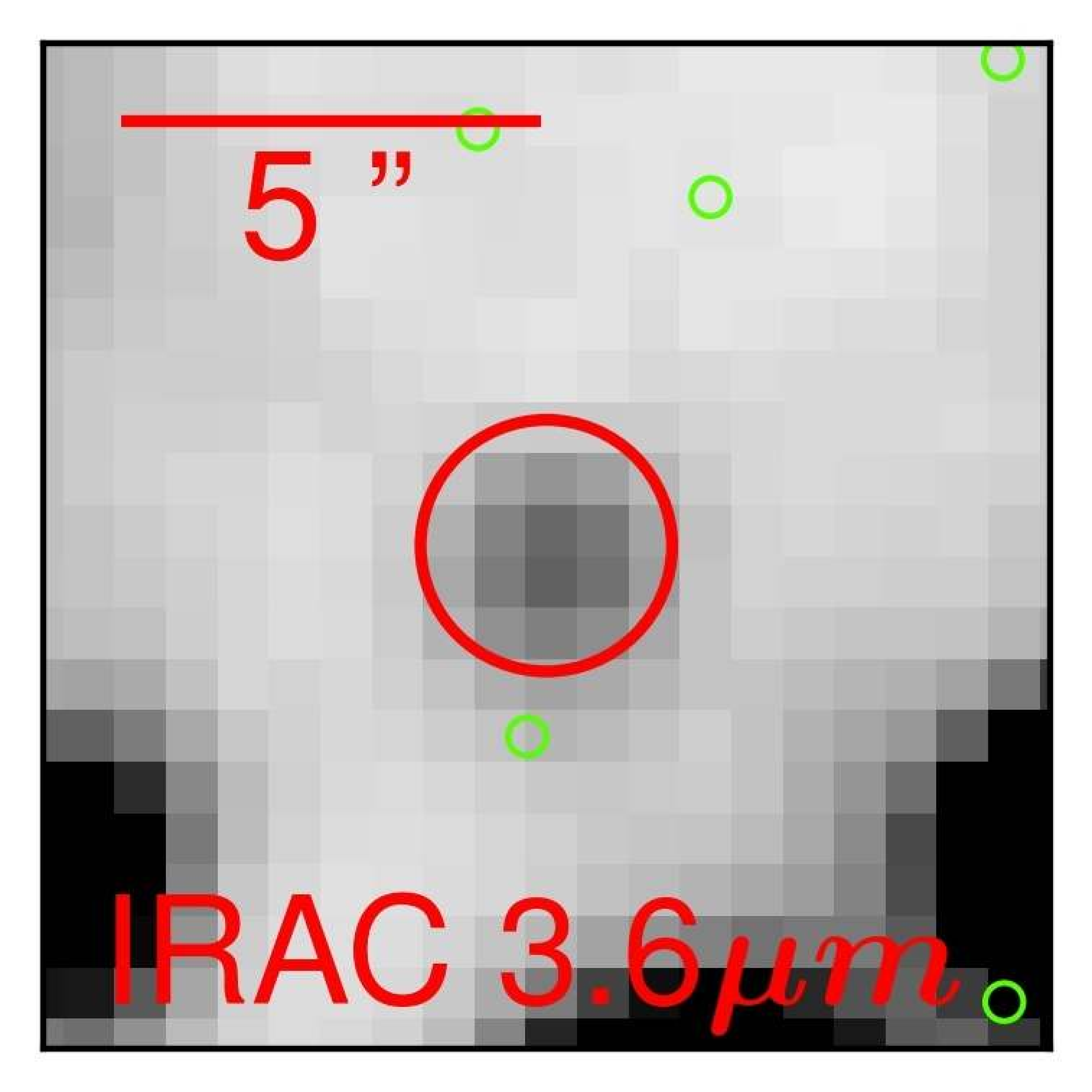}		
		\end{minipage}	
		\begin{minipage}[b]{0.315\linewidth}
			\includegraphics[width=1.\linewidth]{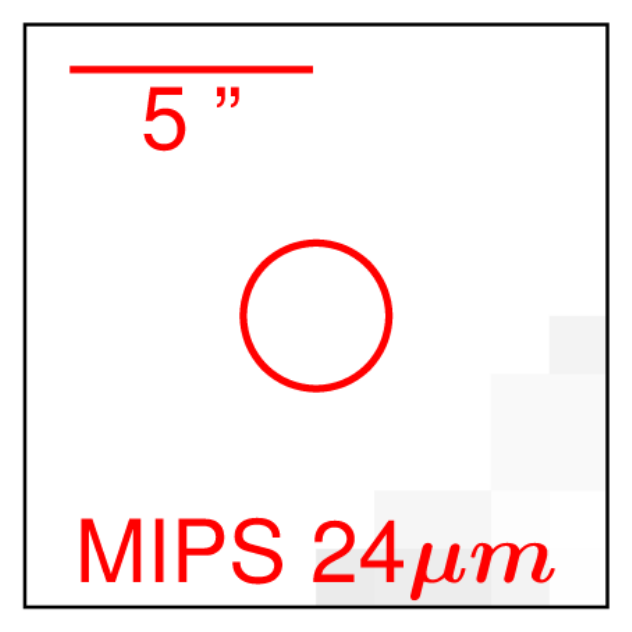}		
		\end{minipage}			
		\includegraphics[width=.49\linewidth]{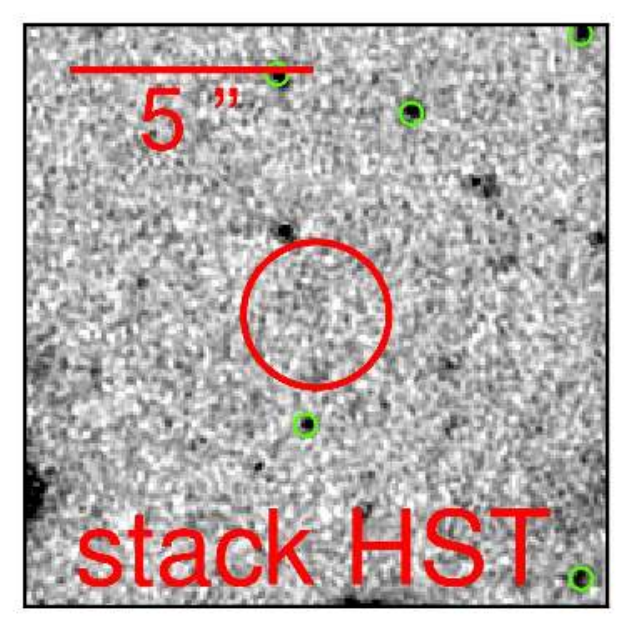}
		\includegraphics[width=.49\linewidth]{SH_ios_gs.pdf}
		\centering
	\end{minipage}
	\quad
	\begin{minipage}[b]{0.52\linewidth}
		\begin{center}
			\includegraphics[width=1.\linewidth]{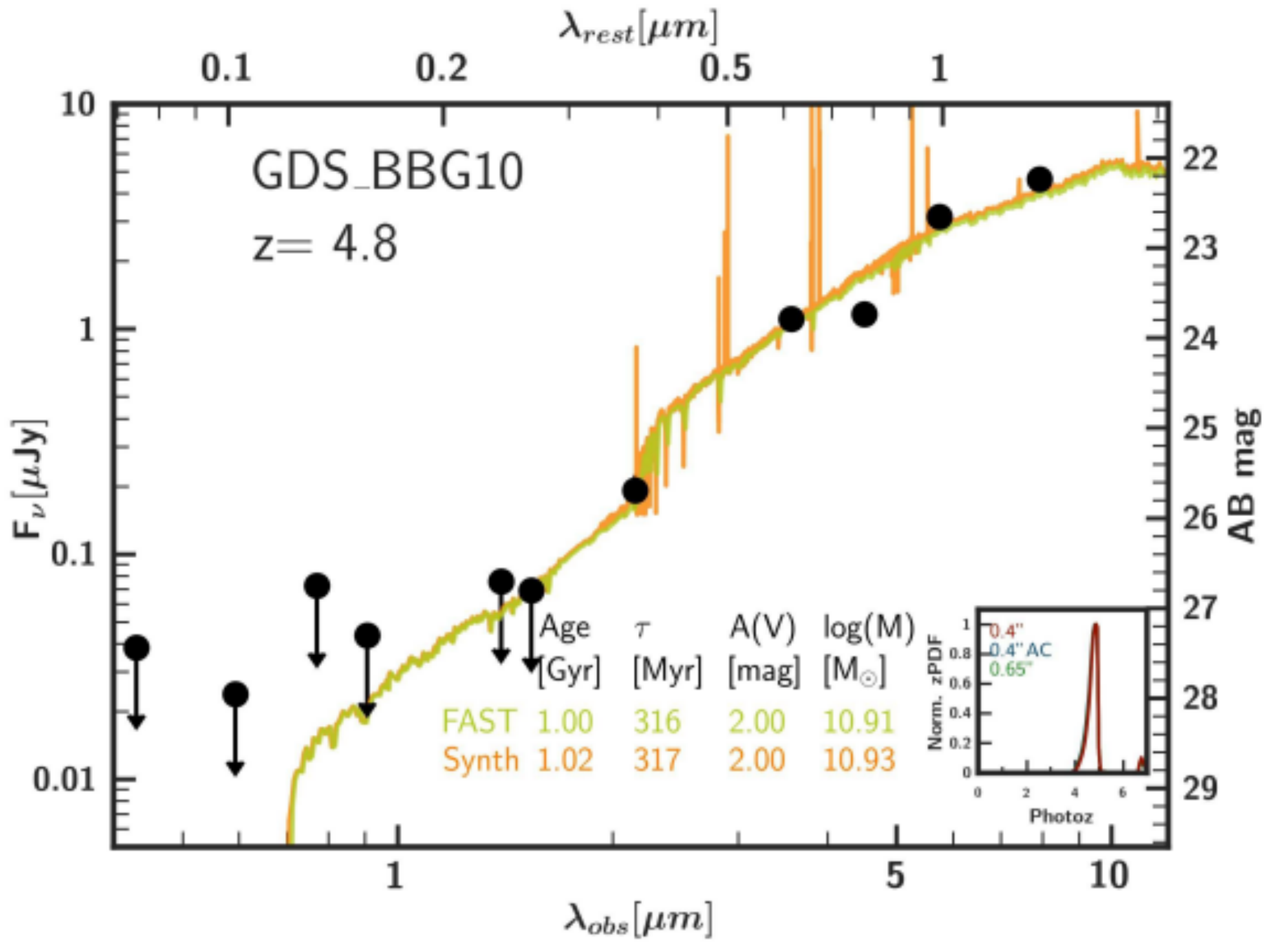}
		\end{center}
	\end{minipage}

% Source 11}

	\begin{minipage}[b]{0.44\linewidth}
		\centering
		\begin{minipage}[b]{0.315\linewidth}
			\includegraphics[width=1.\linewidth]{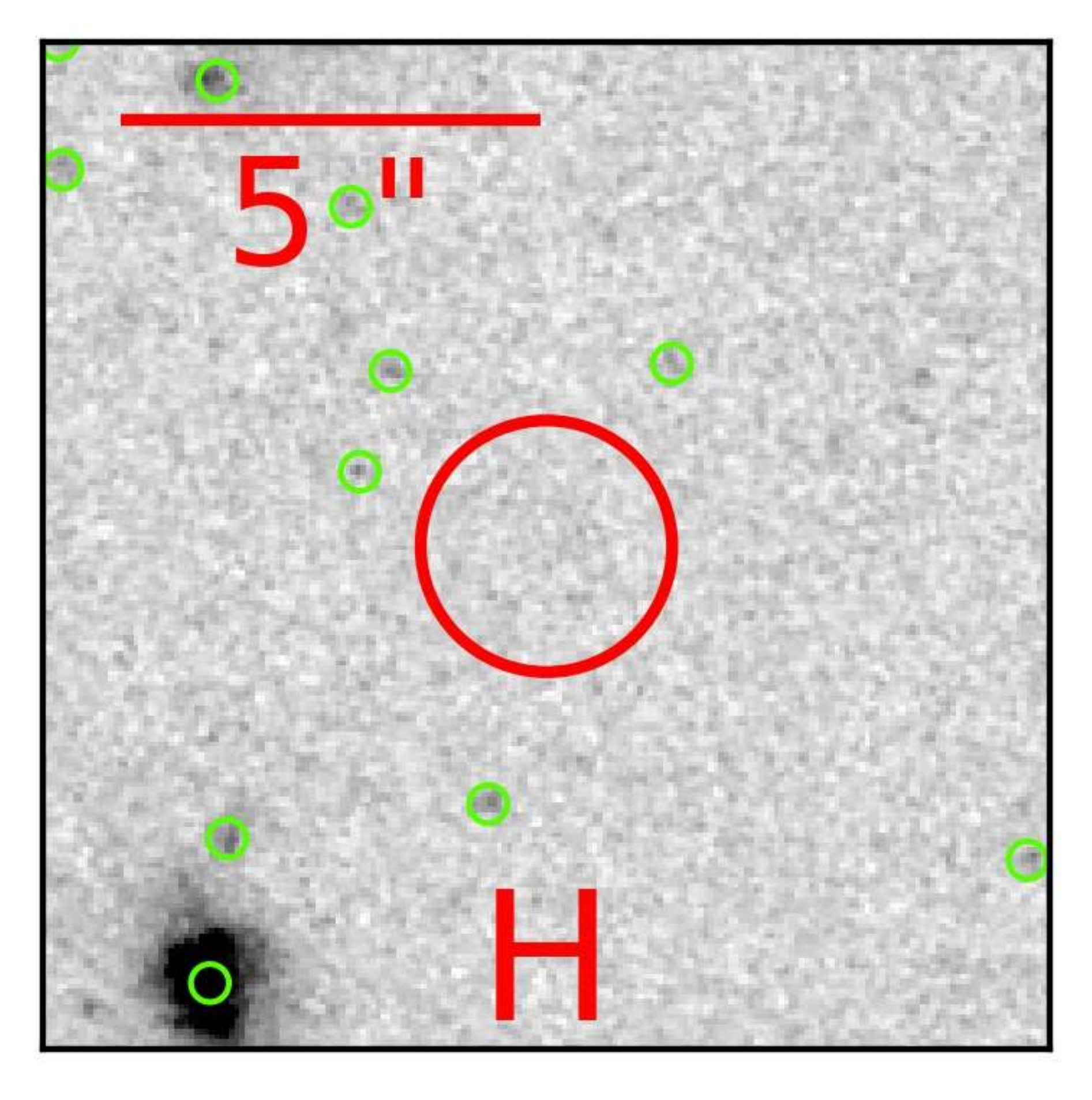}
		\end{minipage}
		\begin{minipage}[b]{0.315\linewidth}
			\includegraphics[width=1.\linewidth]{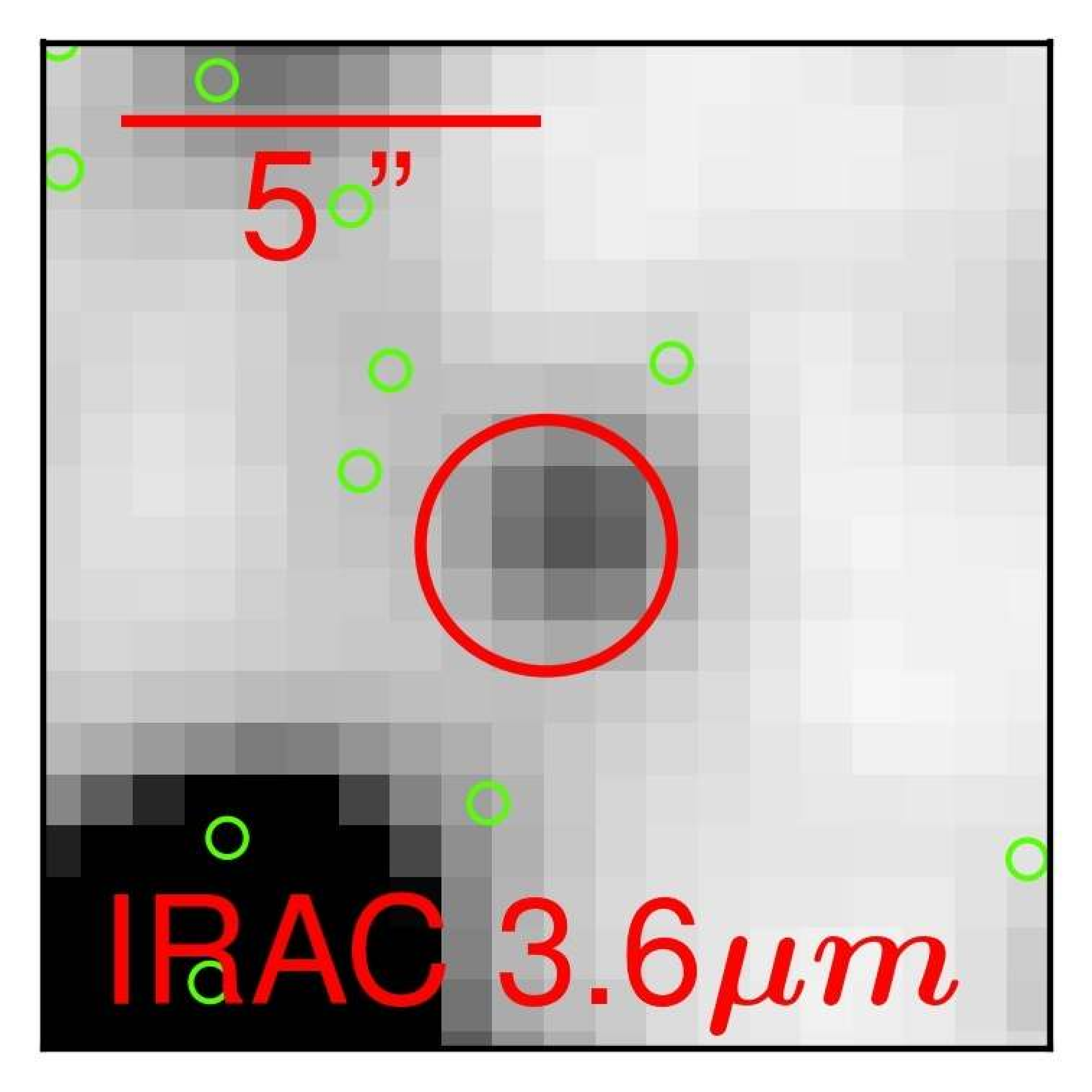}		
		\end{minipage}	
		\begin{minipage}[b]{0.315\linewidth}
			\includegraphics[width=1.\linewidth]{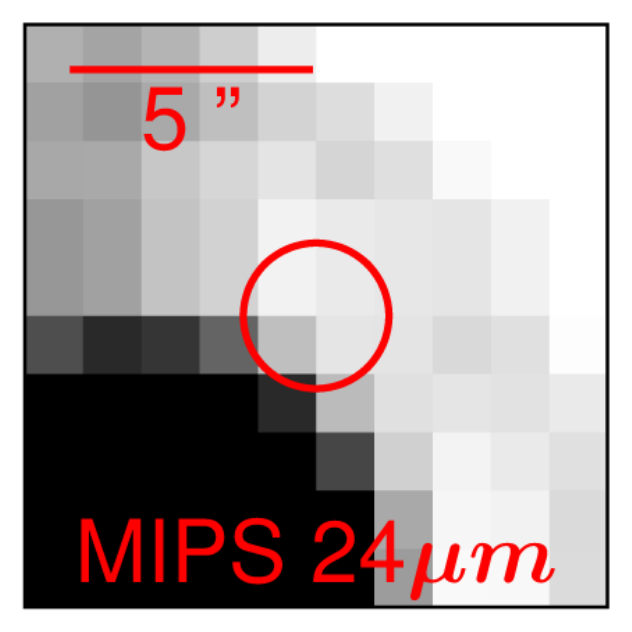}		
		\end{minipage}			
		\includegraphics[width=.49\linewidth]{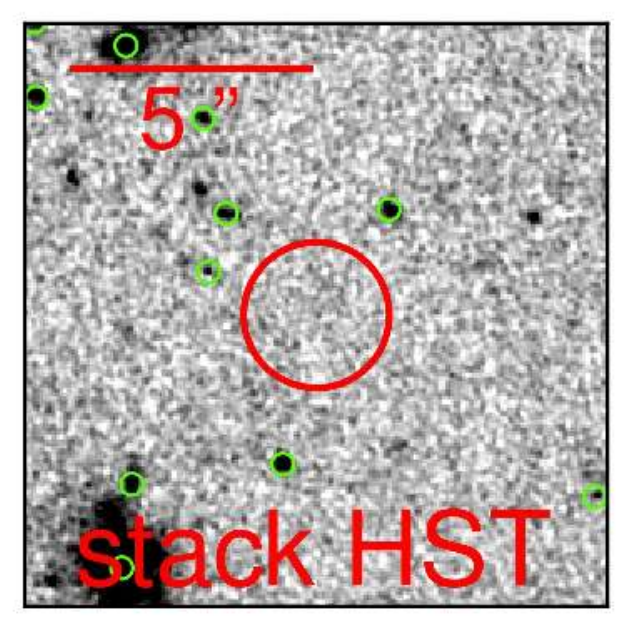}
		\includegraphics[width=.49\linewidth]{SH_ios_gs.pdf}
		\centering
	\end{minipage}
	\quad
	\begin{minipage}[b]{0.52\linewidth}
		\begin{center}
			\includegraphics[width=1.\linewidth]{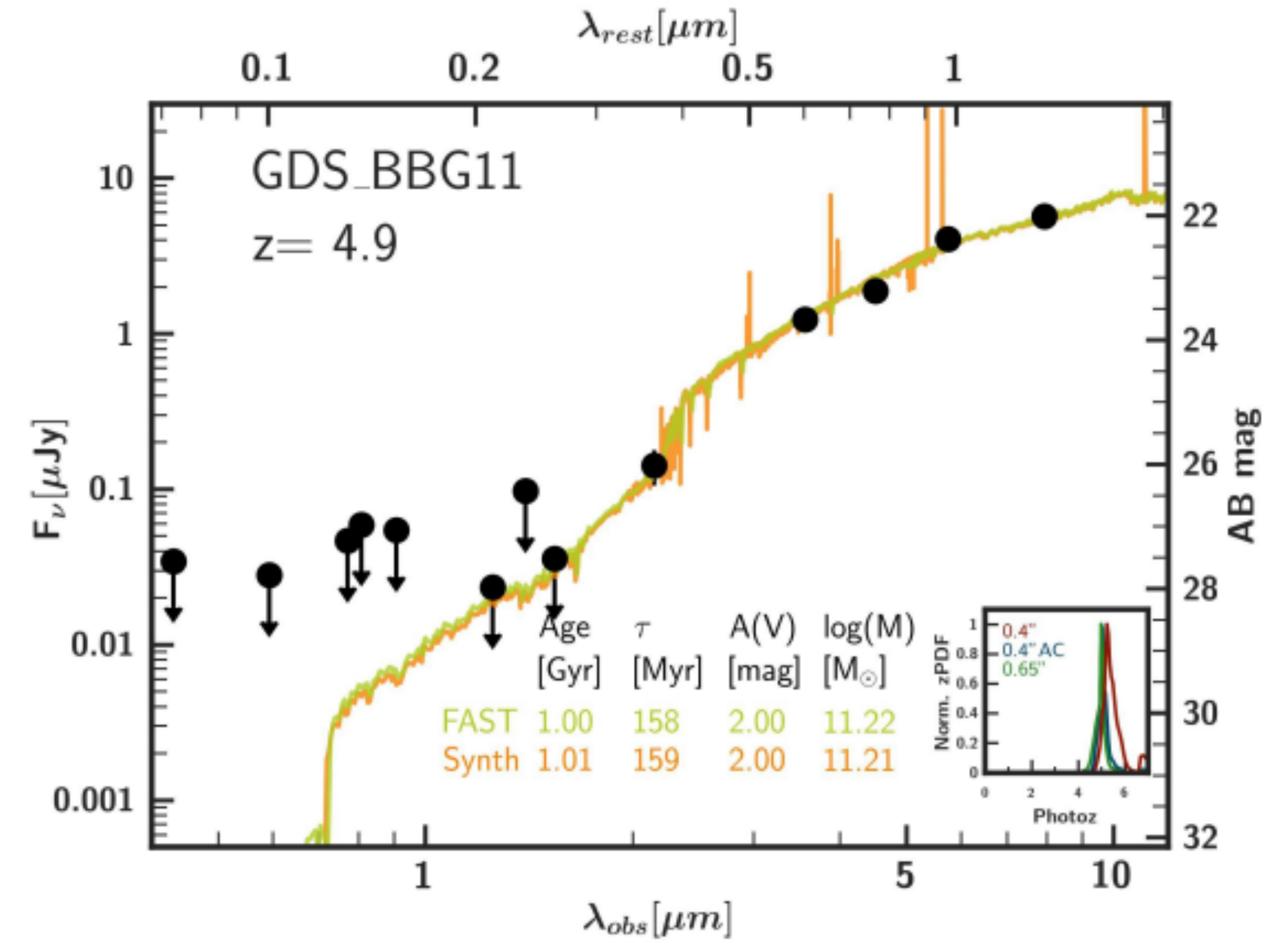}
		\end{center}
	\end{minipage}

% Source 12}

	\begin{minipage}[b]{0.44\linewidth}
		\centering
		\begin{minipage}[b]{0.315\linewidth}
			\includegraphics[width=1.\linewidth]{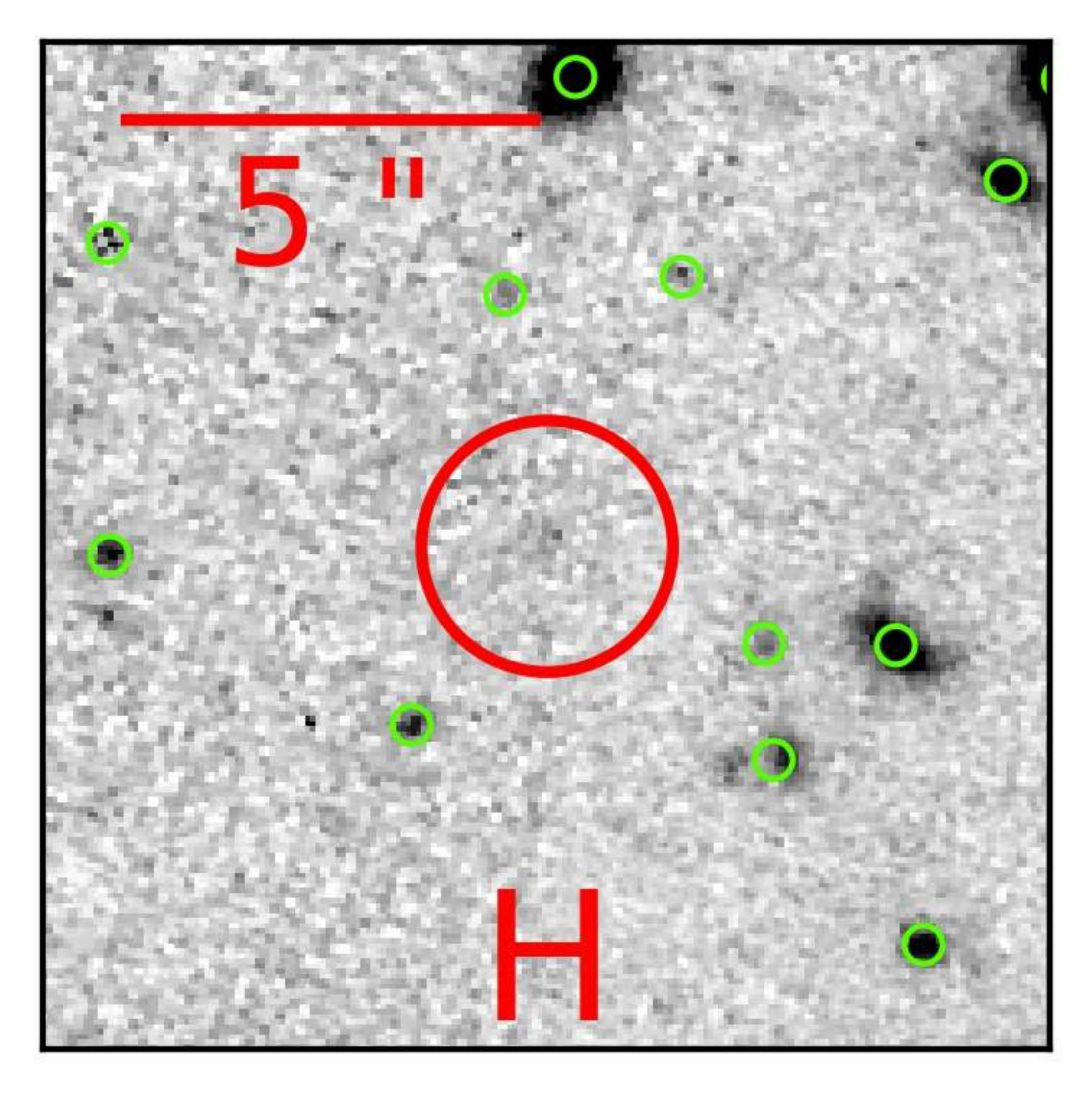}
		\end{minipage}
		\begin{minipage}[b]{0.315\linewidth}
			\includegraphics[width=1.\linewidth]{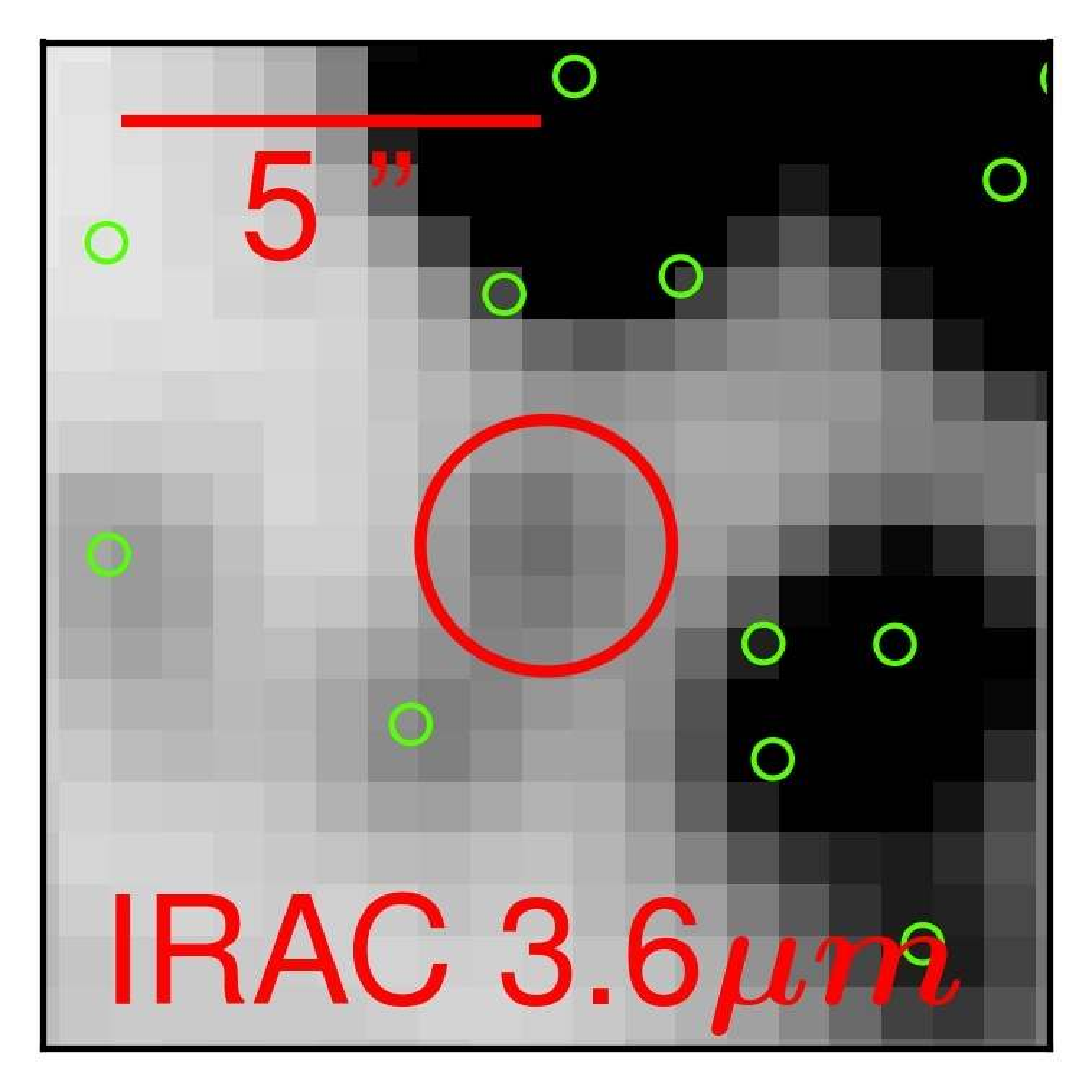}		
		\end{minipage}	
		\begin{minipage}[b]{0.315\linewidth}
			\includegraphics[width=1.\linewidth]{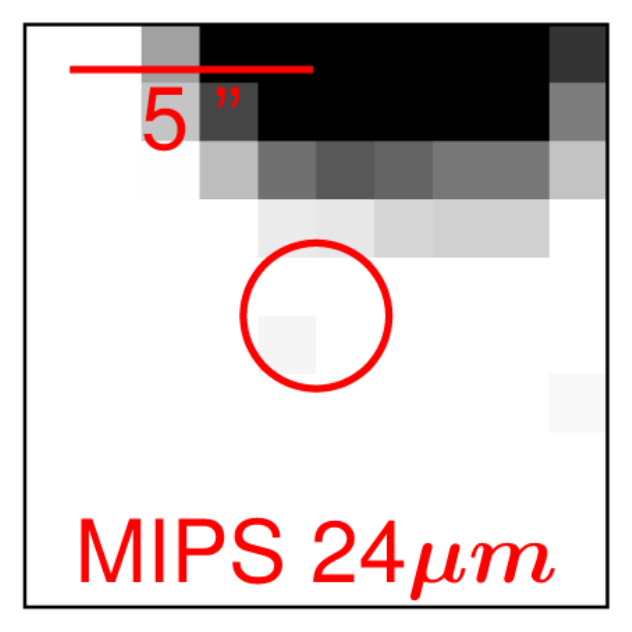}		
		\end{minipage}			
		\includegraphics[width=.49\linewidth]{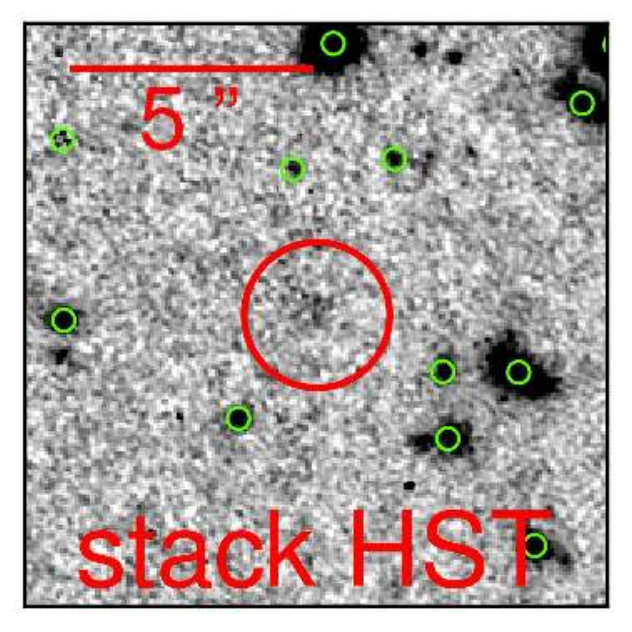}
		\includegraphics[width=.49\linewidth]{SH_ios_gs.pdf}
		\centering
	\end{minipage}
	\quad
	\begin{minipage}[b]{0.52\linewidth}
		\begin{center}
			\includegraphics[width=1.\linewidth]{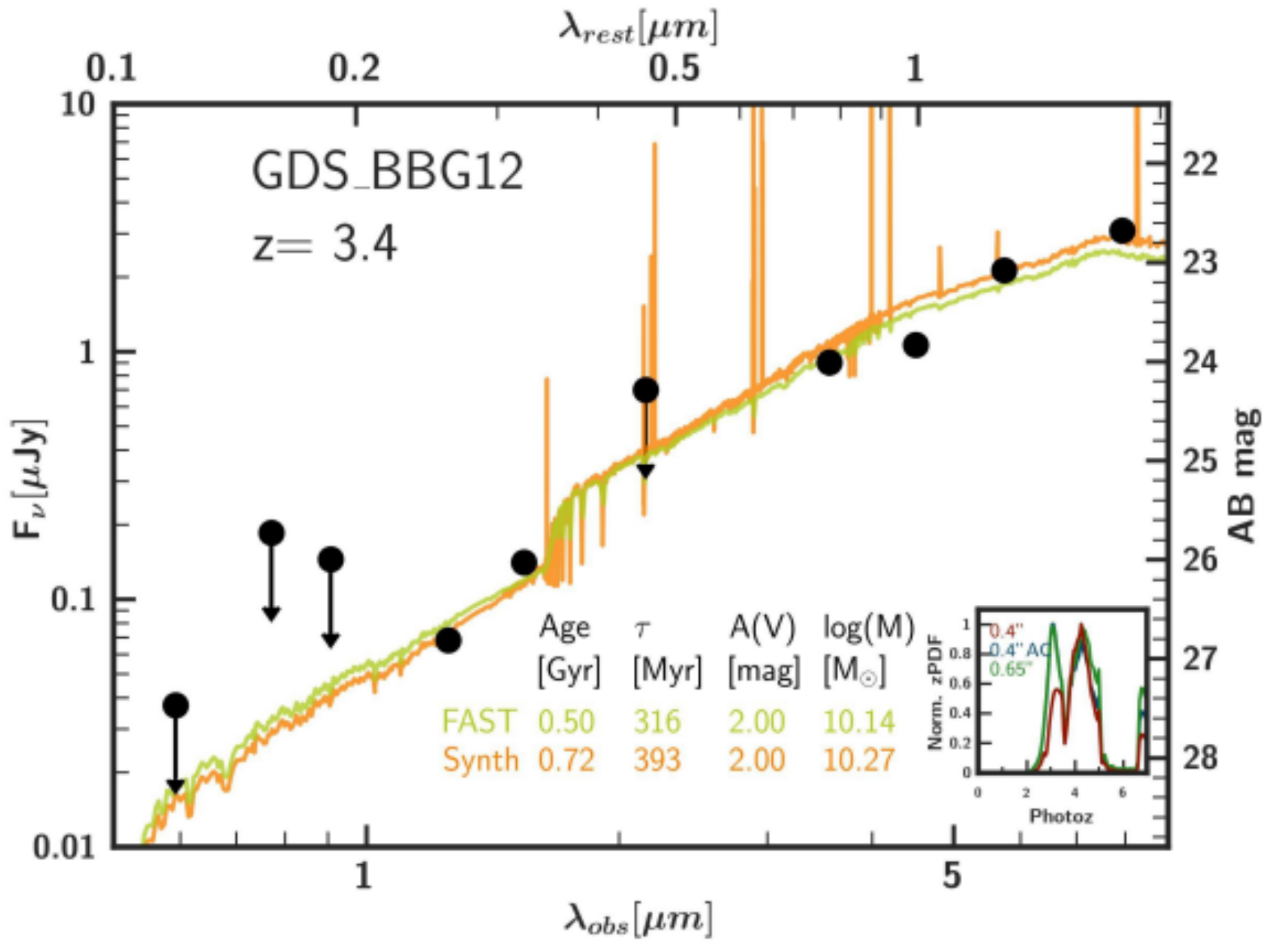}
			\centering
		\end{center}
	\end{minipage}
\end{figure*}

% Source 13}
\begin{figure*}
	\begin{minipage}[b]{0.44\linewidth}
		\begin{minipage}[b]{0.315\linewidth}
			\includegraphics[width=1.\linewidth]{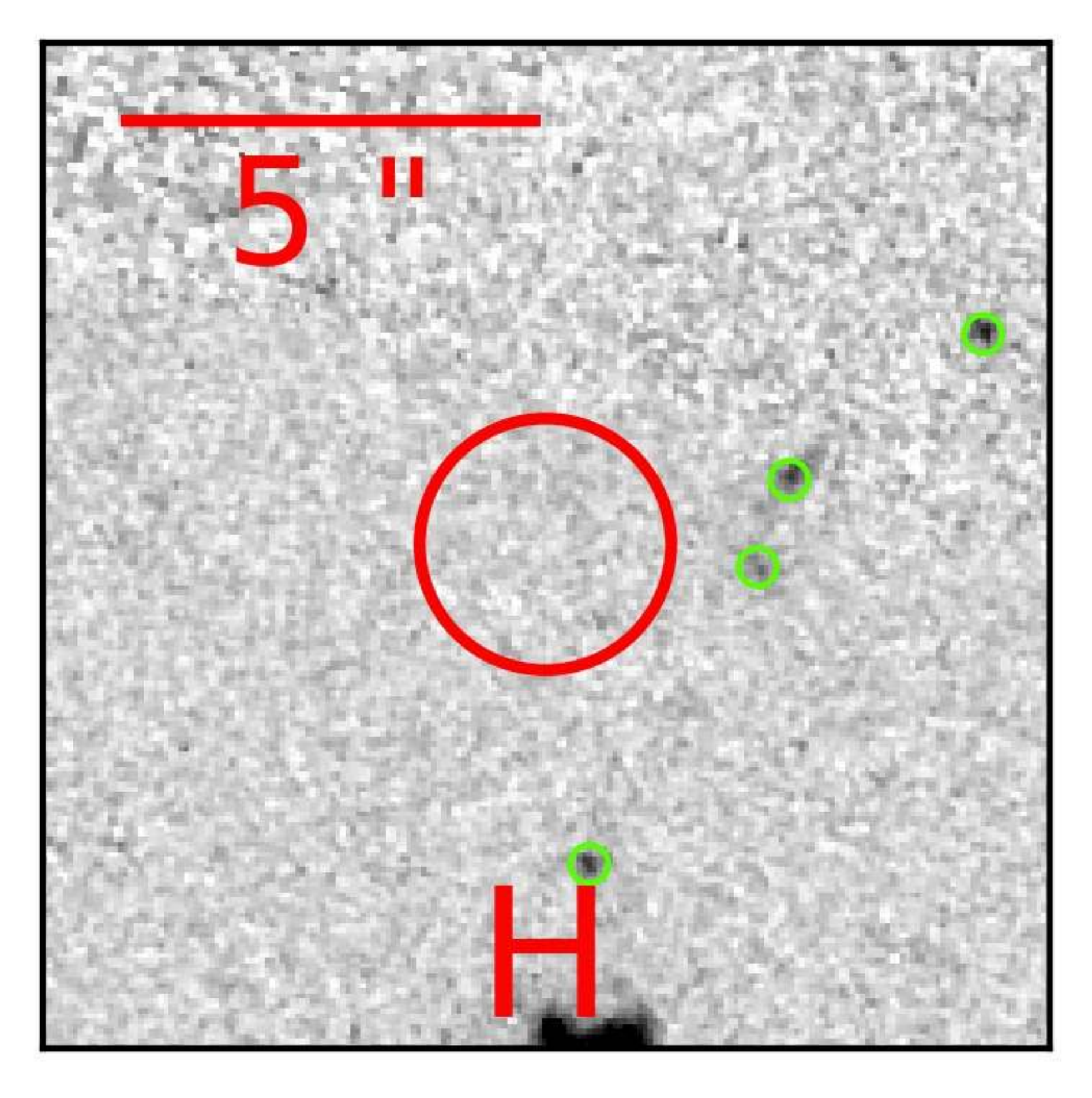}
		\end{minipage}
		\begin{minipage}[b]{0.315\linewidth}
			\includegraphics[width=1.\linewidth]{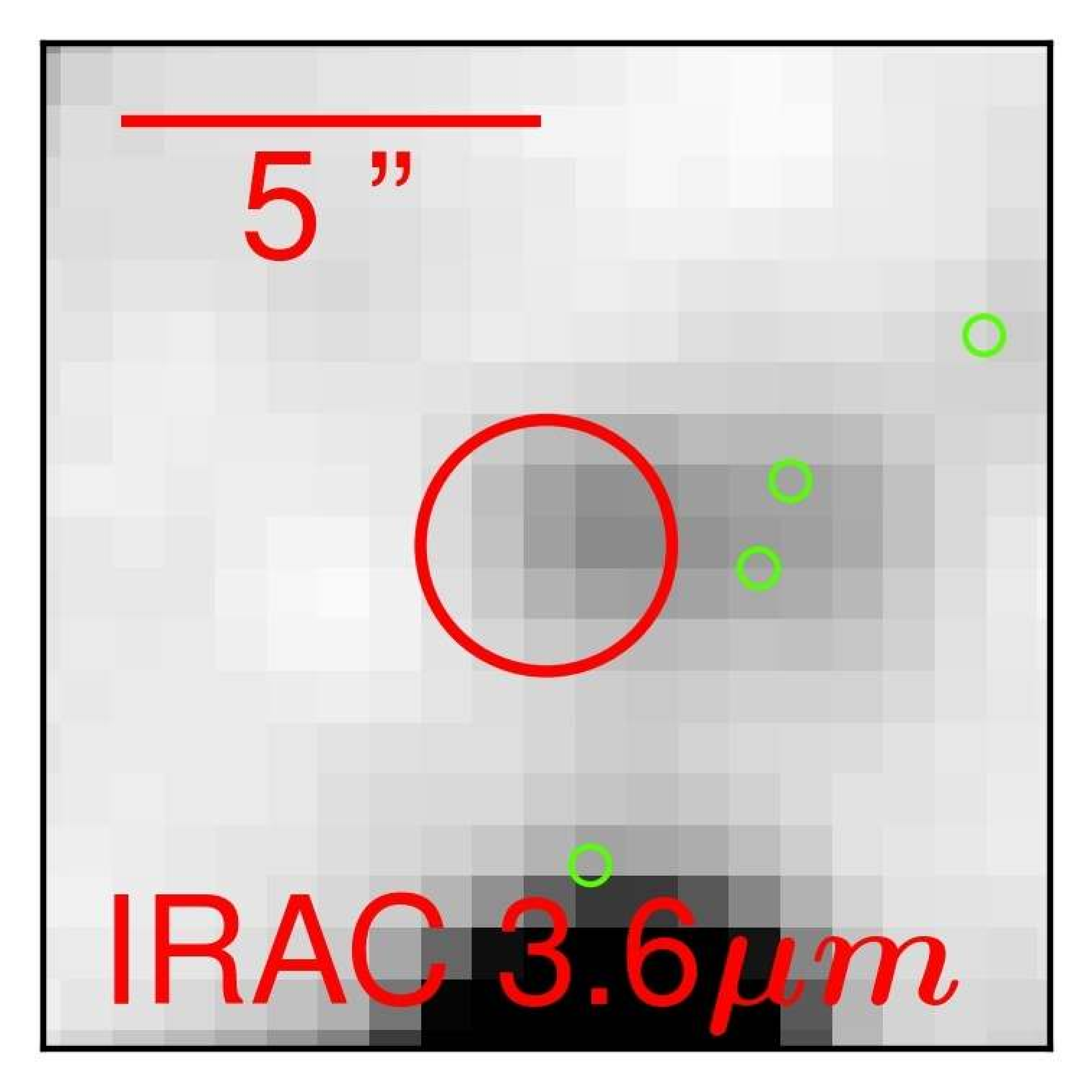}		
		\end{minipage}	
		\begin{minipage}[b]{0.315\linewidth}
			\includegraphics[width=1.\linewidth]{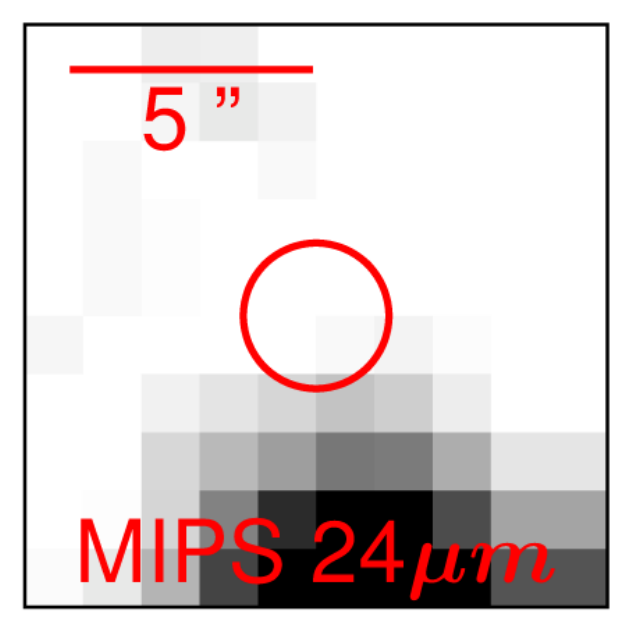}		
		\end{minipage}			
		\includegraphics[width=.49\linewidth]{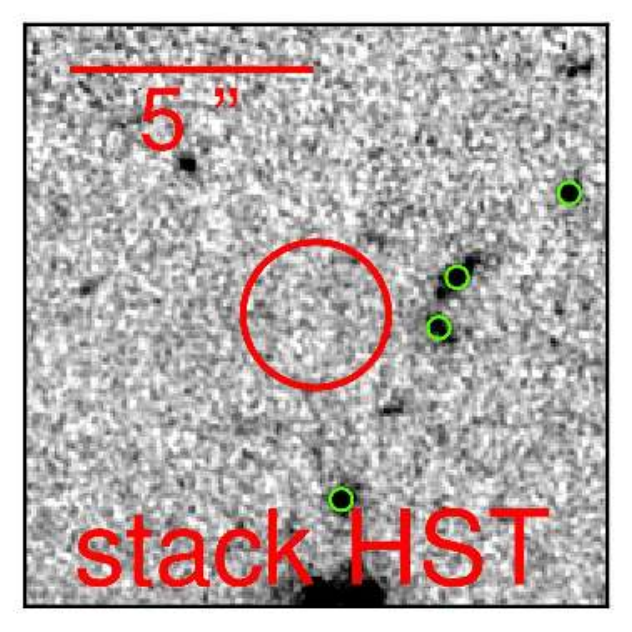}
		\includegraphics[width=.49\linewidth]{SH_ios_gs.pdf}
		\centering
	\end{minipage}
	\quad
	\begin{minipage}[b]{0.52\linewidth}
		\includegraphics[width=1.\linewidth]{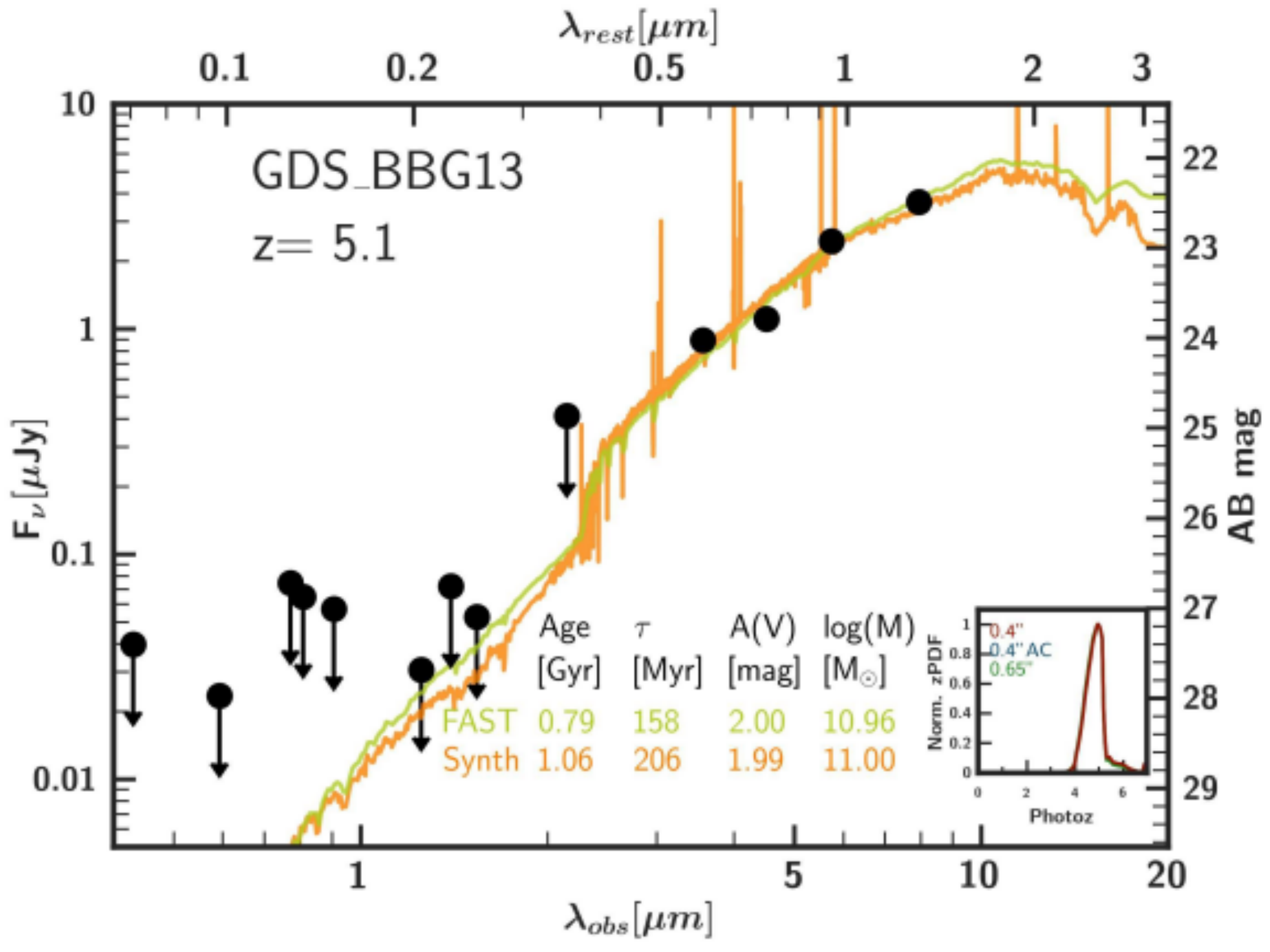}
		\centering
	\end{minipage}

% Source 14}
	\begin{minipage}[b]{0.44\linewidth}
		\centering
		\begin{minipage}[b]{0.315\linewidth}
			\includegraphics[width=1.\linewidth]{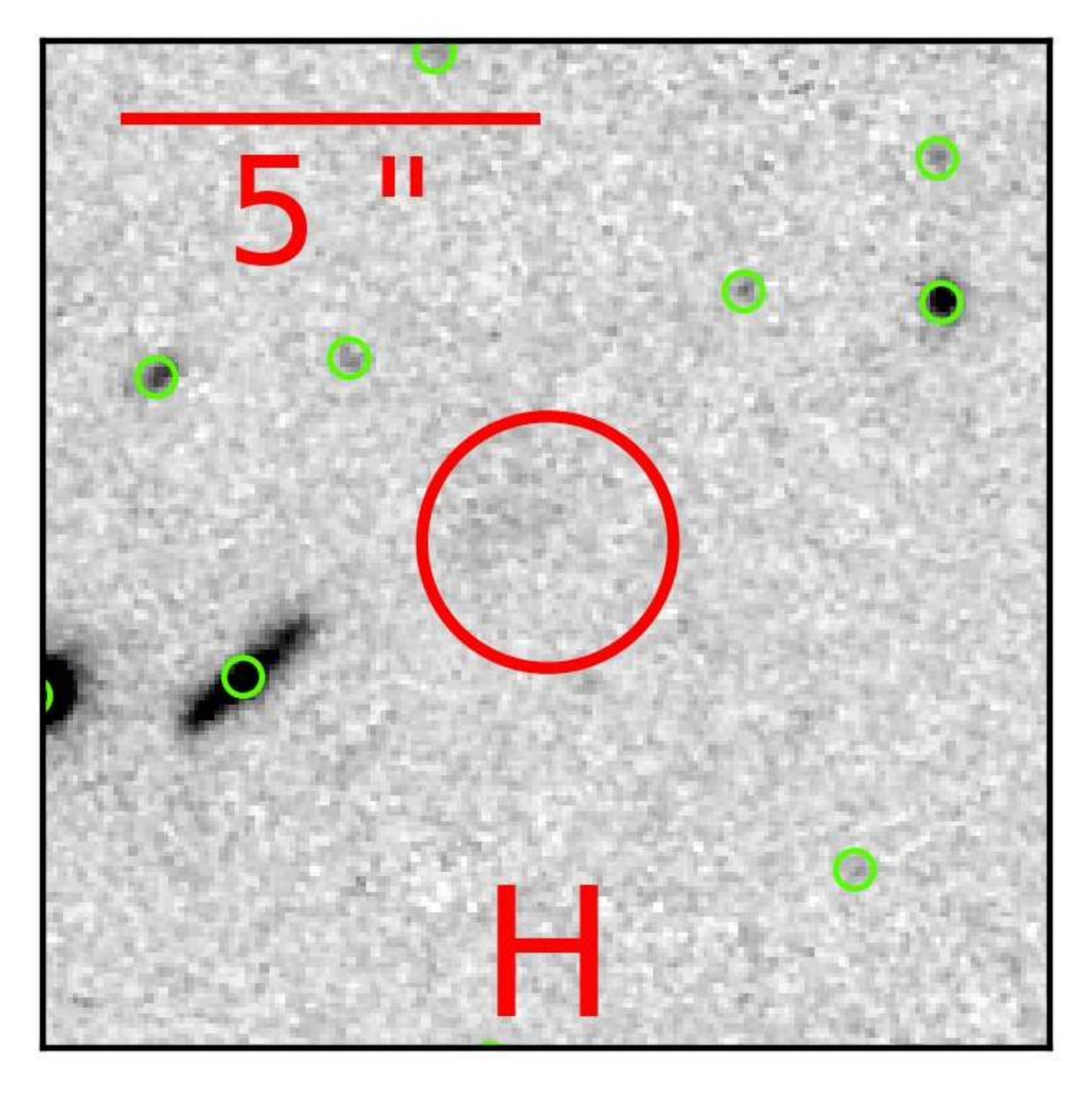}
		\end{minipage}
		\begin{minipage}[b]{0.315\linewidth}
			\includegraphics[width=1.\linewidth]{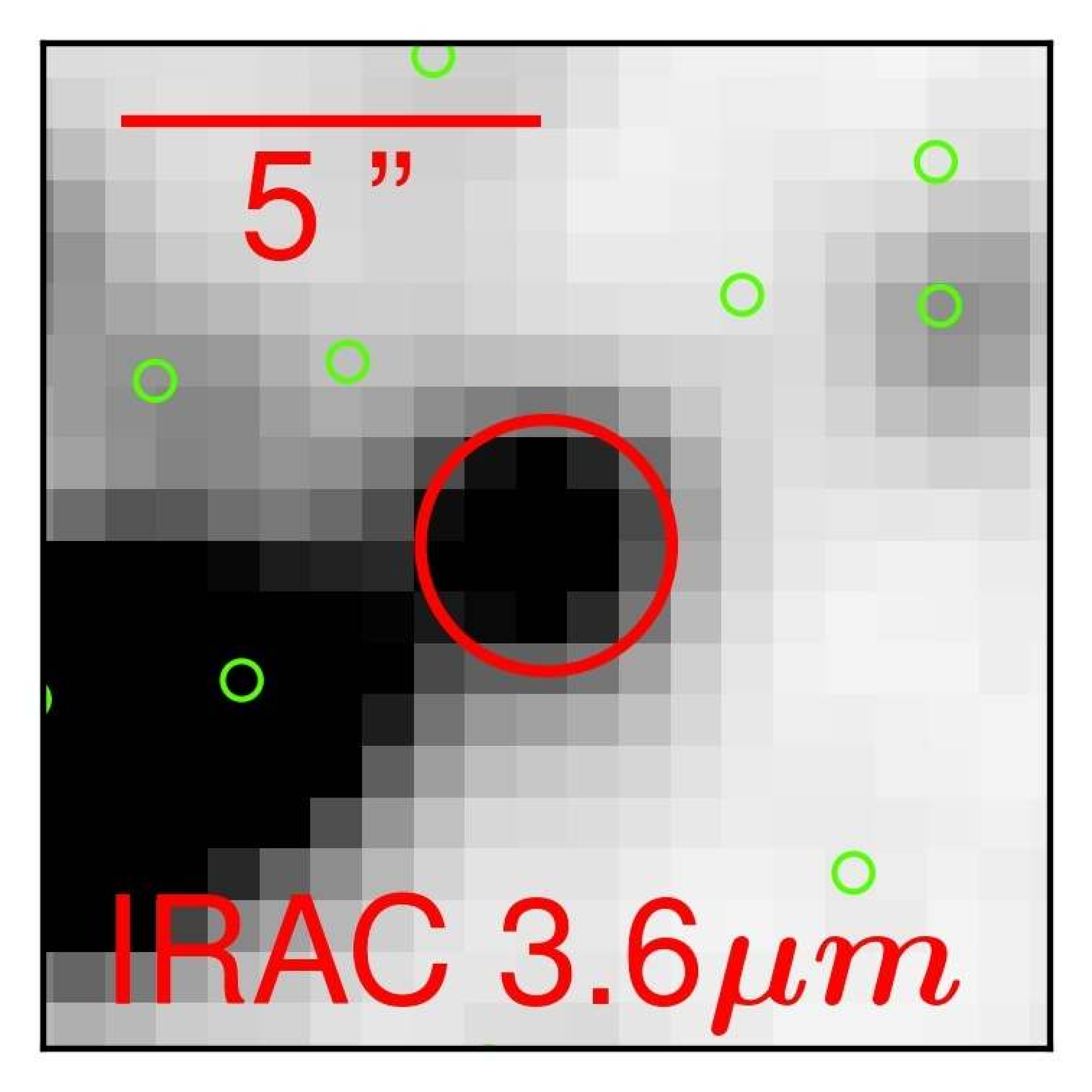}		
		\end{minipage}	
		\begin{minipage}[b]{0.315\linewidth}
			\includegraphics[width=1.\linewidth]{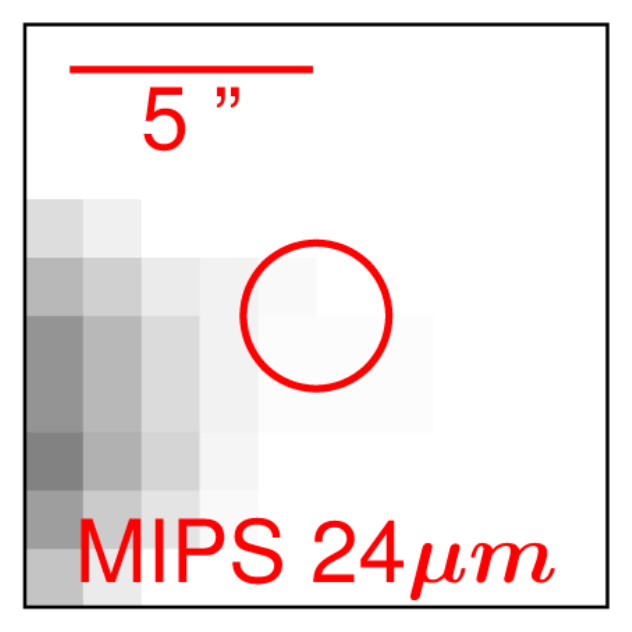}		
		\end{minipage}			
		\includegraphics[width=.49\linewidth]{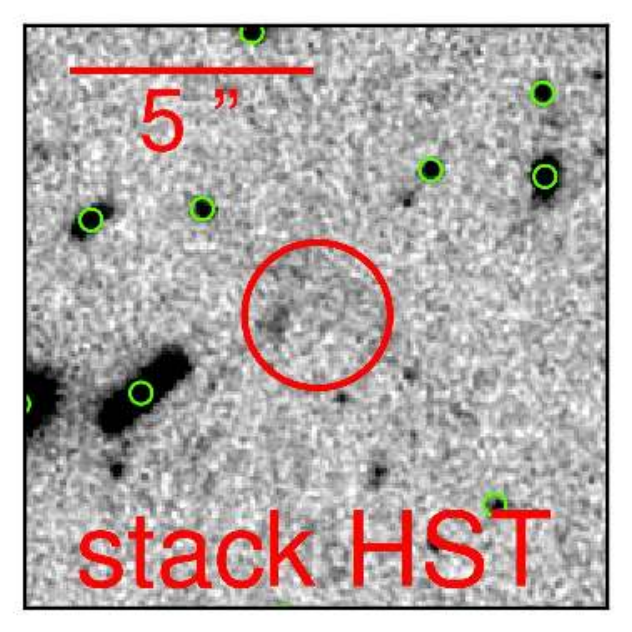}
		\includegraphics[width=.49\linewidth]{SH_ios_gs.pdf}
		\centering
	\end{minipage}
	\quad
	\begin{minipage}[b]{0.52\linewidth}
		\begin{center}
			\includegraphics[width=1.\linewidth]{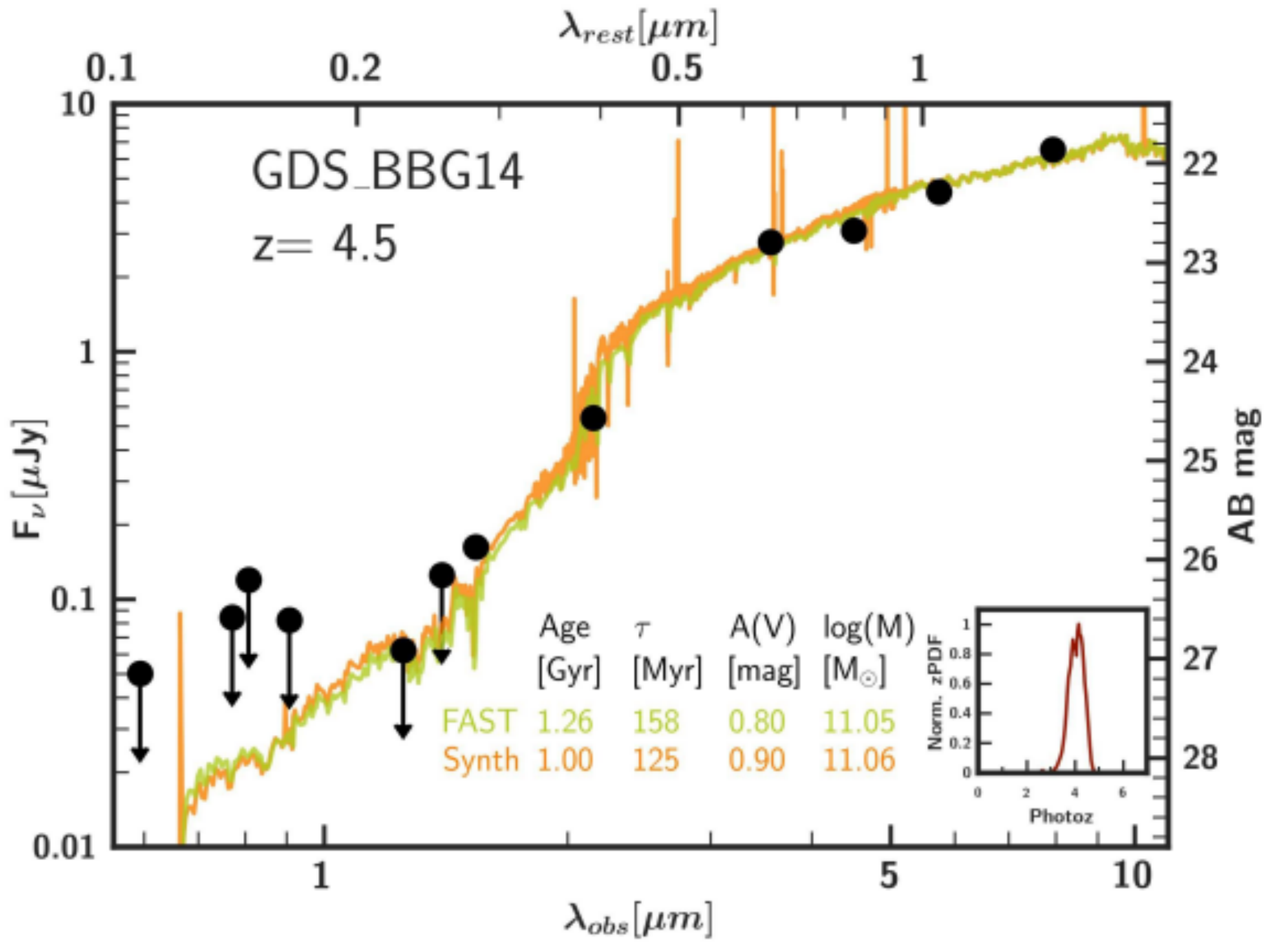}
		\end{center}
	\end{minipage}

% Source 15}

	\begin{minipage}[b]{0.44\linewidth}
		\centering
		\begin{minipage}[b]{0.315\linewidth}
			\includegraphics[width=1.\linewidth]{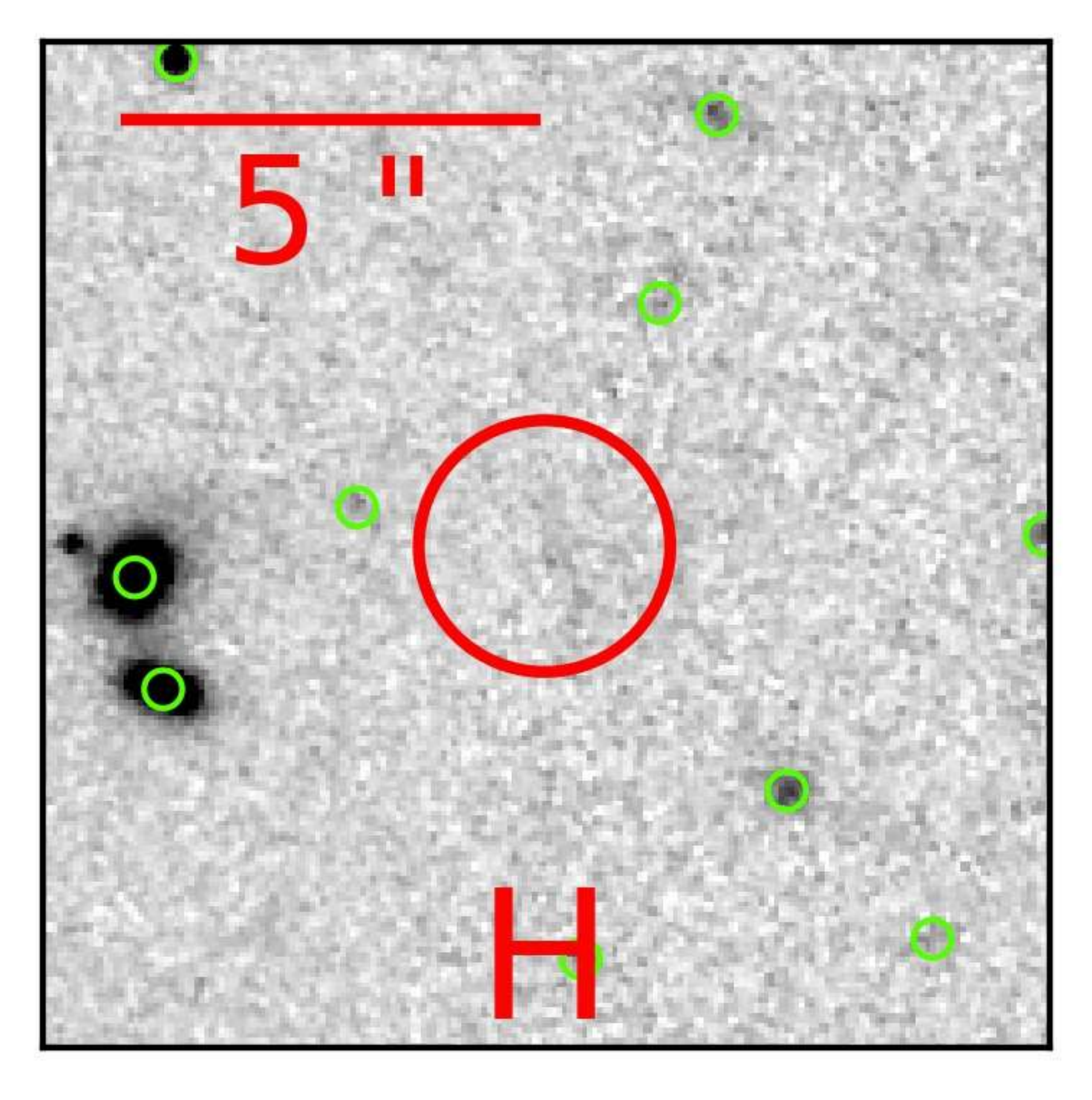}
		\end{minipage}
		\begin{minipage}[b]{0.315\linewidth}
			\includegraphics[width=1.\linewidth]{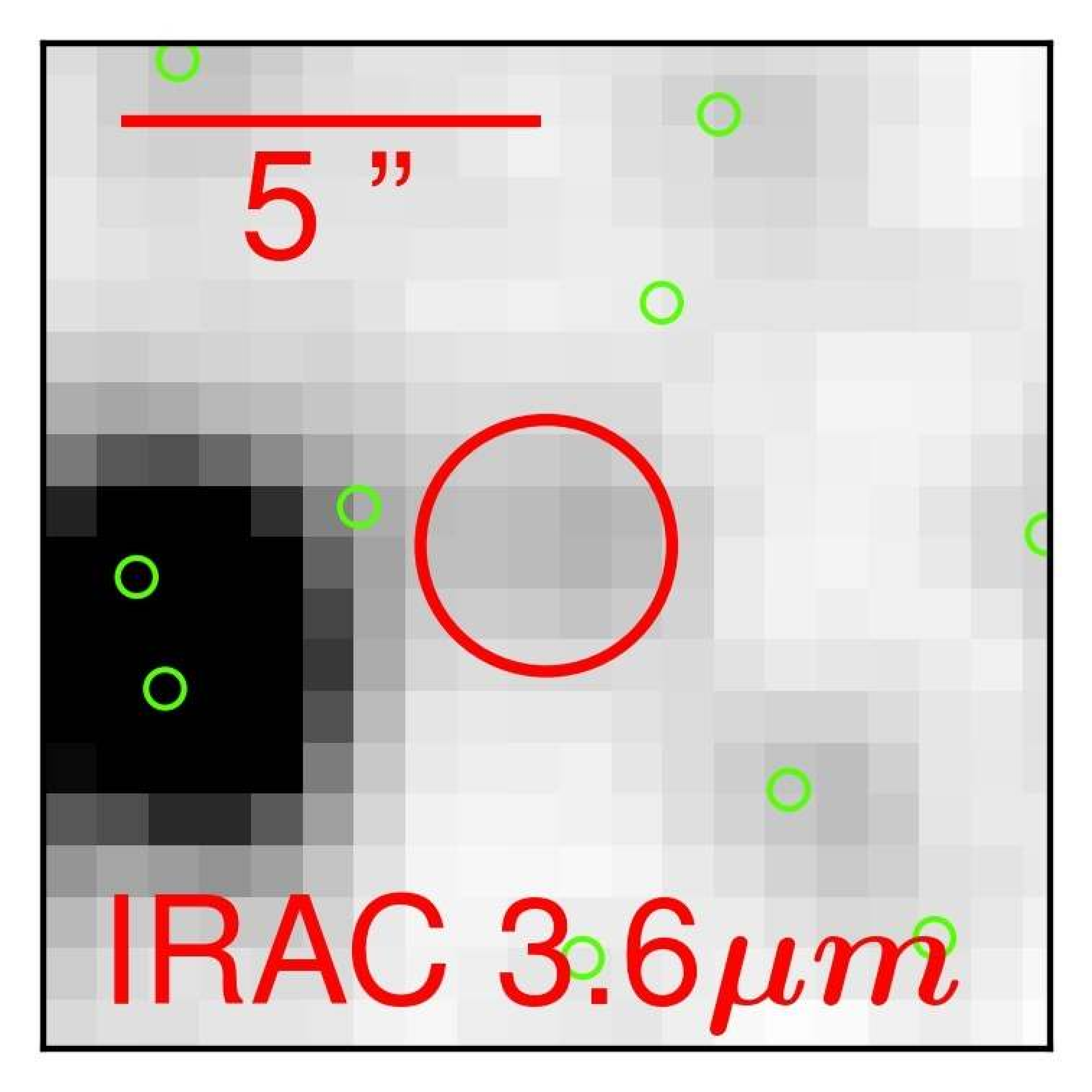}		
		\end{minipage}	
		\begin{minipage}[b]{0.315\linewidth}
			\includegraphics[width=1.\linewidth]{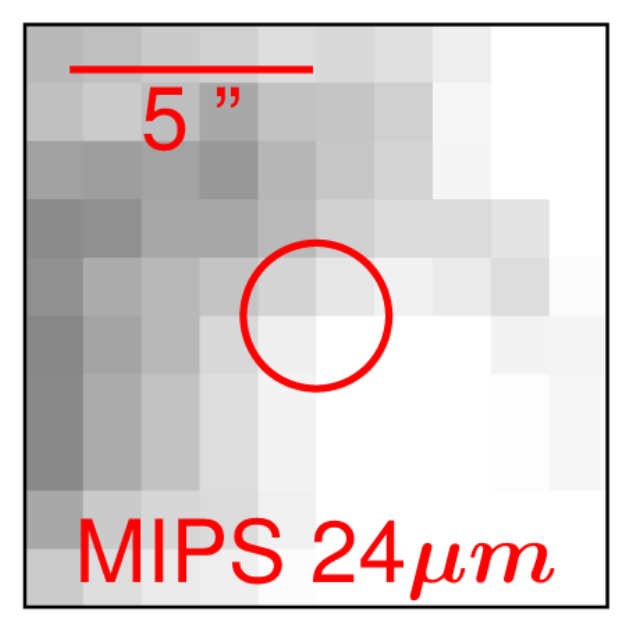}		
		\end{minipage}			
		\includegraphics[width=.49\linewidth]{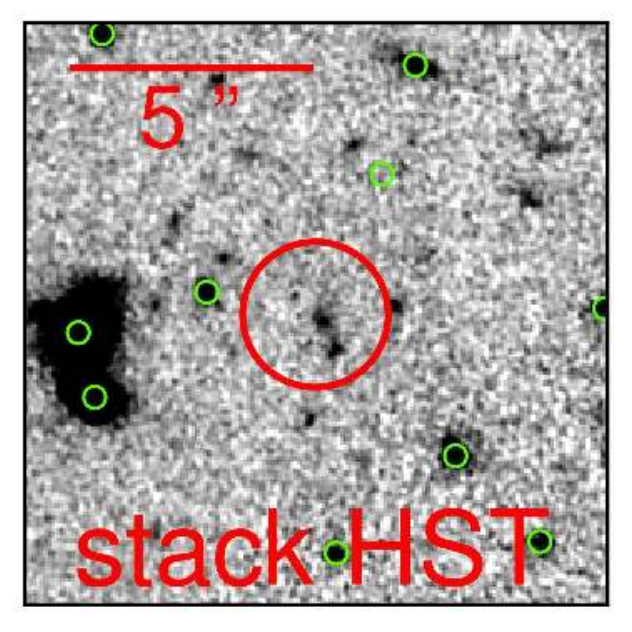}
		\includegraphics[width=.49\linewidth]{SH_ios_gs.pdf}
		\centering
	\end{minipage}
	\quad
	\begin{minipage}[b]{0.52\linewidth}
		\begin{center}
			\includegraphics[width=1.\linewidth]{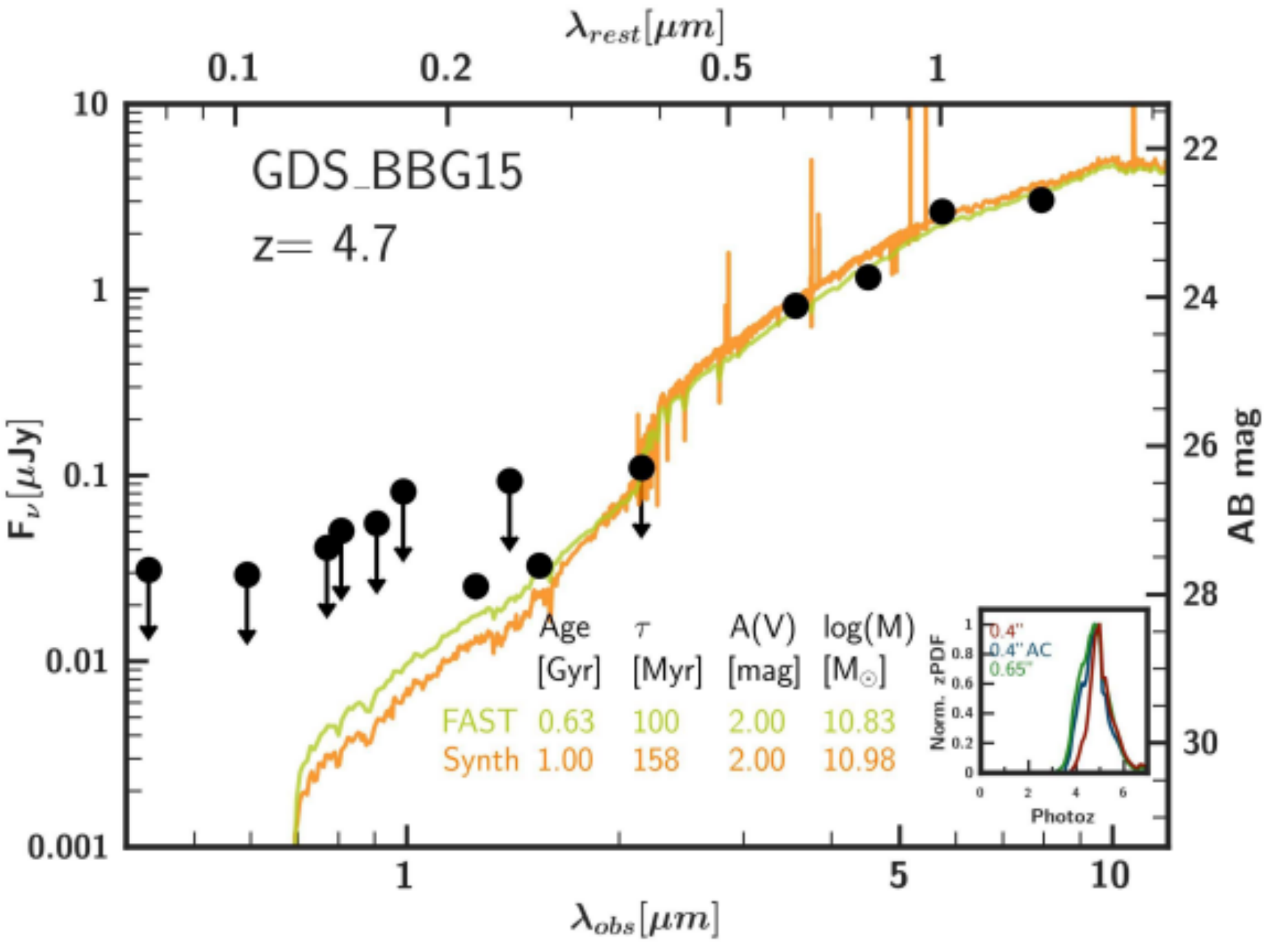}
		\end{center}
	\end{minipage}
\end{figure*}

% Source 16}
\begin{figure*}
	\begin{minipage}[b]{0.44\linewidth}
		\begin{minipage}[b]{0.315\linewidth}
			\includegraphics[width=1.\linewidth]{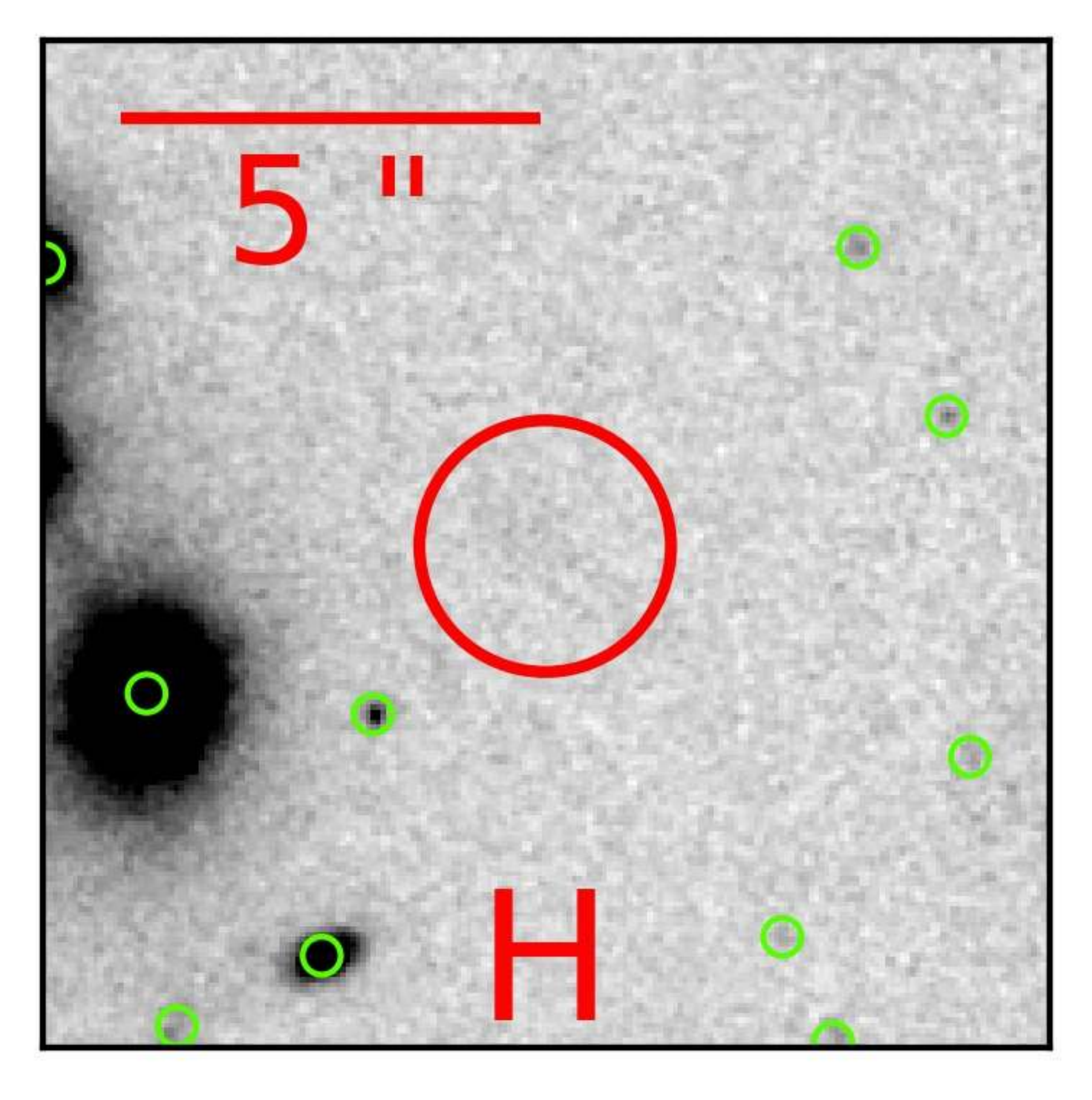}
		\end{minipage}
		\begin{minipage}[b]{0.315\linewidth}
			\includegraphics[width=1.\linewidth]{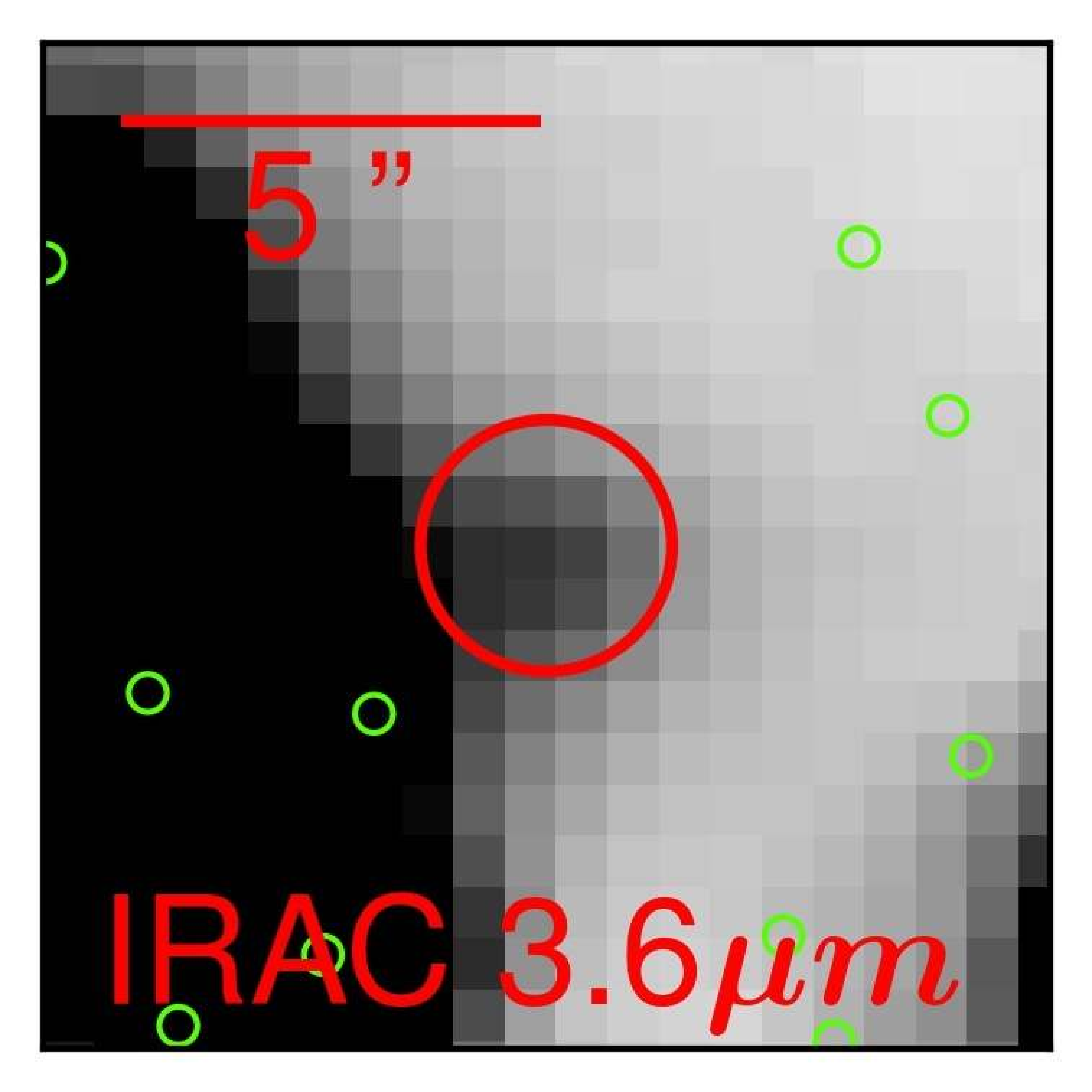}		
		\end{minipage}	
		\begin{minipage}[b]{0.315\linewidth}
			\includegraphics[width=1.\linewidth]{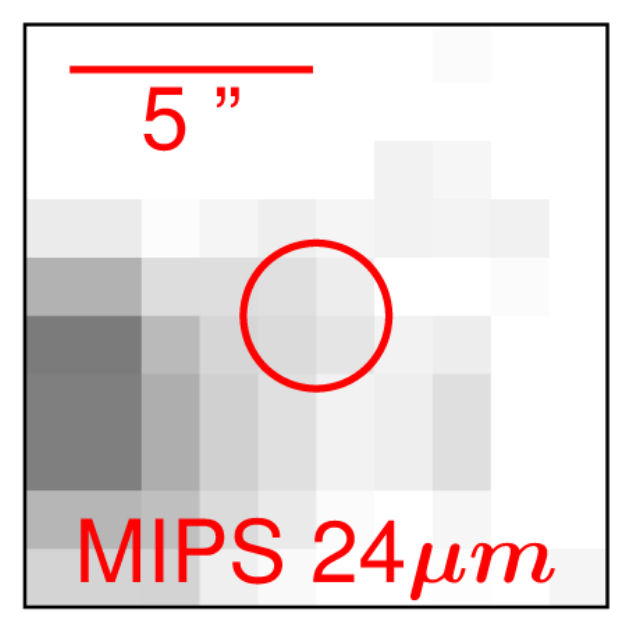}		
		\end{minipage}			
		\includegraphics[width=.49\linewidth]{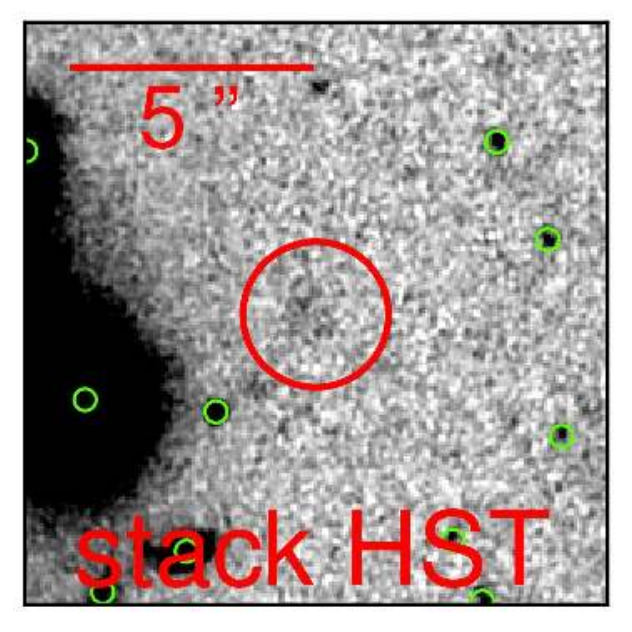}
		\includegraphics[width=.49\linewidth]{SH_ios_gs.pdf}
		\centering
	\end{minipage}
	\quad
	\begin{minipage}[b]{0.52\linewidth}
		\includegraphics[width=1.\linewidth]{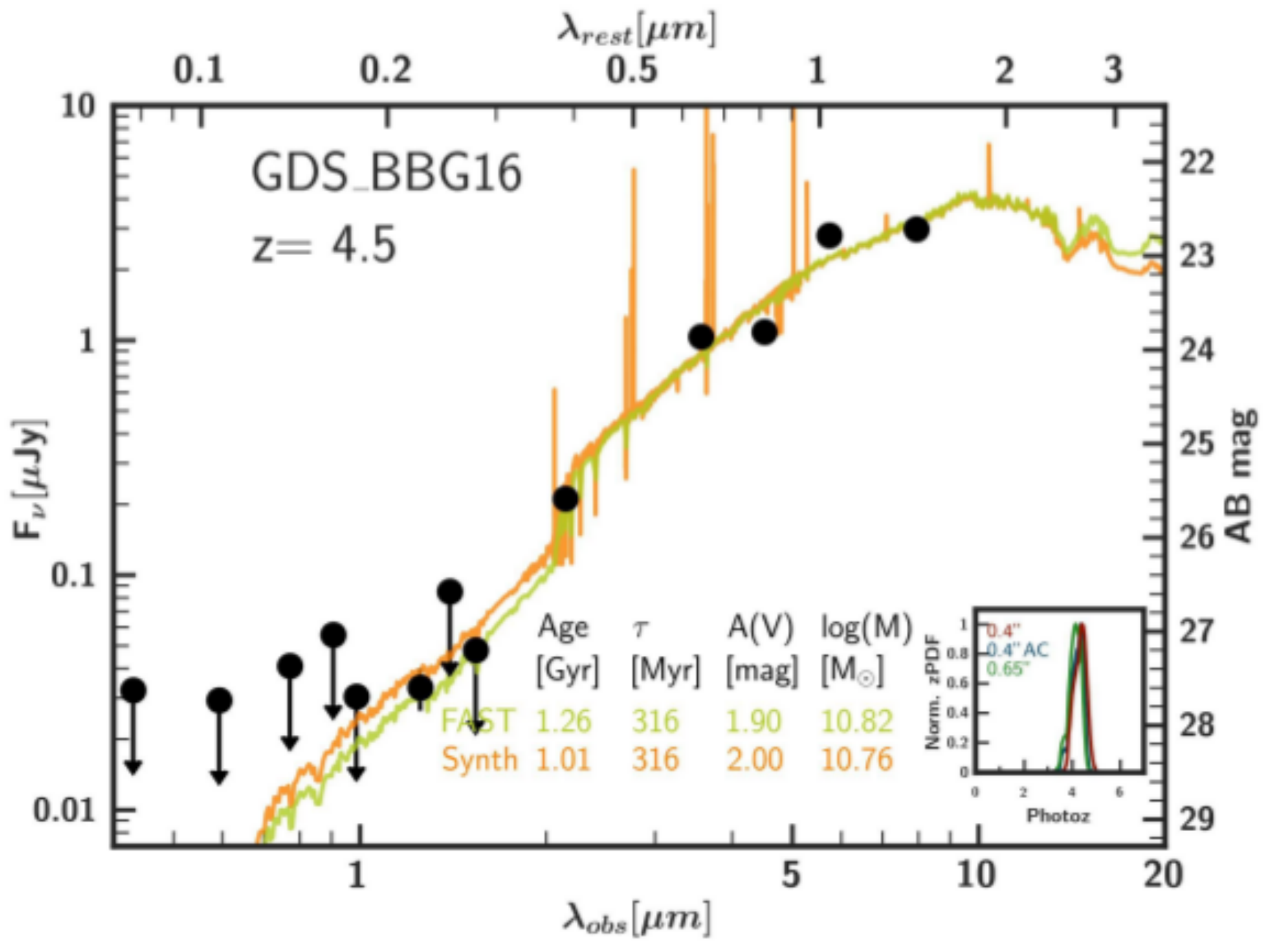}
		\centering
	\end{minipage}
\end{figure*}

\clearpage
% % % % % % % % % % % % % % % % % % % % % % % % % % % % % % % % % % % % % % % % % % % % % % % % % % % % % % % %

\end{appendices}

\end{document}